\documentclass[12pt]{book}

\usepackage{a4wide}
\usepackage{fancyhdr}
\usepackage[dvipdfm]{graphicx}
\usepackage{threeparttable}
\usepackage{color}   
\usepackage[bookmarks]{hyperref}
\usepackage[latin1]{inputenc}
\fontfamily{ptm}\selectfont

\renewcommand{\chaptermark}[1]%
         {\markboth{\thechapter.\ #1}{}}
\renewcommand{\sectionmark}[1]%
         {\markright{\thesection\ #1}}
\newcommand{\fermi}{{\it Fermi}}
\newcommand{\apj}{ApJ}
\newcommand{\nat}{Nature}
\newcommand{\prd}{Physical Review D}
\newcommand{\apjl}{ApJ Letters}
\newcommand{\aap}{A\&A}
\newcommand{\apjs}{ApJS}
\newcommand{\apss}{Astrophysics and Space Science}
\newcommand{\araa}{Annual review of astronomy and astrophysics}
\newcommand{\mnras}{MNRAS}
\newcommand{\LMUTitle}[9]{
  \thispagestyle{empty}
  \vspace*{\stretch{1}}
  {\parindent0cm
   \rule{\linewidth}{.7ex}}
  \begin{flushright}

    \vspace*{\stretch{1}}
    \sffamily\bfseries\Huge
    #1\\
    \vspace*{\stretch{1}}
    \sffamily\bfseries\large
    #2
    \vspace*{\stretch{1}}
  \end{flushright}
  \rule{\linewidth}{.7ex}
  \vspace*{\stretch{5}}
  \begin{center}
    \includegraphics[width=2in]{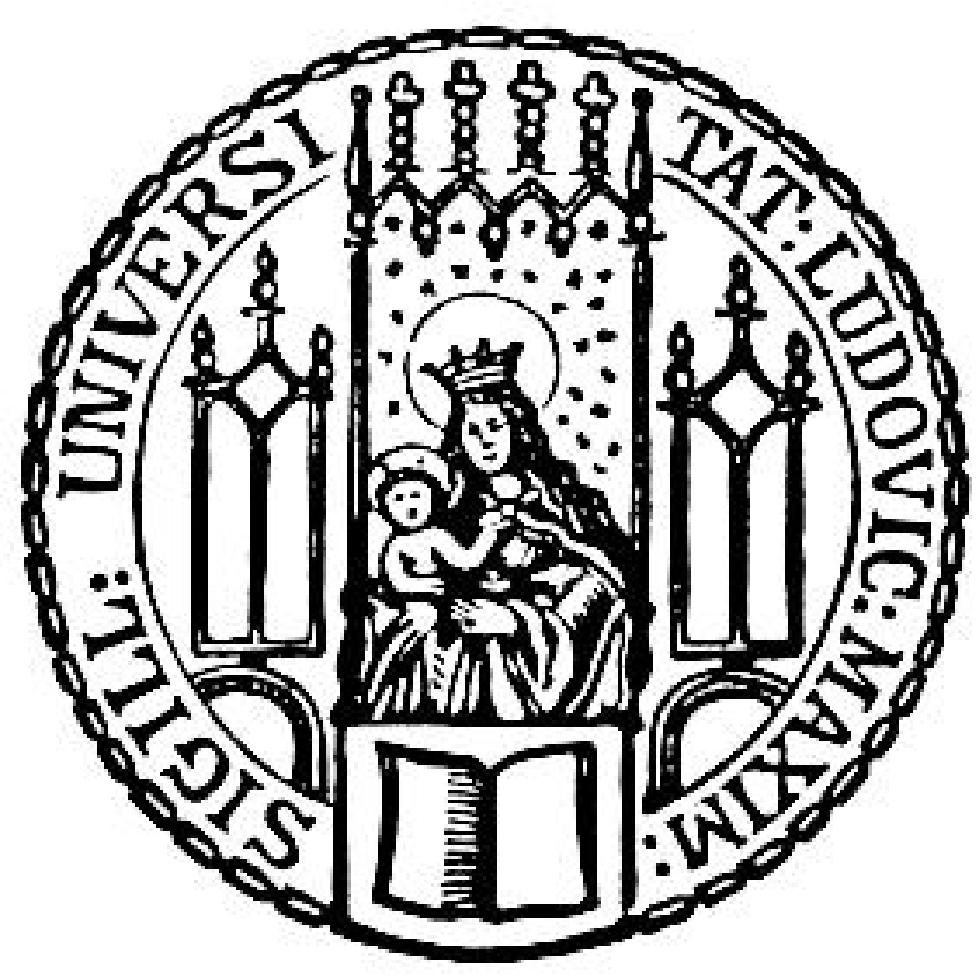}
  \end{center}
  \vspace*{\stretch{1}}
  \begin{center}\sffamily\LARGE{#5}\end{center}
  \newpage
  \thispagestyle{empty}

  \cleardoublepage
  \thispagestyle{empty}

  \vspace*{\stretch{1}}
  {\parindent0cm
  \rule{\linewidth}{.7ex}}
  \begin{flushright}
    \vspace*{\stretch{1}}
    \sffamily\bfseries\Huge
    #1\\
    \vspace*{\stretch{1}}
    \sffamily\bfseries\large
    #2
    \vspace*{\stretch{1}}
  \end{flushright}
  \rule{\linewidth}{.7ex}

  \vspace*{\stretch{3}}
  \begin{center}
    \Large Dissertation\\
    \Large an der #4\\
    \Large der Ludwig--Maximilians--Universi\"at\\
    \Large M\"unchen\\
    \vspace*{\stretch{1}}
    \Large vorgelegt von\\
    \Large #2\\
    \Large aus #3\\
    \vspace*{\stretch{2}}
    \Large M\"unchen, den #6
  \end{center}

  \newpage
  \thispagestyle{empty}

  \vspace*{\stretch{1}}

  \begin{flushleft}
    \large Erstgutachter:  #7 \\[1mm]
    \large Zweitgutachter: #8 \\[1mm]
    \large Tag der m\"undlichen Pr\"ufung: #9\\
  \end{flushleft}

  \cleardoublepage
}

\begin{document}


  \LMUTitle
      {Study of the High Energy Gamma-ray Emission 
	from the Crab Pulsar with \\the MAGIC telescope and \fermi-LAT
}               
      {Takayuki Saito}                       
      {Saitama, Japan}                             
      {Fakult\"at f\"ur Physik}                         
      {M\"unchen 2010}                          
      {30.09.2010}                            
      {Prof. Dr. Christian Kiesling}                          
{Prof. Dr. Masahiro Teshima}
{06.12.2010}

  \chapter*{\vspace{-3.8cm}Abstract}
\vspace{-0.5cm}


 My thesis deals with a fundamental question of high energy gamma-ray astronomy. 
Namely, I studied the cut-off shape of the Crab pulsar spectrum
to distinguish between the leading scenarios for the pulsar models.

   Pulsars are celestial objects, which emit periodic pulsed electromagnetic
radiation (pulsation) from radio to high energy gamma-rays. 
Two major scenarios evolved in past 40 years to explain the pulsation mechanism:
 the inner magnetosphere scenario and the outer magnetosphere scenario. 
Both scenarios predict a high energy cut-off in the gamma-ray energy spectrum,
but with different cut-off sharpness.
An exponential cut-off is expected for the outer magnetosphere scenario
while a super-exponential cut-off is predicted for the inner magnetosphere scenario.
Therefore, one of the best ways to confirm or rule out these scenarios is 
to measure the energy spectrum of a pulsar at around the cut-off energy, i.e.,
 at energies between a few GeV and a~few tens of GeV. 
All past attempts ($>10$) to measure pulsars with ground-based instruments 
have failed while satellite-borne detectors had a too small area to study detailed spectra
 in the GeV~domain.

   In this thesis, the gamma-ray emission at around the cut-off energy from the Crab pulsar
is studied with the MAGIC telescope. 
The public data of the satellite-borne gamma-ray detector, \fermi-LAT, are also analyzed
 in order to discuss the MAGIC observation results in comparison with the adjacent energy band.

   In late 2007, a new trigger system (SUM trigger system) allowed to reduce the threshold energy of
 the MAGIC telescope from 50 GeV to 25 GeV and the Crab pulsar was successfully detected 
during observations from October 2007 and January 2009.
 My analysis reveals that the energy spectrum is 
consistent with a simple power law 
between 25 GeV to 100 GeV.
The extension of the energy spectrum up to 100 GeV rules out 
the inner magnetosphere scenario.

   \fermi-LAT started operation in August 2008. 
The \fermi-LAT data reveal
that a power law with an exponential cut-off at a few GeV can well describe
 the energy spectrum of the Crab pulsar between 100 MeV and 30 GeV.
This is consistent with the outer magnetosphere scenario and again, 
inconsistent with the inner magnetosphere scenario. 

The measurements of both experiments strongly disfavor the inner magnetosphere scenario. 
However, by combining the results of the two experiments, 
it turns out that even the standard outer magnetosphere scenario cannot
 explain the measurements.
 Various assumptions have been made to explain this discrepancy.
 By modifying the energy spectrum of the electrons 
which emit high energy gamma-rays via the curvature radiation,
 the combined measurements can be reproduced but further studies 
with higher statistics and a better energy resolution are needed to support this assumption.

   The energy-dependent pulse profile from 100 MeV to 100 GeV has also been studied in detail. 
Many interesting features have been found, among which the variabilities of the pulse edges 
and the pulse peak phases are the most remarkable. 
More data would allow an investigation of the fine structure of the pulsar magnetosphere 
based on these features.

 Aiming at better observations of pulsars and other sources below 100 GeV, 
a new photosensor, HPD R9792U-40, has been investigated. 
Many beneficial properties, such as a very high photodetection efficiency, 
an extremely low ion-feedback probability and an excellent charge resolution have been found.
  \chapter*{\vspace{-3.8cm}Abstract}
\vspace{-0.5cm}

Diese Dissertation behandelt eine grundlegende Fragestellung der 
Hochenergie-Astrophysik. 
Um herauszufinden, welches der f\"uhrenden Modelle die Emission von Pulsaren
 korrekt beschreibt, untersuche ich das Abknickverhalten des Energiespektrums
 des Pulsars im Krebsnebel zu hohen Energien hin. Pulsare sind astronomische
 Objekte, die durch periodische Abstrahlung ("Pulsation") elektromagnetischer
 Strahlung von Radiowellen bis hin zu Gammastrahlen gekennzeichnet sind.
 Zwei Modellklassen haben sich in den vergangenen 40 Jahren herausgebildet, 
welche den Pulsationsmechanismus mit Teilchenbeschleunigung in der inneren bzw.
\"aussssseren Magnetosph\"are erkl\"aren k\"onnen. 
Beide Szenarien sagen ein Abknicken des Energiespektrums im 
Gammastrahlungsbereich voraus, allerdings mit unterschiedlicher St\"arke. 
Wenn die Strahlung in der \"ausssseren Magnetosph\"are erzeugt wird, erwartet man 
einen exponentiellen Abfall, w\"ahrend f\"ur die Erzeugung in der inneren 
Magnetosph\"are ein noch st\"arkerer Abfall vorhergesagt wird. 
Deswegen ist die Bestimmung eines Pulsar-Energiespektrums in der N\"ahe 
des erwarteten Abknickens, also bei Energien zwischen einigen GeV und einigen
 zehn GeV eine bevorzugte Methode, einem dieser Szenerien Glaubw\"urdigkeit
 zu verleihen oder es andererseits auszuschliesssssen. 
Alle bisherigen Versuche, Pulsarspektren im genannten Energiebereich
 mit bodengebundenen Instrumenten zu vermessen, schlugen fehl; 
gleichzeitig hatten satellitengest\"utzte Detektoren eine zu kleine 
Sammelfl\"ache, als dass die genaue Vermessung von Spektren m\"oglich gewesen w\"are.

Diese Arbeit untersucht die Gammastrahlungsemission des Krebsnebels 
in der Umgebung der Abknick-Energie mit Hilfe des MAGIC-Teleskops. 
Ebenfalls ausgewertet werden \"offentlich zug\"angliche Daten des 
satellitengest\"utzten Gammastrahlungsdetektors Fermi-LAT, 
so dass die MAGIC-Ergebnisse mit dem bei niedrigeren Energien anschliessenden
 Energiebereich von Fermi-LAT verglichen werden k\"onnen. 

Vor kurzem wurde ein neues Triggersystem (ein analoger Summentrigger)
 in Betrieb genommen, welches die Energieschwelle des MAGIC-Teleskopes
 von 50 GeV auf 25 GeV heruntersetzt; damit wurde w\"ahrend Beobachtungen von
 Oktober 2007 bis Januar 2009 der Pulsar im Krebsnebel erstmalig nachgewiesen.
 Meine Untersuchungen zeigen, dass das Energiespektum im Bereich von 25 GeV
 bis 100 GeV mit einem einfachen Potenzgesetz vertr\"aglich ist.
 Allein schon die Tatsache, dass sich das Spektrum bis 100 GeV erstreckt,
 schliessssst eine Erzeugung in der inneren Magnetosph\"are aus.

Die Fermi-LAT Beobachtungen begannen im August 2008. 
Die \"offentlichen Fermi-LAT Daten zeigen, dass ein Potenzgesetz 
mit exponentiellem Abfall bei einigen GeV das Krebspulsarspektrum 
zwischen 100 MeV und 30 GeV gut beschreibt. 
Das ist vertr\"aglich mit einer Erzeugung in der \"ausssseren, aber nicht 
in der inneren Magnetosph\"are. Die Messungen beider Instrumente deuten 
somit stark darauf hin, dass das entsprechende Modell 
f\"ur eine Erzeugung in der inneren Magnetosph\"are die Daten nicht 
korrekt beschreibt. Wenn man nun die Resultate beider Messungen kombiniert,
 zeigt sich, dass ein einfaches Modell zur Gammastrahlungs-Erzeugung in der
 \"ausssseren Magnetosph\"are die Daten nur unzul\"anglich beschreibt. 
Verschiedene Annahmen sind notwendig, um die vorhandenen 
Abweichungen zu erkl\"aren.

Weiterhin wurde die Energieabh\"angigkeit des Pulsprofils zwischen 100 MeV 
und 100 GeV genau untersucht. Diverse interessante Eigenschaften wurden 
gefunden, von denen die Ver\"anderungen in den Pulsflanken und die Entwicklung
 der Phasen der Pulsationsmaxima die erw\"ahnenswertesten sind. Von diesen
 Eigenschaften ausgehend, w\"urde eine gr\"ossssere Menge von Beobachtungsdaten
 erlauben, den Aufbau der Pulsarmagnetosph\"are genau zu untersuchen.

Mit dem Ziel, die Beobachachtungsm\"oglichkeiten fur Pulsare und andere 
Quellklassen unter 100 GeV zu verbessern, wurde eine neuer Photondetekor, 
der HPD R9792U-40, charakterisiert. Viele positive Eigenschaften konnten 
gefunden werden, wie beispielsweise eine sehr hohe Photon-Nachweiseffizienz,
 eine ausssserordentlich niedrige Ionenr\"uckkopplungswahrscheinlichkeit, 
sowie eine ausgezeichnete Ladungsaufl\"osung.
  \tableofcontents

  \cleardoublepage

%

\pagenumbering{arabic}
 \mainmatter\setcounter{page}{1}
 
\addcontentsline{toc}{chapter}{\protect Introduction}

\chapter*{Introduction}

The first detection of the TeV gamma-rays from a celestial object, i.e. 
the discovery of the Crab Nebula by the Whipple telescope in 1989, 
opened a new field in astronomy, which is called very high energy (VHE) gamma-ray astronomy. 
In this thesis, VHE gamma-ray stands for photons with energies above 10 GeV. 
As of now approximately 100 VHE gamma-ray sources are known, which are categorized
into several classes such as active galactic nuclei, supernova remnants, 
pulsar wind nebulae and gamma-ray binaries.
Observations in 
VHE gamma-rays have given first insight into the nature of these extremely dynamical objects
emitting non-thermal radiation.

Pulsars are a class of celestial objects
, which emit periodic pulsed radiation (pulsation) extending from radio up to gamma-rays.
VHE gamma-rays can serve as an important probe for the radiation mechanism of pulsars, too.
Pulsars are explained as rapidly rotating neutron stars 
which possess extremely strong magnetic fields. 
Electrons are accelerated within their magnetosphere
by strong electric field and emit beamed electromagnetic radiation, which will be observed
as pulsation due to the rotation of the neutrons star. 
On top of this general picture of the pulsation mechanism, 
there are two competing major scenarios which specify the acceleration/emission region within the magnetosphere.
One is the inner magnetosphere
 scenario, in which the pulsation originates from near the magnetic pole on the neutron
 star surface. The other is the outer magnetosphere scenario, in which the pulsation comes 
 from a region along the last closed magnetic field lines in the outer magnetosphere. 
Both scenarios could reasonably explain all the features of pulsars observed before 2007.  
 One of the best ways to verify or refute these scenarios is measuring the energy spectrum 
 at around cut-off energy, i.e. at energies between a few GeV and a few tens of GeV.
The reason is as follows: Both scenarios predict a high energy cut-off in the gamma-ray energy spectrum,
but with different cut-off sharpness.
An exponential cut-off is expected for the outer magnetosphere scenario
while a super-exponential cut-off is predicted for the inner magnetosphere scenario.
The sharpness of the cut-off and the highest energy of the observed photons allow one
to constrain the emission region.

Before 2007, a satellite-borne detector, EGRET, detected 7 gamma-ray pulsars
and the energy spectra could well be measured only up to $\sim 5$ GeV. 
On the other hand, ground-based imaging atmospheric Cherenkov telescopes (IACTs)
 could set flux upper limits only above 100 GeV. 
There existed no sufficient measurement 
in the important energy range, i.e., at energies
between a few GeV and a few tens of GeV. 
It was evident that filling this energy gap would lead to 
a clarification between the competing scenarios
and a better understanding of the pulsation mechanism.




 The MAGIC telescope is the IACT that has a largest single dish reflector 
with a 17 m diameter. Accordingly, it has
the lowest energy threshold among IACTs (50 GeV in the case of the standard trigger). 
MAGIC has been the best instrument to fill 
the energy gap from the higher side. Moreover, in October 2007, the new trigger system 
(SUM trigger system) was
implemented, which reduced the energy threshold even further, from 50 GeV
to 25 GeV. 
Nearly at the same time, a new satellite-borne gamma-ray detector,
\fermi-LAT, became operational in August 2008, which has a $\sim 10$ times better sensitivity
than EGRET and could measure the pulsar energy spectrum well beyond 10 GeV. It began filling the
gap from the lower side. 

In this thesis, the observational results on the Crab pulsar
by MAGIC at energies above 25 GeV are presented. 
The public data of the \fermi-LAT on the Crab pulsar are also analyzed from 100~MeV 
to $\sim 30$~GeV. 
The combined analysis of the results from two experiments is also carried out,
carefully taking into account the systematic uncertainties of both experiments.
Then, several constraints in the pulsar model based on the combined energy spectrum 
are discussed. 
In addition to the spectral study, energy-dependent pulse profiles between 100 MeV to 100 GeV
are intensely studied. The possibility to infer the fine structure 
of the emission region based on the pulse profile is also discussed.

Aiming for better observations of some selected pulsars and other sources 
below 100 GeV with MAGIC, 
a new photodetector, HPD R9792U-40, is investigated. 
HPD R9792U-40 is a hybrid photodetector. Since a hybrid photodetector has never been used in any IACTs,
its properties and performance are thoroughly studied.
\\

 The thesis is structured as follows. An introduction to VHE gamma-ray astronomy 
is given in Chapter~1. Pulsars and theoretical models of the pulsation mechanism
are introduced in Chapter~2. Chapter~3 describes the IACT technique and the MAGIC telescope. 
The analysis methods of the MAGIC data and its performance are explained in detail in Chapter~4.
The analysis results of the MAGIC data and the public \fermi-LAT data on the Crab pulsar are 
presented in Chapter~5 and Chapter~6, respectively. In Chapter~7, a combined analysis of the
results of the two experiments is performed. Physics discussions on the results are
presented in Chapter~8. 
The properties and performance of HPD R9792U-40 and the development of associated operation
circuits are presented in
Chapter~9. Conclusions and outlook are added in Chapter~10.

 \chapter{Very High Energy Gamma Ray Astronomy}

see http://wwwmagic.mppmu.mpg.de/publications/theses/TSaito.pdf

 \chapter{Pulsars}

see http://wwwmagic.mppmu.mpg.de/publications/theses/TSaito.pdf

 \chapter{Imaging Atmospheric Cherenkov Telescope Technique and \\
The MAGIC Telescope}

see http://wwwmagic.mppmu.mpg.de/publications/theses/TSaito.pdf
 \chapter{Analysis Method of MAGIC Data}

see http://wwwmagic.mppmu.mpg.de/publications/theses/TSaito.pdf

 \chapter{MAGIC Observations of the Crab Pulsar and Data Analysis}
\label{ChapMAGICObs}

The MAGIC observations are grouped in one-year long cycles and Cycle I started in May 2005.
 In Cycles I and II, the Crab pulsar was observed 
with the standard trigger. 
16 hours of good-condition data showed only a weak signal of pulsation with 2.9 $\sigma$ 
(see Sect \ref{SectCrabObsBefore2007}). Here, I analyze the data recorded in Cycle III and
Cycle IV with the SUM trigger with a much lower energy threshold and a higher
sensitivity below 100 GeV.

\section{Observations}
 The SUM trigger was installed in October 2007 and, subsequently, the Crab pulsar 
was observed 
in Cycles III and IV, for 48 hours (over 47 days) and 78 hours (over 36 days), respectively. 
All the observations were made in ON-mode,  
since the SUM trigger was designed for ON-mode observations.
Although, as described in Sect \ref{SectSignalExtract}, pulsar observations do not require 
OFF observations, 
in order to assure the validity of the analysis chain 
and the quality of the data sets
by the Crab nebula emission, OFF observations were made in October and December 2007 for
$\sim 10$ hours.  

\section{Sum Trigger Sub-patch Malfunction}
\label{SectMalfunc}
\begin{figure}[h]
\centering
\includegraphics[width=0.45\textwidth]{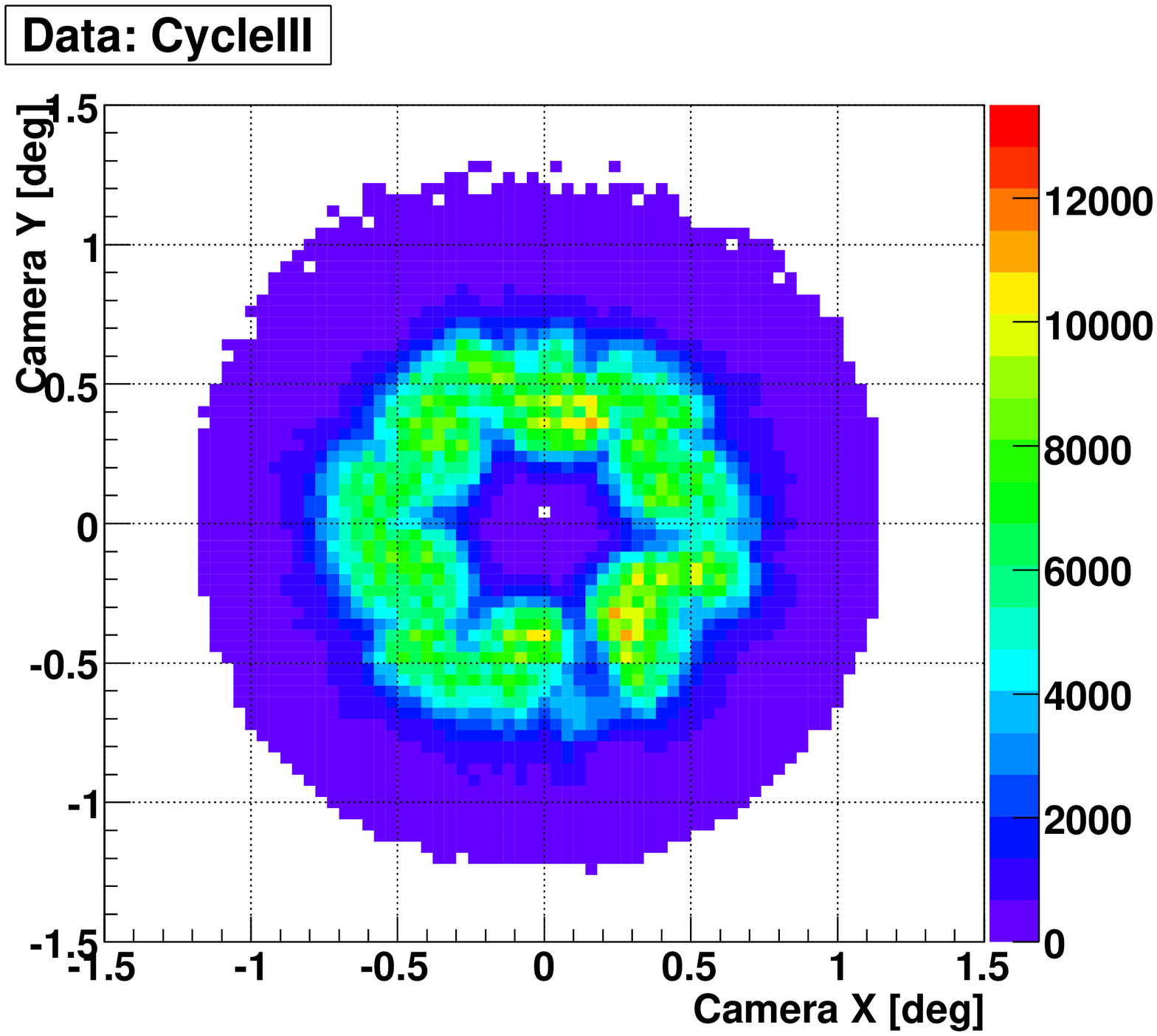}
\includegraphics[width=0.45\textwidth]{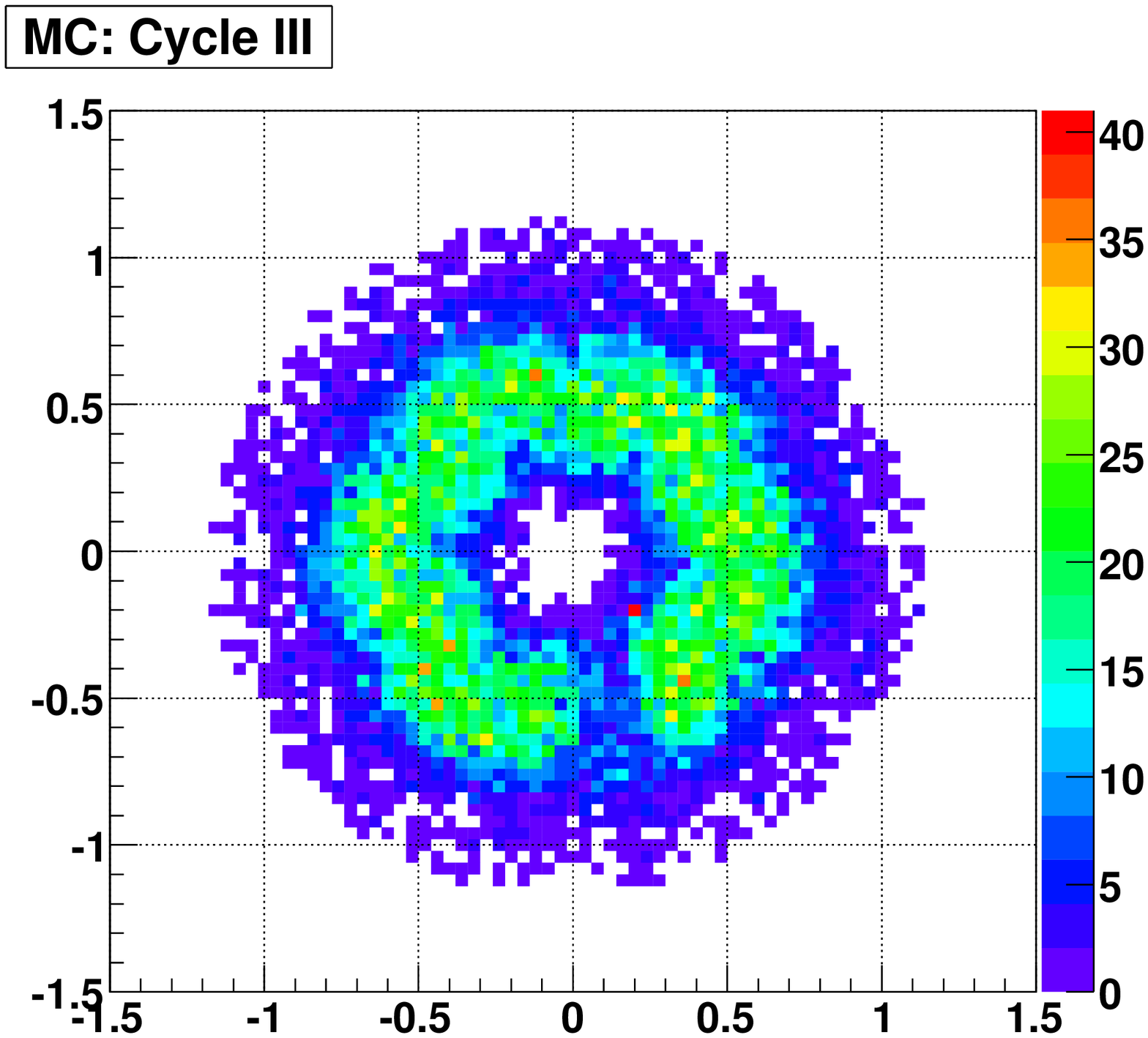}
\includegraphics[width=0.45\textwidth]{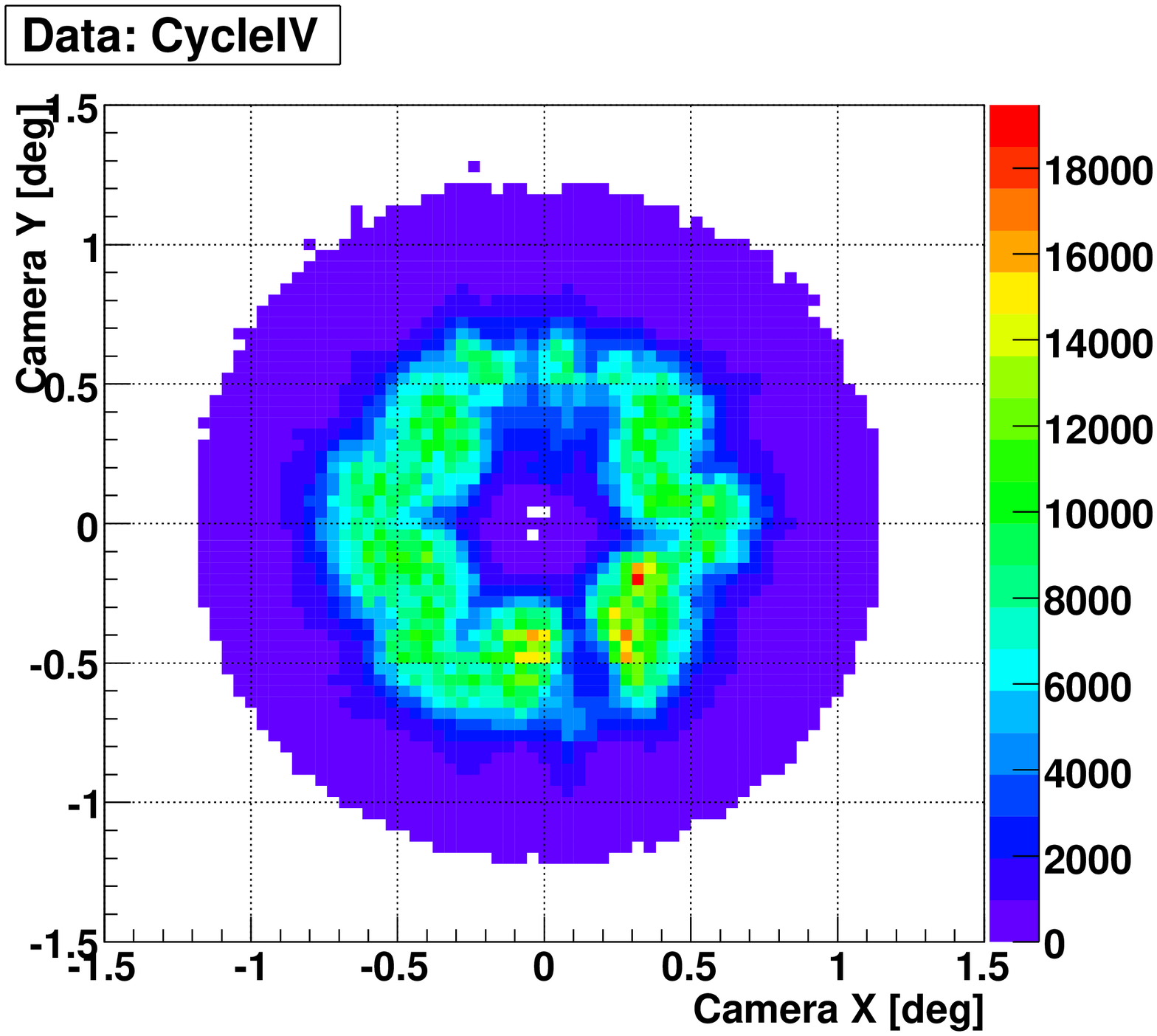}
\includegraphics[width=0.45\textwidth]{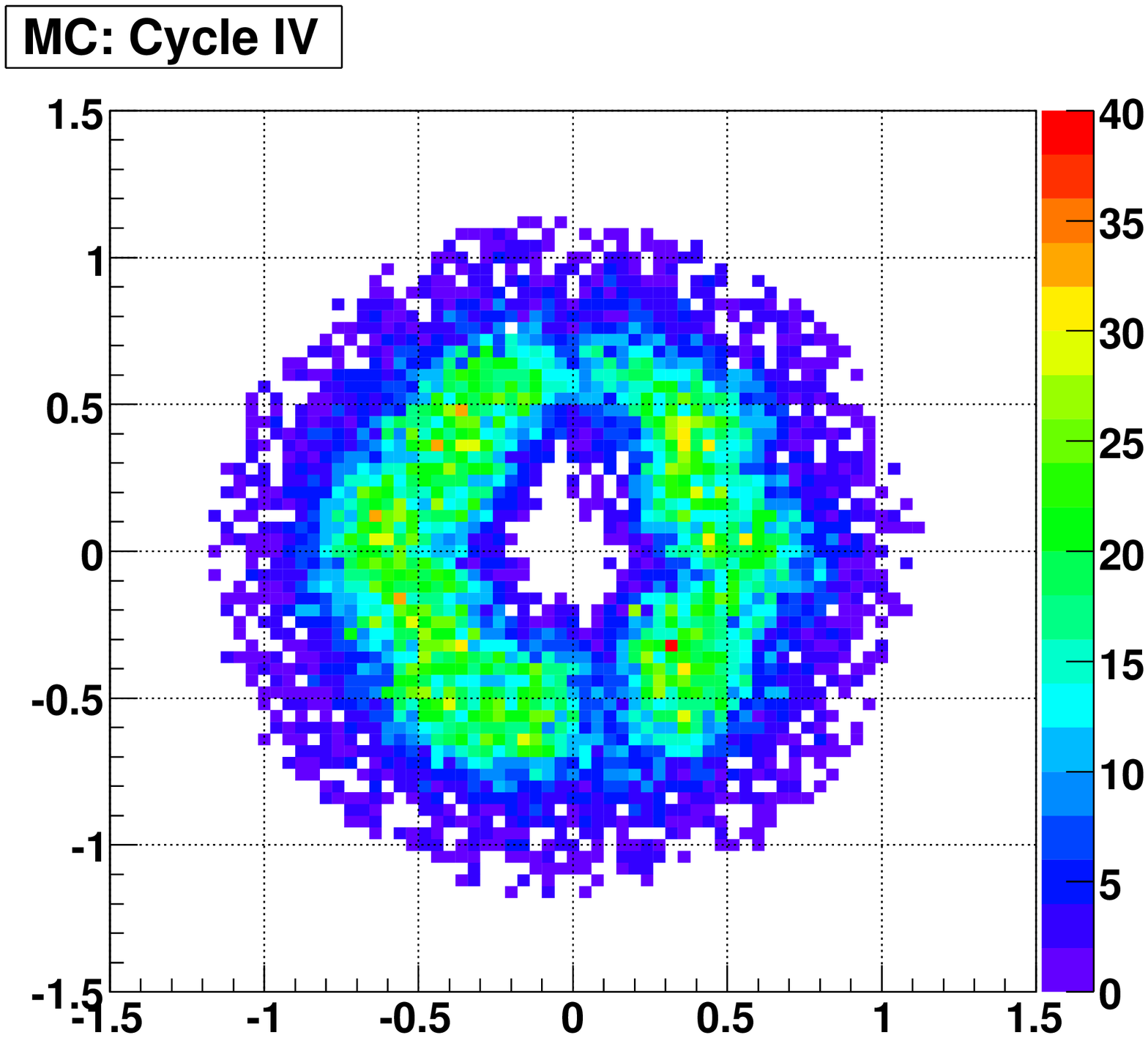}
\includegraphics[width=0.45\textwidth]{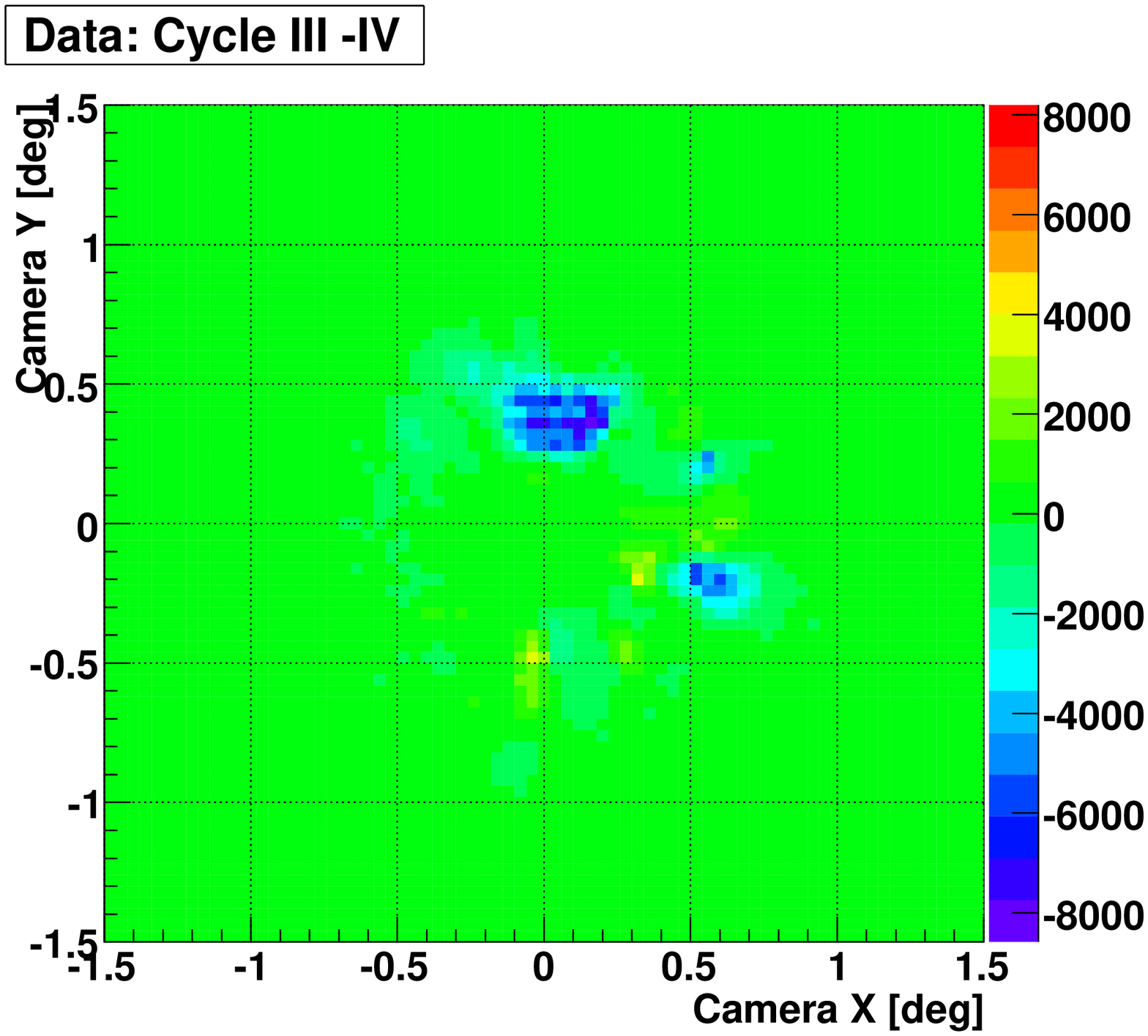}
\includegraphics[width=0.45\textwidth]{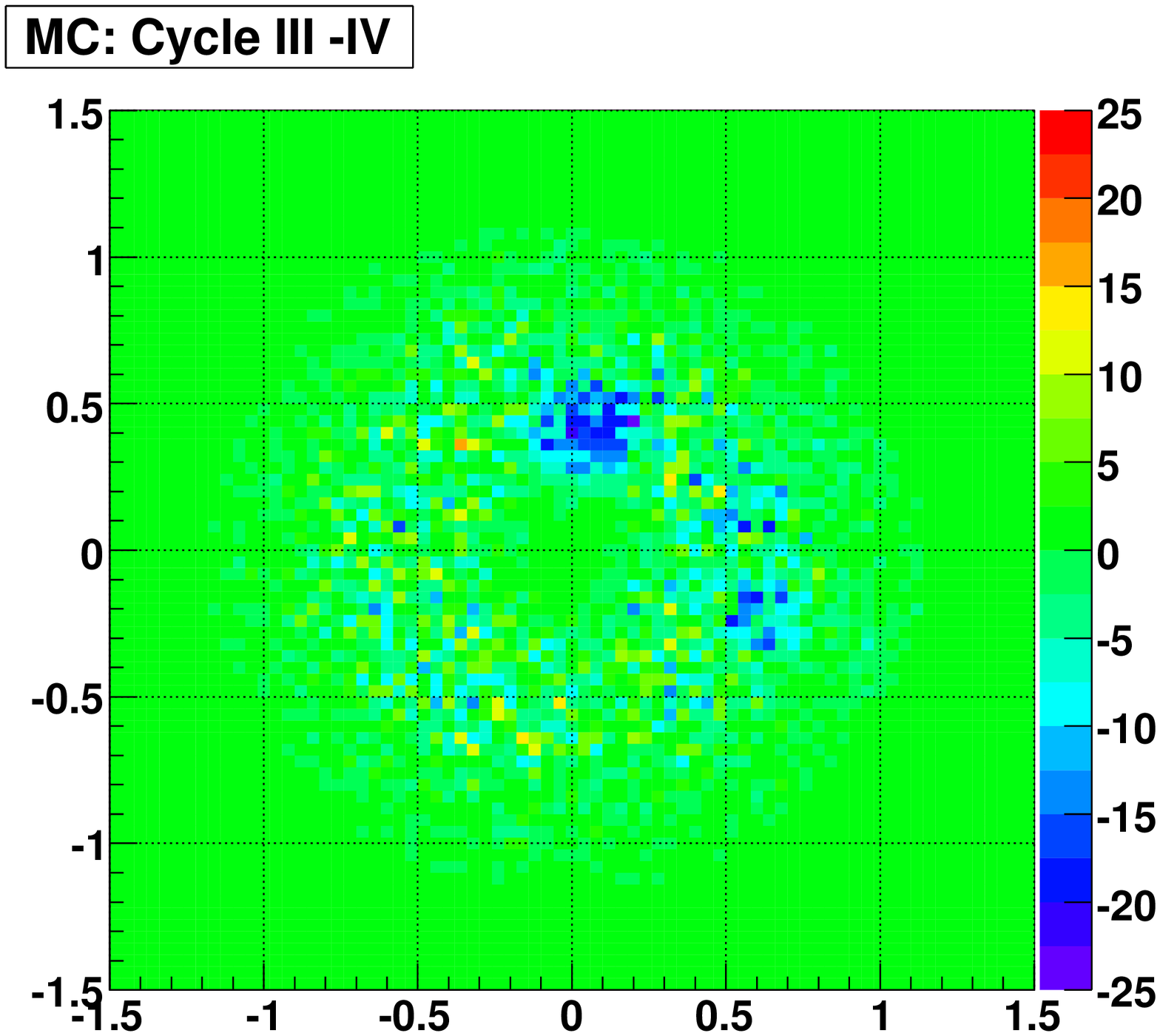}
\caption{COG distributions on the camera. The left and right columns show 
observed data and gamma-ray MC events, respectively. The first, second and 
 third rows show Cycle III, Cycle IV and difference between the two cycles, respectively.
The effect of the subpatch malfunction is seen as holes of COG distributions. 
Simulations reproduce the effect reasonably well. 
}
\label{FigCOG}
\end{figure}

 One and five out of the 36 Sum Trigger sub-patches were malfunctioning during the Crab pulsar
observations in Cycle III and Cycle IV, respectively. 
 The effect can be seen by plotting $COG$ of images on the camera. The regions near the broken patches show a hole in $COG$ distribution, as shown in the top left (Cycle III data) and the middle left (Cycle IV data) panel of Fig.\ref{FigCOG}. When a variability of the Crab pulsar flux
between the two cycles is discussed, this difference must be taken into account. 
By deactivating the broken patches in MC, the effect can be reproduced as shown in the top right (Cycle III MC) and the middle right (Cycle IV MC). 
 The bottom left and the bottom right panel
show the difference between Cycle III and Cycle IV in data and MC, respectively. According to the MC, the differences in gamma-ray detection efficiency between the two cycles for 
$SIZE$ 25-50, 50-100, 100-200 and 200-400 ph.e. are about 21\%, 17\%, 11\% and 7\%. 
This effect will be corrected when the variability is discussed (see Sect. \ref{SectVariability}).
 The calculation of the energy spectra will also be carried out with MCs which reflect
these subpatch malfunctions. 

\clearpage
\section{Data Selection}
\label{SectCuts}
The new SUM trigger system lowers the energy threshold from 50 GeV to 25 GeV
(see Sect. \ref{SectSUMTrigger}). 
It has a big impact on the pulsar observations because most of the signal is expected 
below 50 GeV (see Fig. \ref{FigThreshold}). 
However, such low energy events are easily affected by the observational conditions.
A slightly worse
condition may result in a significant increase
 in the energy threshold and a worse sensitivity. 
Therefore, a more careful data selection than the normal observations above 50 GeV 
is required,
in terms of both hardware and environmental conditions. 

\subsection{Reflector Performance Selection}
\begin{figure}[h]
\centering
\includegraphics[width=0.65\textwidth]{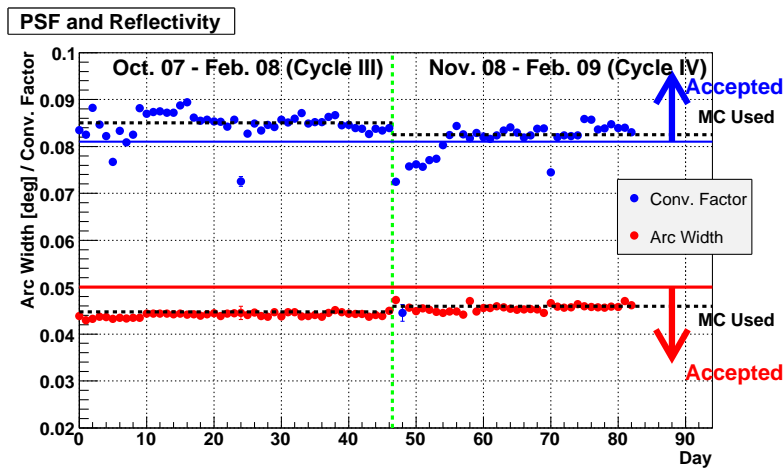}
\caption{Daily monitoring of the light collection efficiency and the PSF of the reflector by muons.
The horizontal axis indicates the observation day and the vertical axis indicates
the mean $ArcWidth$ and the mean conversion factor. 
Data taken in the days when the light collection efficiency (blue points) is below the limit 
(a blue line) are excluded from the analysis. 
}
\label{FigConvPSF}
\end{figure}

The mirror panels of the MAGIC telescope are adjustable (see Sect. \ref{SectReflector}) 
and their alignment
 on some days can be worse than it should be. 
As described in Sect. \ref{SectMuon}, this can be checked by muon events. 
Blue points in Fig. \ref{FigConvPSF} show the light collection efficiency 
estimated by  muon events
(the conversion factor
from the number of photons hitting the reflector to that of the detected photoelectrons)
for each observation day. 
The days when the efficiency is lower than 0.081 (5\% lower than the average)
are excluded from the analysis, since it may affect the trigger efficiency and the energy 
reconstruction.
On average, Cycle IV has 3\% lower efficiency than Cycle III, which is taken into account
in the MC when the energy spectrum is calculated. 
The PSF was checked by fitting a linear function to the $ArcWidth/Radius - Radius$ relation
 (see Sect \ref{SectMuon}). The value at $Radius = 1.15$ degrees is used to evaluate the PSF,
 as shown by the red points in 
Fig. \ref{FigConvPSF}.
 From the study with MC, it was found that the PSF in Cycle IV was $\sim 5\%$ worse than
that in Cycle III, which is also taken into account when the energy spectrum is computed.

\subsection{Zenith Angle Selection}
 As the angular distance from the source to the zenith (Zenith Angle, $ZA$) becomes larger, 
the distance from the shower maximum to the telescope
 increases. The relation between the gamma-ray energy and the number of detected photons 
would change depending on the $ZA$, due to the different Cherenkov photon density on the ground
\footnote{This is a consequence of three effects: a) a higher Cherenkov threshold
leading to fewer photons (see the top right panel of Fig. \ref{FigRefIndex}), b) a wider spread of photons on the ground due to the larger distance
and c) increase in absorption and scattering losses (see Sect. \ref{SectAbsorp}).}.
Needless to say, the threshold energy is also affected.  
Therefore, in order to assure the lowest possible threshold and a uniform $SIZE$ -- energy 
relation, 
I selected data with the $ZA$ below 20
 degrees.
\begin{figure}[h]
\centering
\includegraphics[width=0.5\textwidth]{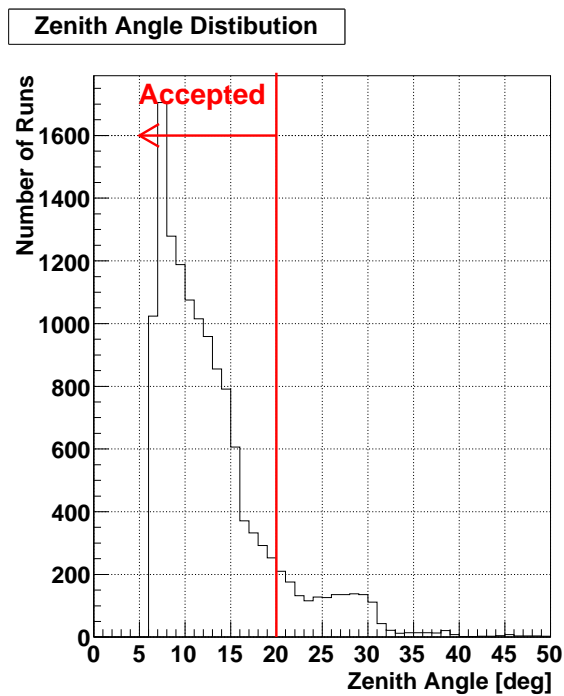}
\caption{$ZA$ distribution of the data. Most of the data are taken below 20 degrees in $ZA$. 
In order to assure the lowest possible energy threshold and the uniform $SIZE$~-~energy
relation, data with $ZA$ below 20 degrees are used in the analysis.
}
\label{FigCrabGeom}
\end{figure}

\subsection{Cloudiness Selection}

\begin{figure}[h]
\centering
\includegraphics[width=0.40\textwidth]{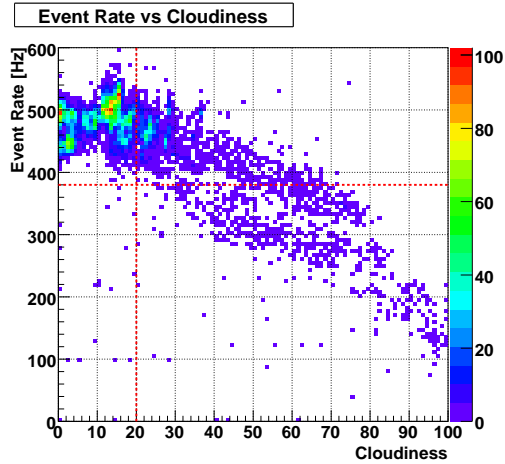}
\caption{The event rate as a function of Cloudiness.
A clear anti-correlation can be seen. Cut values in the event rate and in $Cloudiness$
are indicated by the red dotted lines.
}
\label{FigClvsRate}
\end{figure}

\begin{figure}[h]
\centering
\includegraphics[width=0.65\textwidth]{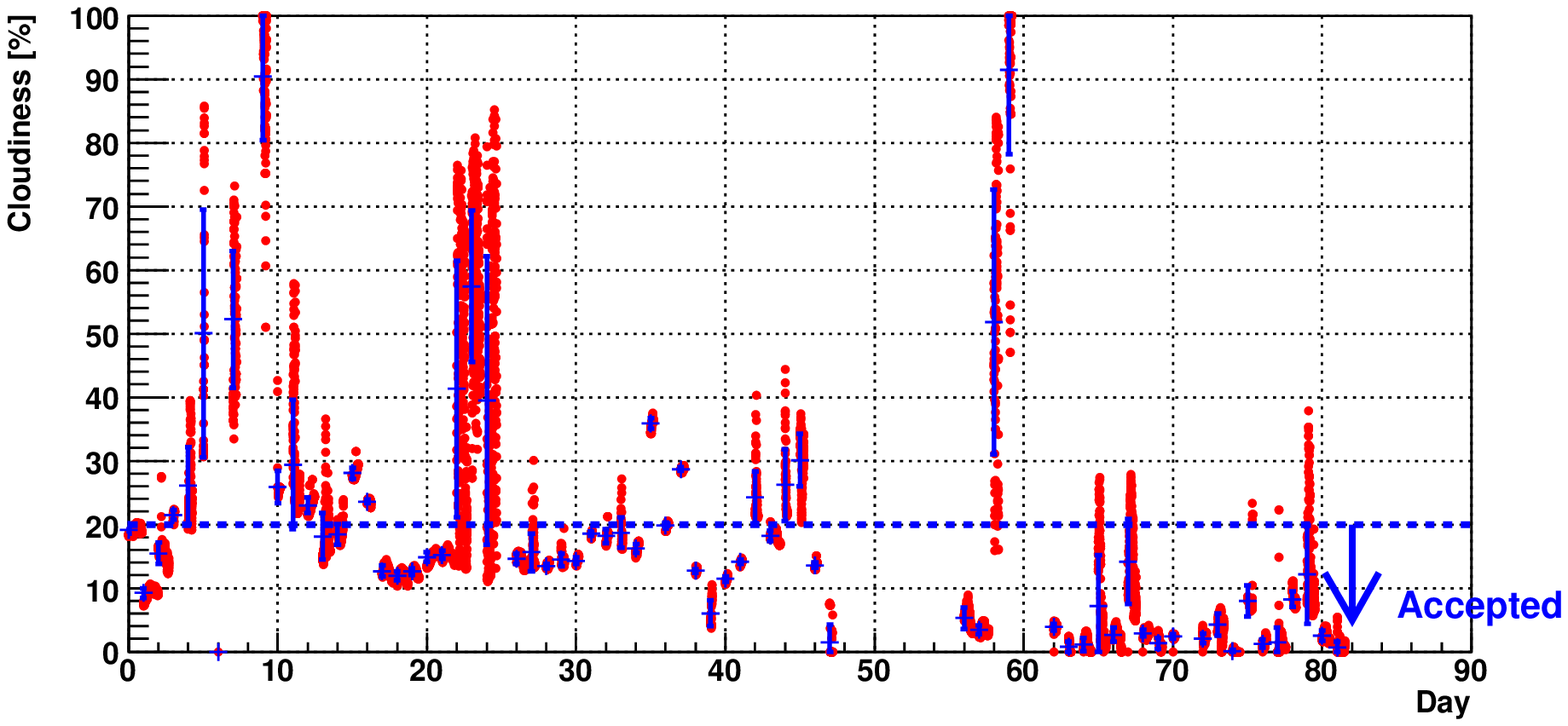}
\includegraphics[width=0.25\textwidth]{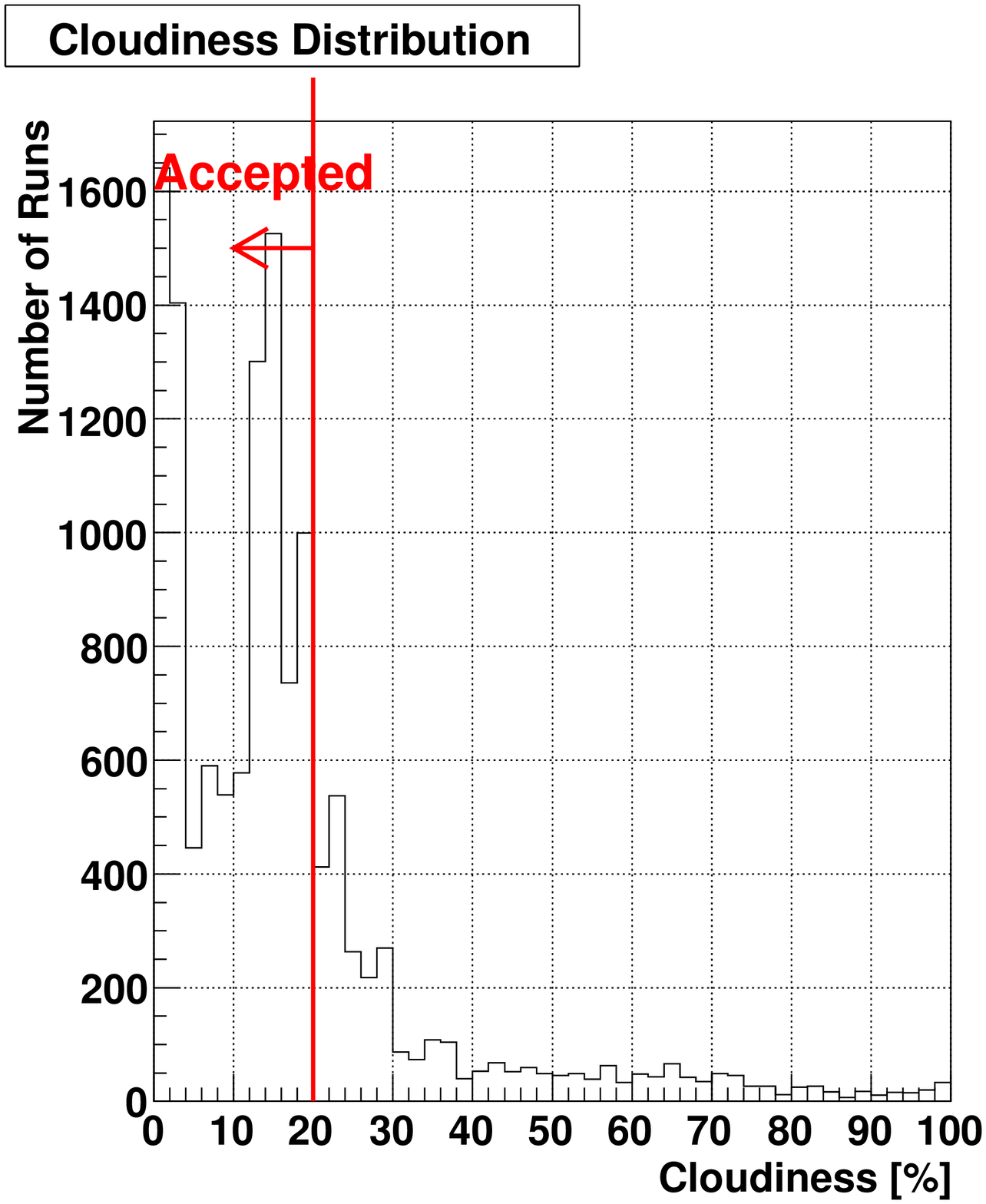}
\caption{Left: $Cloudiness$ for each run. 
The horizontal axis indicates the observation day and the vertical axis indicates
$Cloudiness$. A red dot corresponds to one run ($\sim 1$ minute).
The mean and the RMS of the $Cloudiness$ for each day are denoted by blue crosses. 
The data with $Cloudiness$ larger than 20 are excluded from the analysis as denoted by the 
blue dotted line. 
Right: $Cloudiness$ distribution. Most of the runs have cloudiness below 20 and 
the data with $Cloudiness$ larger than 20 are excluded from the analysis. 
}
\label{FigDailyCloud}
\end{figure}

As described in Chapter \ref{SectPyrometer}, haze and clouds affect the air shower images. 
Especially for the lowest energies below 50 GeV, the  effect would
be significant. Fig. \ref{FigClvsRate} shows the event rate 
\footnote{Not the trigger rate but the rate of the events which are not
completely erased by the image cleaning procedure. Therefore, most of the NSB accidental trigger
events are not included in the rate}
as a function of $Cloudiness$ (see Sect. \ref{SectPyrometer}).
A clear anti-correlation between $Cloudiness$ and the event rate can be seen. 
To assure that no data is affected 
by haze and clouds, I selected data taken with $Cloudiness$ lower than 20, as shown in Fig. \ref{FigDailyCloud}.
If the daily average value is more than 20, all the data taken on that day are excluded. 

\clearpage

\subsection{Event Rate Selection}
\begin{figure}[h]
\centering
\includegraphics[width=0.7\textwidth]{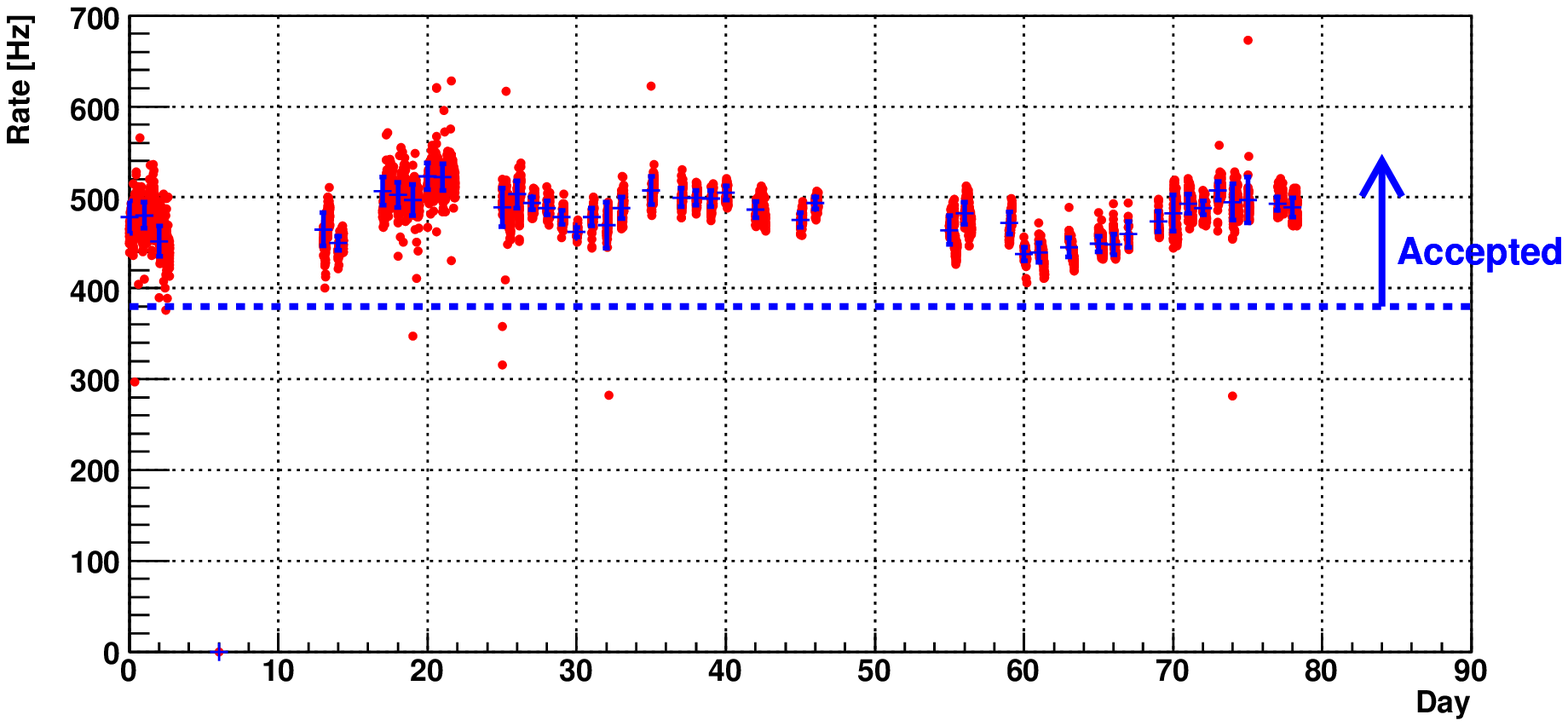}
\includegraphics[width=0.25\textwidth]{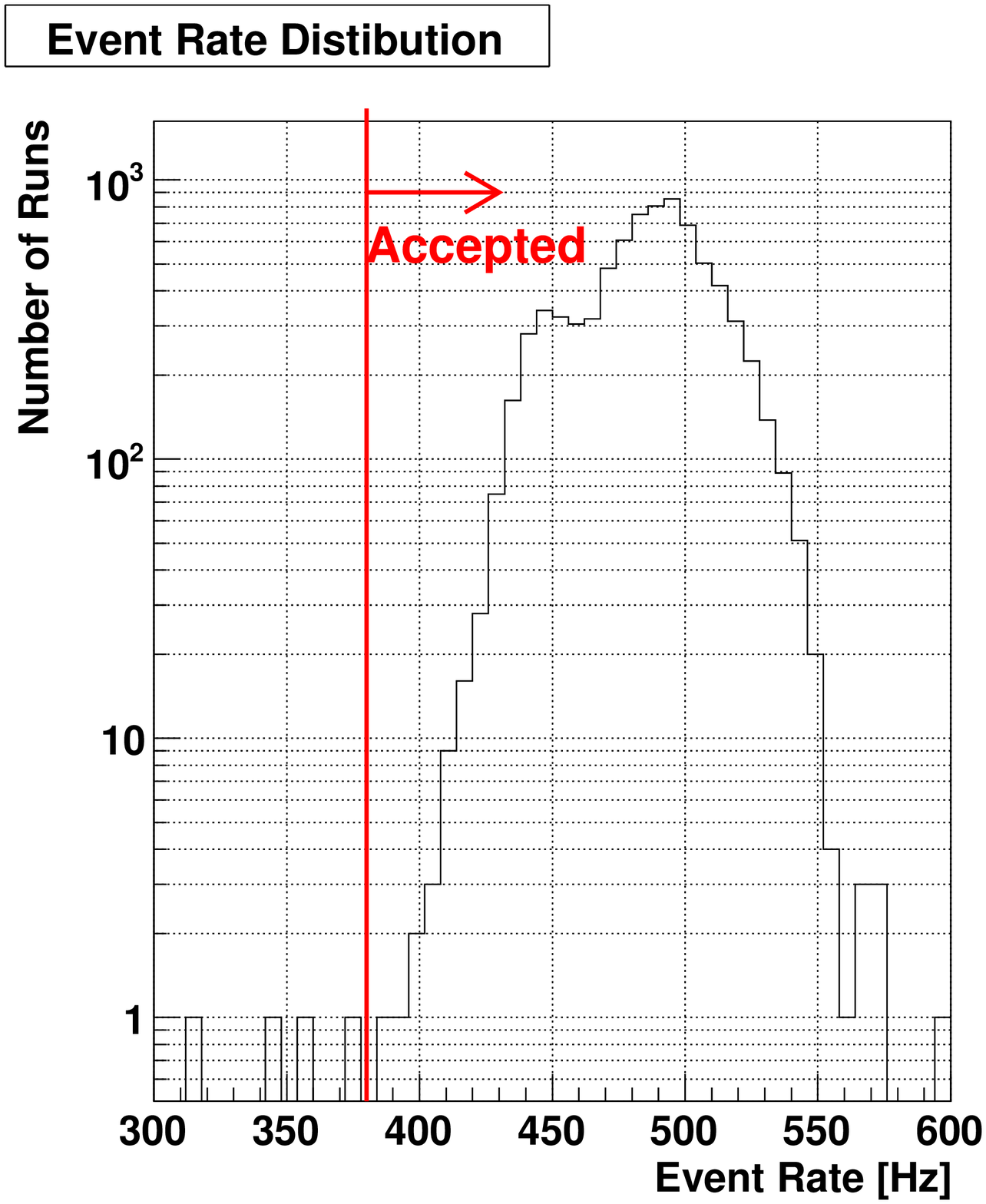}
\caption{Left: The event rate for each run. 
The horizontal axis indicates the observation day and the vertical axis indicates
the event rate. A red dot corresponds to one run ($\sim 1$ minute).
The mean and the RMS of the event rate for each day are denoted by blue crosses. 
The data with the event rate smaller than 380 Hz are excluded from the analysis,
as denoted by the blue dotted line. 
Right: The event rate distribution. Only a few runs 
have an event rate smaller than 380 Hz and they are excluded from the analysis. 
}
\label{FigDailyCR}
\end{figure}

Even if the environmental conditions and the reflector status are good, 
 the telescope performance might not be optimal because of inappropriate
DAQ settings, such as wrong threshold settings and the change in signal transmission
time (length), which can happen during long-term observations. 
Such problems can be identified by checking the event rate. 
For the very low threshold observations like 
pulsar observations, almost half the trigger rate
 is due to NSB + after-pulsing accidental events and, hence, 
the trigger rate may not reflect 
improper DAQ settings. The event rate after the image cleaning would be more indicative because
images of such accidental events would be completely erased by the image cleaning.
As you can see from Fig. \ref{FigDailyCR}, after cutting away
 bad reflector days and cloudy days, 
almost all runs have a good rate, meaning that hardware settings had been fine.
 Runs with the event rate lower than 380 Hz were discarded just in case. 

\subsection{Nebula Measurement Selection}
\label{SectNebulaCut}
From the rate cut described in the previous section, it is almost guaranteed that event selection
has been properly carried out. Just as an additional cross-check, 
the detection efficiency of the Crab nebula emission was examined. 
I analyzed the Crab nebula emission\footnote{As noted in Sect. \ref{SectCrabGeom}, the nebula
and the pulsar cannot be spatially resolved by IACTs. The non-pulsed gamma-ray emission 
above 100 GeV is considered as a nebula emission}
 with $SIZE$ 
above 300, for which $Hadronness$ cut is very powerful,
i.e., the gamma-ray/hadron separation is highly efficient 
 (Sect. \ref{SectGHsep}) and analysis is 
rather easy. 
The $Hadronness$ cut at 0.1 is applied to the data and the excess is evaluated with 
the $ALPHA$ cut at
 10 degrees. The background was estimated by 
fitting the $ALPHA$ distribution from 20 degrees to 80 degrees with a parabolic function
and then extrapolating the fitted function down to 0 degree.
\begin{figure}[h]
\centering
\includegraphics[width=0.4\textwidth]{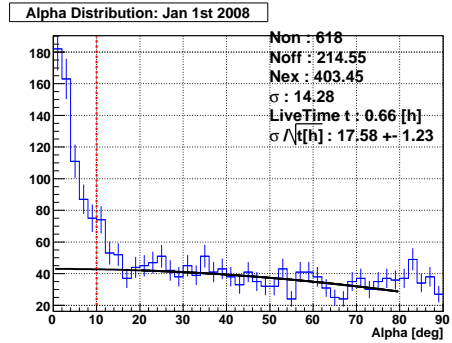}
\caption{An example of the signal from the Crab nebula with $SIZE > 300$
and $Hadronness < 0.1$.
ALPHA analysis is used (see Sect. \ref{SectAlphaDist}).  }
\label{FigNebulaAlpha}
\end{figure}
An example of this analysis is shown in Fig. \ref{FigNebulaAlpha}, which is for the observation
on 1st January, 2008. In order to check the stability of detection efficiency, 
the significance of the excess for each of the observation days was scaled to one hour's
 observation
and plotted in Fig. \ref{FigNebulaExcess}. 
They are very stable and their mean value is 16.1, which is consistent with
the telescope sensitivity. 
\begin{figure}[h]
\centering
\includegraphics[width=0.6\textwidth]{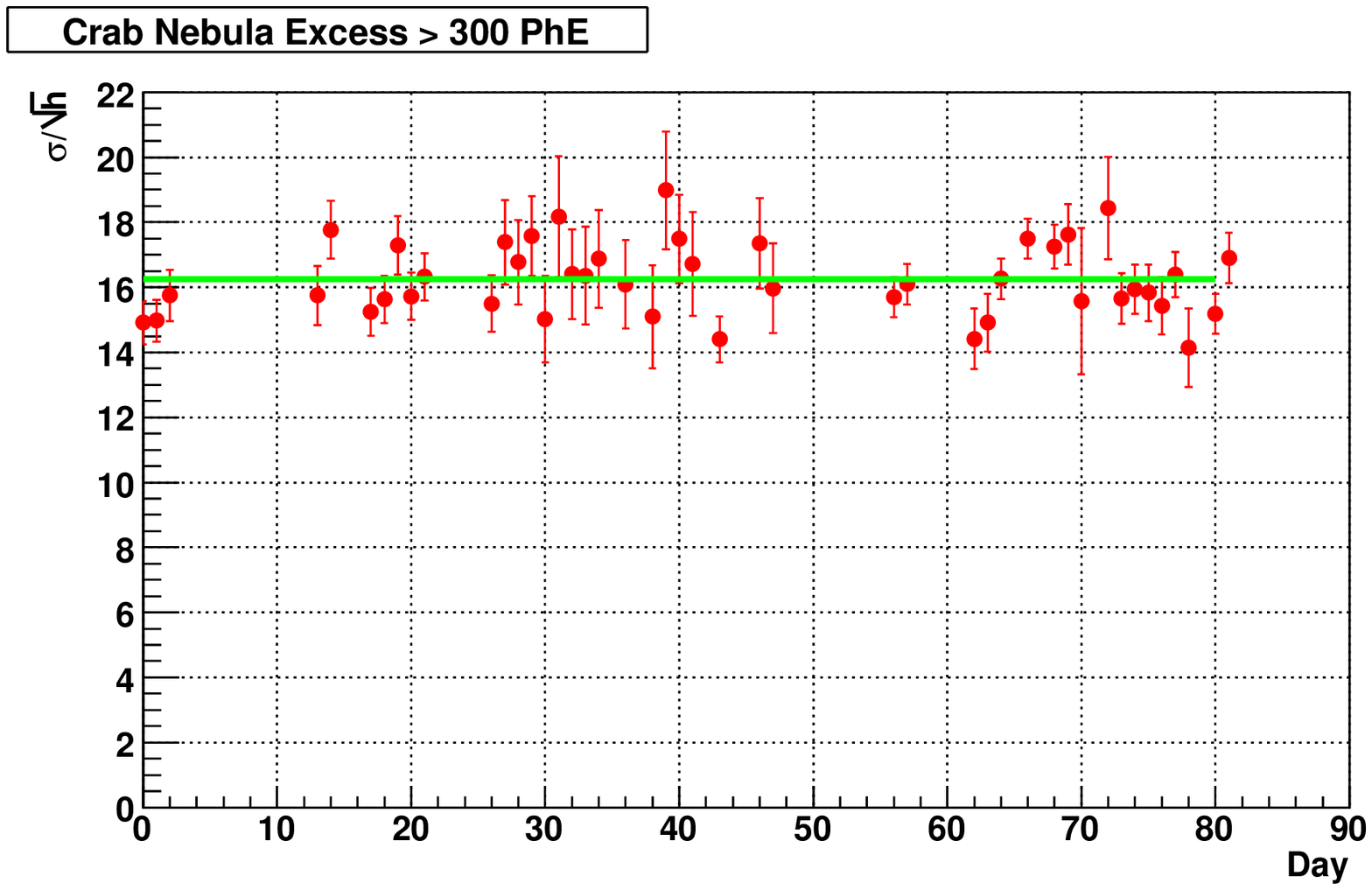}
\includegraphics[width=0.3\textwidth]{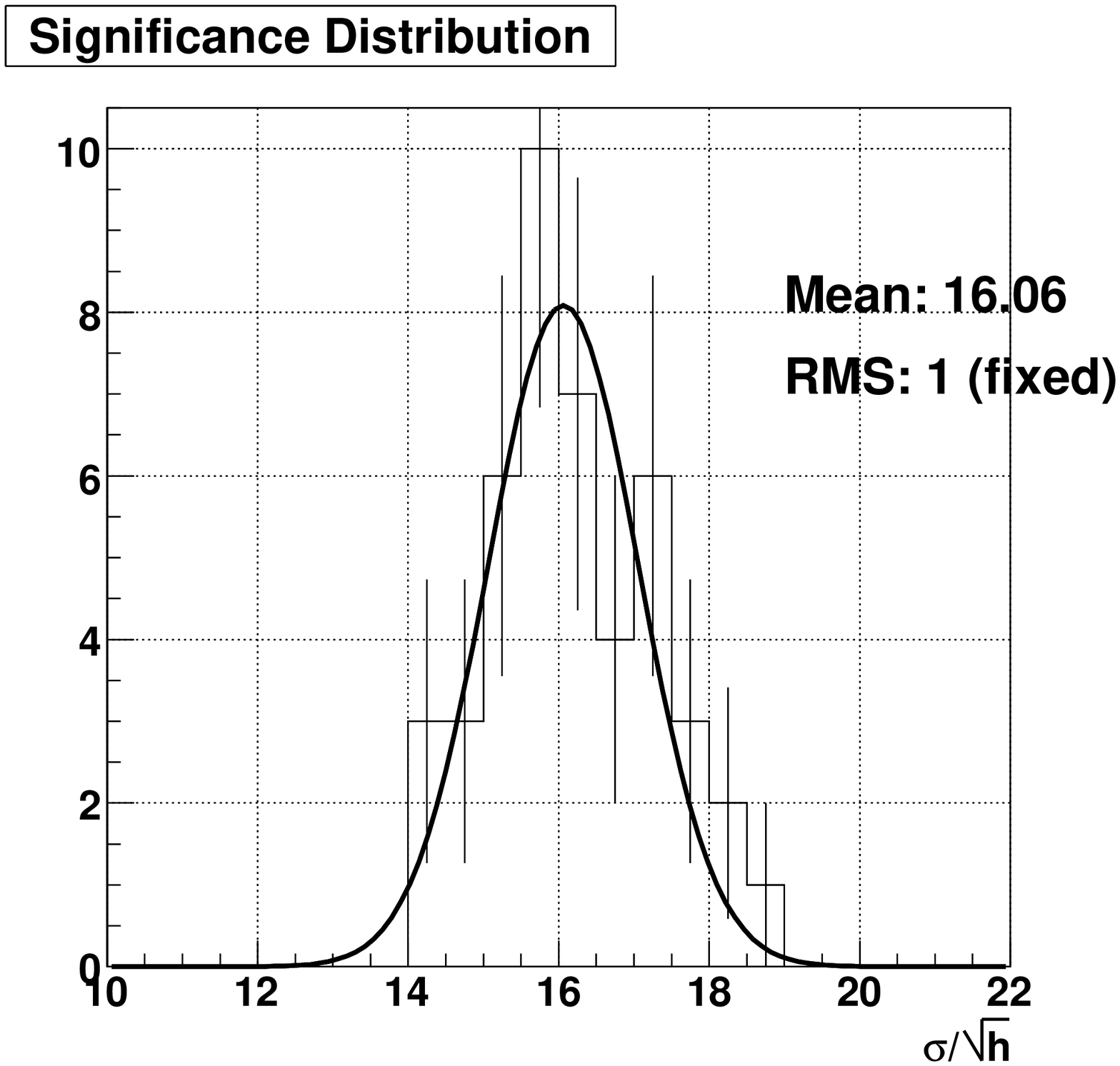}
\caption{Left: Daily monitoring of the detection efficiency of the nebula emission.
The horizontal axis indicates the observation day and the vertical axis indicates
the excess significance normalized to one hour of observation time.
Right: Distribution of the excess significance normalized to one hour. The variation is
compatible with the statistical fluctuations. 
 }
\label{FigNebulaExcess}
\end{figure}

\subsection{Summary of the Data Selection}
The selection of the good observation days are summarized in Table. \ref{TableDataSelection}. 
26 days out of 47 days and 18 days out of 36 days passed all the selection cuts
 for Cycle III and IV, 
respectively. 
Within a good day, some runs were also excluded due to unfavorable $ZA$, $Cloudiness$ or 
the event rate. 25.1 and 34.0 hours of data remained for Cycle III and IV, respectively.

\begin{table}[h]
\centering
\begin{tabular}{|r|c|c|c|c|c|r|c|c|c|c|}
\cline{1-5} \cline{7-11}
Date & PSF/ &   &   & & &Date & PSF/ &  & &  \\
yymmdd & Ref & Z.A. & Cloud & Used & &yymmdd & Ref. & Z.A. & Cloud & Used\\
\cline{1-5} \cline{7-11}
\multicolumn{5}{|c|}{Cycle III} & & 080129 & & & & yes \\	 
\cline{1-5}
071021 & & & &       yes &   & 080131 & & & Cloudy & no \\       
071022 & & & &       yes &   & 080201 & & & Cloudy & no\\       
071105 & & & Cloudy &no  &   & 080202 & & & &  yes \\	   	     
071106 & & & Cloudy &no  &   & 080203 & & & &  yes   \\     
\cline{7-11}                 
071107 & & & Cloudy &no &  & \multicolumn{5}{|c|}{Cycle IV} \\  
\cline{7-11} 
071108 & Bad & & &no & 	& 081105 & Bad & & & no\\	      
071109 & & & Cloudy &no &   & 081107 & Bad & & & no\\	     	  	
071110 & Bad & & & no& 	   & 081126 & Bad & & & no\\	     	  
071111  & & & Cloudy & no& & 081129 & Bad & & &  no\\	     	  
071112  & & & Cloudy & no& & 081130 & Bad & & &  no\\	     	  
071113  & & & Cloudy & no& & 081201 & Bad & & &  no\\	     	  
071114  & & & Cloudy & no& & 081202 & Bad & & &  no\\	     	  
071115  & & & &         yes& & 081203 & Bad & &  & no\\	     	  
071116  & & & &         yes & & 081204 & &  & & yes \\		  
071117  & & & Cloudy &  no& & 081205 & & & &  yes \\		     
071118  & & & Cloudy &  no& & 081206 & & & Cloudy &  no\\	     	     
071205 & & & &  yes & 	   & 081207 & & & Cloudy &  no\\	     	  
071206 & & & &  yes & 	   & 081208 & & $> 20^\circ$& &  no\\      
071207 & & & &  yes & 	   & 081210 & & $> 20^\circ$& &  no\\       
071208 & & & &  yes & 	   & 081219 & & & &  yes \\		  
071209 & & & &  yes & 	   & 081220 & & & &  yes \\		     
071210 & & & Cloudy &  no&  & 081222 & & & &  yes \\		     
071211 & & & Cloudy &  no&  & 081223 & & & Cloudy &  no\\	     	     
071212 & & & Cloudy &  no&  & 081229 & & & &  yes \\		  
071213 &Bad & & &   no& 	   & 090101 & & & Cloudy &  no\\	     	     
071214 & & & &  yes & 	   & 090102 & & & &  yes \\		  
071230 & & & &  yes & 	   & 090103 & & & &  yes \\		     
071231 & & & &  yes & 	   & 090104 & & & &  yes \\		     
080101  & & & &  yes & 	   & 090118 & Bad &  & & no\\	     	     
080102  & & & &  yes & 	   & 090119 & & & &  yes \\		  
080103  & & & &  yes & 	   & 090120 & & & &  yes \\		     
080104 & & & &  yes & 	   & 090121 & & & &  yes \\		     
080105 & & & &  yes & 	   & 090122 & & & &  yes \\		     
080106 & & & &  yes & 	   & 090124 & & & &  yes \\		     
080107 & & & Cloudy  &no &  & 090125 & & & &  yes \\		     
080108 & & & &  yes & 	   & 090126 & & & &  yes \\		     
080109 & & & Cloudy &no &  & 090127 & & & Cloudy &  no\\	     	     
080110 & & & & yes & 	   & 090128 & & & &  yes \\		  
080111 & & & & yes & 	   & 090130 & & & &  yes \\		     
080112 & & & &  yes & 	   & 090131 & & $> 20^\circ$& &  no\\         
080113 & & & &  yes & 	   & 090201 & & $> 20^\circ$& &  no\\      
080126 & & & Cloudy & no & & & & & & \\
\cline{1-5} \cline{7-11}
\end{tabular}
\caption{\it Selection of the data on daily basis based on the reflector performance,
the zenith angle and $Cloudiness$. 
} 
\label{TableDataSelection}
\end{table}

\clearpage

\section{$ALPHA$ Cut Optimization for Pulsar Analysis}
\label{SectDynAlphaCut}
\begin{figure}[h]
\centering
\includegraphics[width=0.4\textwidth]{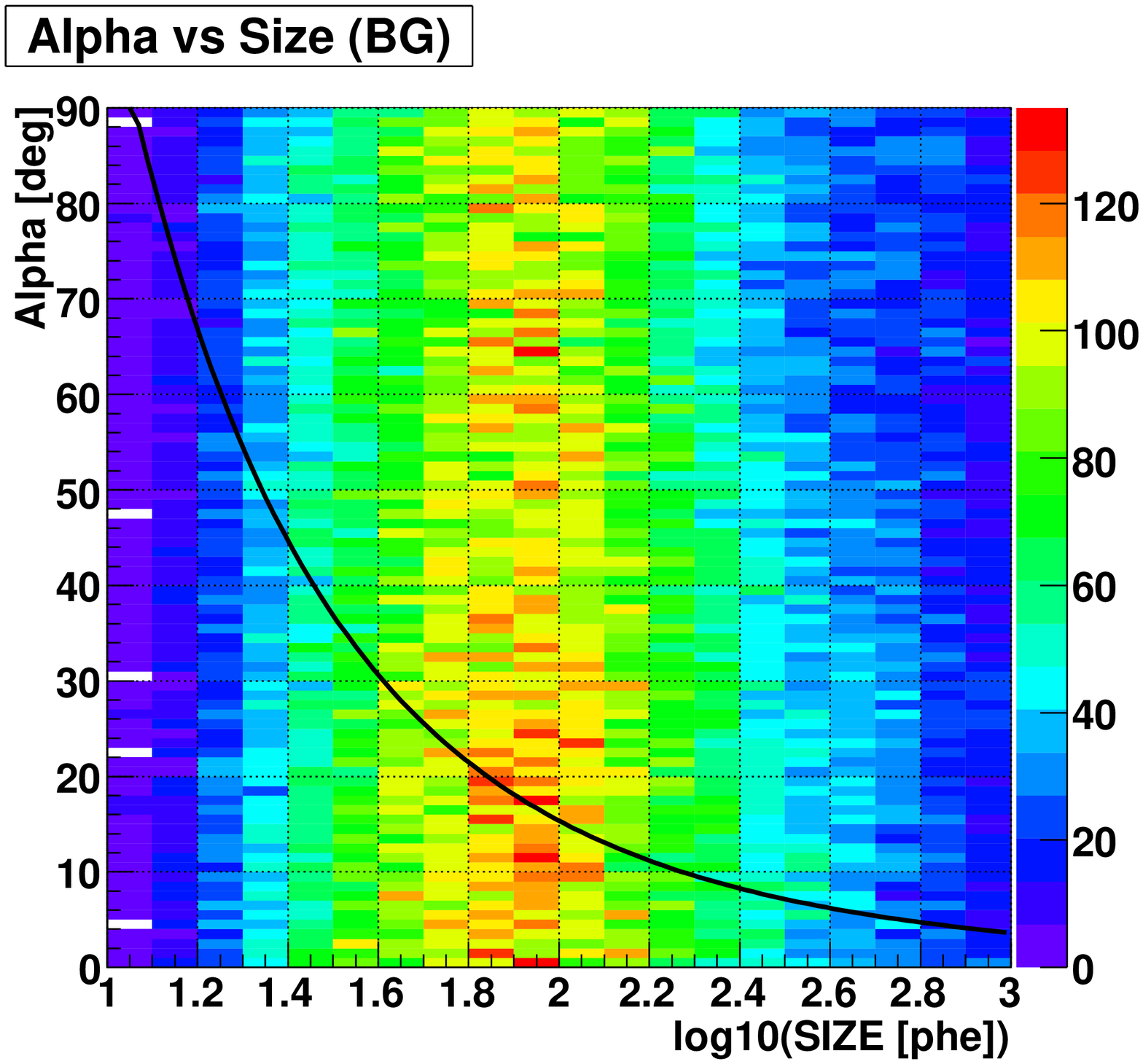}
\includegraphics[width=0.4\textwidth]{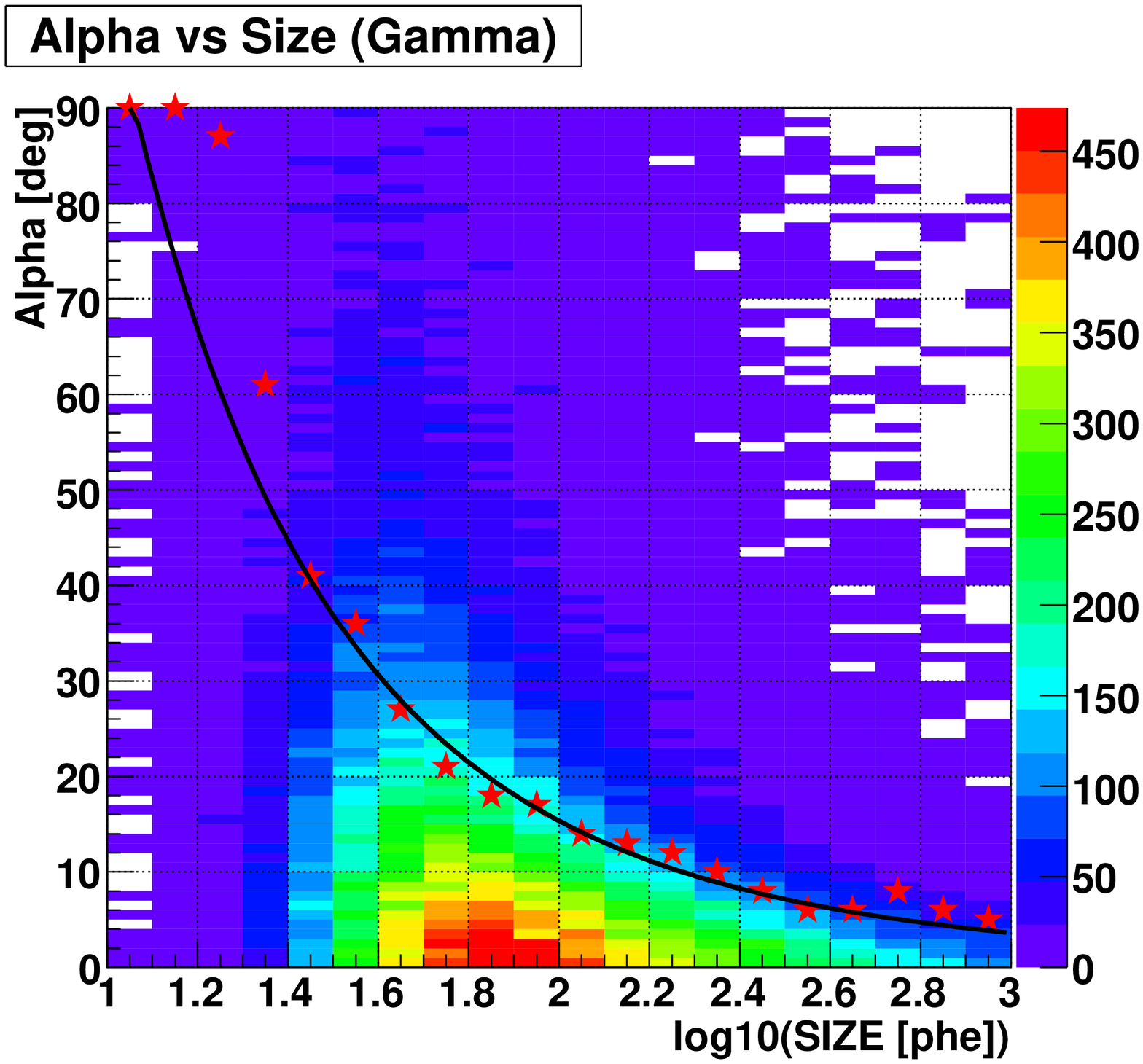}
\includegraphics[width=0.4\textwidth]{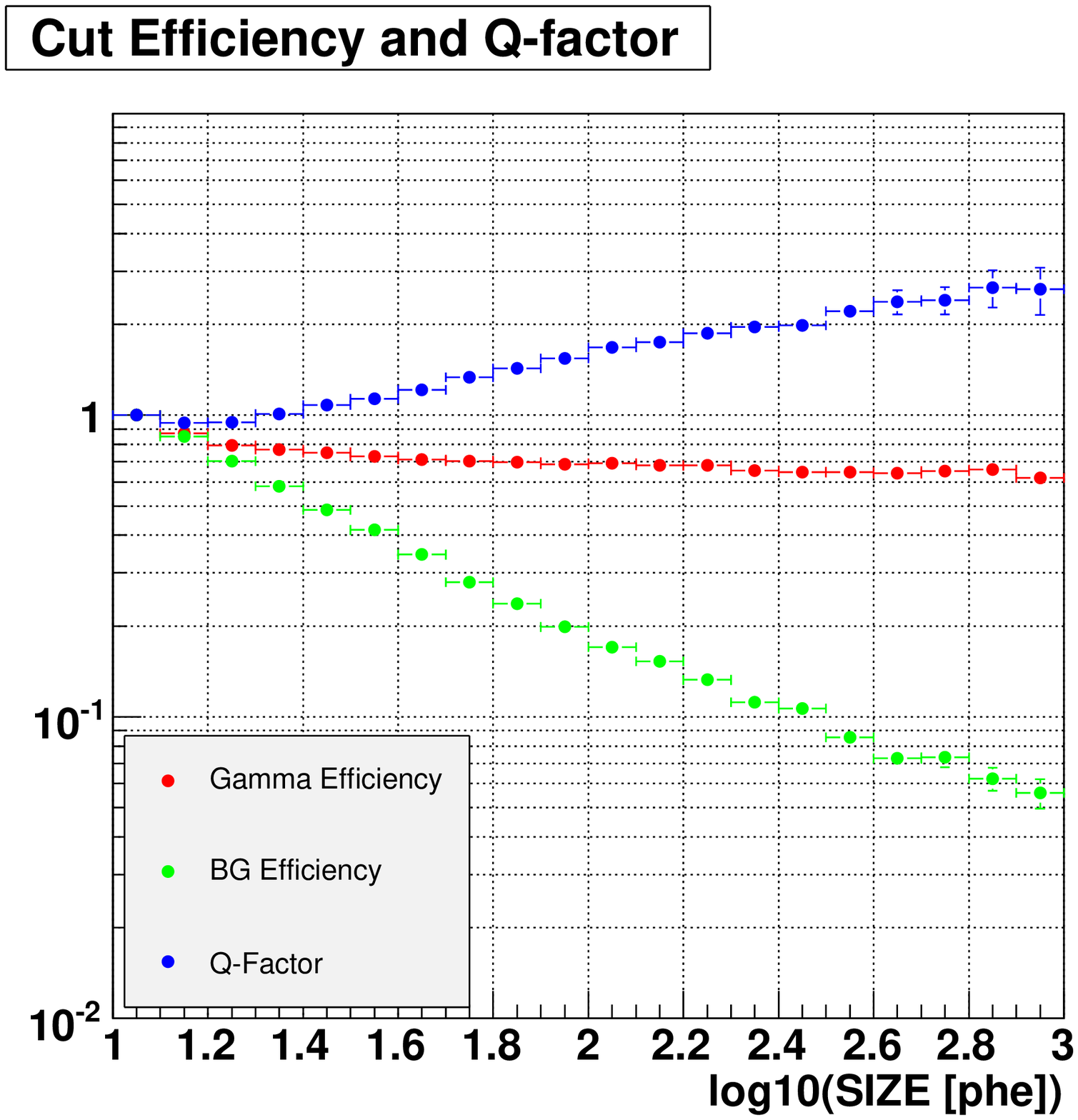}
\caption{Top left: $ALPHA$ as a function of log$_{10}$(SIZE) for an observed data sample. 
Since the data sample is dominated by the hadron background events, $ALPHA$ distributes uniformly
from 0 to 90 degree. A black line indicates the $SIZE$-dependent cut values.
Top right: $ALPHA$ as a function of log$_{10}$(SIZE) for an MC gamma-ray event sample. 
$ALPHA$ is concentrated around 0, whereas the concentration gets weaker as $SIZE$ decreases.
Red stars indicate the best cut values which maximize the Q-factor.
A black line indicates the $SIZE$-dependent cut values.
Bottom: The Q-factor (blue) and the survival ratios for hadrons (green) and gamma-rays (red)
as a function of log$_{10}$(SIZE). Q-factor is about 1.5 at $SIZE = 100$ and 1 at $SIZE = 25$.
 }
\label{FigAlphaOptim}
\end{figure}

For the pulsar analysis, the signal extraction can be carried out
using the light curve (see Sect. 
\ref{SectSigExtLC}). The image parameter $ALPHA$ is additionally used as an event
selection parameter in order to increase the signal-to-noise ratio.
The best cut value on $ALPHA$ changes with $SIZE$ because the larger the $SIZE$,
the better the shower direction estimation. The dependency
is especially strong at $SIZE < 100$,
as one can see from the top right panel of Fig. \ref{FigAlphaOptim}.
Since most of the signal from the Crab pulsar is expected at $SIZE < 100$,
a $SIZE$-dependent $ALPHA$ cut is applied to the Crab pulsar data.
It is optimized as follows:
First, MC gamma-ray and data (mostly hadron) samples are divided into 20 of log$_{10}$($SIZE$)
 bins from 1 (10 ph.e.) to 3 (1000 ph.e.), as shown in Fig. \ref{FigAlphaOptim}.
For each of the bins,
the best $ALPHA$ cut is calculated which maximizes the so-called Q-factor $Q = \epsilon_g/\sqrt{\epsilon_h}$,  where $\epsilon_g$ and $\epsilon_h$ is the fraction of events which survive
 the $ALPHA$ cut for gamma-ray and hadron samples, respectively. 
 Red stars in the top right panel of Fig. \ref{FigAlphaOptim} indicate the best $ALPHA$ cut values for each bin.
Then, those best values as a function of log$_{10}(SIZE)$ are fitted by a function
$ A_{cut}(SIZE) = a ({\rm log}_{10}(SIZE) + b )^c$, obtaining $a = 3.7 \times 10^4, b = 1.674, c=-5.988$ as the best parameters. 
The function is shown in the same panel as a black line. 
In the data analysis, the events which fulfill $ALPHA < A_{cut}(SIZE)$ are considered as
 gamma-ray candidates.
At size 25 ph.e. the cut is at 45 degrees, while at 250 ph.e.
it is at 8 degrees. $\epsilon_h$, $\epsilon_\gamma$ and $Q$ are shown in the bottom panel of Fig.  \ref{FigAlphaOptim}. 
$Q$ is approximately 1.5 at 100 ph.e. and lower for smaller $SIZE$s.

 It should be noted that a cut in $Hadronness$ was found not to improve the signal-to-noise ratio
as much as in $ALPHA$ at $SIZE <100$ where most of the signal from the Crab pulsar is expected.
 In order to avoid systematic errors in analysis, a $Hadronness$ cut
is not applied. 

\clearpage

\section{Analysis of the Energy Spectrum of the Crab Nebula}

The Crab nebula is generally used as a standard candle for the calibration of the IACTs
(see Sect. \ref{SectCrabNebula}). 
It is the brightest steady point-like source above 100 GeV. 
Actually that is why it could conveniently be used to verify the quality of 
the data for the Crab pulsar analysis (see Sect. \ref{SectNebulaCut}). 
It would also be very important to calculate its energy spectrum 
in order to assure that the analysis tool and the MC samples are
appropriate before starting the Crab pulsar analysis.
\subsection{Data Sample for the Crab Nebula Analysis}
 I used 17.1 hours of data (ON-data) and 5.2 hours of OFF observation data (OFF-data),
 both of which were taken in October and December 2007 and passed all the selection criteria described
in Sect. \ref{SectCuts}. 
The conditions of the observations are summarized in Table \ref{TableNebulaOnOff}. The pointing position of the OFF-data is on the same declination 
as for the ON-data but differs by 2 hours in right ascension, which  results in the same trajectory 
on the sky
between ON and OFF with a 2-hour time lag. 

\begin{table}[h]
\centering
\begin{tabular}{|r|r|r|}
\hline
  & ON & OFF \\
\hline 
\hline
Dec. [deg.] & 22.014 & 22.014 \\
R.A. [hour] & 5.5756 & 3.5756 \\
Period & October 2007 & October 2007  \\
& and December 2007 & and December 2007 \\
Eff. Time [hours] & 17.1 &  5.2\\
Zenith Range [deg] & 6 to 20 & 6 to 20 \\
\hline
\end{tabular}
\caption{The observation conditions for the Crab nebula data.} 
\label{TableNebulaOnOff}
\end{table}

\subsection{Energy Range of the Crab Nebula Analysis}
For the energy range below 50 GeV, a precise background estimation is not easy in the case of the 
nebula observations. The reason is as follows:
For the nebula analysis, the light curve cannot be used for the signal extraction and, hence,
the $ALPHA$ distribution is used instead (see Sect. \ref{SectSignalExtract}).
However, since the images with a small $SIZE$ are easily affected by the sky conditions,
$ALPHA$ distribution may slightly vary depending on the conditions, 
which may lead to a difference between ON-data and OFF-data.
Since the observed events are strongly dominated by cosmic-ray background events, 
even a tiny mismatch in $ALPHA$ distributions may result in a large systematic 
error in the background estimation.  
 It should be stressed that this is not the case for the pulsar analysis. Since the light curve 
can be
 used for the pulsed signal extraction, it does not require any OFF-observation. 
Unless the sky condition changes significantly 
in the time scale of a pulsar period (34 ms for the Crab pulsar), 
which is very hard to imagine, 
the background estimation can be properly carried out.
 For this reason, the nebula emission was analyzed only above 50 GeV. 

 The MC samples were generated from 6 GeV to 2 TeV. They are meant to be used for the analysis of the Crab pulsar,  which is known not to have significant emission above 100 GeV. With 
these MC samples, it is very hard to analyze the spectrum above 1 TeV properly because no 
information above 2 TeV is present in MC and energy resolution is limited ($\sim 20\%$ at 1 TeV).
In order to avoid any analytical bias due to the limited energy range in MC production, the spectrum was analyzed up to 700 GeV.

\begin{figure}[h]
\centering
\includegraphics[width=0.8\textwidth]{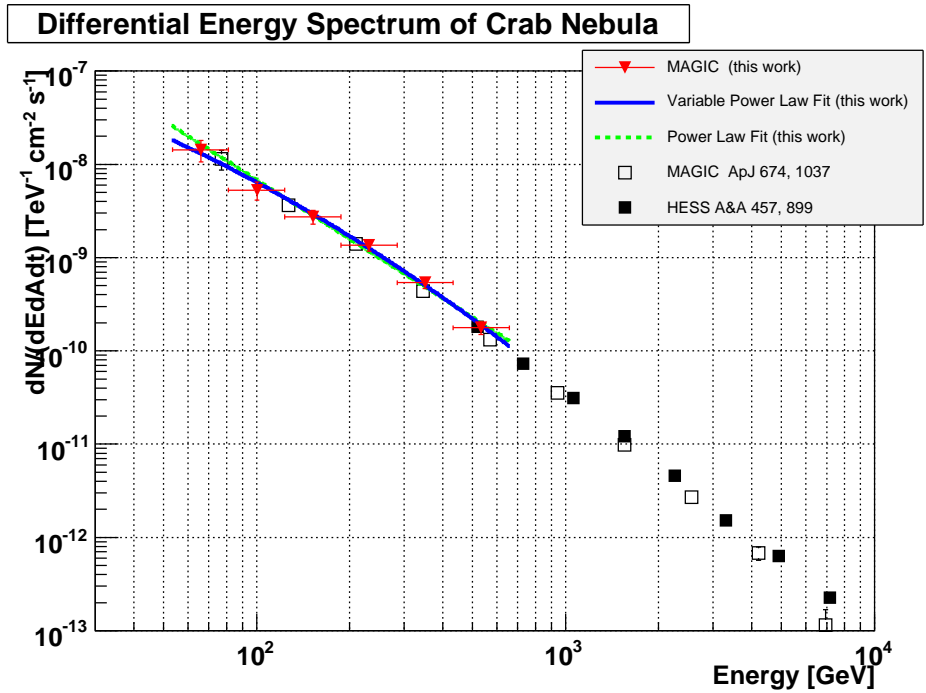}
\caption{The energy spectrum of the Crab nebula. Red triangles, 
open squares and filled squares indicate 
the MAGIC observation results with the SUM trigger (this work), the published MAGIC result 
with the standard trigger taken in 2006 (taken from \cite{MAGICCrab1}), 
and the published HESS results (taken from \cite{HESSCrab}). They are consistent
in the overlapped energy region. The power law fit (a green line) and the variable power law fit
(a blue line) to the
MAGIC results with the SUM trigger are also shown.
}
\label{FigMAGICNebula}
\end{figure}

 The results are shown in Fig. \ref{FigMAGICNebula}. The spectrum was unfolded using the Tikhonov 
regularization method. It is known that the spectrum in this energy range is better fitted by a variable power law 
\begin{eqnarray}
\frac{{\rm d}^3F}{{\rm d}E{\rm d}A{\rm d}t} = f_0(E/300 {\rm GeV})^{[\Gamma_1+\Gamma_2{\rm log}_{10}(E/300{\rm GeV})]}
\end{eqnarray}
than a simple  power law.
\begin{eqnarray}
\frac{{\rm d}^3F}{{\rm d}E{\rm d}A{\rm d}t} = f_0(E/300 {\rm GeV})^\Gamma
\end{eqnarray}

 The power law fitting gives the best parameters of $f_0 = (6.6 \pm 0.7) \times 10^{-10}$ [cm$^{-2}$s$^{-1}$TeV$^{-1}$] and
$\Gamma = -2.11 \pm 0.10$, while $\chi^2 = 4.74$ with the degree of freedom 4 ($\chi^2$ probability = 31.5 \%). On the other hand, the variable power law gives $f_0 = (7.0 \pm 0.8) \times 10^{-10}$[cm$^{-2}$s$^{-1}$TeV$^{-1}$],
 $\Gamma_1 = -2.21 \pm 0.15$ and $\Gamma_2 = -0.45 \pm 0.47$, while $\chi^2 = 3.55$ 
with the degree of freedom 3 ($\chi^2$ probability = 31.4\%). In this limited energy range,
 both functions fit well. It should be noted that the fits take into account the correlation 
between the spectral points which 
is introduced by the unfolding procedure.
 The spectrum measured by HESS telescopes above 500 GeV (see \cite{HESSCrab})
and the previous MAGIC measurements above 60 GeV (see \cite{MAGICCrab1})
are also shown in the same figure. They are consistent with one another in the overlapping
 range, 
verifying the validity of the MC samples and analysis tools used for the pulsar analysis. 

\clearpage

\section{Analysis of Optical Pulsation}
\label{SectOptPulsation}

For the pulsar analysis, the proper barycentric correction and the accurate pulsar period
 information
are essential. In order to check if the barycentric correction and the pulse phase calculation
are correctly done, 
the optical pulsation of the Crab pulsar is very useful
and helpful since it can be clearly detected within 10 minutes of observation thanks to
the large reflector of MAGIC. 

\begin{figure}[ht]
\centering
\includegraphics[width=0.95\textwidth]{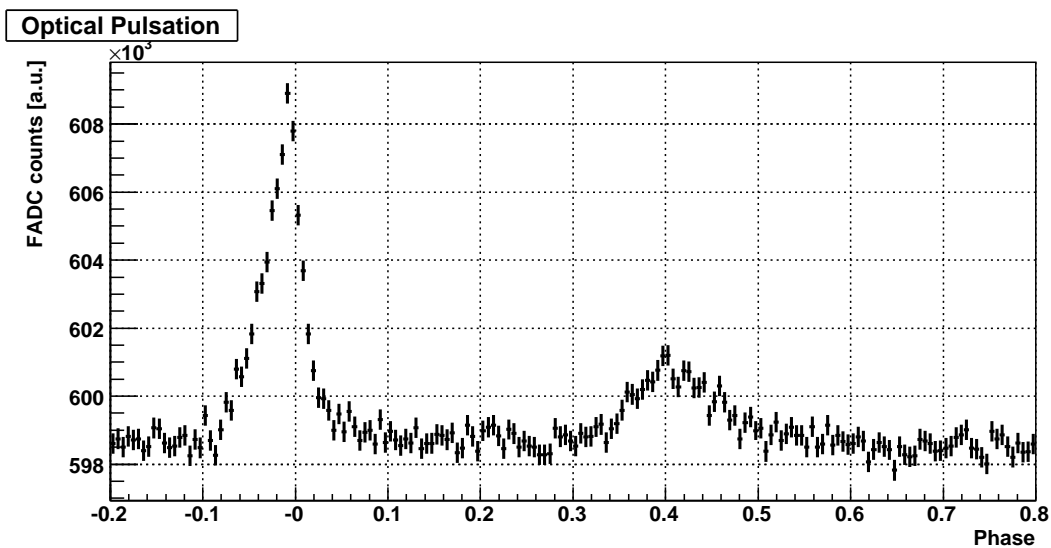}
\caption{The light curve of the optical pulsation from the Crab pulsar observed by MAGIC
with the central pixel (see Sect. \ref{SectCPIX}).
$\sim 30$ hours of the observation data are used. 10 minutes is enough to see the two peaks
 clearly.}
\label{FigMAGICOptical}
\end{figure}

Fig. \ref{FigMAGICOptical} shows optical pulsation detected by the central pixel of MAGIC 
(see Sect. \ref{SectCPIX}). 
$\sim 30$ hours of the observation data from both cycles are used.
The barycentric correction was done by $Tempo$ and 
the period information i.e. $\nu$, $\dot \nu$ and $t_0$, are
taken from the ``Jodrell Bank Crab Pulsar Monthly Ephemeris'',
 as described in Sect. \ref{SectPhaseCal}. 
The phase is calculated by Eq. \ref{EqPhaseCalc}.
A glitch (see \ref{SectGlitch}) occurred in May 2008, which is between 
the Cycle III and the Cycle IV observations. Since there is no Crab pulsar observation in that
month and $\nu$, $\dot \nu$ and $t_0$ are updated monthly, the glitch does not affect
the phase calculations.

The peak phase of P1 is slightly shifted earlier with respect to the radio peak phase
by $\sim 0.01$ in phase, corresponding to $\sim 300$ $\mu$s,
which is known and consistent with other observations (see e.g. \cite{Oosterbroek2008}). 
Pulse shapes are also in good agreement with other observations, although a small time variability 
has been reported and quantitative comparison is not easy (see e.g. \cite{Karpov2007}).
Exactly the same method of phase calculation is applied to gamma-ray signals. 

\section{Detection of the Very High Energy Gamma-ray Pulsation from the Crab Pulsar}

After all the selection described in Sect. \ref{SectCuts}, 59.1 hours 
(Cycle III~+~IV) of good data remained. 
The quality of the data sets, MC samples, analysis tools and the pulsar phase calculations
were verified, 
as described in the previous sections. In this section, 
the analysis of the pulsed gamma-ray signal from the Crab pulsar
is described.

\subsection{The Pulsed Gamma-ray Signal}
\begin{figure}[ht]
\centering
\includegraphics[width=0.95\textwidth]{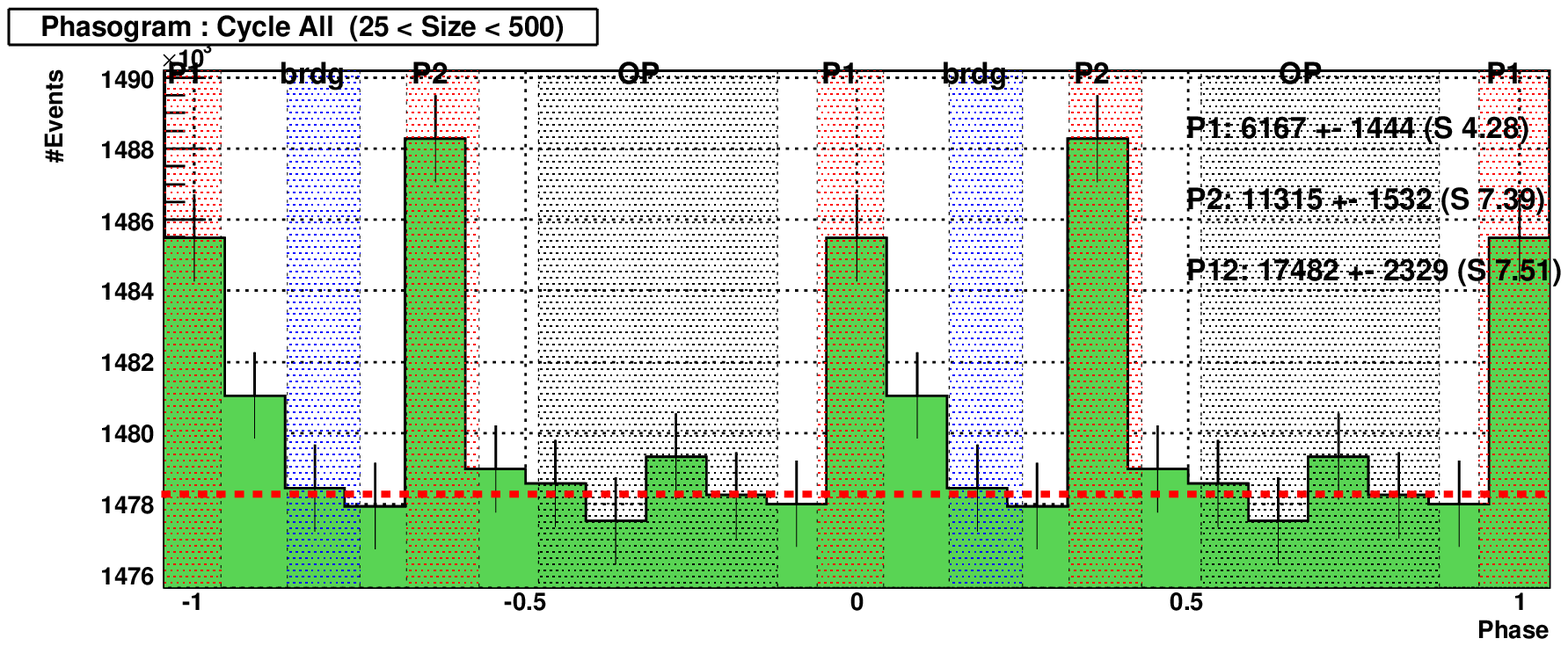}
\includegraphics[width=0.95\textwidth]{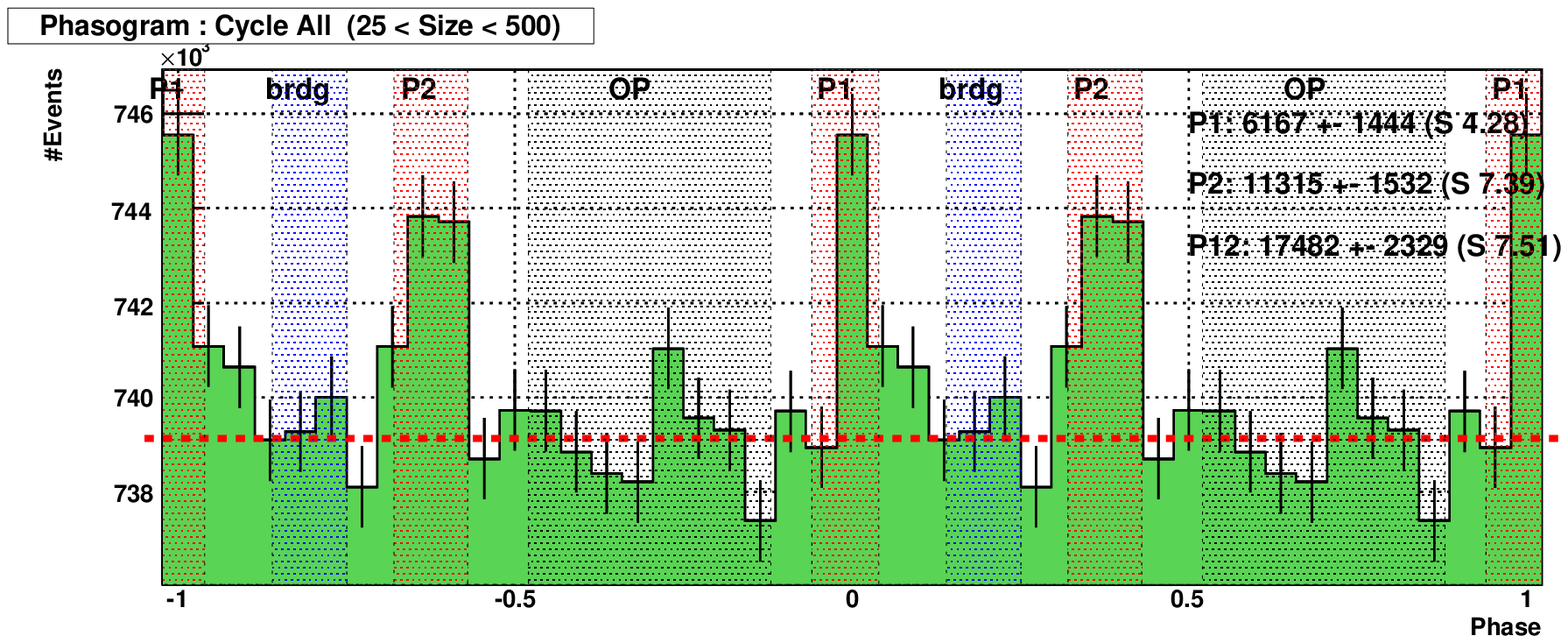}
\caption{The light curves of the Crab pulsar detected by MAGIC.
59.1 hours of the observation data are used. 
Following the convention in pulsar analyses, the light curves are plotted 
from phase -1 to 1.
The upper and lower panels show the light curves 
with 11 bins (3 milliseconds per bin) and  with 
22 bins (1.5 milliseconds per bin) per period. 
The red, shaded regions indicate the phases of P1 and P2, the blue shaded regions indicate
the Bridge emission phases and the black shaded regions indicates the OP (off-pulse) phases
(see Sect. \ref{SectPhaseNaming} for the definition of these phases).
The statistical significance of the excess for P1, P2, and P1~+~P2 are 
4.28 $\sigma$, 7.39 $\sigma$ and 7.51 $\sigma$, respectively.
}
\label{FigIntExAllnoHad}
\end{figure}

\clearpage

The light curve of all events with $SIZE$ range from 25 to 500 are shown in Fig. \ref{FigIntExAllnoHad}. A $SIZE$-dependent $ALPHA$ cut described in \ref{SectDynAlphaCut} was also applied.
The background was estimated using the OP (off-pulse) region (phase 0.52 to 0.88).
 P1 (phase -0.06 to 0.04), P2 (phase 0.32 to 0.43) and the sum of P1 and 
P2 have $ 6267 \pm 1444$, $11315 \pm 1532$
and  $17482 \pm 2329$ excess events with statistical significance of 4.28$\sigma$
, 7.39$\sigma$ and $7.51 \sigma$, respectively. 
For the definition of the phase names, see Sect. \ref{SectPhaseNaming}.
The flux of P2 is twice as high as that of P1.
As can be seen in Fig. \ref{FigCrabPulses}, at 1 GeV, P1 has a higher flux than P2. 
The energy dependence of P2/P1 ratio will be discussed in Sect. \ref{SectP1P2Bridge}.
The so-called bridge emission, which is seen in some energy bands, is not visible in the MAGIC data. This will also be discussed in Sect. \ref{SectP1P2Bridge} 
As one can see the bottom panel of Fig. \ref{FigIntExAllnoHad}, although P1 is conventionally
 defined as 0.32 to 0.43, most of the excess is concentrated in
a narrower phase interval. The precise discussion of the pulse shape will 
take place in Sect. 
\ref{SectEdgeStudy}.

\subsection{Further Investigation of the Signal}

\begin{figure}[ht]
\centering
\includegraphics[width=0.45\textwidth]{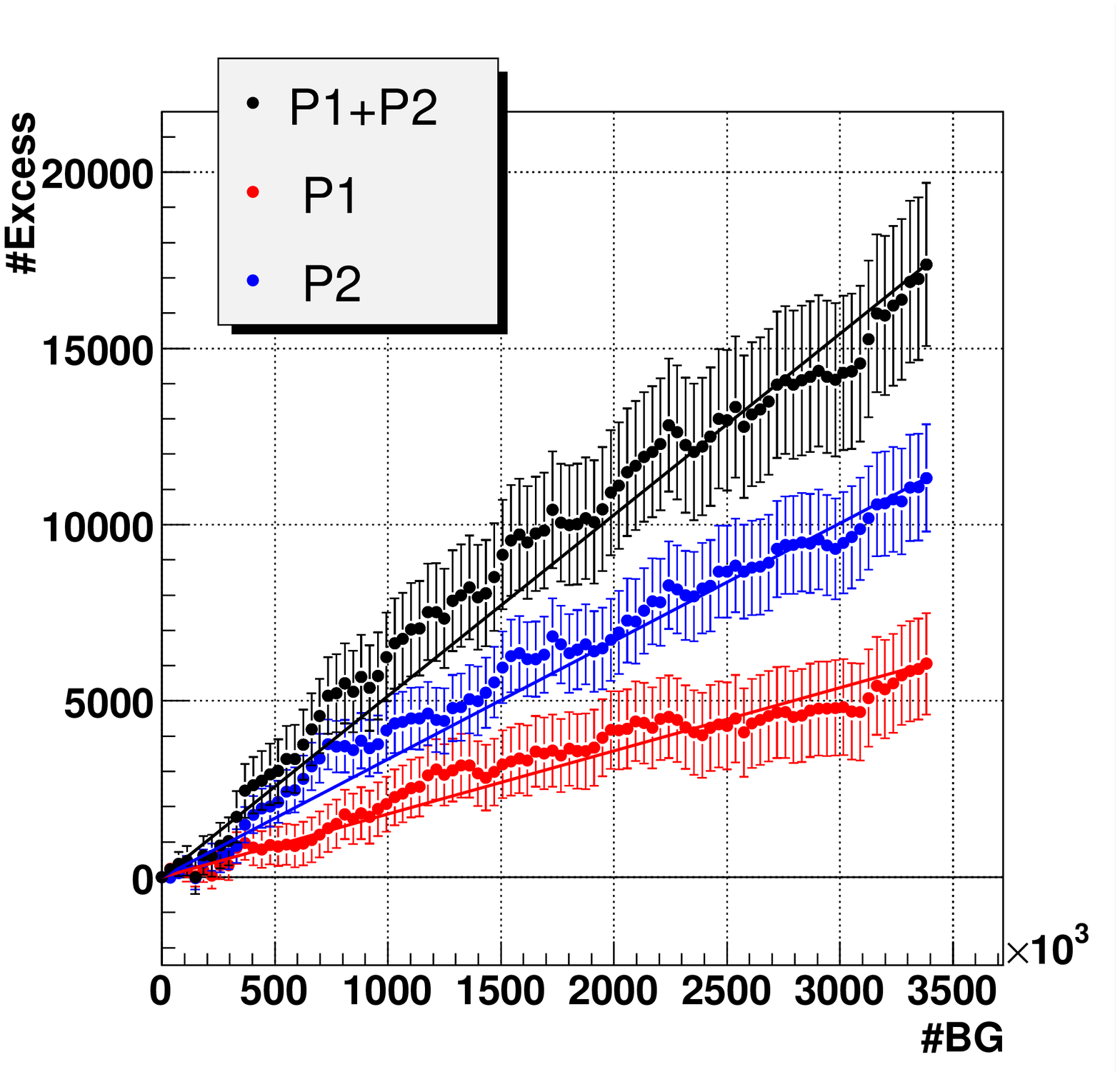}
\includegraphics[width=0.45\textwidth]{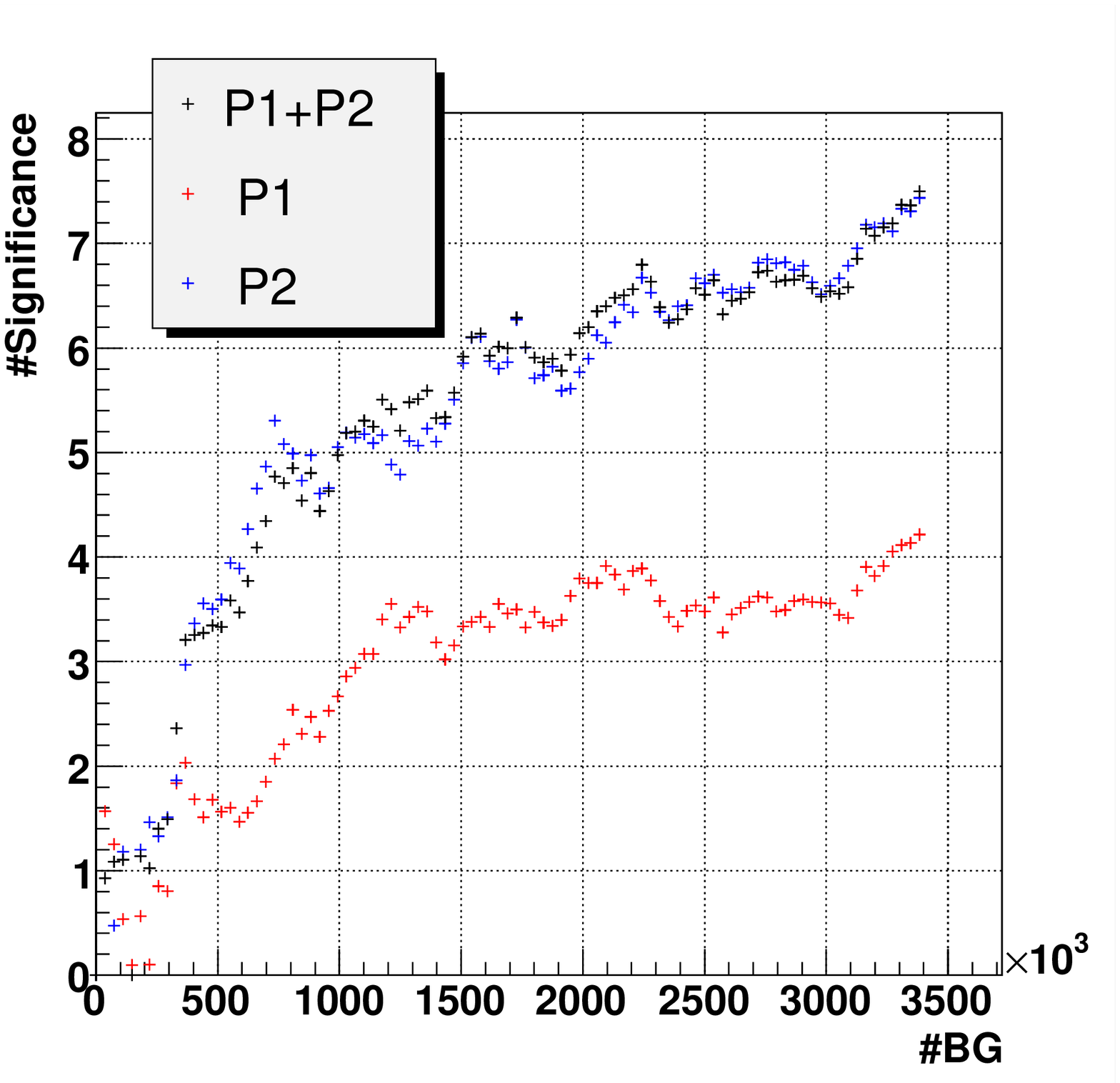}
\caption{Left: the growth of the number of excess events as a function of the number of 
background events. Red, blue and black points indicate P1, P2, and P1~+~P2, respectively.
Right: The growth of the statistical significance as a function of the number of
background events. Red, blue, and black points indicate P1, P2, and P1~+~P2, respectively.
}
\label{FigExGrowth}
\end{figure}

 Since this is the first detection of a gamma-ray pulsar by an IACT
\footnote{The discovery of the Crab pulsar with MAGIC was first achieved by the efforts of 
my colleagues listed in Sect. \ref{SectChoice} and reported in \cite{MAGICCrab2}},
it is important to 
assure that the signal is not an artifact of the analysis or of the instrument. 
A useful check is 
the growth of the number of excess events as a function of the number of background events. 
Since pulsar emission is thought to be stable in time, the excess should grow linearly.
 The results are shown in the left panel of Fig. \ref{FigExGrowth} and indeed the excess grows linearly. 
The growth of statistical significance is also shown in the right panel.

\begin{figure}[ht]
\centering
\includegraphics[width=0.85\textwidth]{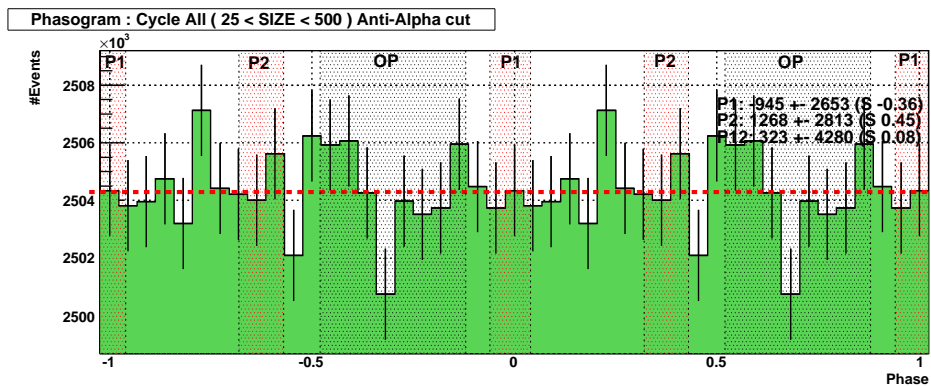}
\caption{The light curve of the Crab pulsar for data which do not pass the $ALPHA$ cut.
No pulsed signal is visible, as expected.}
\label{FigAntiAlpha}
\end{figure}

 Another useful check is the ``inverse selection'' of events. If the excess is due to gamma-ray
signals, events discarded by the $ALPHA$ cuts should not contain a significant excess. 
Figure \ref{FigAntiAlpha} shows the light curve produced with events which do NOT pass the $ALPHA$ cut. Excesses are compatible with the background fluctuation.

\section{SIZE Dependence of the Pulsation}
In order to roughly estimate from which energies the excess events come from, 
the energy dependence of the light curves should be examined. 
Instead of the energy reconstructed with Random Forest (see Sect. \ref{SectEneRec}),
which suffers a trigger-bias effect at these low energies as discussed
in Sect. \ref{SectEneRec}, 
I use $SIZE$ for this study.
$SIZE$ is the total number of photoelectrons in an image and a good indicator of the primary gamma-ray energy, especially for this data set, for which the $ZA$ range is limited up to 20 degrees.
As one can see in Fig. \ref{FigEvsS}, $SIZE$ in ph.e. corresponds roughly to two times the energy
 in GeV. 

\begin{figure}[h]
\centering
\includegraphics[width=0.63\textwidth]{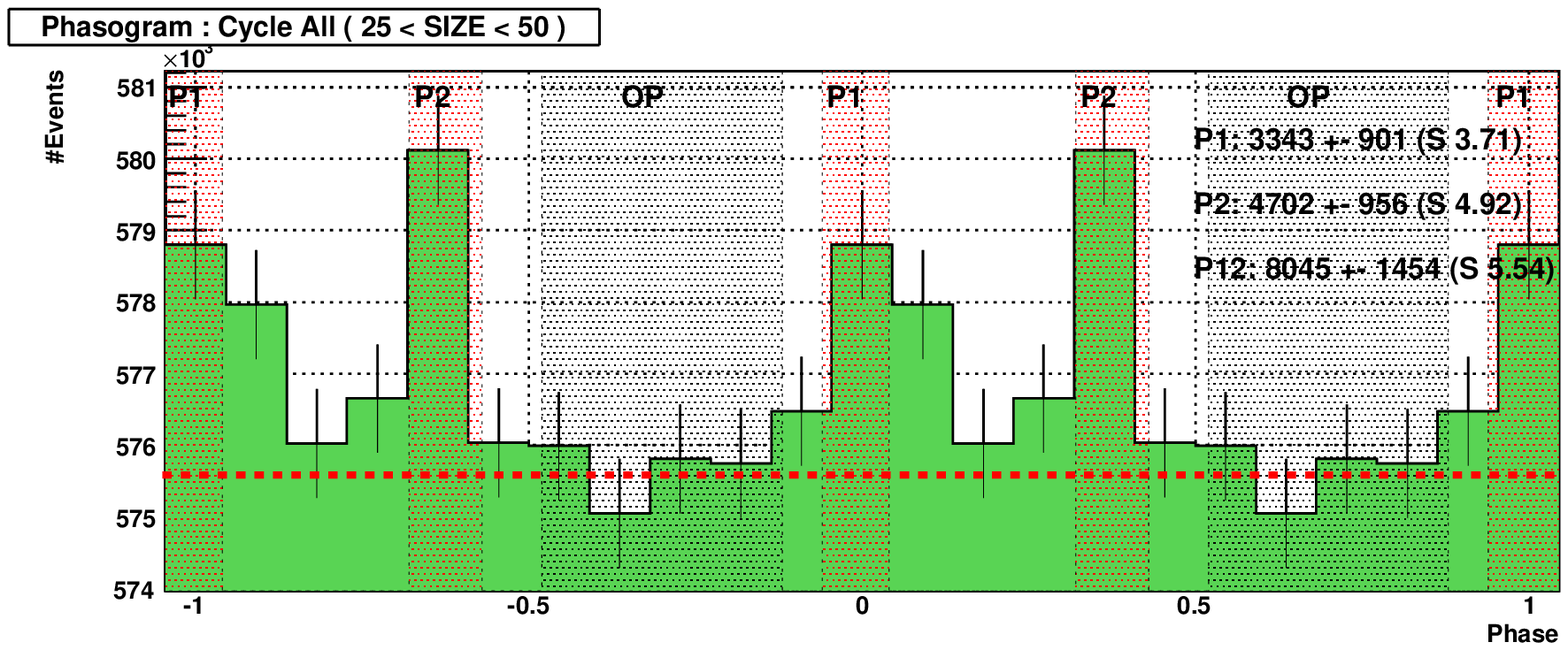}
\includegraphics[width=0.63\textwidth]{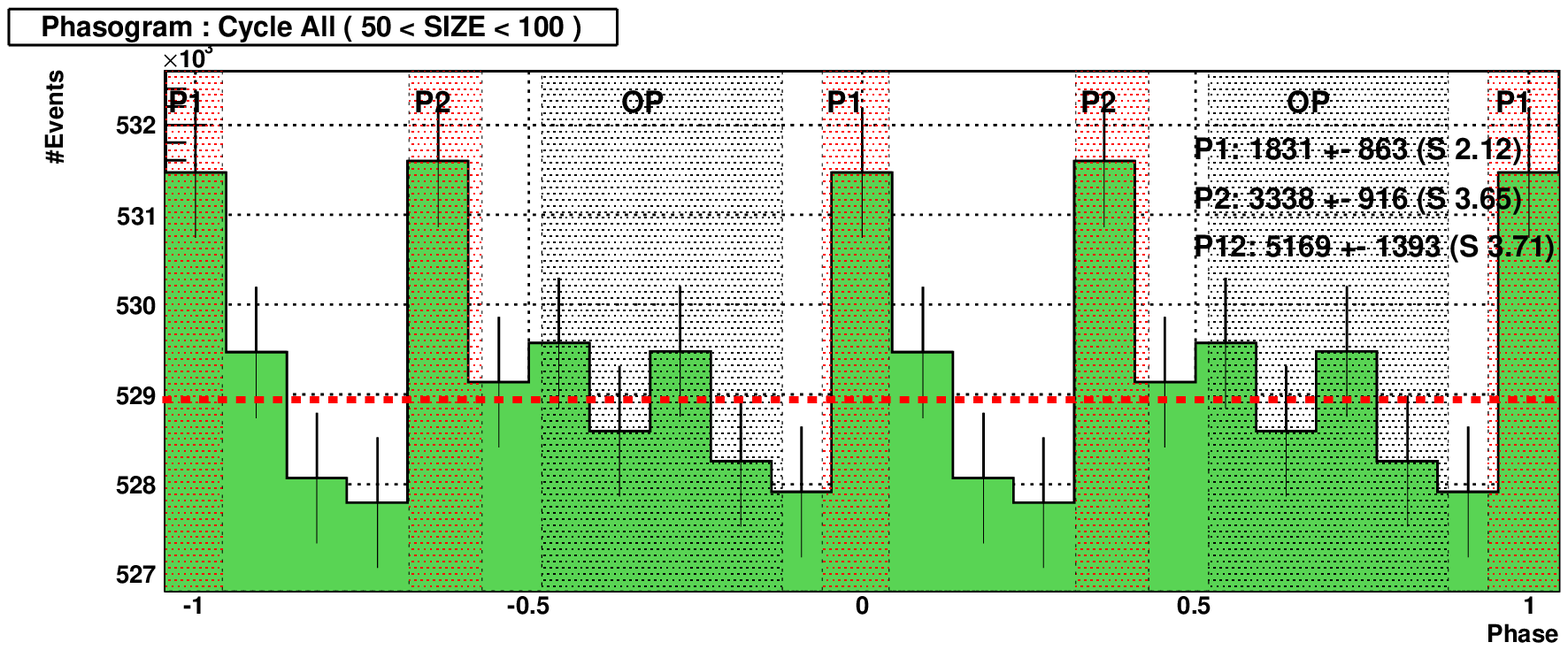}
\includegraphics[width=0.63\textwidth]{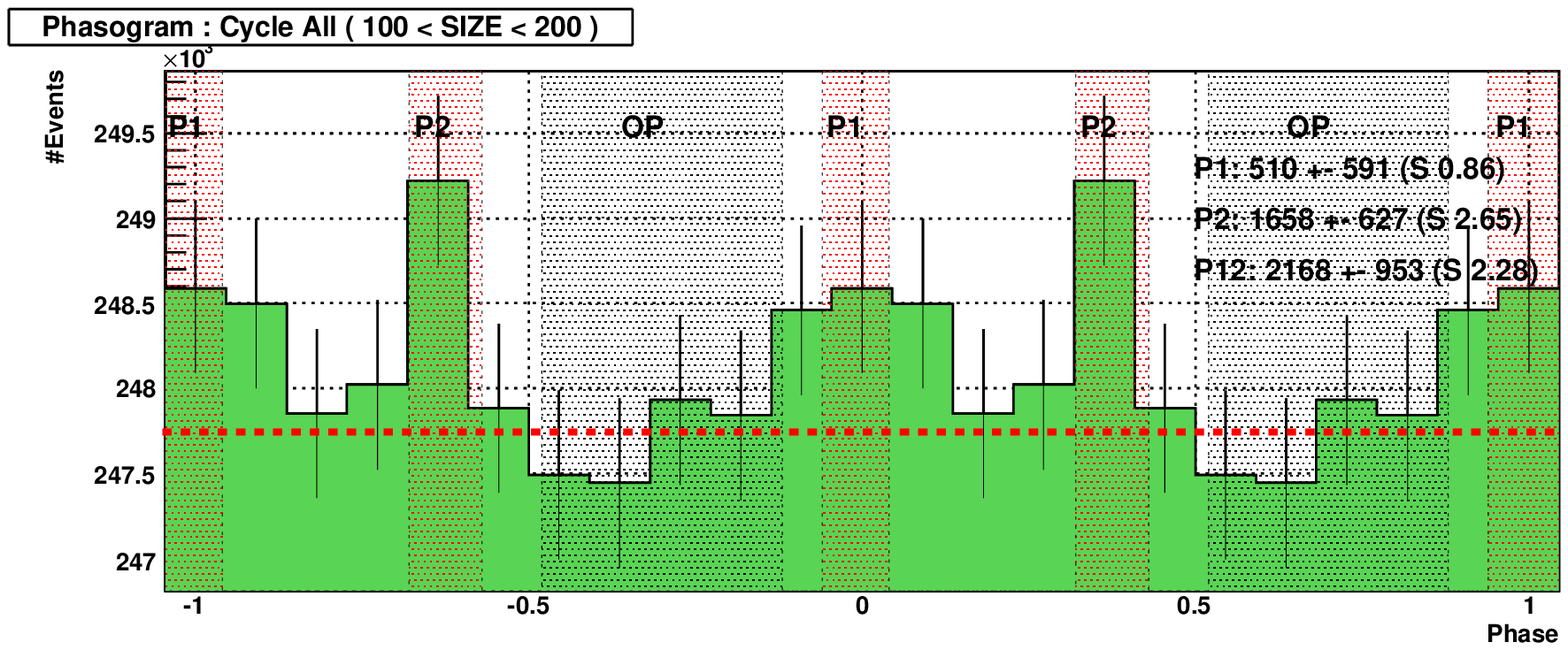}
\includegraphics[width=0.63\textwidth]{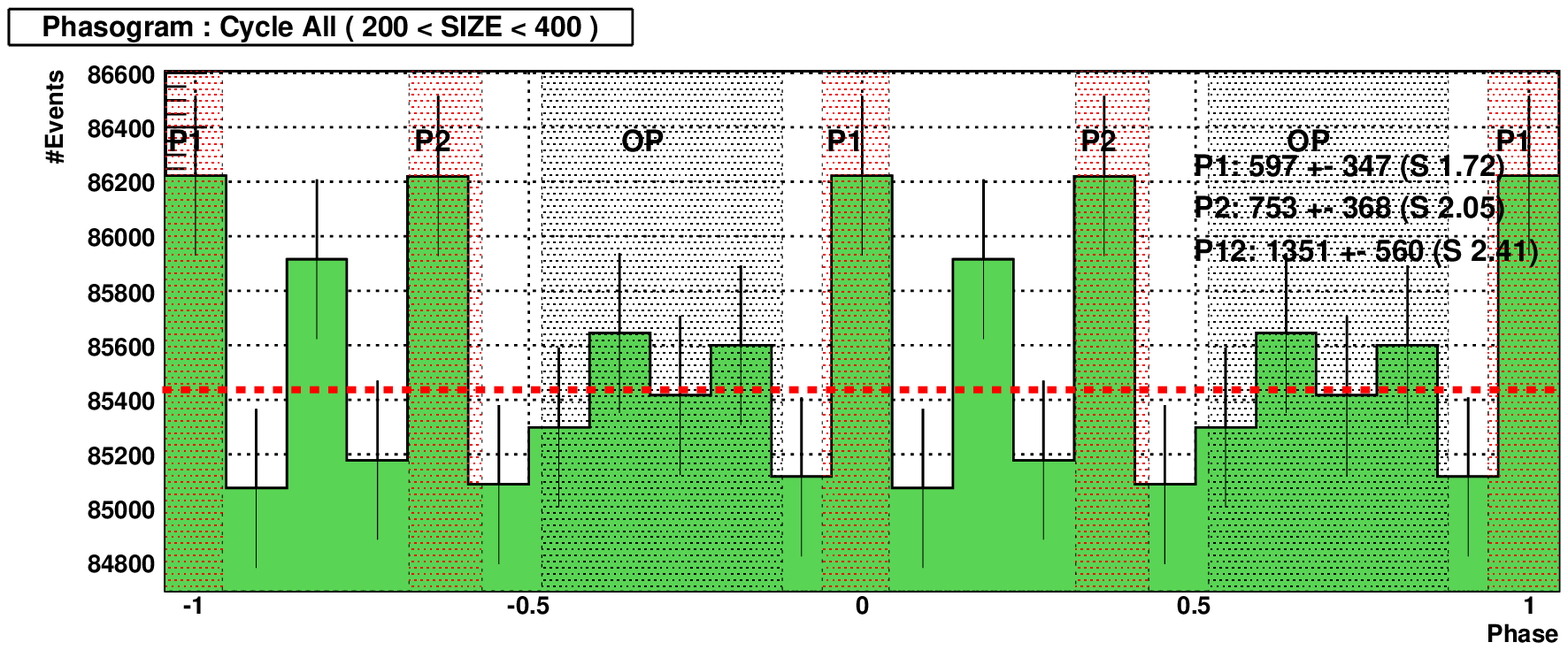}
\includegraphics[width=0.63\textwidth]{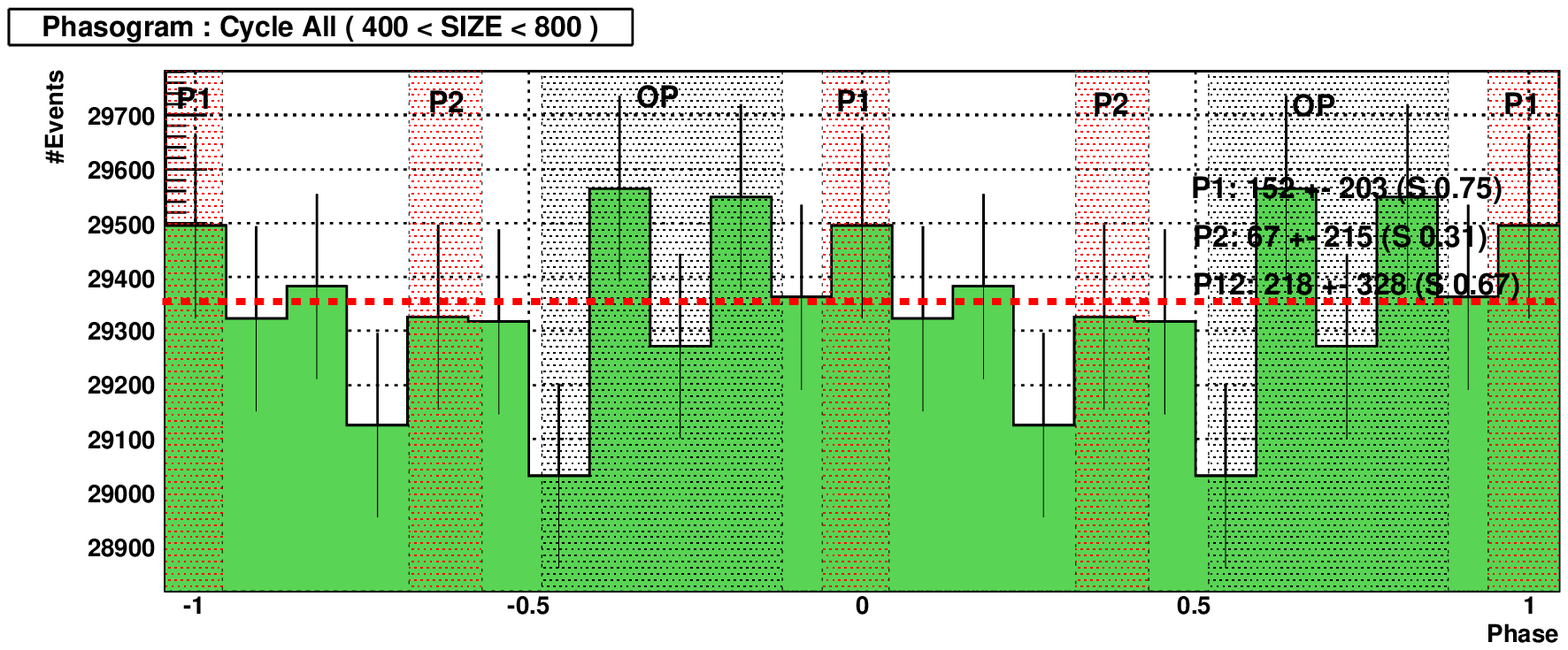}
\includegraphics[width=0.63\textwidth]{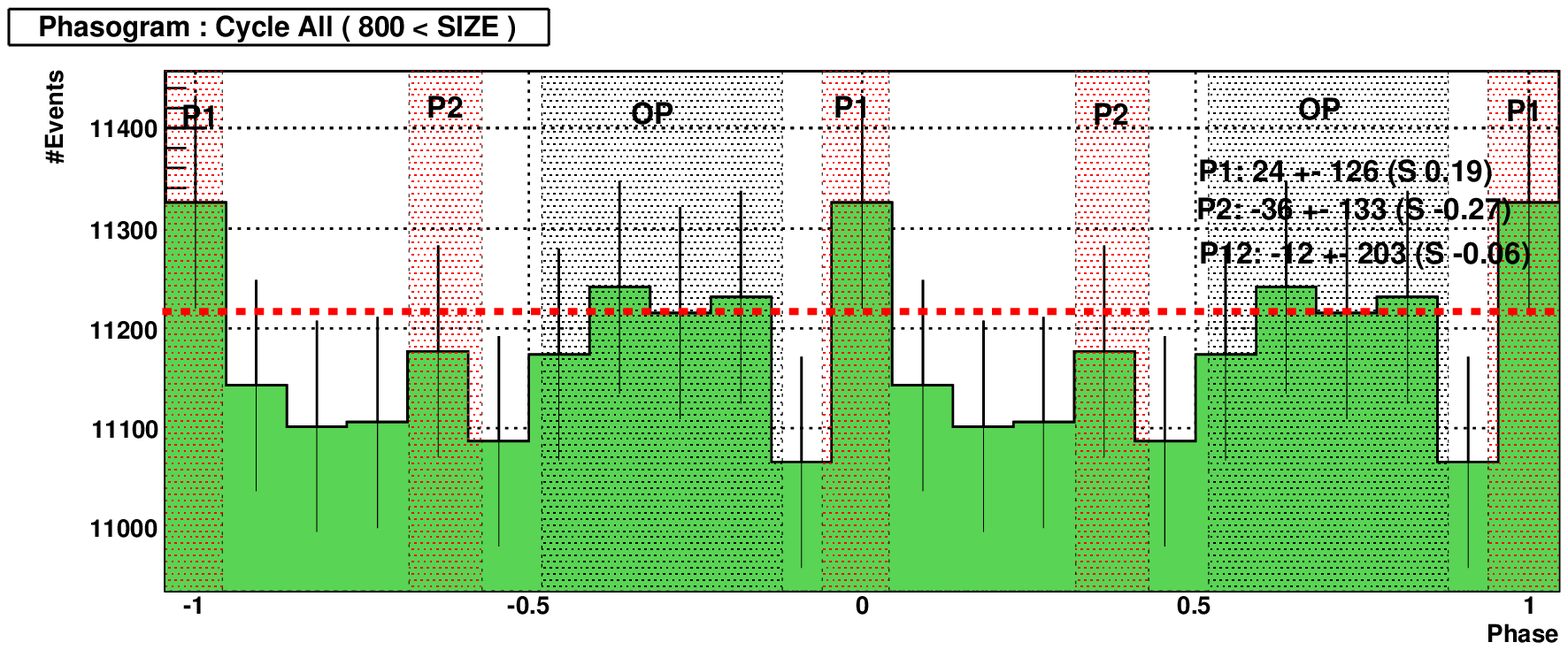}
\caption{The light curves for different $SIZE$ intervals. From the top: 25 to 50, 50 to 100,
100 to 200, 200 to 400, 400 to 800 and above 800 are shown.
Most of the excess events are seen at $SIZE < 100$, while some are still visible 
at $SIZE > 100$.
} 
\label{FigSizeDepPhasogram}
\end{figure}

\begin{figure}[h]
\centering
\includegraphics[width=0.6\textwidth]{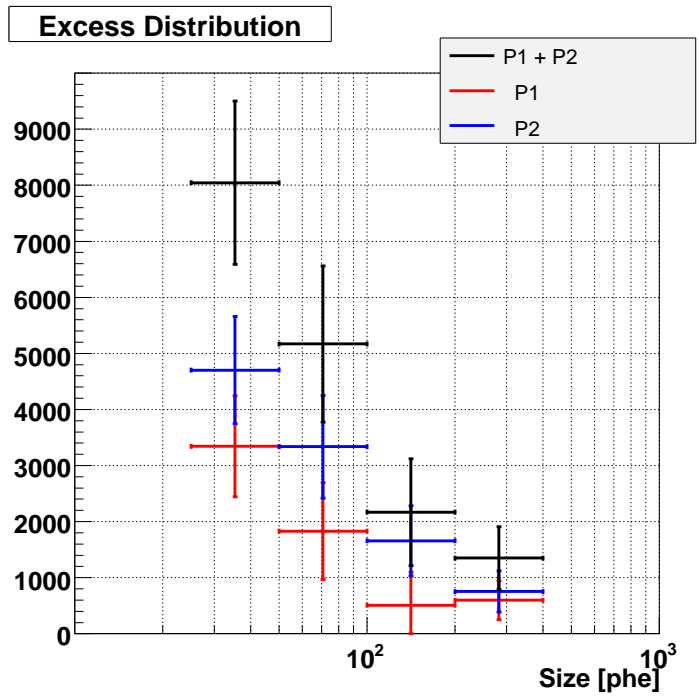}
\caption{The numbers of excess events as a function of $SIZE$. Red, blue and black lines
indicate P1, P2 and P1~+~P2, respectively.
Most of the excess events are at $SIZE < 100$, while some excess events are also visible at 
$SIZE > 100$.
} 
\label{FigExcessVsSize}
\end{figure}

The data with $SIZE$ from 25 to 800 were divided into 5 bins in log$_{10}(SIZE)$.
Data with $SIZE$ above 800 were also analyzed.
The light curves of these six sub-samples are shown in Fig. \ref{FigSizeDepPhasogram}.
The numbers of excess events for P1, P2 and P1~+~P2, shown in the right upper corner of each panel,
 were calculated by estimating the background level with the OP region (0.52 to 0.58). 
Most of the excess events are in the two lowest $SIZE$ bins. The third (100 -200) and 
fourth (200-400) bins also show a 2 $\sigma$ level
excess. Above 400 ph.e., no more excess is visible.
The size-dependence of the excess is shown graphically in Fig \ref{FigExcessVsSize}.

A detail calculation of the energy spectrum 
of the pulsed gamma-ray signal
will take place in Sect. \ref{SectMAGICspectrum}.

\clearpage

\section{Time Variation of the Pulsation from the Crab Pulsar}
\label{SectVariability}

The previous study by L. Kuiper et al. (see \cite{Kuiper2001}) shows the flux of the Crab pulsar
at 1-10 MeV 
 is
 stable in a time scale of years.
However, it is still interesting to check if the flux measured by MAGIC is stable or not,
 especially because 
it is the flux beyond the cut-off energy. The growth of the number of excess events as a function
of the number of background events already showed that there is no drastic change in flux
(see Fig. \ref{FigExGrowth}). Here I examine the stability of the flux and the light curve
more quantitatively.
Due to the limited statistical significance of the excess, dividing data sample into too many 
subsets does not make much sense.
Therefore, I only compared Cycle III and Cycle IV to search for a possible yearly variability.
\subsection{Variability in Light Curve}

\begin{figure}[h]
\centering
\includegraphics[width=0.9\textwidth]{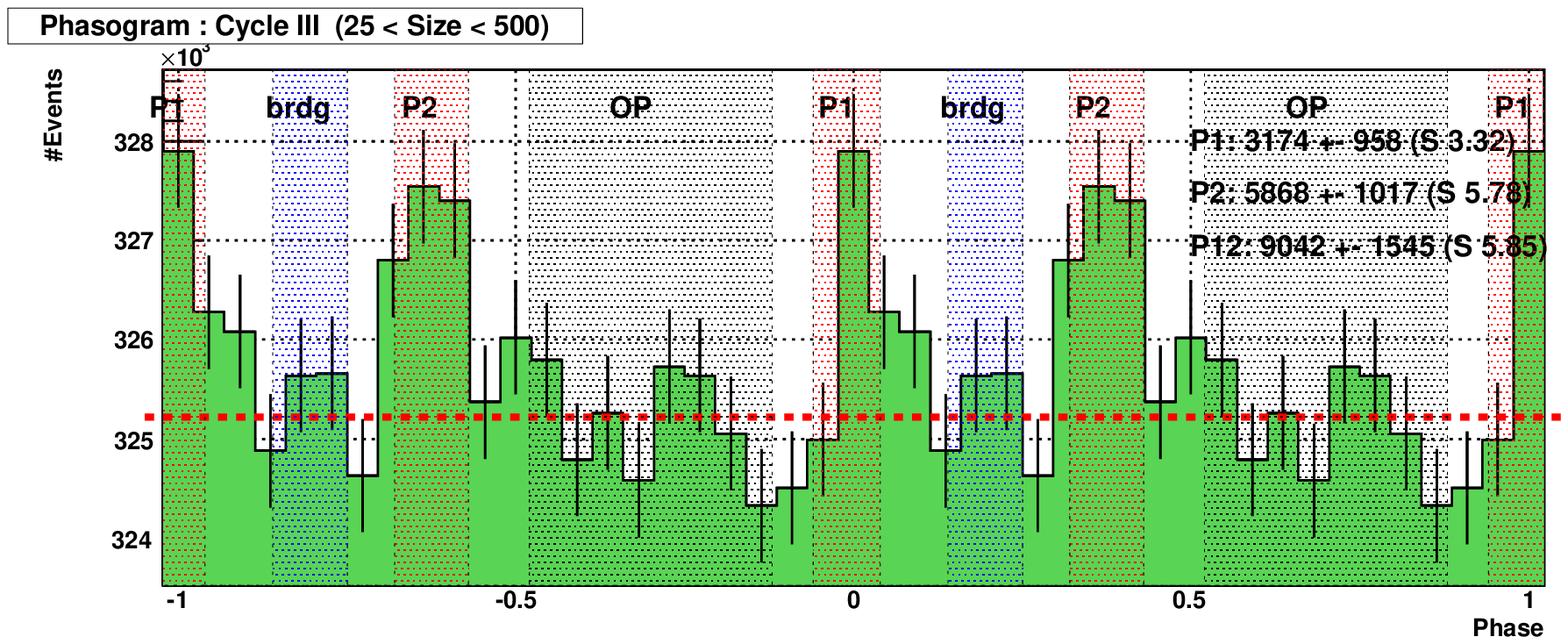}
\includegraphics[width=0.9\textwidth]{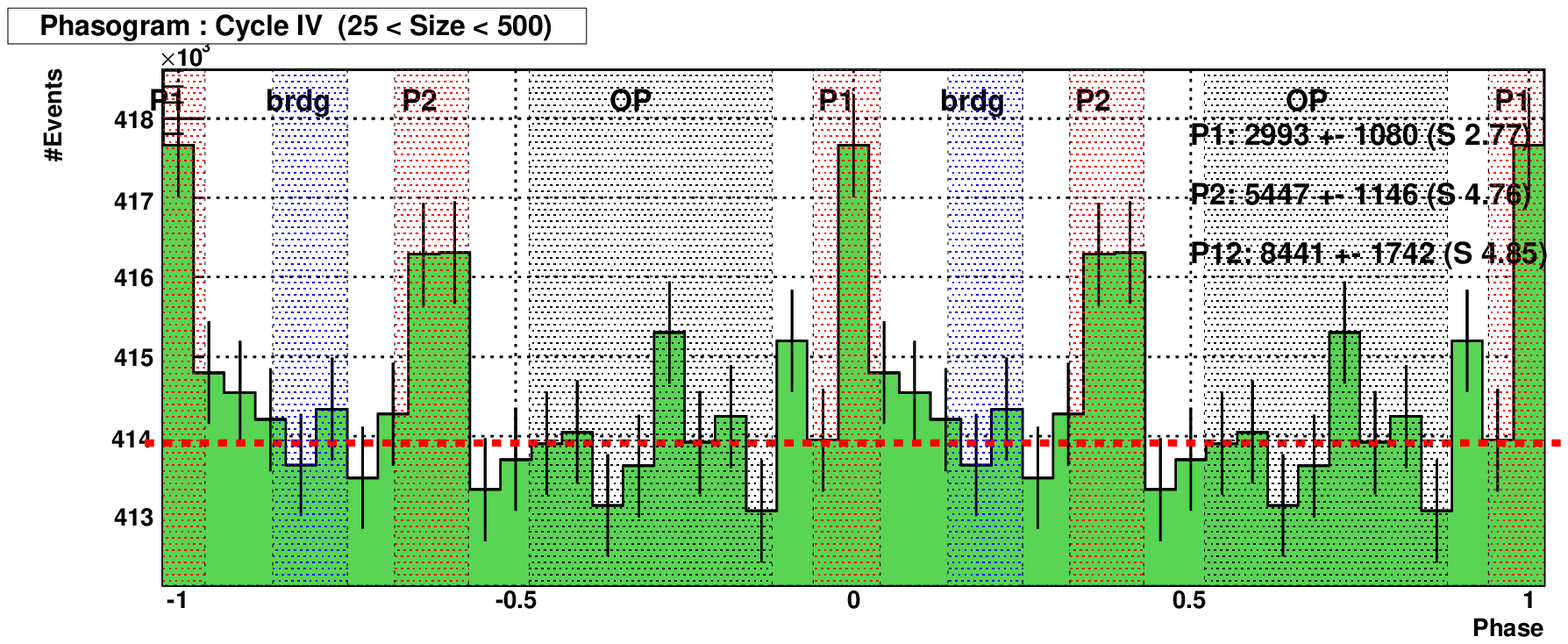}
\caption{The comparison of the light curves between Cycle III and IV. 
Significant variation of the light curve cannot be seen.}
\label{FigPhasogramIIIandIV}
\end{figure}

 Fig. \ref{FigPhasogramIIIandIV} shows the light curves of Cycle III (top) and Cycle IV (bottom). 
The $SIZE$ range is from 25 to 500.
In order to evaluate the variation in the light curve, the $\chi2$ test was performed for
the 11 bins starting from -0.0682 to 0.432 in the histograms,
which is roughly from the beginning of P1 to the end of P2. The $\chi2$ was 5.00 while the 
degree of freedom was 10, indicating no significant difference between the two light curves.

\subsection{Variability in Flux}

\begin{figure}[h]
\centering
\includegraphics[width=0.3\textwidth]{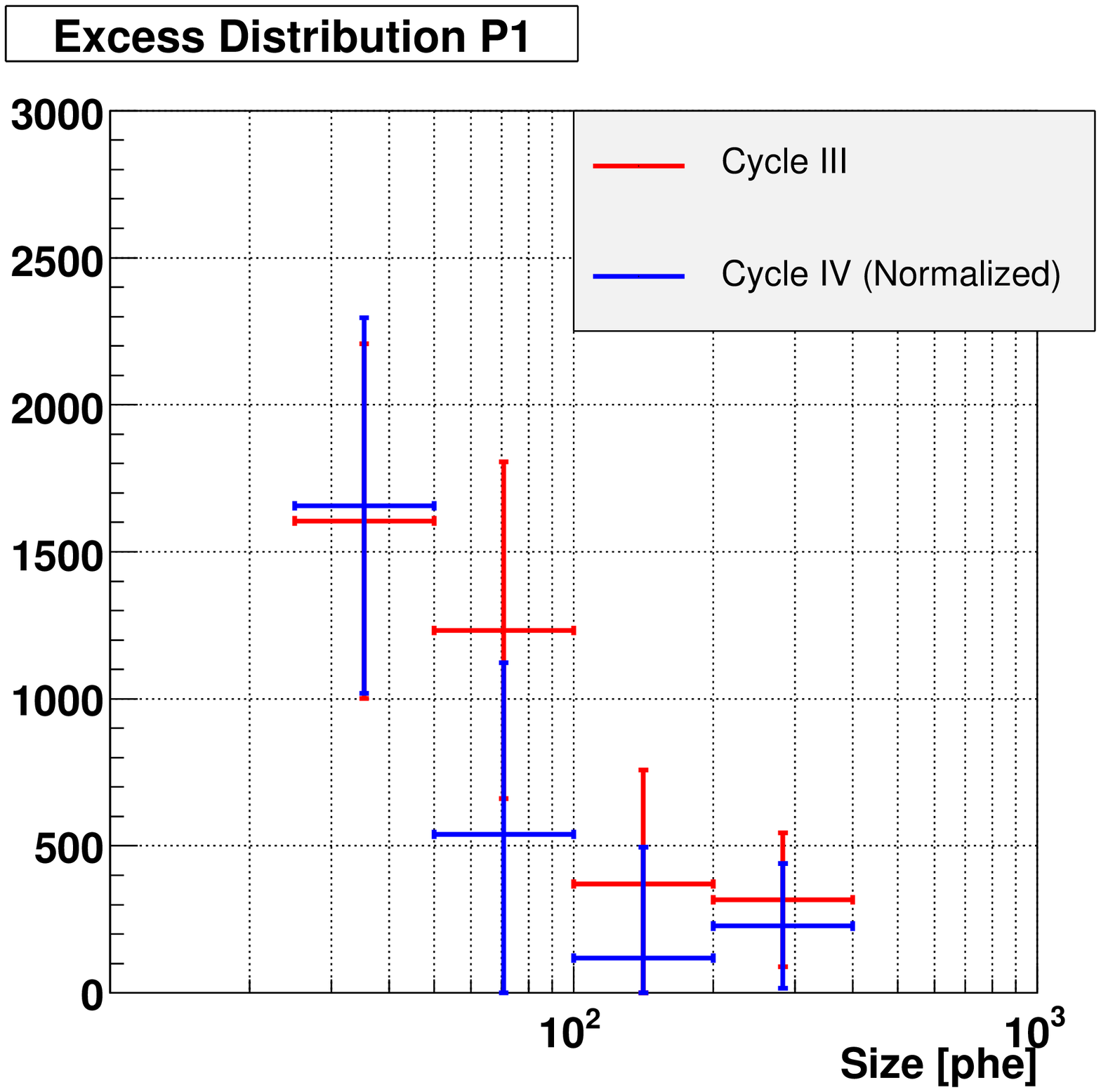}
\includegraphics[width=0.3\textwidth]{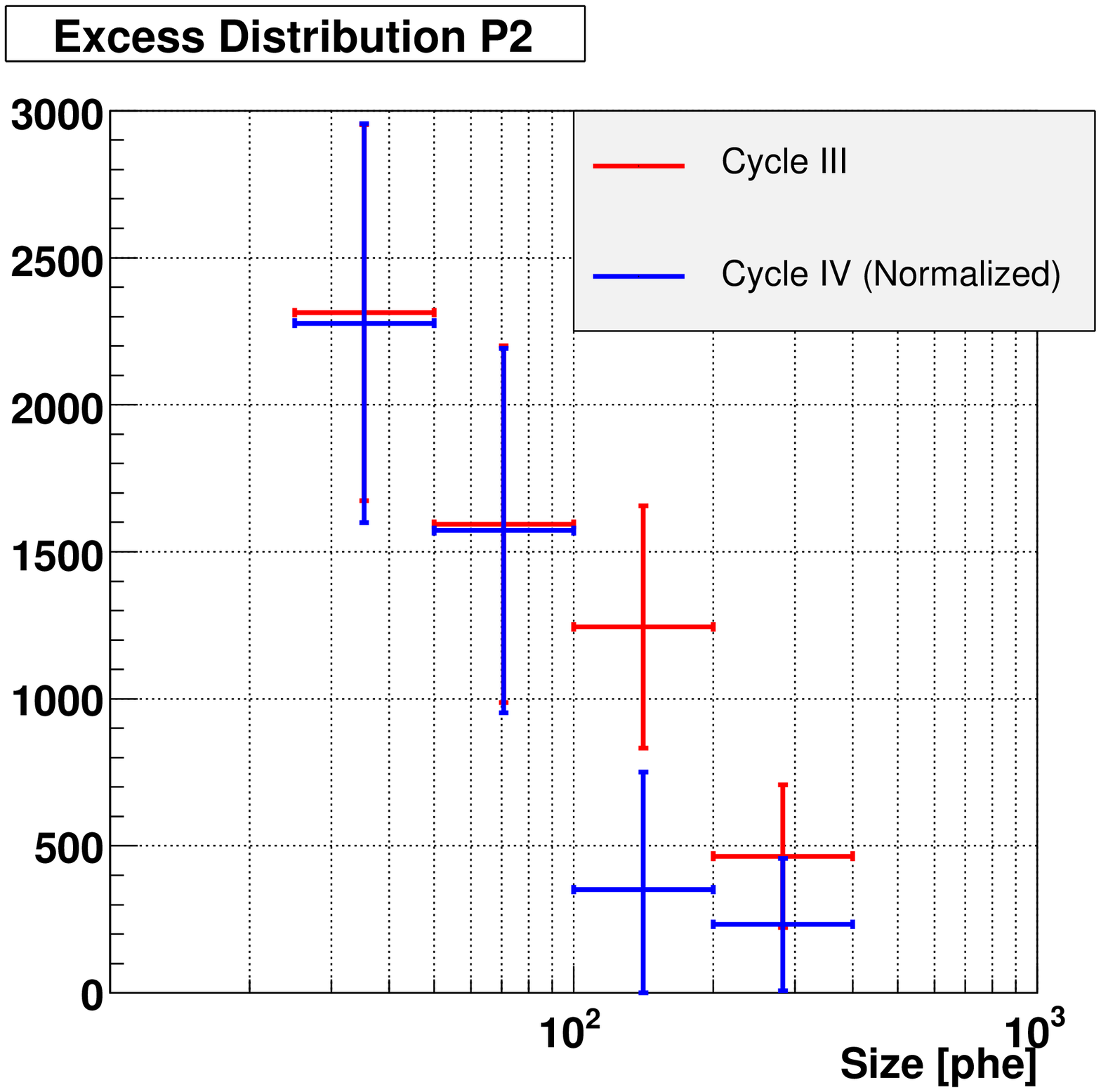}
\includegraphics[width=0.3\textwidth]{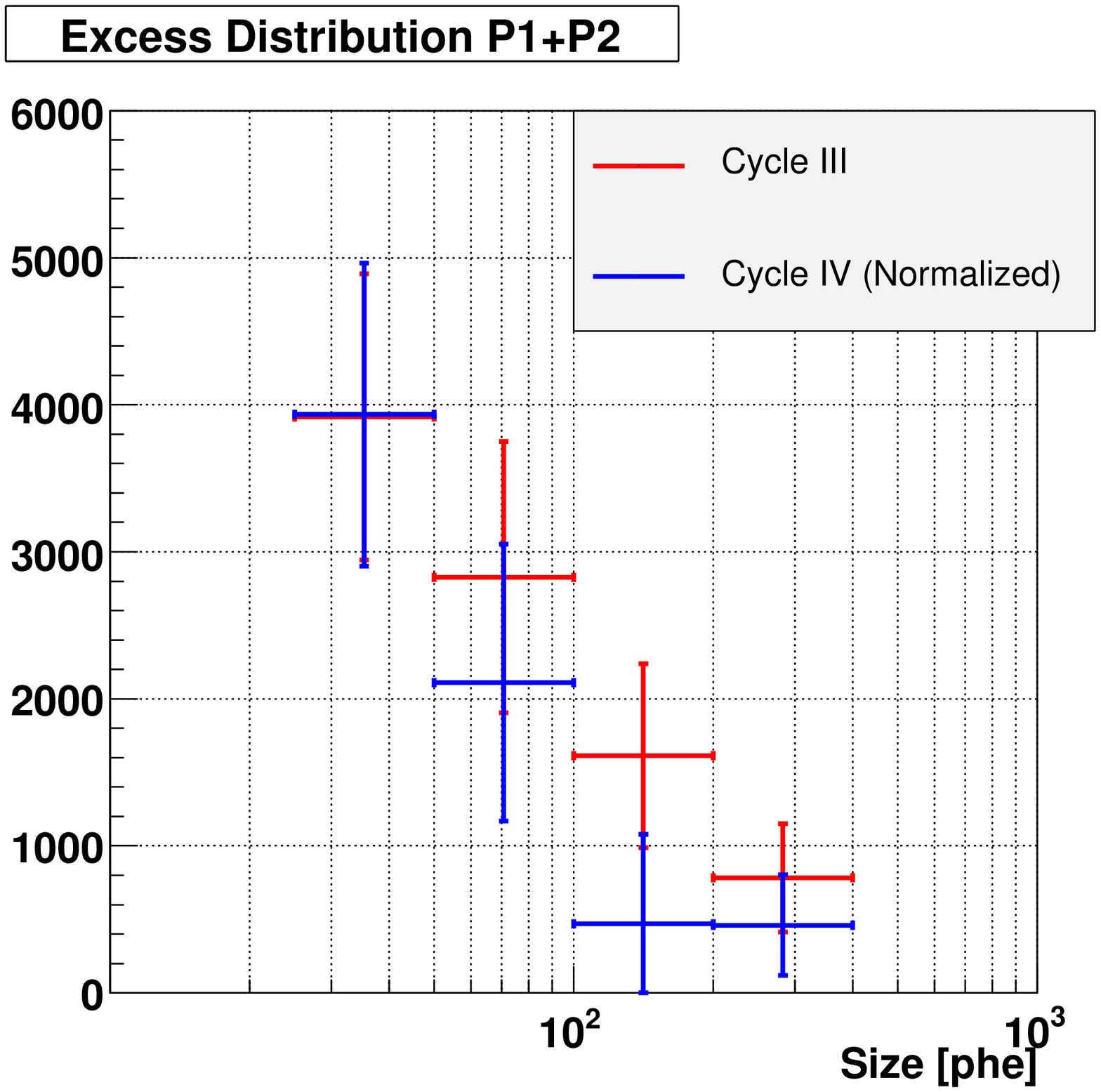}
\caption{The comparison of the number of excess events as a function of $SIZE$
between Cycle III and IV. The correction for the effect of the subpatch malfunction described 
in \ref{SectMalfunc} is applied for each $SIZE$ bin. In addition, the numbers of excess 
events for Cycle IV are multiplied by 25.1/34.0 in order to correct for the difference of 
the observation time.
}
\label{FigExComp}
\end{figure}

As one can see from Fig. \ref{FigPhasogramIIIandIV}, 
although Cycle IV has 30\% longer observation time, 
it shows less excess events than Cycle III. This can be explained only by statistical fluctuation
but the hardware malfunction described in Sect. \ref{SectMalfunc}
may also have played a role. 
$SIZE$-dependence of the number of excess events
 are also compared in Fig. \ref{FigExComp}. Left, middle
 and right panels are for P1, P2 and P1~+~P2. 
The difference in observation time between the two cycles is corrected. 
The effect of broken sub-patches
estimated by MC (see Sect. \ref{SectMalfunc}) is also corrected, such that Cycle III 
and IV have the same gamma-ray detection efficiency. 
$\chi2$s are 1.04, 3.14, 2.46 for P1, P2 and P1~+~P2, while
the number of {\it dof} is 4 for all phase intervals. 
Signals observed in Cycle III and IV are statistically consistent and no significant variability
 is seen between Cycle III and IV.

\section{Energy Spectra of the Pulsation from the Crab Pulsar}
\label{SectMAGICspectrum}
Here I show the energy spectra for the pulsation of P1, P2 and P1~+~P2.
Since the effective area is increasing rapidly from 20 GeV to 200 GeV (see Sect. \ref{SectStdSumComp}) and the energy resolution is rather poor in this energy region (see Sect. \ref{SectUnFold}),
the spectra must be calculated with great care. 
Events with $SIZE$ lower than 30 will be excluded from the spectrum calculation in order to
avoid a possible mismatch between MC and data (see Sect. \ref{SectMAGICSyst}). 
A few different methods will be tried in order to estimate
 the analytical uncertainty of the results.
\subsection{Excess Distribution in the Reconstructed Energy and Size}
\label{SectExcessDist}
\begin{figure}[h]
 \centering
 \includegraphics[width=0.49\textwidth]{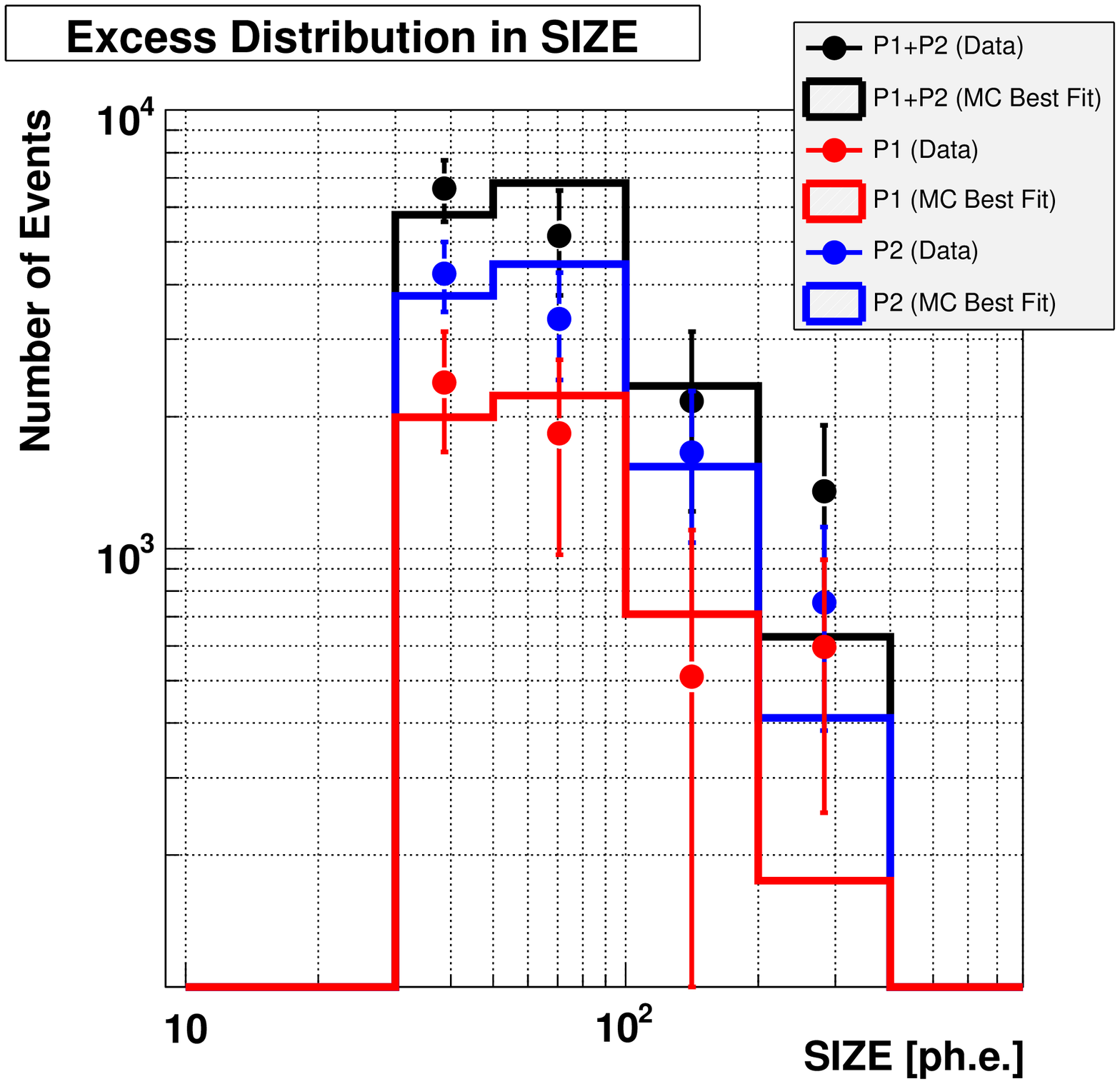}
 \includegraphics[width=0.49\textwidth]{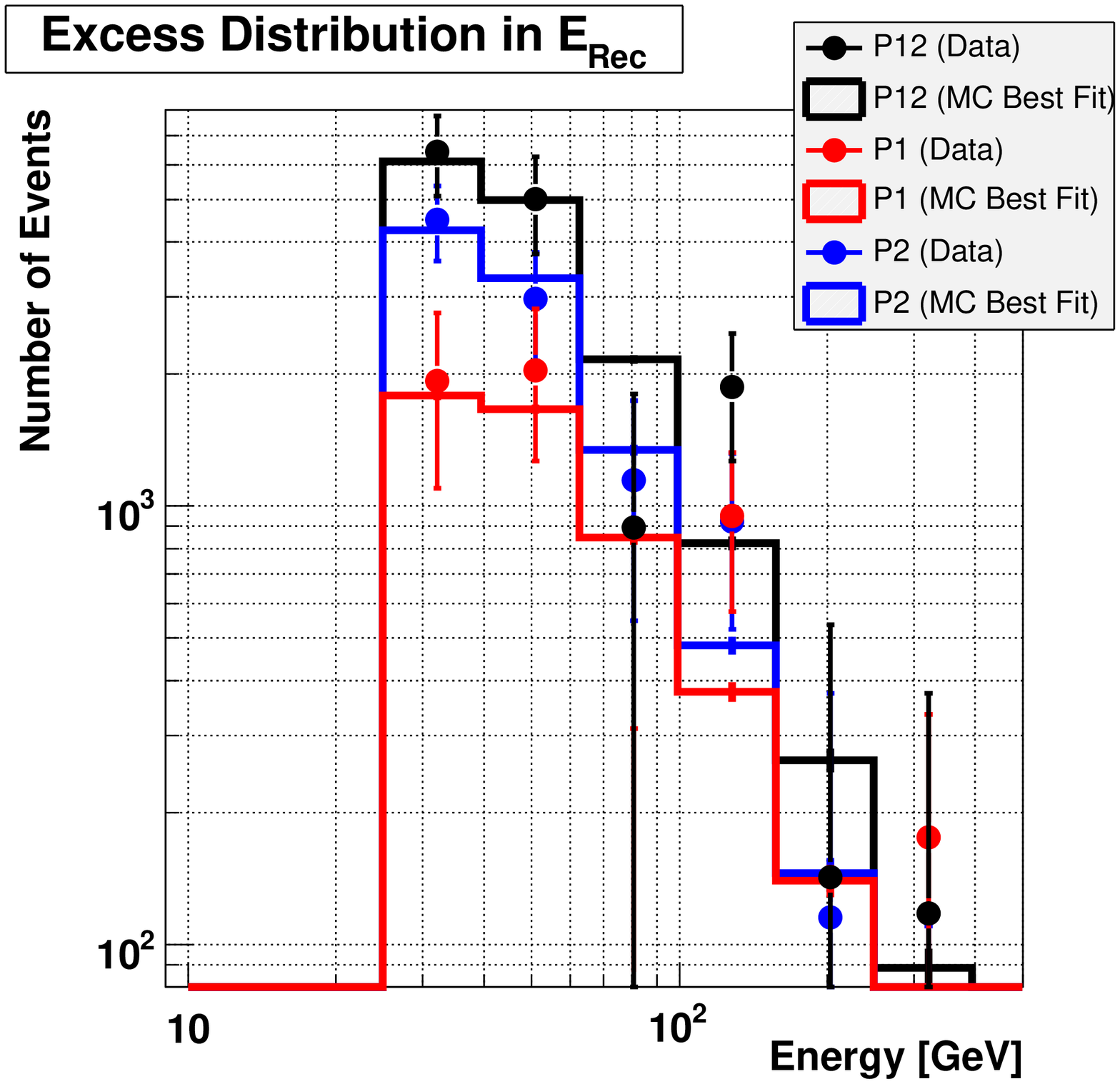}
 \caption{
The numbers of excess events 
as a function of $SIZE$ (the left panel) or the reconstructed energy (the right panel). 
Points indicate the observed data used for the spectrum calculations, while 
solid lines indicate the expected number of 
excess events from the spectra obtained by the forward unfolding.
In order to reduce the possible systematic error for the spectrum,
the events with $SIZE < 30$ are excluded. 
Red, blue and black lines are for P1, P2 and P1~+~P2, respectively.
}
 \label{FigExcessDist}
 \end{figure}

Fig. \ref{FigExcessDist} shows the distributions of the number of excess events
 as a function of $SIZE$ and the reconstructed
energy. It should be emphasized again that the reconstructed energy is strongly biased by the trigger
effect (see Sect. \ref{SectUnFold}). 
The energy spectrum will be 
calculated from these distributions. $SIZE$ (total charge in a shower image) has less systematic uncertainty than 
the reconstructed energy,
whereas the resolution of the energy estimation by $SIZE$ is poorer than that by 
the reconstructed energy.
Therefore, it is important to analyze the spectrum with both and
compare the results.

\subsection{ Spectrum Calculation by Forward Unfolding Assuming a Power Law }
As long as the assumption of the spectral shape is valid,
the forward unfolding method should provide the most robust and reliable result,
 as described in Sect. \ref{SectUnFold}. 
Here, I assumed a power law spectrum with two free parameters,
namely,
differential flux at 30 GeV $f_{30}$ and the spectral index $\Gamma$:
\begin{equation}
\frac{{\rm d}^3F}{{\rm d}E{\rm d}A{\rm d}t} = f_{30} (E/30{\rm GeV})^\Gamma
\label{EqPowerLaw}
\end{equation}
Due to the relatively poor statistics and the narrow energy range, 
even if the true spectrum is slightly curved, the power law assumption should be valid.

The results based on the $SIZE$ distribution (the left panel of Fig. \ref{FigExcessDist})
and the reconstructed energy distribution (the right panel of Fig. \ref{FigExcessDist})
are shown in Table \ref{TableCrabPulsarSpectrum}.
The obtained spectra are also graphically shown in Fig. \ref{FigUnfoldedSpectra}
by green (based on $SIZE$) and red (based on the reconstructed energy) lines.
Unfolded excess distributions, i.e.,
the expected excess distributions from the obtained spectra,
 are overlaid in Fig. \ref{FigExcessDist}.
The $\chi^2$ values between the unfolded excess distribution and the observed excess distribution
are shown in the fifth column of Table \ref{TableCrabPulsarSpectrum}. 
They are sufficiently small, assuring the validity of the power law assumption.

\subsection{ Spectrum Calculation by Unfolding with Different Regularization Methods}
\begin{figure}[h]
 \centering
 \includegraphics[width=0.55\textwidth]{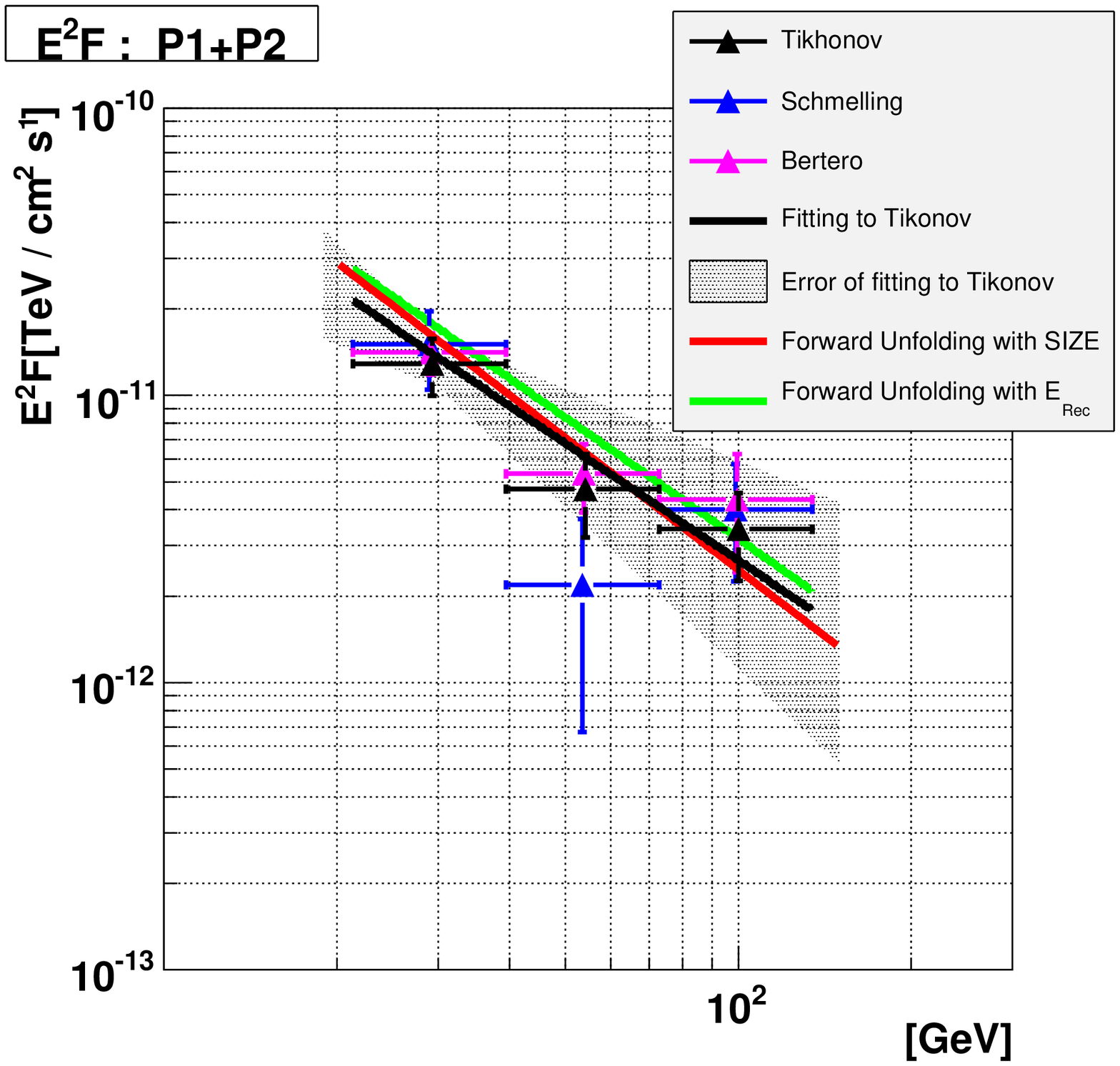}
 \includegraphics[width=0.49\textwidth]{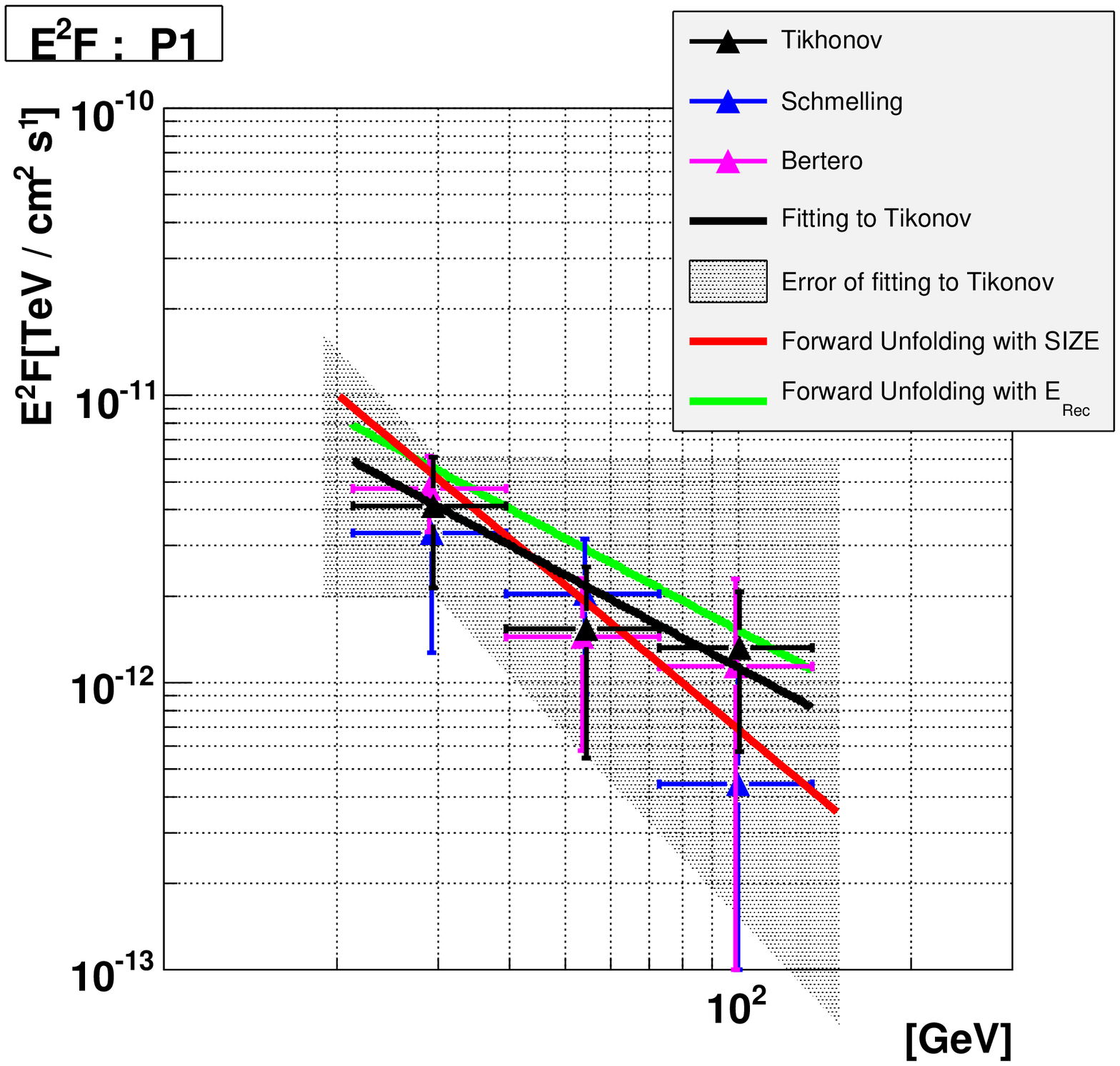}
 \includegraphics[width=0.49\textwidth]{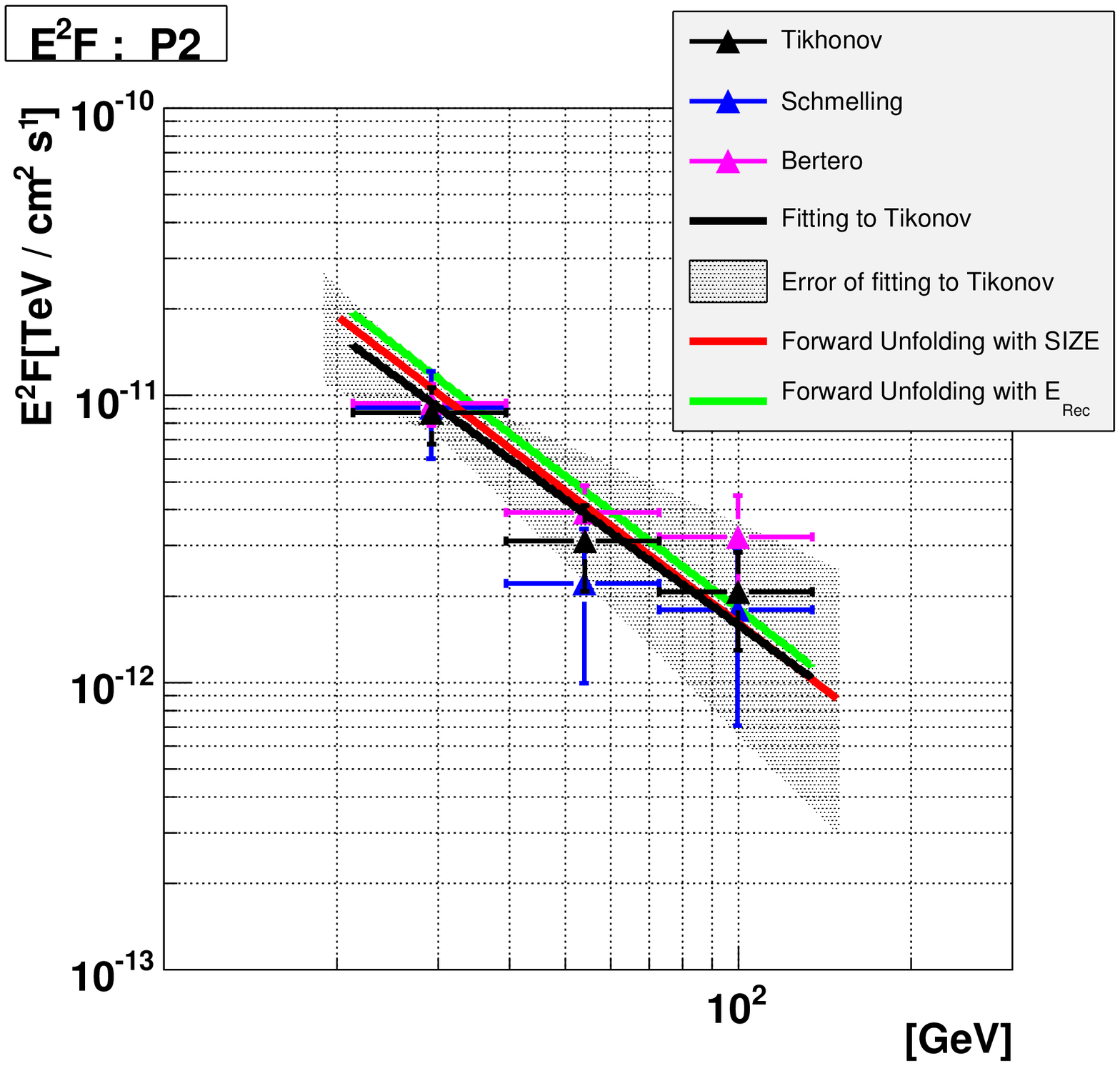}

 \caption{The energy spectra calculated with the various methods. 
Red and green lines show the power law spectra obtained by forward unfolding 
based on $SIZE$ distribution and the reconstructed energy distribution, respectively. 
Black, blue and pink points are the spectral points obtained by the unfolding with
Tikhonov, Schmelling and Bertero regularization methods, respectively. Black lines
indicate the power law fitting to the Tikhonov results while the shaded areas
indicate the error of the fitting.
}
 \label{FigUnfoldedSpectra}
 \end{figure}

(Backward) unfolding does not require an assumption of the spectral shape a priori. 
However, as described in Sect \ref{SectUnFold}, different regularization methods might give
 different results. The results are reliable only when all the regularization methods 
show consistency. 
Tikhonov (see \cite{Tikhonov}), Schmelling (see \cite{Schmelling}) and Bertero (see \cite{Bertero}) regularization
methods were used for unfolding the reconstructed energy distribution.
 The results are shown in Fig \ref{FigUnfoldedSpectra}
as black (Tikhonov), blue (Schmelling) and pink (Bertero) points. 
All the results are consistent. 
The results with the Tikhonov method were fitted by a power law function
 (Eq. \ref{EqPowerLaw}),
shown as a black line in the figure.
It should be noted that points are correlated with each other 
because of the unfolding procedure 
but that the correlation is taken into 
account when fitting is performed. The best fit parameters and $\chi^2$ values
are summarized in Table 
\ref{TableCrabPulsarSpectrum}.


\begin{table}[h]
\centering
\begin{tabular}{|r|r|r|r|r|}
\hline
Phase & Method & $f_{30}$ [10$^{-9}$cm$^{-2}$s$^{-1}$TeV$^{-1}$] & $\Gamma$ & $\chi^2 /dof$ (prob.) \\
\hline
        & Forward with Size     &   17.3 $\pm$ 2.1 & -3.53 $\pm$ 0.41  & 3.7/2 (15.7\%) \\ 
P1~+~P2 & Forward with E$_{rec}$&   18.8 $\pm$ 2.4 & -3.42 $\pm$ 0.34 &  5.1 /4 (27.7\%) \\
        & Tikhonov               &   14.9 $\pm$ 2.9 & -3.35 $\pm$ 0.52 & 2.90/1 (8.9\%) \\
\hline	
        & Forward with Size     &   5.7 $\pm$ 1.4 &  -3.67 $\pm$ 0.80  & 2.1/ 2 (35.0\%)\\
P1      & Forward with E$_{rec}$&   6.0 $\pm$ 1.5 & -3.06 $\pm$ 0.59 &  7.2 /4 (12.6\%)\\
        & Tikhonov               &   4.5 $\pm$ 2.3 & -3.07 $\pm$ 1.04 & 1.20/1 (27.3\%) \\
 \hline
        & Forward with Size     &   11.3 $\pm$ 1.5 & -3.53 $\pm$ 0.39  & 2.7/2 (25.9\%)\\
P2      & Forward with E$_{rec}$&   12.6 $\pm$ 1.6 & -3.54 $\pm$ 0.32 & 2.0/4 (73.6 \%)\\
        & Tikhonov               &   10.0 $\pm$ 1.9 & -3.45 $\pm$ 0.54 & 2.19/1 (13.9\%) \\
\hline
\end{tabular}
\caption{The parameters obtained for power law spectra (see Eq. \ref{EqPowerLaw}) for different phase intervals.}
\label{TableCrabPulsarSpectrum}
\end{table}

\subsection{Discussion of the Results}
As summarized in Table \ref{TableCrabPulsarSpectrum}, the three methods, namely, 
the forward unfolding
with the $SIZE$ distribution, the forward unfolding 
with the reconstructed energy distribution 
and the unfolding with reconstructed energy distribution 
using the Tikhonov regularization method, 
show consistent
results. The energy spectra can be well described by a simple power law, which is partially
due to the limited statistics and the narrow energy range.  
The flux of P2 at 30 GeV is twice as large as that of P1, being consistent with
the light curve. On the other hand, no difference in spectral 
indices between them is visible.
The indices are $-3.35 \pm 0.52$, $-3.07 \pm 1.04$ and $-3.45 \pm 0.54$
for P1~+~P2, P1 and P2, respectively, in the case of the unfolding with 
the Tikhonov regularization method.

\section{Concluding Remarks}
After many careful checks, 59.1 hours of high quality data have been obtained. 
All the analysis tools have been carefully examined as well. 
From them, $ 6267 \pm 1444$, $11315 \pm 1532$ and  $17482 \pm 2329$ gamma-ray signal events 
have been detected for P1, P2 and P1~+~P2, respectively, corresponding to 
4.28$\sigma$, 7.39$\sigma$ and $7.51 \sigma$ in statistical significance.
The light curves show some interesting features compared to lower energy bands, such as
very narrow P1 peak and the absence of the bridge emission. These features will be further
 discussed in 
Chapter \ref{ChapCombAna} together with lower energy observations. 
Most of the excess events are concentrated on $SIZE < 100$ whereas some excess events are still
visible at $SIZE > 100$. The energy spectra have been calculated with a few different methods
and all of them have given consistent results. All of P1, P2 and P1~+~P2 can be 
described by a power law from 25 GeV to 100 GeV and
P2 has twice as high a flux as P1 at 30 GeV. 
The power law indices of P1 and P2 are compatible 
and they are approximately $-3.5$. 
Significant time variation of the pulsation between Cycles III and IV
is not seen. These results will be discussed in more detail in Chapter \ref{ChapCombAna},
by comparing with Fermi-LAT data from 100 MeV to $\sim 30$ GeV.

 \chapter{Analysis of \fermi-LAT public Data}
\label{ChapFermi}

MAGIC could observe gamma-rays from the Crab pulsar only above 25 GeV, 
which is apparently beyond the spectral cut-off point. 
In order to make progress in understanding the emission mechanism,
the MAGIC results need to be discussed in connection with the lower energies.
However, even after 9 years (April 1991 to May 2000) of operation of EGRET, 
which was the only GeV gamma-ray detector that could detect the Crab pulsar before 2007, 
only $\sim 20$ photons above 5 GeV had been detected from the Crab pulsar
 (see Fig. \ref{Fig2.9sigma}). 
This gap in energy coverage between MAGIC and EGRET was soon to be filled 
by data from a new instrument.

The Large Area Telescope (LAT) aboard the {\it Fermi} Gamma-ray Space Telescope
was successfully launched on June 11, 2008. 
It can observe gamma-rays above 100 MeV and clearly saw the Crab pulsar up
to $\sim $30 GeV after 8 months of operation (see \cite{FermiCrab}). 
It is certain that the data of \fermi-LAT help to interpret
 the results of MAGIC observations discussed in the previous chapter.
Therefore, I analyzed one year of its data,
which was made public in August 2009. 



\section{Detector Design of \fermi-LAT}

\begin{figure}[h]
\centering
\includegraphics[width=0.3\textwidth]{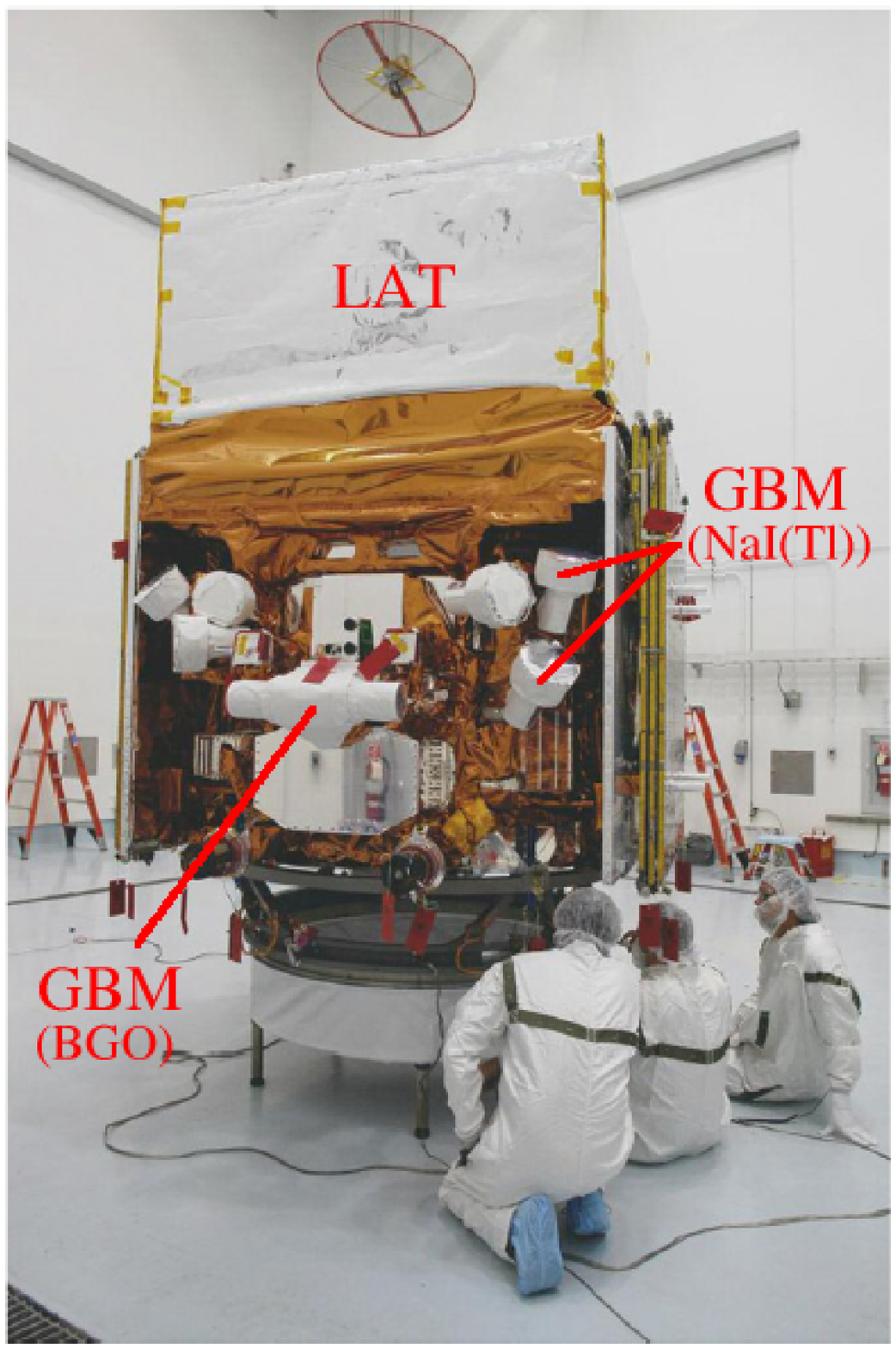}
\includegraphics[width=0.55\textwidth]{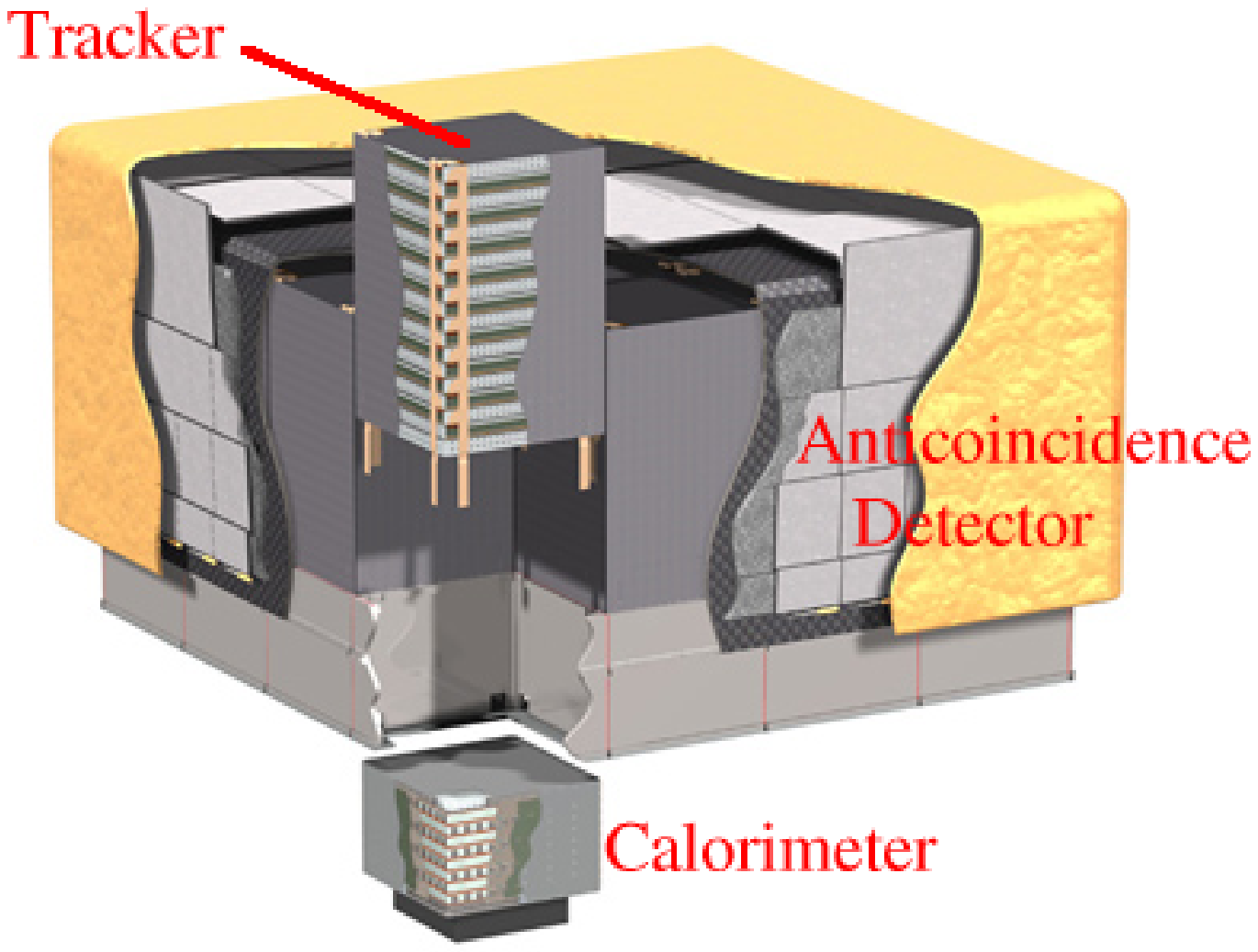}
\caption{Left: A photograph of the \fermi~Gamma-ray Space Telescope, adopted from \cite{FermiGBM}. The high energy gamma-ray
($ > 100$ MeV) detector LAT is seen at the top of the photograph.
Six of the NaI(Tl) scintillation counters (for 8 keV to 1 MeV) and one of 
the BGO scintillation counters
(for 200 keV to 40 MeV) which are for the gamma-ray burst monitor can also be seen.
Right: The LAT detector. It consists of the tracker, the calorimeter and the anti-coincidence
detector. Figure adopted from \cite{FermiLAT}. See text for details.
  }
\label{FigLAT}
\end{figure}

The {\it Fermi} Gamma-ray Space Telescope is equipped with the Gamma-ray Burst Monitor (GBM) and the
Large Area Telescope (LAT) (see the left panel of Fig. \ref{FigLAT}).
The GBM consists of 12 thallium-activated sodium iodide (NaI(Tl)) scintillation counters and two bismuth germanate (BGO) scintillation counters (see \cite{FermiGBM}).
Each counter has an area of 126 cm$^2$ and energy ranges are 8 keV to 1 MeV and 200 keV to 40 MeV, for the NaI(Tl) counter and the BGO counter, respectively. 
The primary aim of the GBM is to detect gamma-ray bursts and its data
 have not been made public.
The
LAT comprises trackers, calorimeters, and an anti-coincidence detector (see \cite{FermiLAT}). 
The LAT estimates the incoming direction and the energy of a gamma-ray by converting it into 
an electron-positron pair, which subsequently cause electromagnetic cascades inside the detector.

\begin{description}
\item[Tracker:] 
There are 16 tracker modules (see the right panel of Fig. \ref{FigLAT}) and each of them has 18 ($x,y$) tracking planes. A tracking plane consists of two layers ($x$ and $y$) of single-sided silicon strip detectors (35 cm long). The 16 planes at the top of the tracker are interleaved with high-$Z$ converter material (tungsten) in which gamma-rays can convert to 
an electron-positron pair 
(see \cite{FermiLAT}). The total vertical depth of the tracker including support material amounts to 1.5 radiation lengths.
\item[Calorimeter: ]
The primary purpose of the calorimeter is twofold:  
1) to measure the energy deposition due to the electromagnetic particle shower that results from the 
electron-positron pair produced by the incident photon 
and 2) to image the shower development profile, thereby providing an important background discriminator and an estimator of the shower energy leakage fluctuations.
 There are 16 calorimeter modules under the trackers (see the right panel of Fig. \ref{FigLAT}). Each of them has 96 cesium iodide crystals doped with thallium (CsI(Tl)) with a size of 2.7 cm $\times$ 2.0 cm $\times$ 32.6 cm. The crystals are optically isolated from each other and are arranged horizontally in eight layers of 12 crystals each. The total vertical depth of the calorimeter is 8.6 radiation lengths, i.e. 10.1 radiation lengths for the total instrument (see \cite{FermiLAT}).

\item[Anti-coincidence detector: ] The purpose of the anti-coincidence detector is to provide charged-particle background rejection.
\footnote{The efficiency of charged particle detection is $>0.9997$ 
(0.99999 when combined with the other subsystems). 
This is one of the key advantages compared to ground-based instruments, where 
it is impossible to install a primary hadron veto.
 }
 It surrounds the tracker modules (see the right panel of Fig. \ref{FigLAT}). 
High energy gamma-rays may cause a so-called ``backsplash'' effect in the massive calorimeter, i.e.
 isotropically distributed secondary particles (mostly 100-1000 keV photons) from the electromagnetic cascade can hit the anti-coincidence detector (the photons cause Compton scattering), creating false veto signals. 
The effect was present in EGRET and lowered the detection efficiency above 10 GeV by at least a factor of two. To minimize the false veto, the anti-coincidence detector is made up 
of 89 segmented plastic scintillators: A  $5 \times 5$ array on the top and $4 \times 4$ arrays on each of the four sides. Each tile is read out by two photomultipliers coupled to wavelength shifting fibers. Tiles near the incident candidate photon trajectory may be considered for background rejection (see \cite{FermiLAT}).

\end{description}

\section{Detector Performance of {\it Fermi}-LAT}
\label{SectFermiPerformance}
The basic performance of the LAT is summarized in Table \ref{TabFermiLAT}.
Parameters are taken from \cite{FermiLAT}.
The energy resolution is dependent on the energy and the incident angle (see the table)
but it is roughly $10\%$, which is better than 
MAGIC (35\% at 30 GeV, see Fig. \ref{FigEneEst}). The large FoV (2.4 sr) would also be of great advantage to compensate
for the small effective area. The timing accuracy of $<10$ $\mu$s is also good enough
to study the Crab pulsar.
On the other hand, the angular resolution is rather poor (0.6 degree at 1 GeV),
leading to large contamination of the galactic diffuse emission or nearby source emissions
to the target source.

\begin{table*}[t]
\centering
 \begin{tabular}{lr}
\hline
\hline
Parameter & Value or Range \\
\hline

Energy range & 20 MeV - 300 GeV\\
Effective Area at normal incidence & 9,500 cm$^2$\\
Energy resolution (equivalent Gaussian 1$\sigma$) & \\
\hspace{1cm}100 MeV-1GeV (on-axis) & 9\%-15\% \\
\hspace{1cm}1 GeV-10GeV (on-axis) & 8\%-9\% \\
\hspace{1cm}10 GeV-300GeV (on-axis) & 8.5\%-18\% \\
\hspace{1cm}$>$10 GeV ($>60^\circ$ incidence) & $\leq 6$\% \\
Single Photon Angular resolution & \\
\hspace{0.5cm} on-axis, 68\% containment radius $\theta_{68\%}$: & \\
\hspace{1.0cm} $>10$ GeV & $\leq$0.15$^\circ$ \\
\hspace{1.0cm} 1 GeV & $0.6^\circ$ \\
\hspace{1.0cm} 100 MeV & 3.5$^\circ$ \\
\hspace{0.5cm} on-axis, 95\% containment radius: & 3 $\times \theta_{68\%}$\\
\hspace{0.5cm} off-axis, containment radius at 55$^\circ$ & 1.7 $\times$ on-axis value\\
Field of View (FoV) & 2.4 sr \\
Timing accuracy & $< 10 \mu$s \\
Event read-out time (dead time) & 26.5 $\mu$s \\
\hline
\hline
 \end{tabular}
 \caption{Summary of LAT Instrument Parameters and Estimated Performance.}
 \label{TabFermiLAT}
 \end{table*}
The systematic uncertainty in the energy scale was conservatively estimated to be
$<$ 5\% for 100 MeV to 1 GeV and $<$ 7\% above 1 GeV, 
from the comparison between electron beam tests and their simulation
(see \cite{FermiAreaSyst} and \cite{FermiEneSyst}).
The systematic uncertainty in effective area was evaluated by comparing
Vela observation results and the simulation for them (see \cite{FermiAreaSyst}). It is 10\% below
100 MeV, decreasing to 5\% at 560 MeV and increasing to 20\% at 10 GeV and above. 

\section{Data Sample} 
After one year of operation, all the Fermi-LAT data and its analysis tools were made public
in August 2009. 
I analyzed one year of data from 4th August 2008 to 3rd August 2009. 
Events with an energy between 100 MeV to 300 GeV and with an arrival direction of 20 degrees
 around the Crab pulsar were downloaded from the public {\it Fermi} website \cite{FermiDataBase}. 
In order to have solid results, the events with a zenith angle smaller than 105 degrees
 and with the highest quality ``Diffuse class'', which means a high probability of being 
a photon, were selected. This event selection was performed by the Fermi official tool {\sf gtselect} (see \cite{FermiDataBase}).  
  Events with imperfect spacecraft information and events taken when the satellite was in the South Atlantic Anomaly were also rejected. These selections were carried out by {\sf gtmktime} (see \cite{FermiDataBase}). 

\begin{figure}[h]
\centering
\includegraphics[width=0.43\textwidth]{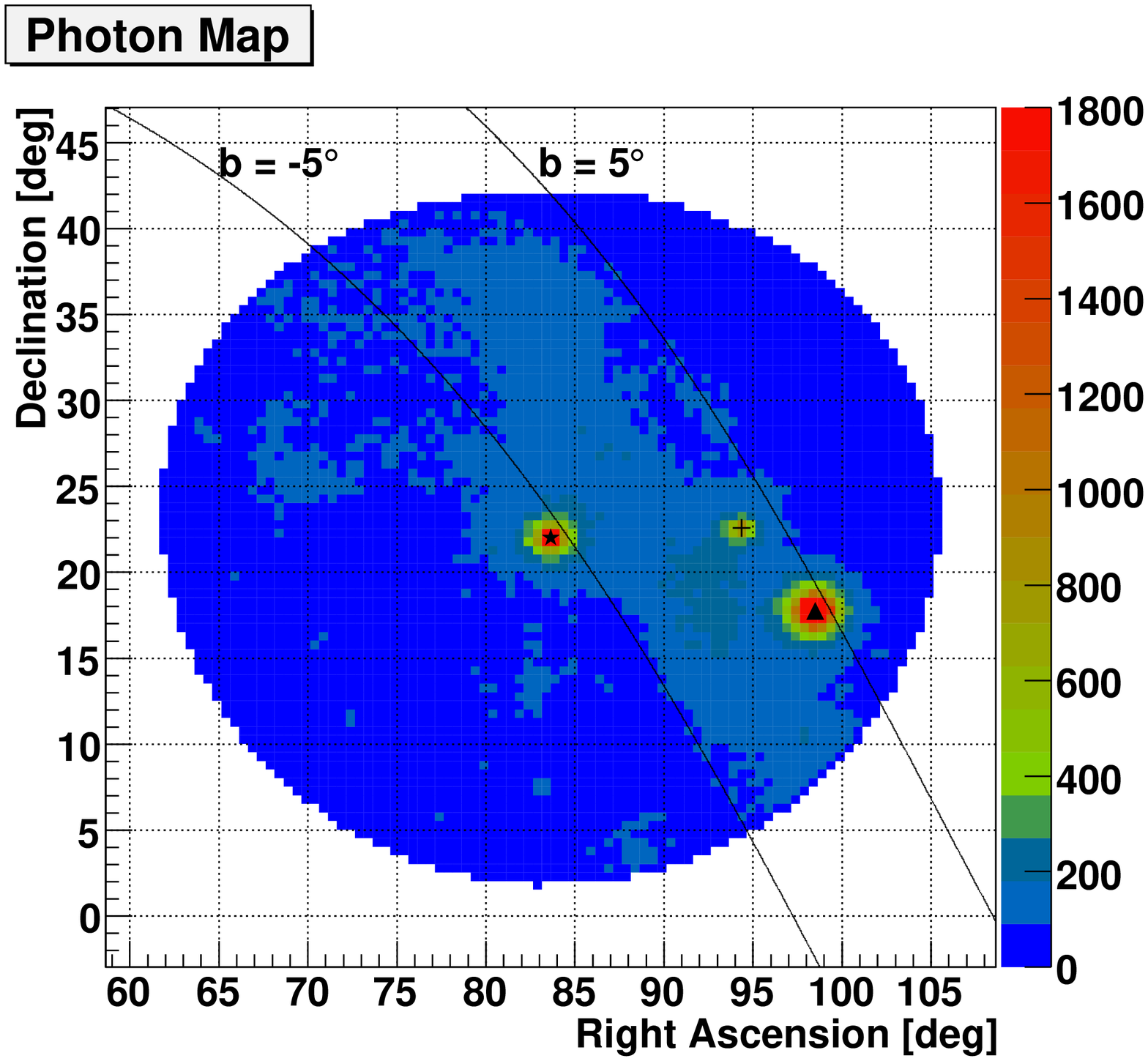}
\includegraphics[width=0.5\textwidth]{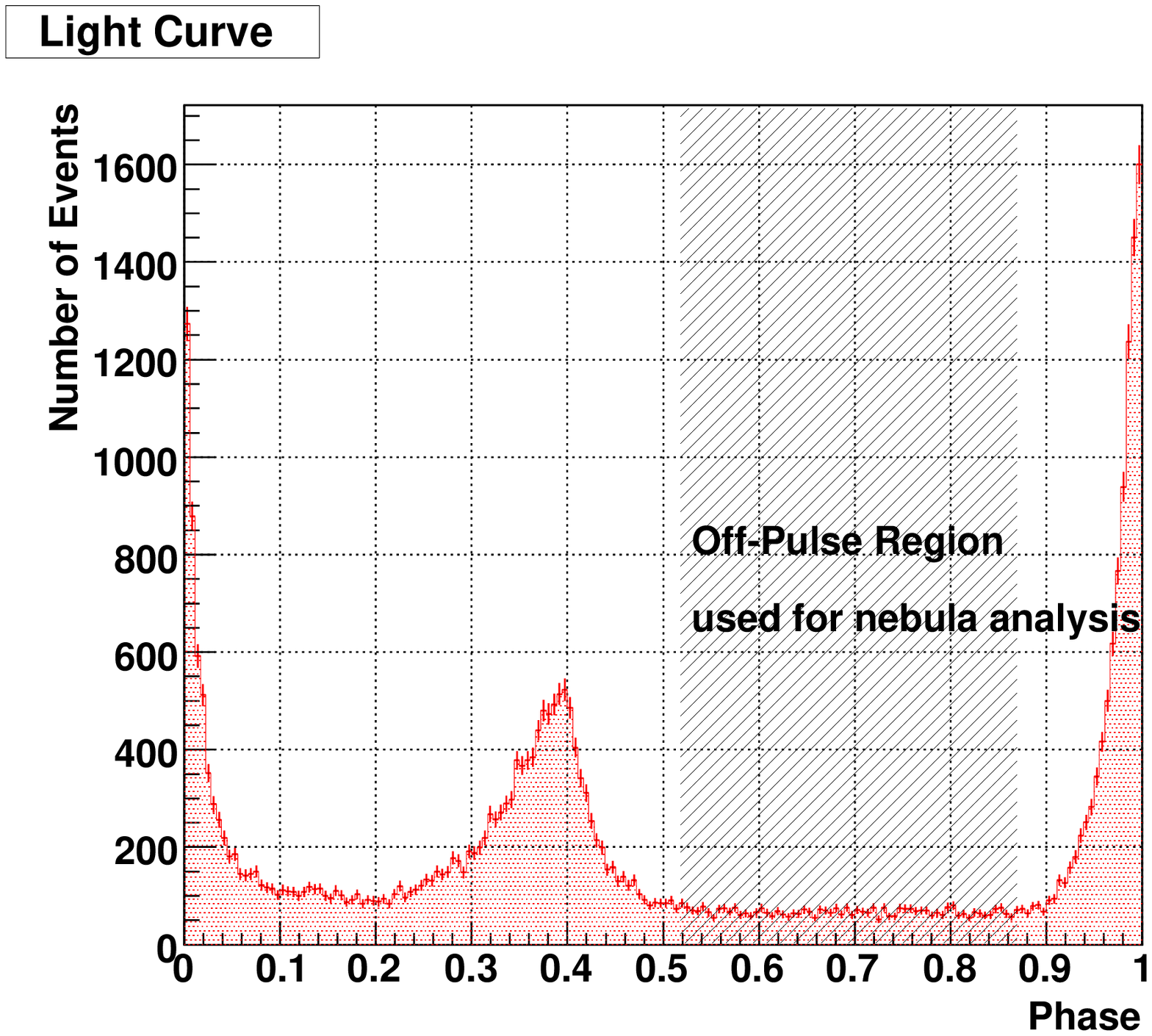}
\caption{Left: An event distribution of all the {\it Fermi}-LAT data used in the analysis 
(100 MeV to 300 GeV) in equatorial coordinates. 
A star, a cross and a triangle indicate the positions of the Crab pulsar, IC 443 and the
Geminga pulsar. Emissions from these sources are clearly seen. In addition, galactic diffuse
emission is also visible.
Right: A light curve of the Crab pulsar based on the {\it Fermi}-LAT data from 100 MeV to 300 GeV.
The data between phase 0.52 and 0.87 (shaded area) are used for the nebula analysis.
The nebula emission is a major background component for the pulsar analysis.
}
\label{FigFermiAllMap}
\end{figure}

 The sky map of all the data used is shown in the left panel of Fig. \ref{FigFermiAllMap}. 
The Crab pulsar is at the center of the map (star) and two more bright sources, namely, IC 443 (cross) and Geminga (triangle), can also be seen. In addition, the galactic diffuse gamma-ray emission is visible. 
Since the angular resolution of the LAT detector is $\sim 3.5$ degree at 100 MeV 
(see \cite{FermiLAT}), 
contamination from the nearby sources and the galactic diffused emission must be taken into account when the Crab pulsar is analyzed. Moreover, emission from the Crab nebula must also be subtracted by using the pulse phase information.

\section{The Light Curve}
\label{SectFermiLightCurve}
To make a light curve (phase diagram) of the Crab pulsar, first of all, 
the pulse phase must be assigned to each individual event. This is done 
by the official {\it Fermi} analysis tool, {\sf gtpphase} (see \cite{FermiDataBase}). 
It requires the pulsar ephemeris information and I used the 
``Jodrell Bank Crab Pulsar Monthly Ephemeris'' for that, as I did 
 for the MAGIC analysis (see Sect \ref{SectPhaseCal}). 
Some of phase-resolved sky maps are shown in Fig. \ref{FigPulsarTopPage}. 
Then, events from the direction around
the Crab pulsar were extracted from the data set. 
Because the angular resolution has a strong energy-dependency,
the extraction radius $R$ [degree] should also be energy-dependent. I used the following radius $R$,
which was also used in the {\it Fermi} official publication (see \cite{FermiCrab}):
\begin{equation}
R= {\rm Max}(6.68-1.76*{\rm log}_{10}(E),1.3) )
\end{equation}
 where $E$ is the estimated energy which is already assigned for each event in the public data.
$R$ decreases linearly to log$_{10}$($E$) until 1.14 GeV and stays constant at 1.3 degrees
 above this energy. It should be noted that the emissions from both the Crab pulsar and
 the Crab nebula are included within $R$. The angular resolution of \fermi-LAT does not
enable a spatial resolution for the pulsar and the nebula.

\begin{figure}[h]
\centering
\includegraphics[width=0.9\textwidth]{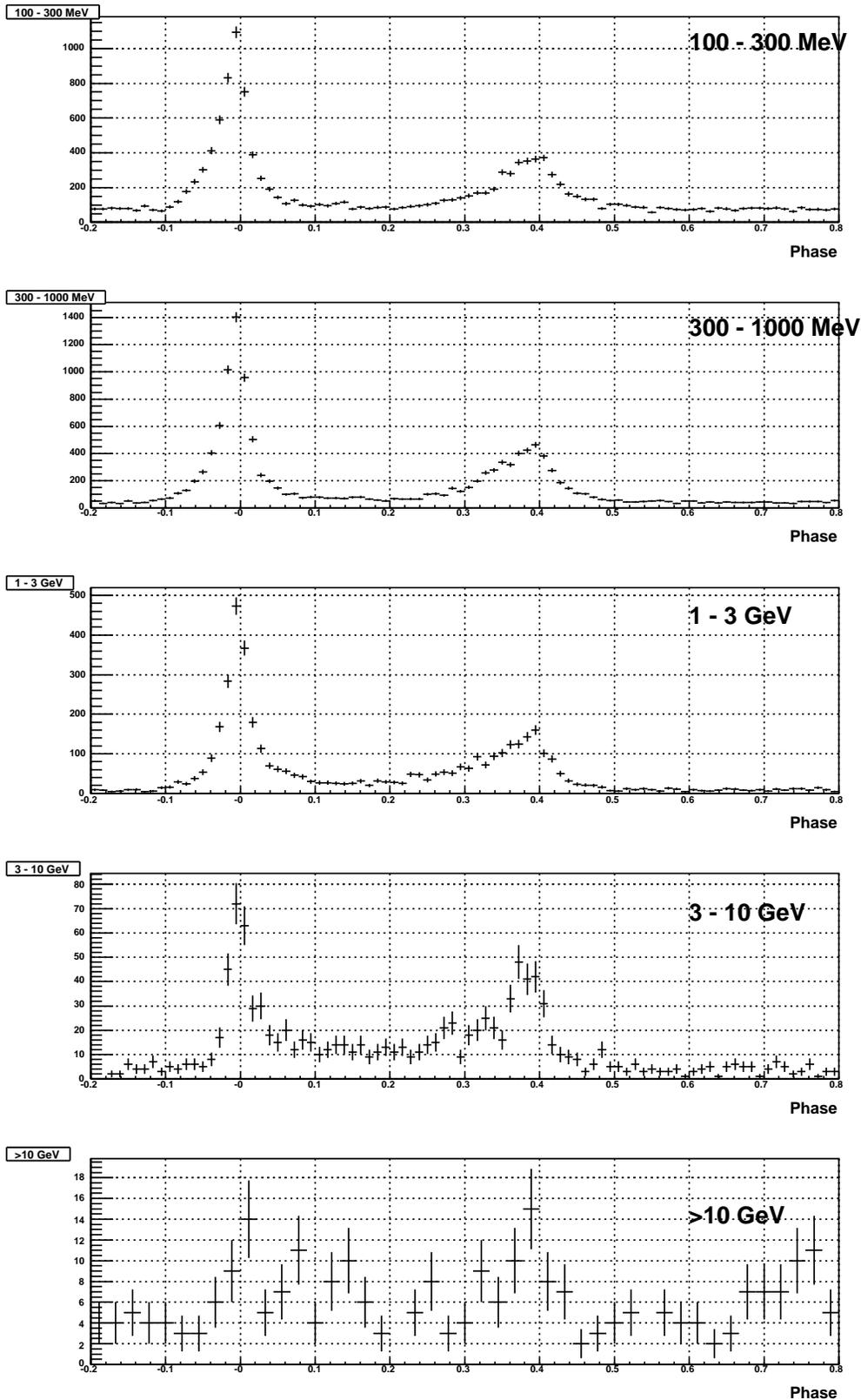}
\caption{The light curves for different energies. From the top: 100 - 300 MeV,
300 - 1000 MeV, 1-3 GeV, 3-10 GeV and above 10 GeV. The energy dependence of P1/P2 ratio
and the pulse width are visible. The possible third peak is also seen
at a phase around 0.75, only above 10 GeV. 
A flat continuum in the light curves 
is mainly from the continuous Crab nebula emission.
}
\label{FigFermiPhaseEne}
\end{figure}

The light curve obtained 
by all the LAT data used is shown in the right panel of Fig. \ref{FigFermiAllMap}. 
Energy-dependent light curves from 100 MeV to above 10 GeV 
are shown in Fig. \ref{FigFermiPhaseEne}.
Below 10 GeV, the pulsations are seen with good precision. 
A flat continuum in the light curves 
is mainly from the continuous Crab nebula emission.
Above 10 GeV, although the two pulses are clearly
visible, the statistical uncertainties are larger than for lower energies. 
P1 (phases -0.06 to 0.04)
and P2 (phases 0.32 to 0.43) have $12.1 \pm 6.4$ and $20.9 \pm 7.3$ excess events above 10 GeV
with a significance of 
2.0 and 3.2, respectively, with the background level (mainly from the nebula emission) estimated using the phases between 0.52 and 0.87. 

 There are several features visible in these light curves: The flux ratio between
P1 and P2 is changing with the energy.
 The widths of the two pulses are decreasing as the energy
goes higher. A hint of a possible third peak is visible at a phase around 0.75, but 
only above 10 GeV.
These detailed features of the light curves
 will be discussed in Chapter \ref{ChapCombAna}, together with the MAGIC results
and lower energy observations.

\clearpage

\section{The Crab Nebula Analysis}
As is the case with MAGIC, the rather poor angular resolution of {\it Fermi}-LAT 
does not enable a spatial
 resolution for the Crab pulsar and the Crab nebula. Therefore, 
 the nebula emission is the dominant background of the Crab pulsar emission for 
{\it Fermi}-LAT data. 
It must be properly analyzed by using the pulse phase information
 and, then, must be subtracted from the Crab pulsar emission.
Fortunately, the analysis of the nebula has an important by-product: 
one can make sure of the analysis method by checking if the obtained energy
 spectrum is smoothly connected to the IACT measurements (see Sect. \ref{SectCrabNebula}). 

In order to analyze the nebula component, 
photons with the pulse phases from 0.52 to 0.87\footnote{
It is almost the same as the OP (off-pulse) phases defined in Sect. \ref{SectPhaseNaming}
but not exactly. Since the official {\it Fermi} publication used these phases for the 
nebula analysis (see \cite{FermiCrab}), I followed their example.}
 (see the right panel of Fig. \ref{FigFermiAllMap}), where no pulsed emission is seen in lower energies,
 are assumed to be from the nebula (and other background photons such as galactic diffuse
emission).
 The selection of the right phase events is carried out
 by the \fermi-LAT analysis tool {\sf gtselect}. 
The effective observation time and the collection area are calculated by 
{\sf gtltcube} and {\sf gtexpmap} (see \cite{FermiDataBase}).  
 The spectrum is determined by means of the likelihood method, using the official 
tool {\sf gtlike} (see \cite{FermiDataBase}).  
It is done in the following way:
The spectral shapes with several
parameters of the sources in the FoV, the galactic and extragalactic diffuse emission models 
and the detector response function are assumed a priori. Then, the best parameters 
that maximize the likelihood of the observed data sets are determined. 
 {\it P6\_V3\_Diffuse}, which is officially provided by the \fermi-LAT collaboration,
was used for the detector response function,
A simple power law spectrum was assumed for the IC 443 while a power law spectrum with an exponential cut-off was assumed for Geminga. 
For the extragalactic and galactic diffuse emission,
{\it isotropic\_iem\_v02.txt} and {\it gll\_iem\_v02.fit}, 
which are included in the \fermi-LAT analysis tool package as a standard model, 
were used.
  For the Crab nebula, the spectrum based on the sum of the two power laws
is assumed aiming for the synchrotron and the inverse Compton emission components, which have been suggested by the previous EGRET (see \cite{Kuiper2001}) and IACT
measurements (see e.g. \cite{MAGICCrab1}, \cite{HESSCrab} and Fig. \ref{FigNebulaSpectrum}).

\begin{figure}[h]
\centering
\includegraphics[width=0.8\textwidth]{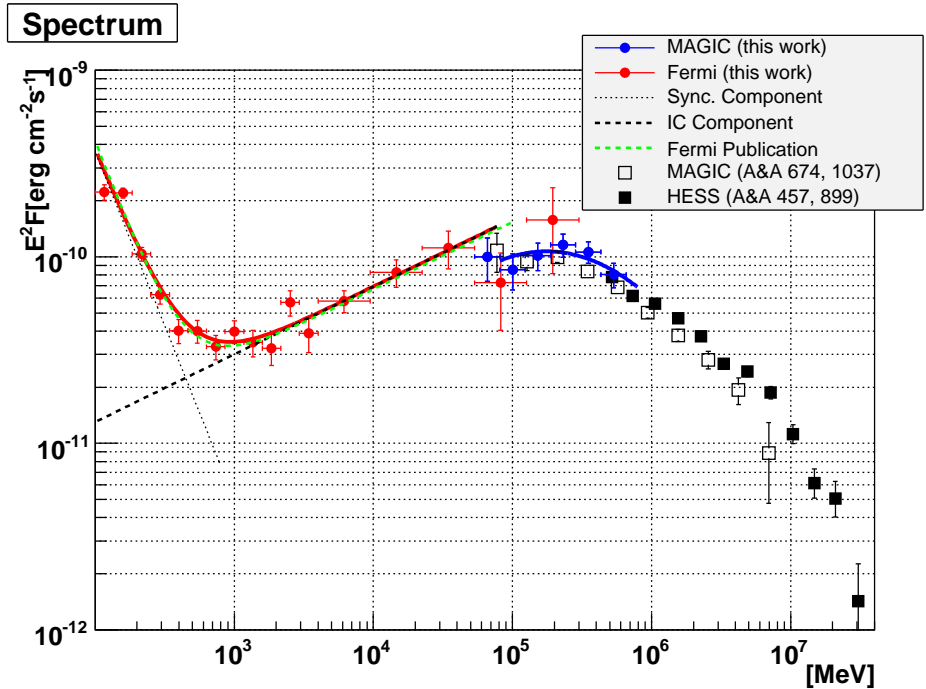}
\caption{The energy spectrum of the Crab nebula. The red line and circles indicate
the spectrum calculated by myself using one year of {\it Fermi}-LAT data.
The green dashed line indicates the published spectrum from the ${\it Fermi}$ collaboration
(see \cite{FermiCrab}). Blue circles, black open squares and black filled squares indicate 
the spectrum calculated by myself with the MAGIC observation data with the SUM trigger,
the published spectrum from the MAGIC collaboration before the SUM trigger was installed 
(see \cite{MAGICCrab1}) and the published spectrum from the HESS collaboration 
(see \cite{HESSCrab}), respectively.
The synchrotron and the inverse Compton components below 100 GeV are indicated by
black dotted and black dashed lines, respectively.
}
\label{FigFermiNebula}
\end{figure}

 The spectrum of the Crab nebula calculated based on the {\it Fermi}-LAT data
is shown in Fig.~\ref{FigFermiNebula} as a red line. 
One can see that synchrotron spectrum is steeply falling from 200 MeV to 500 MeV and the inverse Compton component becomes dominant above 800 MeV. They can be described as
\begin{eqnarray}
\frac{{\rm d}N}{{\rm d}E{\rm d}A{\rm d}T}_{sync} &=& (8.6 \pm 1.4) \times 10^{-11} (E/300{~\rm MeV})^{-4.14 \pm 0.47} \\
\frac{{\rm d}N}{{\rm d}E{\rm d}A{\rm d}T}_{IC} &=& (7.3 \pm 0.7 \times 10^{-12} (E/1{~\rm GeV})^{-1.68 \pm 0.05} \\
\end{eqnarray}

 In order to make sure that the assumption of the spectral shape of the Crab pulsar is valid, 
the same data sets were divided into many subsets according to the energy. Then, 
the likelihood analysis was applied to each subset, assuming a simple power law in each
small energy range.
 The red points in the figure indicate the results for the
divided subsets. Instead of showing many short truncated lines, the value at the bin center and its error are shown. All the points are very well aligned along the line,
 showing the validity of the assumption.

The spectrum published by the {\it Fermi}-LAT collaboration with smaller data samples (
8 months of data)
(see \cite{FermiCrab}) are also shown in the same figure  as a green dashed line, 
which is consistent with my analysis. The measurements by MAGIC (the published one
and the one I calculated with the data samples used for the pulsar analysis)
and by HESS are also shown in the same figure. The spectra are smoothly connected
from 100 MeV to above 10 TeV. 
\footnote{
The true spectrum should not be a simple power law from 10 to 100 GeV. 
Therefore, the $\simeq 50\%$ difference 
between the fitting line of \fermi-LAT data and that of MAGIC data at 100 GeV 
in Fig. \ref{FigFermiNebula} is not very meaningful.
}

It should be noted that the poor statistics of \fermi-LAT data in the overlapped energy region from 50 GeV to 300 GeV
does not allow a relative flux scale calibration 
between \fermi-LAT and MAGIC 
with a precision better than $\simeq 60\%$, which is larger than
the systematic uncertainties of both experiments 
(see Sect. \ref{SectMAGICSyst} and Sect. \ref{SectFermiPerformance}).

\section{The Crab Pulsar Spectrum}
\label{SectFermiSpectrum}
 From the nebula analysis, all the background components (the nearby source emissions,
the diffuse emissions and the nebula emission itself) have been determined. 
Keeping the spectral parameters for background components fixed, 
the pulsed component is analyzed with the same likelihood method.
Analyses are made for four different phase intervals; TP (total pulse, phase 0.00 to 1.00), P1 (-0.06 to 0.04), P2 (0.32 to 0.43) and P1 + P2 (sum of P1 and P2). These definitions are the
same as those described in Sect. \ref{SectPhaseNaming}.

\subsection{Power Law + Cut-off Assumption for the Crab Pulsar Spectrum}
\label{SectFermiPowerLawWithCutoff}
Previous EGRET measurements show that the energy spectrum of the Crab pulsar 
can be well described by a power law between 100 MeV to a few GeV (see \cite{Kuiper2001}), 
whereas non-detection by IACTs above 100 GeV 
(see \cite{MAGICCrab1}) imply the cut-off to be around 10 GeV as mentioned in Sect. \ref{SectCrabObsBefore2007}.
 Therefore, I assumed the spectral shape as 
\begin{eqnarray}
\frac{{\rm d}^3F}{{\rm d}E{\rm d}A{\rm d}t} = f_1(E/1{~\rm GeV})^{\Gamma_1}{\rm exp}( -(E/E_c)^{\Gamma_2} )
\label{EqCutoffSpectra}
\end{eqnarray}
 There are 4 free parameters, namely, the flux at 1 GeV $f_1$, the cut-off energy $E_c$, 
the power law index $\Gamma_1$ and the sharpness of the cut-off $\Gamma_2$
\footnote{As can be seen in Eq. \ref{EqCutoffSpectra}, although the flux above the cut-off energy
is suppressed, a low flux can still be expected above the cut-off energy.
A word ``cut-off'' does not mean extinction of a gamma-ray flux.
}. 
As discussed in Sect \ref{SectCutoffSteepness}, in the case where the emission region is 
close to the pulsar surface, 
the cut-off should be sharp, i.e. $\Gamma_2$ should be significantly larger than 1, while,
in the case where emission comes from the outer magnetosphere, $\Gamma_2$ should be 1.
 Therefore, estimation of $\Gamma_2$ is important
for the investigation of the emission mechanism.
However, due to the lack of statistics above 10 GeV, the likelihood analysis with four free parameters sometimes gives unstable results.
Therefore, I made five analyses with three free parameters, i.e.
with $f_1$, $E_c$, and $\Gamma_1$ being free parameters and with $\Gamma_2$ 
fixed to be 0.66, 1, 1.33, 1.66 and 2. 

\begin{figure}[h]
\centering
\includegraphics[width=0.9\textwidth]{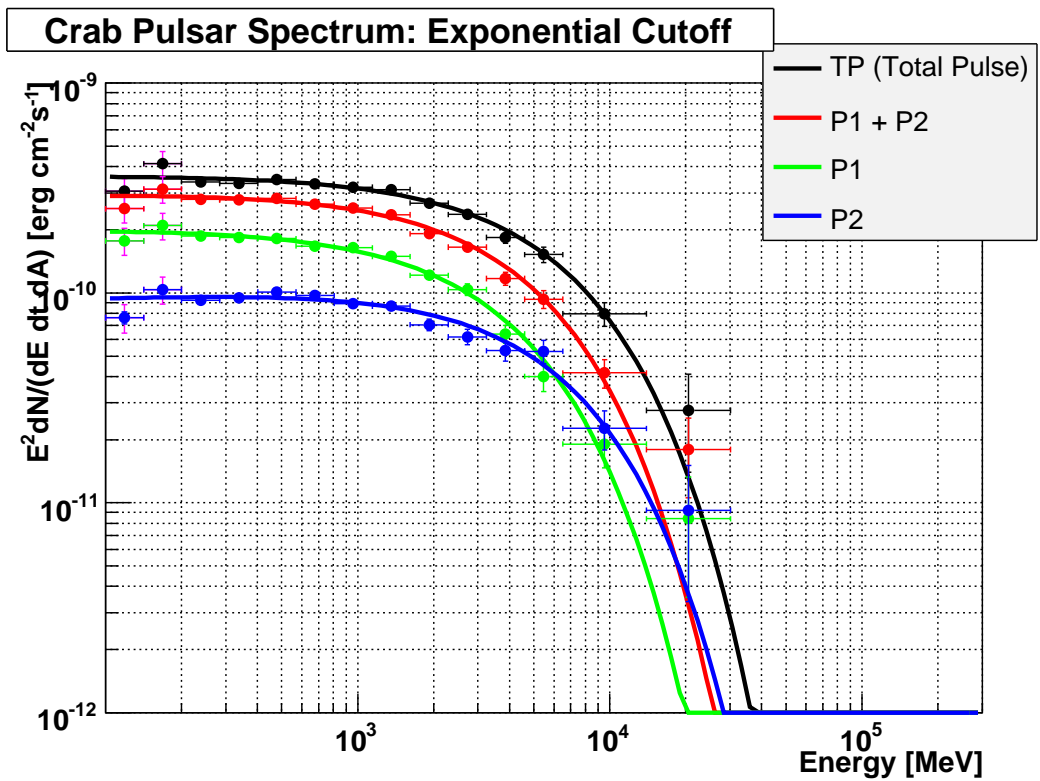}
\caption{The energy spectra of the Crab pulsar for different phase intervals. 
The thick solid lines are determined by the likelihood analysis, assuming the spectral shape to
be a power law with an exponential cut-off. The points are obtained by applying the same
likelihood analysis to the small energy range, assuming a power law energy spectrum
within the range
(see text for details). For the the first two points below 200 MeV, where
the systematic error strongly dominates the statistical error, the quadratical convolution
of the systematic and statistical errors are indicated as pink lines.
}
\label{FigFermiPulsarSpectrum}
\end{figure}

\begin{figure}[h]
\centering
\includegraphics[width=0.49\textwidth]{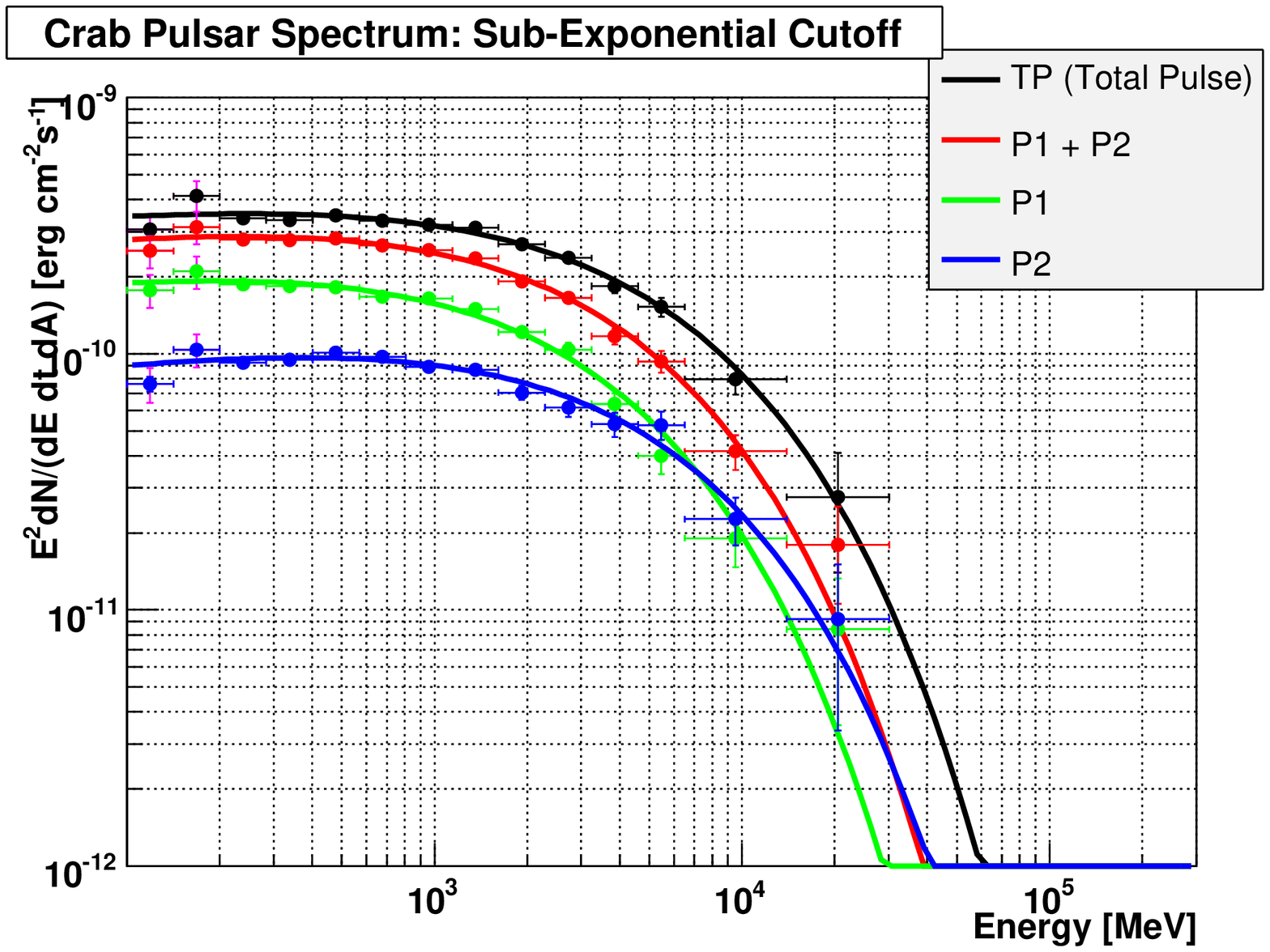}
\includegraphics[width=0.49\textwidth]{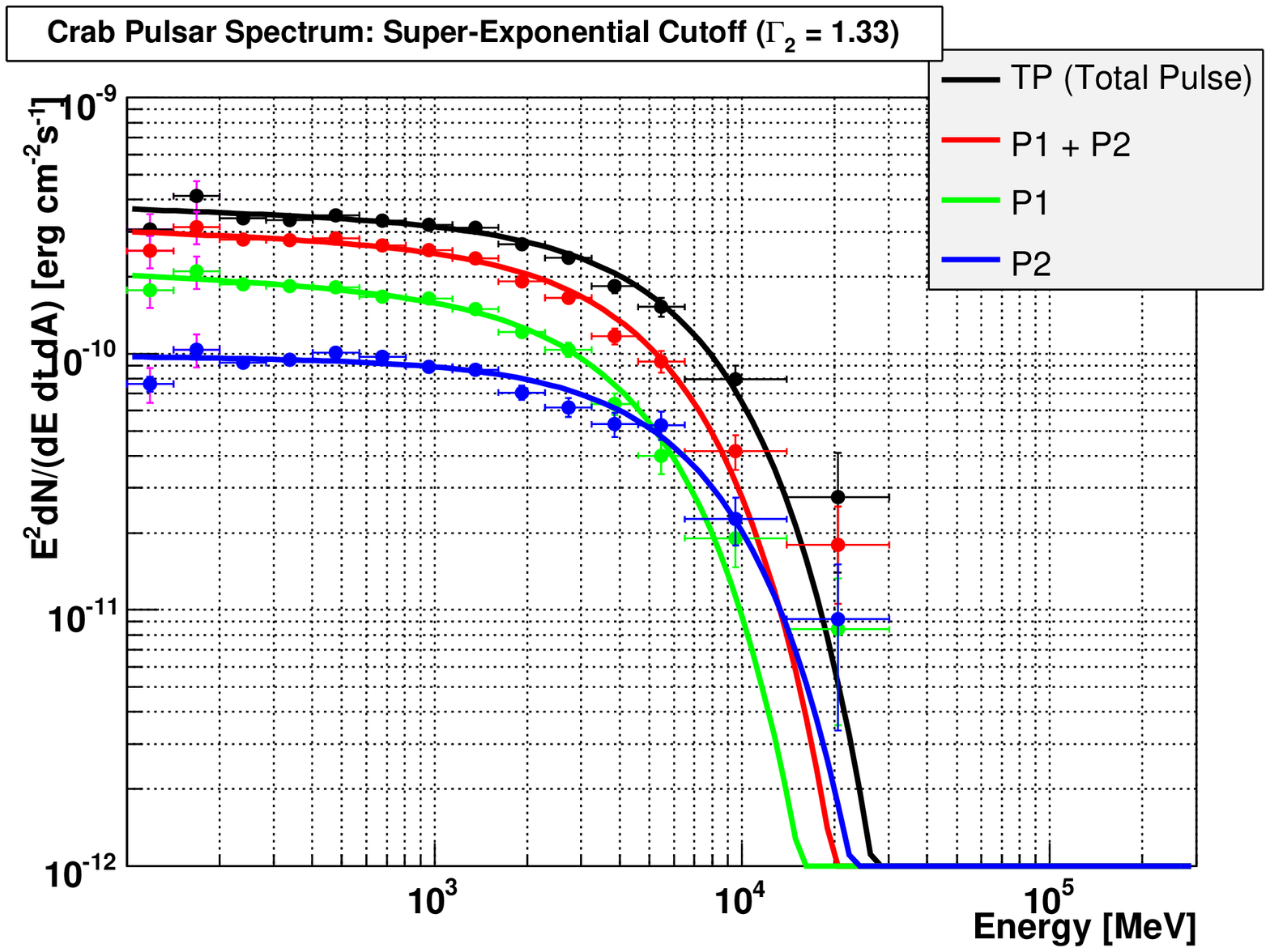}
\includegraphics[width=0.49\textwidth]{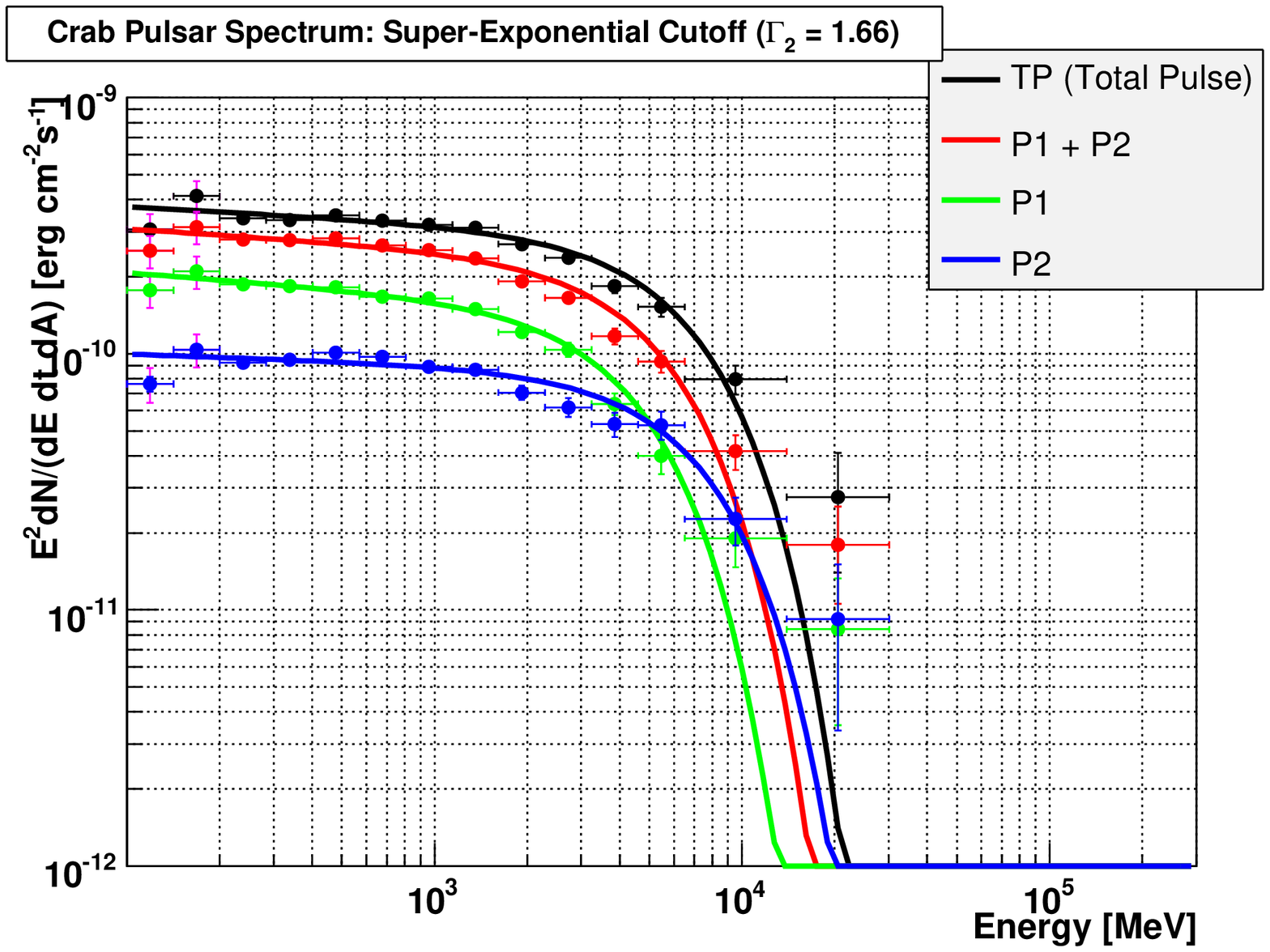}
\includegraphics[width=0.49\textwidth]{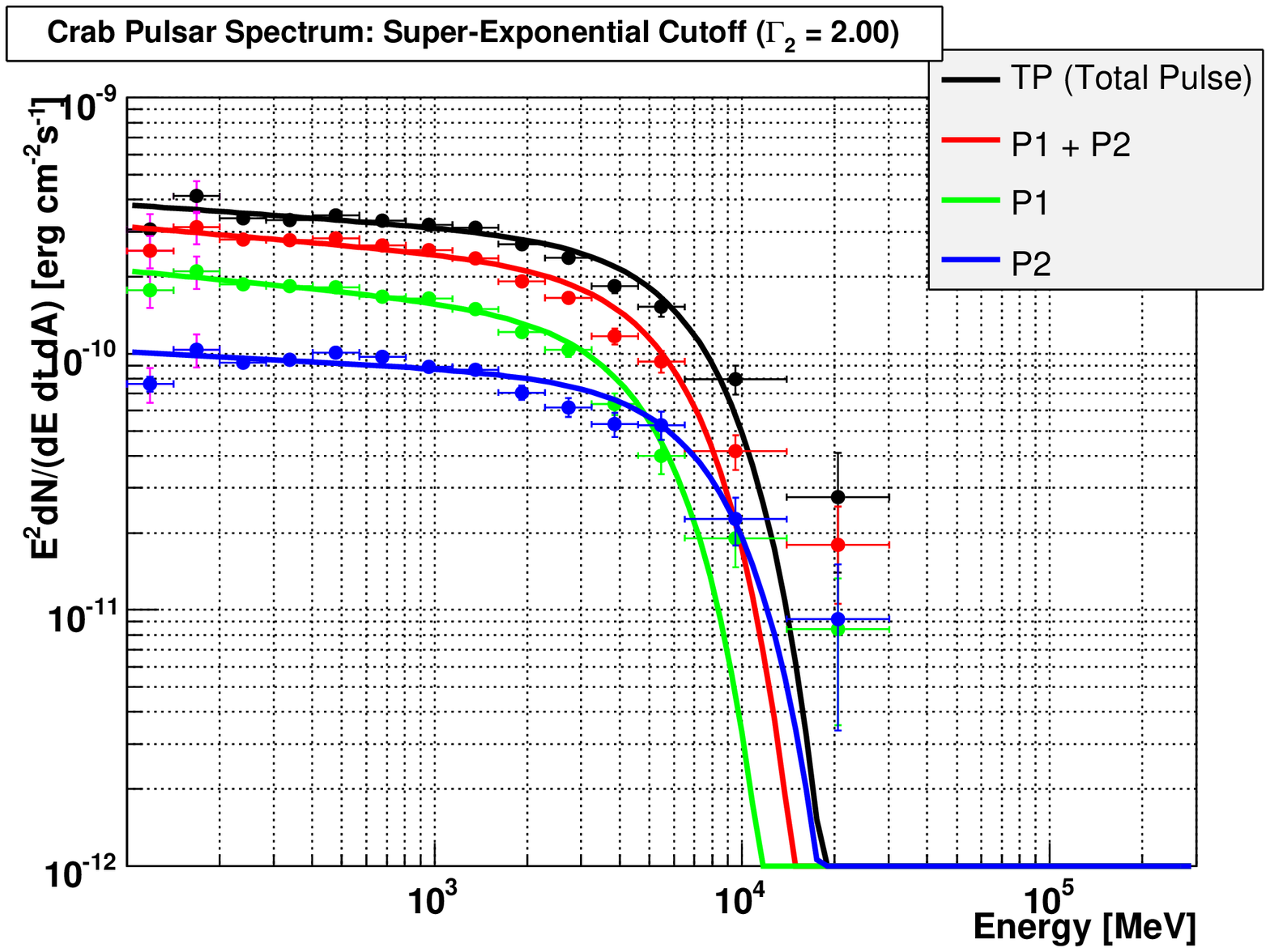}
\caption{The same as Fig. \ref{FigFermiPulsarSpectrum} but with different spectral shape 
assumptions, namely, sub-exponential cut-off (top left), 
super-exponential cut-off with $\Gamma_2 = 1.33$ (top right), 
super-exponential cut-off with $\Gamma_2 = 1.66$ (bottom left), and
super-exponential cut-off with $\Gamma_2 = 2.00$ (bottom right)
}
\label{FigFermiPulsarSpectrum2}
\end{figure}

The results are shown in Fig. \ref{FigFermiPulsarSpectrum}, Fig. \ref{FigFermiPulsarSpectrum2}
 and Table \ref{TableFermiPulsarSpectrum}. Hereafter, when $\Gamma_2 = 1$, 
the spectral cut-off shape will be called the ``exponential cut-off'', 
while when $\Gamma_2 = 0.66$, it will
be called the ``sub-exponential cut-off''. The rest (with $\Gamma_2 > 1$) will be called 
the ``super-exponential cut-off''. 
In order to evaluate which assumption is more appropriate, the likelihood ratio 
$-\log(L/L_{ex})$ is calculated and shown in the seventh column 
of Table \ref{TableFermiPulsarSpectrum}, where $L$ and $L_{ex}$ are the likelihood value 
for a given assumption and that for the exponential cut-off assumption, respectively.
The eighth column shows the corresponding probability.
The spectral parameters published by \fermi-LAT collaboration 
for TP under the exponential cut-off assumption
\footnote{\fermi-LAT collaboration reported these spectral parameters only under the exponential
cut-off assumption in their publication \cite{FermiCrab}. 
In addition to the TP spectrum, they analyzed the spectrum of many narrow phase intervals
(width $\simeq 0.01$). However, they did not publish the spectrum of P1 or P2,
which I want to compare with MAGIC results in Chapter 7. 
}
 are also shown in the last row of
Table \ref{TableFermiPulsarSpectrum}. They are consistent with my analysis
(compare with the first row).

 What one can see from Fig. \ref{FigFermiPulsarSpectrum},
Fig. \ref{FigFermiPulsarSpectrum2}
 and Table \ref{TableFermiPulsarSpectrum} is the following:

\begin{itemize} 
\item No significant difference is seen in the power law index $\Gamma_1$ among different phase intervals (although $\Gamma_1$ is dependent on the sharpness of the cut-off $\Gamma_2$ 
due to a mathematical effect). For the exponential cut-off assumption, 
$\Gamma_1 \simeq -2.0$.
\item The super-exponential assumptions are significantly worse than the exponential cut-off one.
On the other hand, the sub-exponential cut-off assumption is as good as the exponential cut-off one.
\item 80\% of the total flux is from P1~+~P2 at 1 GeV. P1 has twice as high a flux as
P2 at 1 GeV.

\item The cut-off energy is higher for P2 than for P1. 
The difference in the cut-off energy between TP and
 P1 + P2 implies a higher cut-off energy for the bridge emission than for P1 + P2.
\item The flux of P1 and P2 become comparable at around 5 GeV because of the higher cut-off
 of P2. P2 dominates above 5 GeV.
\end{itemize}

The fact that super-exponential assumptions lead to worse fitting 
than the exponential one
suggests that the emission region of gamma-rays is not close to the neutron star surface. 
This will be discussed further in Chapter \ref{ChapCombAna}.
 For the exponential cut-off model, the cut-off energies are estimated to be 6.1 $\pm$ 0.5,
 $4.5 \pm 0.3$, $3.7 \pm 0.3$ and $5.9 \pm 0.7$ GeV for TP, P1 + P2, P1 and P2 respectively.

\clearpage

\begin{table*}
\begin{threeparttable}
\begin{tabular}{|c|c|c|c|c|c|c|c|}\hline
 &  & $f_{1}$ [10$^{-10}$cm$^{-2}$&  &  &  & & \\
Model & Phase &  s$^{-1}$ MeV$^{-1}$] & $\Gamma_1$ & $\Gamma_2$ & $E_{c}$ [MeV] & $LR$\tnote{a} & Rejection Power\tnote{b}\\
\hline
\hline
 & TP & 2.32 $\pm$ 0.05  & -1.99 $\pm$ 0.02  & 1  & 6111 $\pm$ 496  & -- & --  \\ \cline{2-8}
Exponential & P1 + P2 & 1.94 $\pm$ 0.05  & -1.98 $\pm$ 0.02  & 1  & 4452 $\pm$ 307  & -- & --  \\ \cline{2-8}
Cut-off & P1 & 1.29 $\pm$ 0.04  & -1.99 $\pm$ 0.02  & 1  & 3682 $\pm$ 292  & -- & --  \\ \cline{2-8}
 & P2 & 0.67 $\pm$ 0.02  & -1.95 $\pm$ 0.03  & 1  & 5856 $\pm$ 740  & -- & --  \\ \cline{2-8}
\hline
\hline
Sub- & TP & 3.10 $\pm$ 0.13  & -1.88 $\pm$ 0.02  & 0.66  & 3359 $\pm$ 379  & 1.6 & 1.9e-01 (0.9 $\sigma$) \\ \cline{2-8}
exponential & P1 + P2 & 2.80 $\pm$ 0.12  & -1.85 $\pm$ 0.02  & 0.66  & 2198 $\pm$ 205  & -4.8 & -- \\ \cline{2-8}
Cut-off & P1 & 1.94 $\pm$ 0.11  & -1.85 $\pm$ 0.03  & 0.66  & 1791 $\pm$ 195  & 0.2 & 8.0e-01 (0.0 $\sigma$) \\ \cline{2-8}
$\Gamma_2 =0.66$& P2 & 0.92 $\pm$ 0.06  & -1.83 $\pm$ 0.04  & 0.66  & 3001 $\pm$ 511  & $ -3.1$ & -- \\ \cline{2-8}
\hline
\hline
Super- & TP & 2.11 $\pm$ 0.03  & -2.04 $\pm$ 0.01  & 1.33  & 7163 $\pm$ 493  & 2.7 & 6.6e-02 (1.5 $\sigma$) \\ \cline{2-8}
exponential & P1 + P2 & 1.71 $\pm$ 0.03  & -2.04 $\pm$ 0.01  & 1.33  & 5536 $\pm$ 325  & 9.2 & 1.0e-04 (3.7 $\sigma$) \\ \cline{2-8}
Cut-off & P1 & 1.13 $\pm$ 0.03  & -2.05 $\pm$ 0.02  & 1.33  & 4581 $\pm$ 303  & 3.3 & 3.7e-02 (1.8 $\sigma$) \\ \cline{2-8}
$\Gamma_2 =1.33$ & P2 & 0.60 $\pm$ 0.02  & -2.01 $\pm$ 0.02  & 1.33  & 7272 $\pm$ 786  & 3.9 & 2.0e-02 (2.0 $\sigma$) \\ \cline{2-8}
\hline
\hline
Super- & TP & 2.01 $\pm$ 0.03  & -2.07 $\pm$ 0.01  & 1.66  & 7577 $\pm$ 472  & 7.8 & 4.2e-04 (3.3 $\sigma$) \\ \cline{2-8}
exponential & P1 + P2 & 1.61 $\pm$ 0.02  & -2.08 $\pm$ 0.01  & 1.66  & 6085 $\pm$ 325  & 20.4 & 1.4e-09 (6.0 $\sigma$) \\ \cline{2-8}
Cut-off & P1 & 1.05 $\pm$ 0.02  & -2.09 $\pm$ 0.01  & 1.66  & 5010 $\pm$ 296  & 8.6 & 1.8e-04 (3.6 $\sigma$) \\ \cline{2-8}
$\Gamma_2 =1.66$ & P2 & 0.57 $\pm$ 0.01  & -2.04 $\pm$ 0.02  & 1.66  & 8025 $\pm$ 786  & 7.9 & 3.8e-04 (3.4 $\sigma$) \\ \cline{2-8}
\hline
\hline
Super- & TP & 1.96 $\pm$ 0.02  & -2.08 $\pm$ 0.01  & 2  & 7761 $\pm$ 458  & 14.1 & 7.6e-07 (4.8 $\sigma$) \\ \cline{2-8}
exponential & P1 + P2 & 1.55 $\pm$ 0.02  & -2.10 $\pm$ 0.01  & 2  & 6411 $\pm$ 325  & 32.4 & 8.9e-15 (7.7 $\sigma$) \\ \cline{2-8}
Cut-off & P1 & 1.01 $\pm$ 0.02  & -2.11 $\pm$ 0.01  & 2  & 5248 $\pm$ 286  & 15.1 & 2.9e-07 (5.0 $\sigma$) \\ \cline{2-8}
$\Gamma_2 =2.00$ & P2 & 0.55 $\pm$ 0.01  & -2.06 $\pm$ 0.02  & 2  & 8496 $\pm$ 784  & 11.7 & 8.7e-06 (4.3 $\sigma$) \\ \cline{2-8}
\hline 
\hline 
\multicolumn{8}{|c|}{Fermi Publication} \\
\hline
Exponential & & & & & & & \\
Cut-off & TP & 2.36 $\pm$ 0.06  & -1.97 $\pm$ 0.02  & 1  & 5800 $\pm$ 500  & -- & --  \\ \cline{2-8}
\hline
\hline 
 \end{tabular}
\begin{tablenotes}
{\footnotesize
\item[a] Likelihood ratio defined as $LR = -\log(L/L_{ex})$, where $L$ and $L_{ex}$
are the likelihood value for a given assumption and that for the exponential cut-off assumption.
\item[b] The probability corresponding to the $LR$ value. It is also expressed as the 
corresponding deviation in the Gaussian distribution. When $LR$ is negative, the probability
is not calculated.
}
\end{tablenotes}
\end{threeparttable}
 \caption{
The results of the likelihood analyses for different spectral shape assumptions.
For the definition of the spectral parameters, see Eq. \ref{EqCutoffSpectra}.
The corresponding spectra are graphically shown in Fig. \ref{FigFermiPulsarSpectrum} and 
Fig. \ref{FigFermiPulsarSpectrum2}.  
The last row shows the spectral parameters from the \fermi-LAT publication
for TP phase under the exponential cut-off assumption, taken from \cite{FermiCrab}.
For other cut-off assumptions and for other phases (P1, P2 or P1~+~P2), 
the spectral parameters were 
not reported in \cite{FermiCrab}.
}
\label{TableFermiPulsarSpectrum}
 \end{table*}

\clearpage

\subsection{Power Law Extension Assumption for the Crab Pulsar Spectrum above the Cut-Off}
\label{SectFermiPL}
It should be noted that the sharpness of the cut-off $\Gamma_2$ is already determined 
below $\sim 7$ GeV and the spectral shape above 7 GeV is not well determined
 due to the lack of statistics, 
as can be seen from the error bars in 
Fig. \ref{FigFermiPulsarSpectrum} and Fig. \ref{FigFermiPulsarSpectrum2}. 
Considering that the MAGIC results show power law spectra with an index of
$\sim -3.5$ above 25 GeV (see Sect. \ref{SectMAGICspectrum}), I examine a power law assumption 
for the spectra above the cut-off energy. The fact that
the last point at 20 GeV is upwardly deviated from the exponential cut-off spectrum 
by  $\simeq 1 \sigma$ for all phase intervals (see Fig. \ref{FigFermiPulsarSpectrum})
may support this assumption.

I selected data above 4 GeV and 
made a spectral analysis assuming a power law; 
\begin{equation}
\frac{{\rm d}^3F}{{\rm d}E{\rm d}A{\rm d}t} = f_{10}(E/10{~\rm GeV})^{\Gamma} 
\label{EqFermiPL}
\end{equation}

\begin{figure}[h]
\centering
\includegraphics[width=0.75\textwidth]{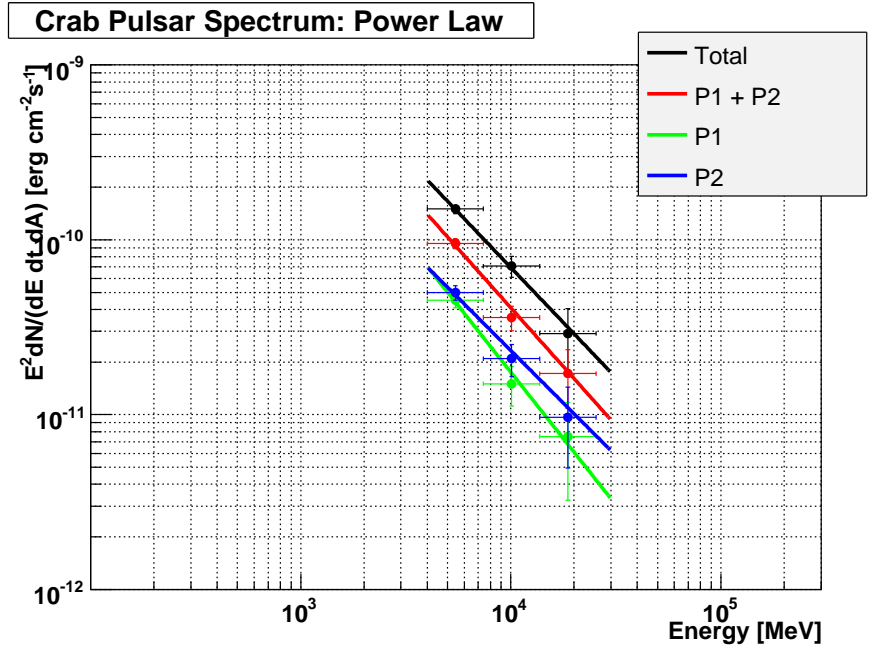}
\caption{The energy spectra of the Crab pulsar above 4 GeV for different phase intervals.
The thick solid lines are determined by the likelihood analysis assuming a power law 
spectrum. The points are obtained by applying the same
likelihood analysis to the small energy range.
Above 4 GeV, the spectra can be well described by a power law.
}
\label{FigFermiPL}
\end{figure}

The results are shown in Fig. \ref{FigFermiPL} and Table \ref{TableFermiPL}.
In order to perform the likelihood ratio test
with respect to the exponential cut-off assumption,
 the exponential cut-off assumption is also applied to
the same data set. This time $\Gamma_1$ is fixed to the best value obtained by the 
previous analyses (see Table \ref{TableFermiPulsarSpectrum}), so that the number of free parameters are the same (two) for the two models.
The likelihood ratio ($LR$) and the corresponding probability are shown in the fifth 
and sixth columns of the table. 
None of the $LR$s are significantly large, which means that above 4~GeV the power law assumption is as good
as the exponential cut-off assumption.
The obtained spectral indices are $\sim -3.3 \pm 0.2$ and consistent with the ones obtained from 
the MAGIC data,
as described in Sect. \ref{SectMAGICspectrum} (see Table \ref{TableCrabPulsarSpectrum}). 
The spectra obtained with MAGIC data and \fermi-LAT data will be compared in more detail
in the next chapter.

\begin{table*}[h]
\begin{threeparttable}
\begin{tabular}{|c|c|c|c|c|c|}\hline
Model & Phase & $f_{10}$ [10$^{-13}$cm$^{-2}$s$^{-1}$ MeV$^{-1}$] & $\Gamma$ & $LR$ \tnote{a}& Rejection Power\tnote{b}\\
\hline
\hline
         & TP & 4.34 $\pm$ 0.42  & -3.26 $\pm$ 0.15  & 1.3 & 2.8e-01 (0.6 $\sigma$) \\ \cline{2-6}
Power Law & P1 + P2   & 2.55 $\pm$ 0.27  & -3.35 $\pm$ 0.16  & 0.0 & 9.9e-01 (0.0 $\sigma$) \\ \cline{2-6}
         & P1    & 1.09 $\pm$ 0.19  & -3.52 $\pm$ 0.26  & -0.8 & -  \\ \cline{2-6}
         & P2    & 1.45 $\pm$ 0.20  & -3.20 $\pm$ 0.21  & 0.6 & 5.5e-01 (0.0 $\sigma$) \\ \cline{2-6}
\hline
\hline
 \end{tabular}
\begin{tablenotes}
{\footnotesize
\item[a] Likelihood ratio defined as $LR = -\log(L/L_{ex})$, where $L$ and $L_{ex}$
are the likelihood value for the power law extension assumption and that for the exponential cut-off assumption.
\item[b] The probability corresponding to the $LR$ value. It is also expressed as the 
corresponding deviation in the Gaussian distribution. When $LR$ is negative, the probability
is not calculated.
}
\end{tablenotes}
\end{threeparttable}
 \caption{
The results of the likelihood analyses 
assuming a power law spectrum (see Eq. \ref{EqFermiPL}) for different phase intervals.
The corresponding spectra are graphically shown in Fig. \ref{FigFermiPL}
}
\label{TableFermiPL}
 \end{table*}

\section{Concluding Remarks}
By using one year of the public \fermi-LAT data, 
the Crab pulsar was analyzed from  100 MeV to $\sim 30$ GeV.
Energy-dependent light curves show a few remarkable features, such
as energy dependence of the pulse width and the P2/P1 ratio. 
The possible third peak is also seen only above 10 GeV.
These features will be discussed in detail, combined with
the MAGIC results and lower energy observations, in Chapter \ref{ChapCombAna}. 
The energy spectrum of the Crab pulsar 
can be well described by a power law with an exponential cut-off,
indicating that the emission region is not close to the neutron star
surface. The cut-off energies are estimated to be 6.1 $\pm$ 0.5 GeV, $4.5 \pm 0.3$ GeV,
 $3.7 \pm 0.3$ GeV, 
and $5.9 \pm 0.7$ GeV, for the TP, P1 + P2, P1 and P2, respectively.
Due to the lack of statistics, the spectral shape above 7 GeV is not well determined.
If data only above 4 GeV are analyzed, a power law function with an index of $\sim -3.3 \pm 0.2$ 
fits the data too, suggesting 
{\bf the possibility that the energy spectrum is extending
by a power law after the cut-off}. Actually, the spectral indices are consistent 
with the MAGIC results.
 These spectra obtained with \fermi-LAT data will be compared with 
the ones obtained by MAGIC in Chapter \ref{ChapCombAna}.

 \chapter{Analysis of the Energy Spectrum and the Light Curve Combining MAGIC and {\it Fermi}-LAT results}
\label{ChapCombAna}

MAGIC could measure the energy spectrum of the Crab pulsar from 25 GeV to 100 GeV,
whereas {\it Fermi}-LAT could measure from 100 MeV to $\sim 30$ GeV. In this chapter, 
the energy spectrum and the light curve of the Crab pulsar from 100 MeV to 100 GeV is further
investigated, by combining the results of the two experiments.

\section{Energy Spectra of P1~+~P2, P1 and P2}
\begin{figure}[h]
\centering
\includegraphics[width=0.65\textwidth]{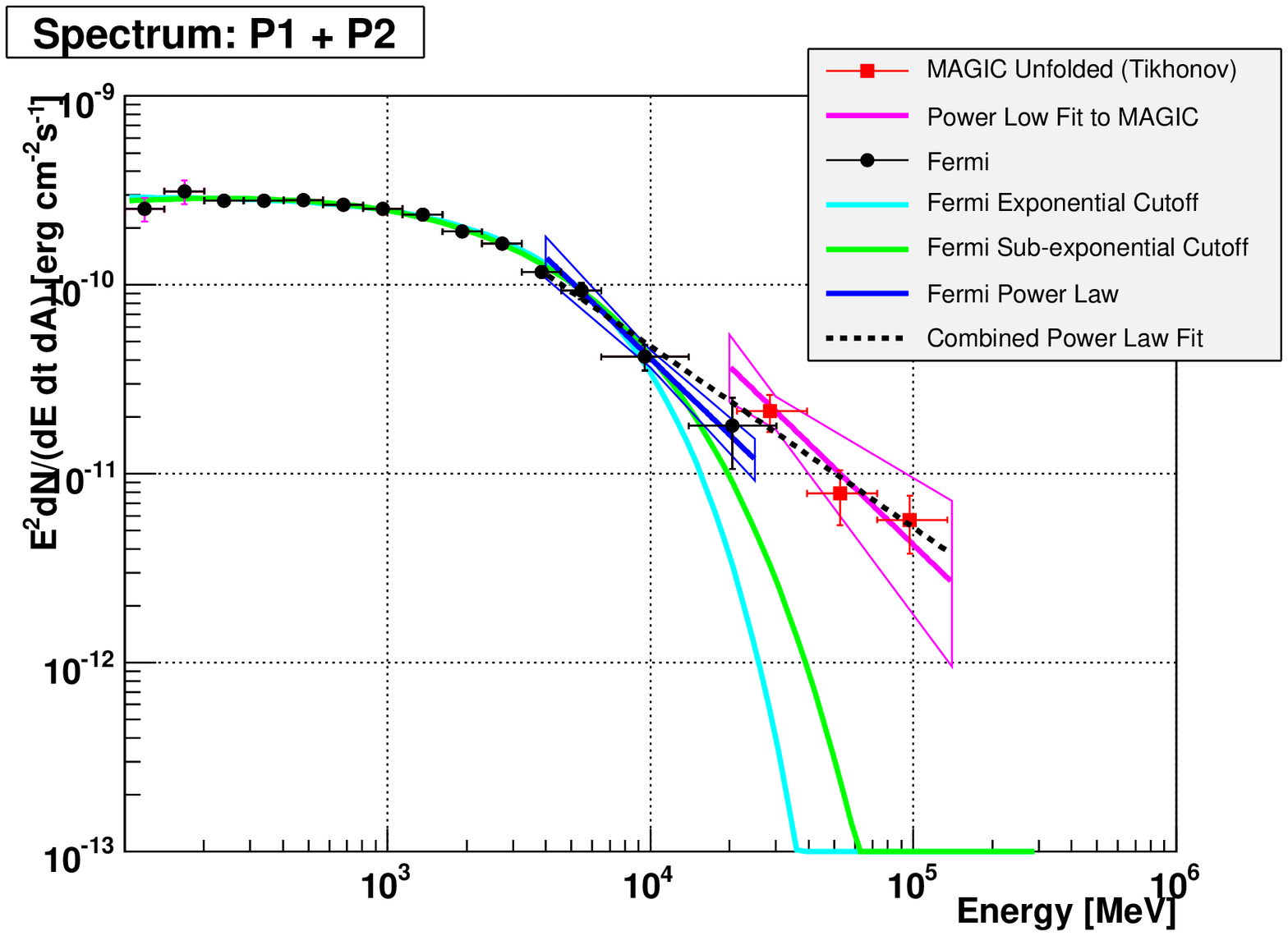}
\includegraphics[width=0.45\textwidth]{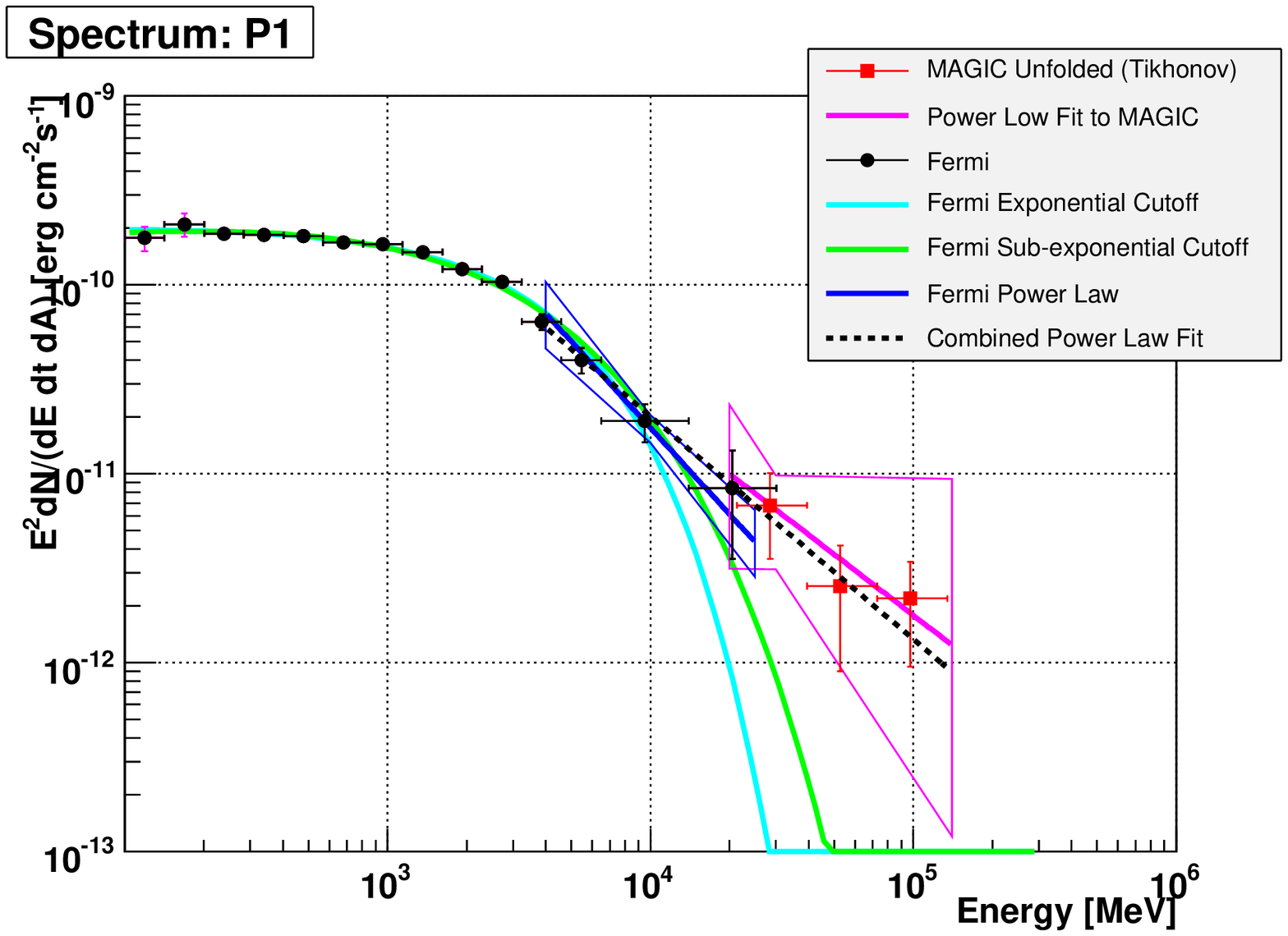}
\includegraphics[width=0.45\textwidth]{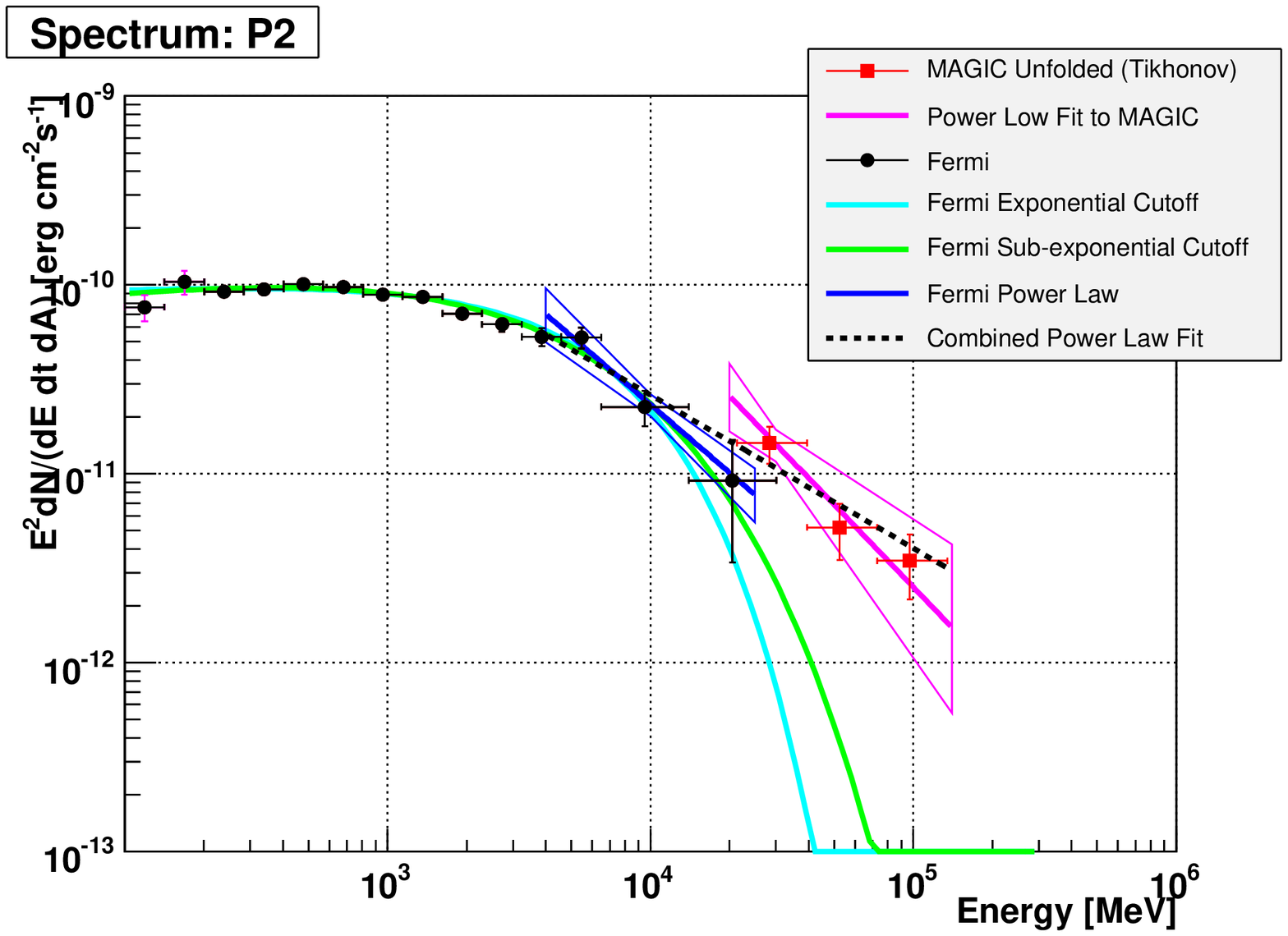}
\caption{The comparison of the energy spectra of the Crab pulsar between \fermi-LAT measurements
and MAGIC measurements for P1 + P2 (top), P1 (bottom left) and P2 (bottom right). The black and the red points
indicate \fermi-LAT and MAGIC results, respectively. The sky blue and the green lines
indicate the best fit spectra based on \fermi-LAT data,
assuming a power law with an exponential cut-off and with a sub-exponential cut-off,
respectively. The blue and the pink lines show the power law fits to the \fermi-LAT data
above 4 GeV and MAGIC data, respectively. The statistical uncertainties of the fits
are also shown by butterfly shape boxes with the corresponding color. 
Power law fits to the \fermi-LAT-MAGIC combined data above 4 GeV are shown
by black dotted lines (see Sect. \ref{SectPL4GeV}).
}
\label{FigSpectraComp}
\end{figure}

The energy spectra measured by MAGIC and \fermi-LAT are shown in Fig. \ref{FigSpectraComp}.
The top, the bottom left and the bottom right panels show those for P1 + P2, P1 and P2, 
respectively (see Sect. \ref{SectPhaseNaming} for the definition of P1 and P2).
 Although the \fermi-LAT results are well described by 
a power law with
an exponential cut-off or with a sub-exponential cut-off,
the MAGIC results are apparently deviated from them.
  In Sect. \ref{SectHowDifficult} and Sect. \ref{SectHowDifficult2}, these deviations 
will be quantitatively examined, 
taking into account the systematic uncertainties of both experiments. 

On the other hand, a power law can well describe both \fermi-LAT results above 4 GeV
(see Sect. \ref{SectFermiPL}) and MAGIC results between 25 GeV and 100 GeV (see Sect. \ref{SectMAGICspectrum}).
The obtained spectral slopes from the two experiments seem similar.
 The power law assumption will be examined in Sect. \ref{SectPL4GeV}, also taking into account
the systematic uncertainties of both experiments.

\clearpage
\subsection{How Much Do the MAGIC Measurements Deviate from an Exponential Cut-off Spectrum?}
\label{SectHowDifficult}
In general, the theoretical calculations based on the outer magnetosphere emission hypothesis 
(the OG model or the SG model)
predict an exponential cut-off at 1 - 10 GeV in the Crab pulsar energy spectrum,
as discussed in Sect. \ref{SectCutoffSteepness}, 
with which \fermi-LAT results are indeed consistent. 
However, as can be seen from Fig. \ref{FigSpectraComp}, the MAGIC measurements seem to be
deviated from the extrapolation from the \fermi-LAT measurements assuming that 
the spectra of P1, P2, and P1~+~P2 above 25 GeV follow a power law with an exponential cut-off.
Here, I evaluate these deviations.

\subsubsection{Method: $\chi^2$ Test on $SIZE$ Distributions}

\begin{figure}[h]
\centering
\includegraphics[width=0.7\textwidth]{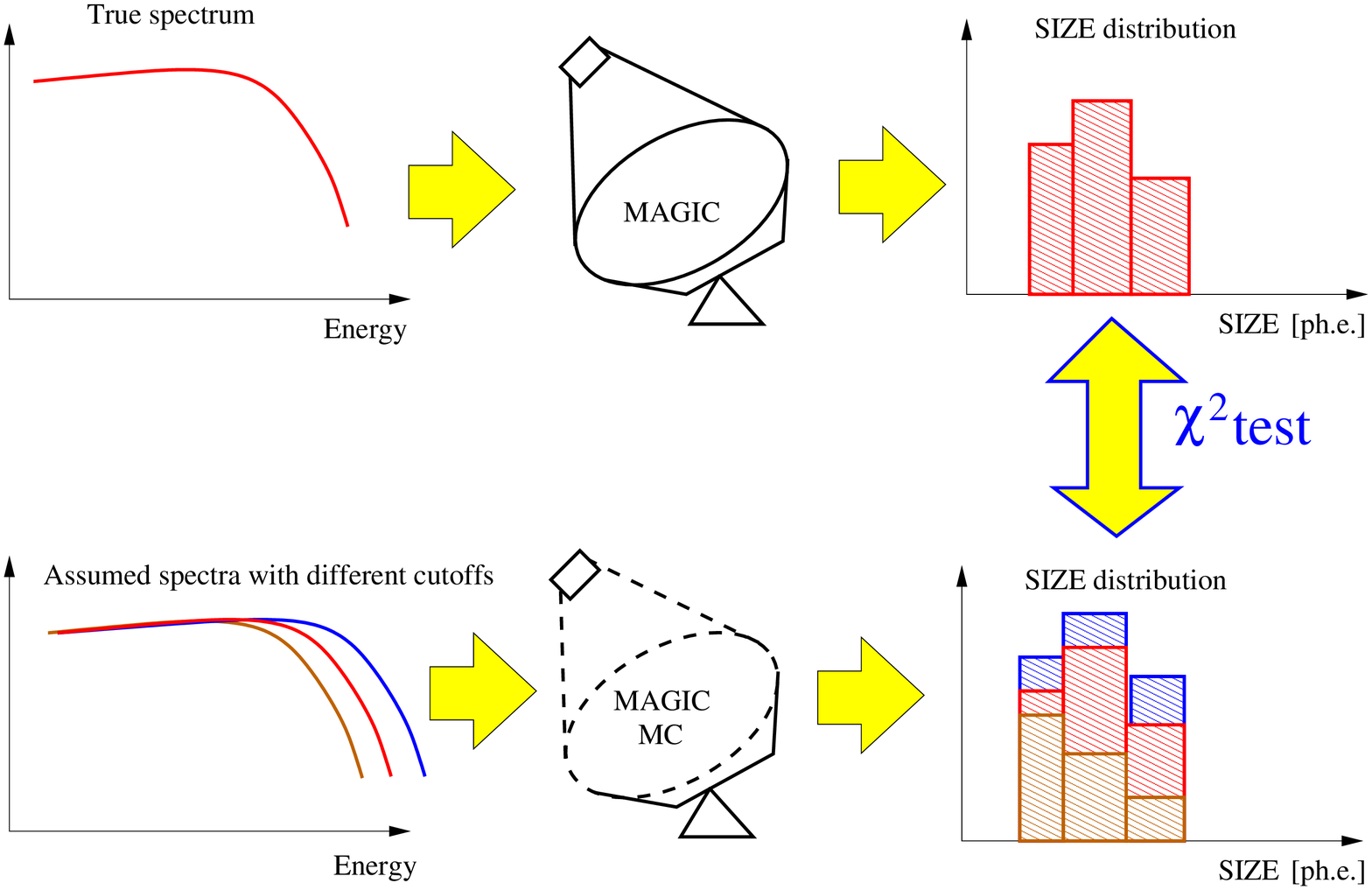}
\caption{Schematical explanation of the method for evaluating the inconsistency between
MAGIC and \fermi-LAT under the exponential cut-off spectrum assumption.
A power law with an exponential cut-off is assumed for the spectral shape in MC. 
A power law part is taken from the \fermi-LAT measurements,
while the cut-off energy is varied from 1 GeV to 25 GeV.
A $\chi^2$ test is performed between the $SIZE$ distribution of the observed data
and that of the MC for each cut-off energy. See text for details.
} 
\label{FigMethod}
\end{figure}

 The power law spectra with an exponential cut-off
for P1, P2 and P1~+~P2 obtained 
by \fermi-LAT data (see Table \ref{TableFermiPulsarSpectrum}) are assumed to be also valid 
above 25 GeV.
 By means of MC with these energy spectra, 
expected $SIZE$ distributions of excess events for P1, P2 and P1~+P2
 in the MAGIC data are computed.
Then, $\chi^2$ tests are performed between 
these MC predictions and the actual observed distributions.
The binning of the $SIZE$ distribution is 
$30 - 50$, $50-100$ and $100-400$, which is the same as for the spectrum calculations
(see Sect. \ref{SectExcessDist}) except that the highest two bins are combined so that
all bins have a meaningful number of excess events (with respect to the statistical uncertainty)
in the data.
Among the free parameters in the \fermi-LAT likelihood analysis, the uncertainty of the
 cut-off energy
 has the largest effect on the expected $SIZE$ distribution in the MAGIC data. 
Therefore, the analysis is repeated
while changing the cut-off energy from 1 GeV to 25 GeV.
This method is schematically shown in Fig. \ref{FigMethod}.

 It should be noted that this method does not involve the energy reconstruction,
which suffers the poor energy resolution and the trigger bias effect below 100 GeV 
(see Sect.\ref{SectUnFold}). $SIZE$ (total charge in a shower image)
is one of the most reliable image parameters and 
a good indicator of the gamma-ray energy. Therefore, a robust evaluation
of the deviation can be expected for this method. 
\clearpage

\subsubsection{Results of the $\chi^2$ Tests on $SIZE$ Distributions}
\begin{figure}[h]
\centering
\includegraphics[width=0.49\textwidth]{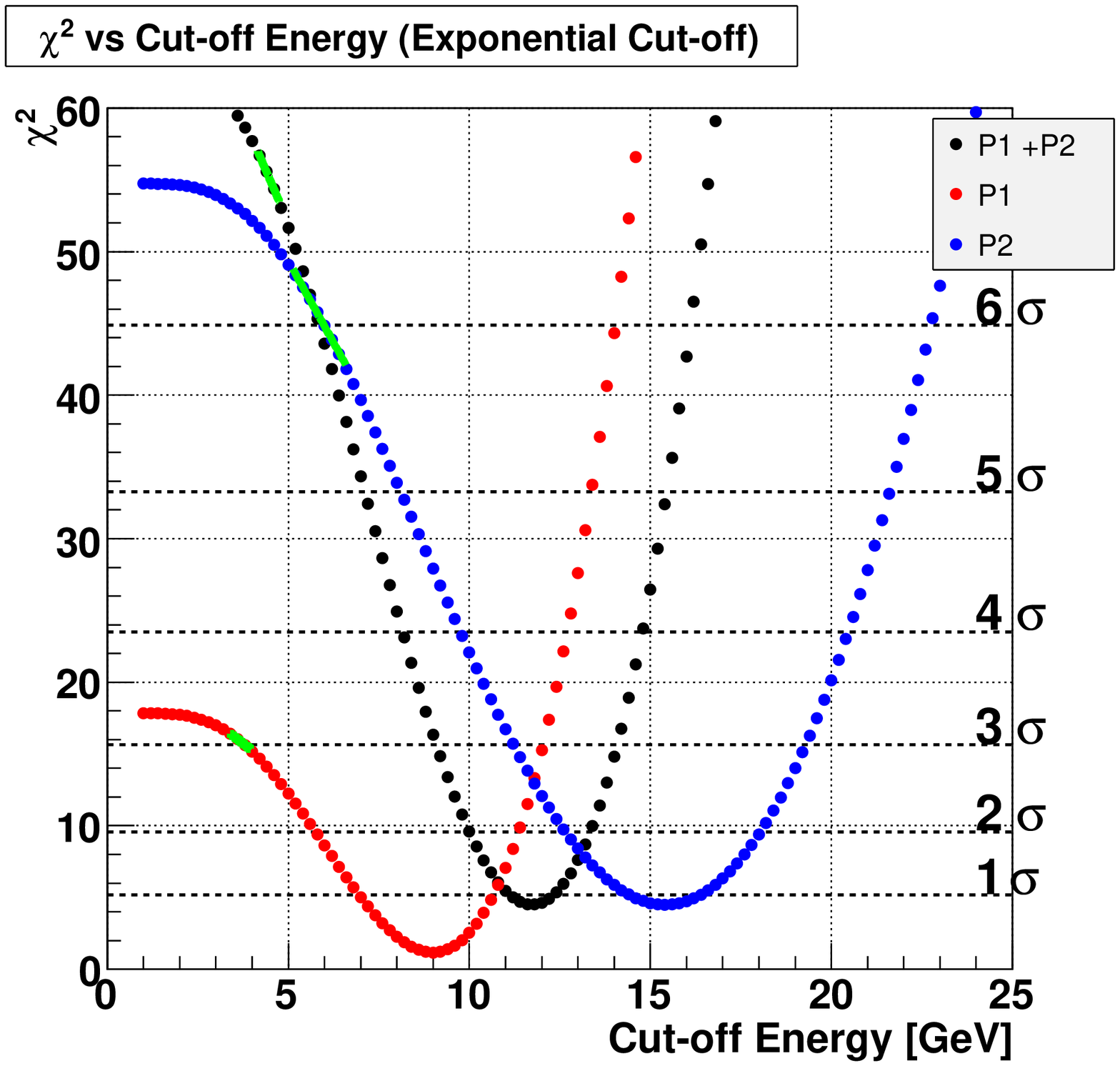}
\includegraphics[width=0.49\textwidth]{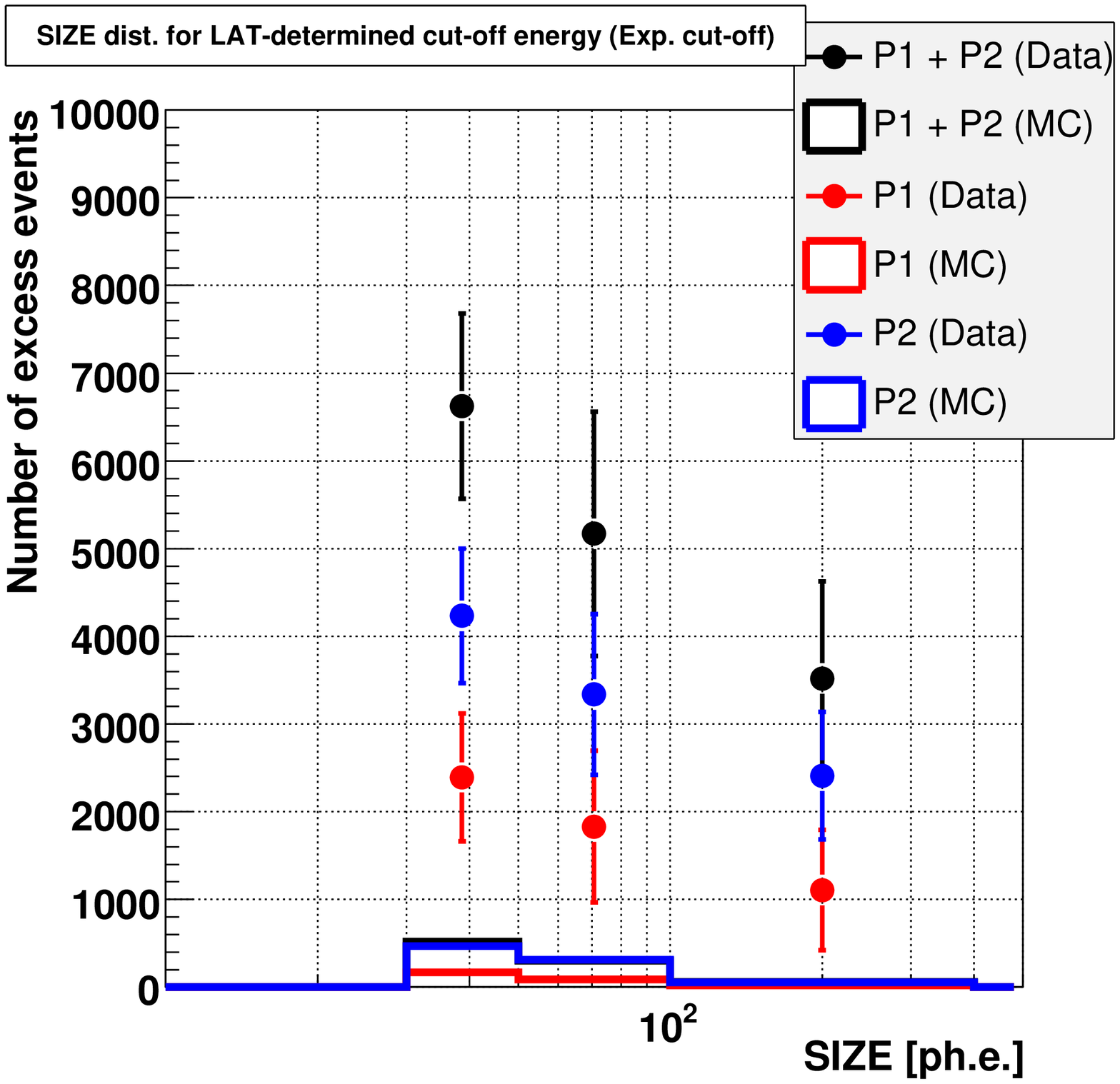}
\includegraphics[width=0.35\textwidth]{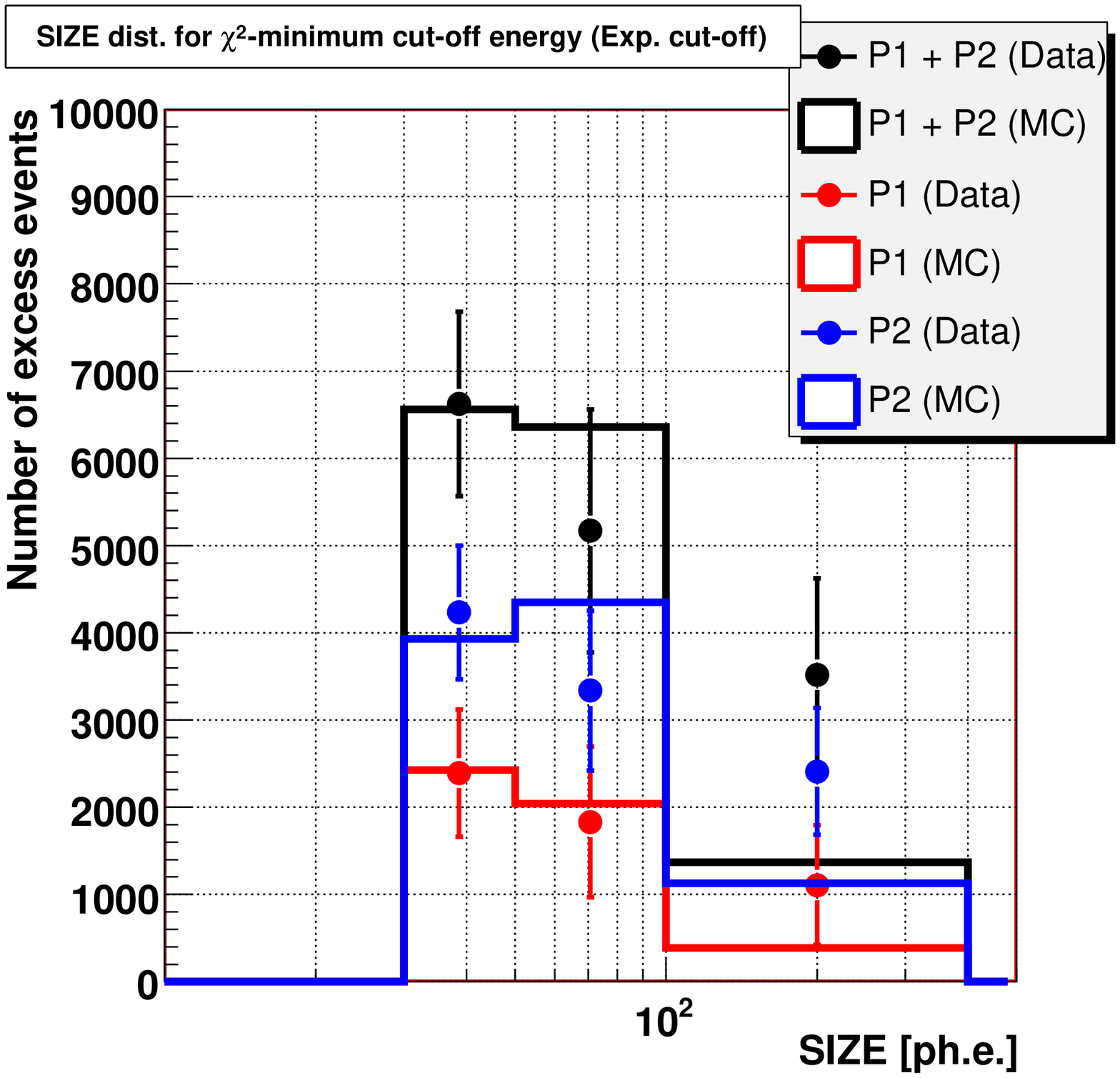}
\caption{The evaluation of the consistency between the \fermi-LAT 
and the MAGIC results under the exponential cut-off assumption. 
Top left: $\chi^2$ value based on the $SIZE$ distribution of the MC prediction and 
that of the observed data, as a function of the cut-off energy. 
The number of degree of freedom is three.
The corresponding upper probability is expressed with the corresponding Gaussian deviation
and is indicated by black dotted lines.
The black, the red and the blue points indicate P1 + P2, P1 and P2, respectively.
Green bars on the plot indicate the cut-off energy and its statistical uncertainty
obtained with the \fermi-LAT data.
Top Right: The comparison of the $SIZE$ distributions between the observed data and
MC predictions. The cut-off energies obtained from the \fermi-LAT data are used for MC.
Bottom: The same as the top right panel but the cut-off energies which minimize
the $\chi^2$ are used for MC.
} 
\label{FigSizeDistComp}
\end{figure}

The top left panel of Fig. \ref{FigSizeDistComp} shows the $\chi^2$ value as a function of
the cut-off energy. The number of degree of freedom is three.
The corresponding upper probability is expressed with the corresponding Gaussian deviation
and is indicated by black dotted lines.
The green lines on the plot indicate
the cut-off energies with statistical errors obtained from the \fermi-LAT data,
i.e.  (4.45 $\pm$ 0.31) GeV , (3.68 $\pm$ 0.29) GeV and  (5.86 $\pm$ 0.74) GeV
 for P1~+~P2, P1 and P2, respectively (see Table \ref{TableFermiPulsarSpectrum}).
 The top right panel of Fig. \ref{FigSizeDistComp} shows the comparison of 
$SIZE$ distributions between the observed data and 
the MC predictions for the \fermi-LAT-determined cut-off energies.
Based on these $SIZE$ distributions,
the inconsistencies between the exponential cut-off spectra determined by \fermi-LAT
and the MAGIC measurements are estimated to be
at $(6.77~\pm~0.13) \sigma$, $(3.01~\pm~0.06) \sigma$ and $(6.04~\pm~0.26) \sigma$ levels for P1~+~P2, P1 and P2, respectively.
The cut-off energies that minimize the $\chi^2$ values are estimated to be 
$(11.7 \pm 0.7)$ GeV, $(8.9 \pm 0.9)$ GeV and $(15.4 \pm 1.2)$ GeV for
P1~+~P2, P1 and P2, respectively, 
which are also in clear contradiction with the \fermi-LAT-determined cut-off energies.
The bottom panel shows the comparison of $SIZE$ distributions between the observed data and the 
MC predictions for the cut-off energies that minimize the $\chi^2$ values.
These results are summarized in Table \ref{TableConsEval}
\clearpage

\clearpage

\subsubsection{Effect of  Systematic Uncertainties}
As mentioned in Sect. \ref{SectMAGICSyst} and \ref{SectFermiPerformance}, 
the systematic uncertainties in the energy scale of MAGIC and \fermi-LAT 
above 10 GeV are 20\% and 7\%, respectively, while 
the systematic uncertainties in the effective area (including the effective observation time)
of MAGIC and \fermi-LAT above 10 GeV are 10\% and 20\%, 
respectively.
The apparent contradiction between \fermi-LAT and MAGIC measurements under the exponential 
cut-off assumption could be not due to the wrong assumption 
but due to these systematic uncertainties of the two experiments. 

In order to examine this possibility, I made the same analysis but scaling down 
the energy for MAGIC by 30\%, i.e., the spectrum $F_{MAGIC}(E)$ used in the MC is 
defined as 
\begin{equation}
\label{EqEScaleDown}
F_{MAGIC}(E) = F_{LAT}(0.7 E) 
\end{equation}
where $F_{LAT}(E)$ is the power law with an exponential cut-off
whose power law part is based on the \fermi-LAT measurements and whose
cut-off energy is a parameter ranging from 1 GeV to 25 GeV.
This 30\% would be very conservative compared to
the systematic uncertainties of both experiments.
On the other hand, no correction for the effective area is applied 
because the uncertainty of the energy scale should dominate
the effect on the results, due to the steep fall-off of the spectrum at MAGIC energies.

\begin{figure}[h]
\centering
\includegraphics[width=0.48\textwidth]{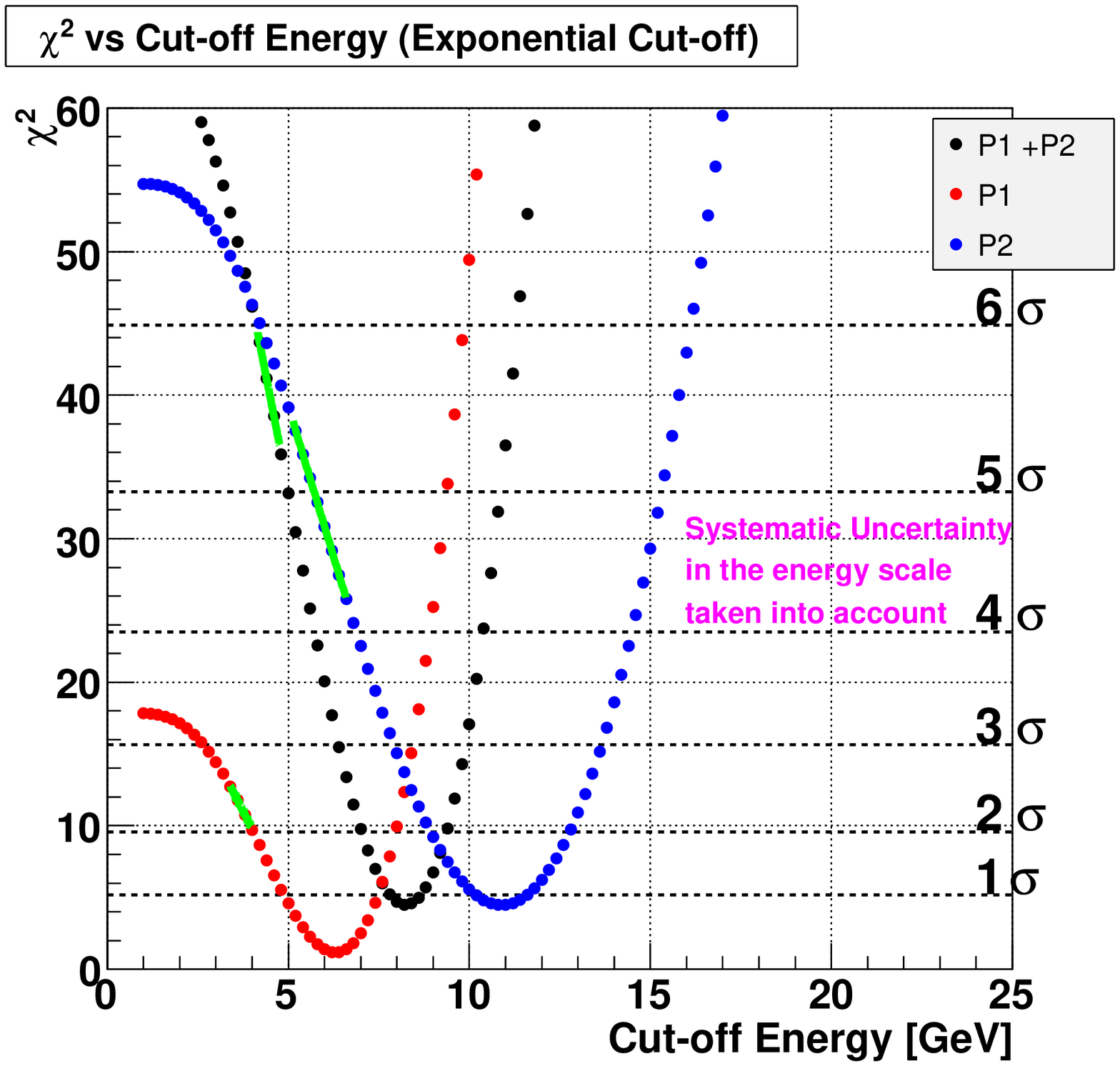}
\includegraphics[width=0.48\textwidth]{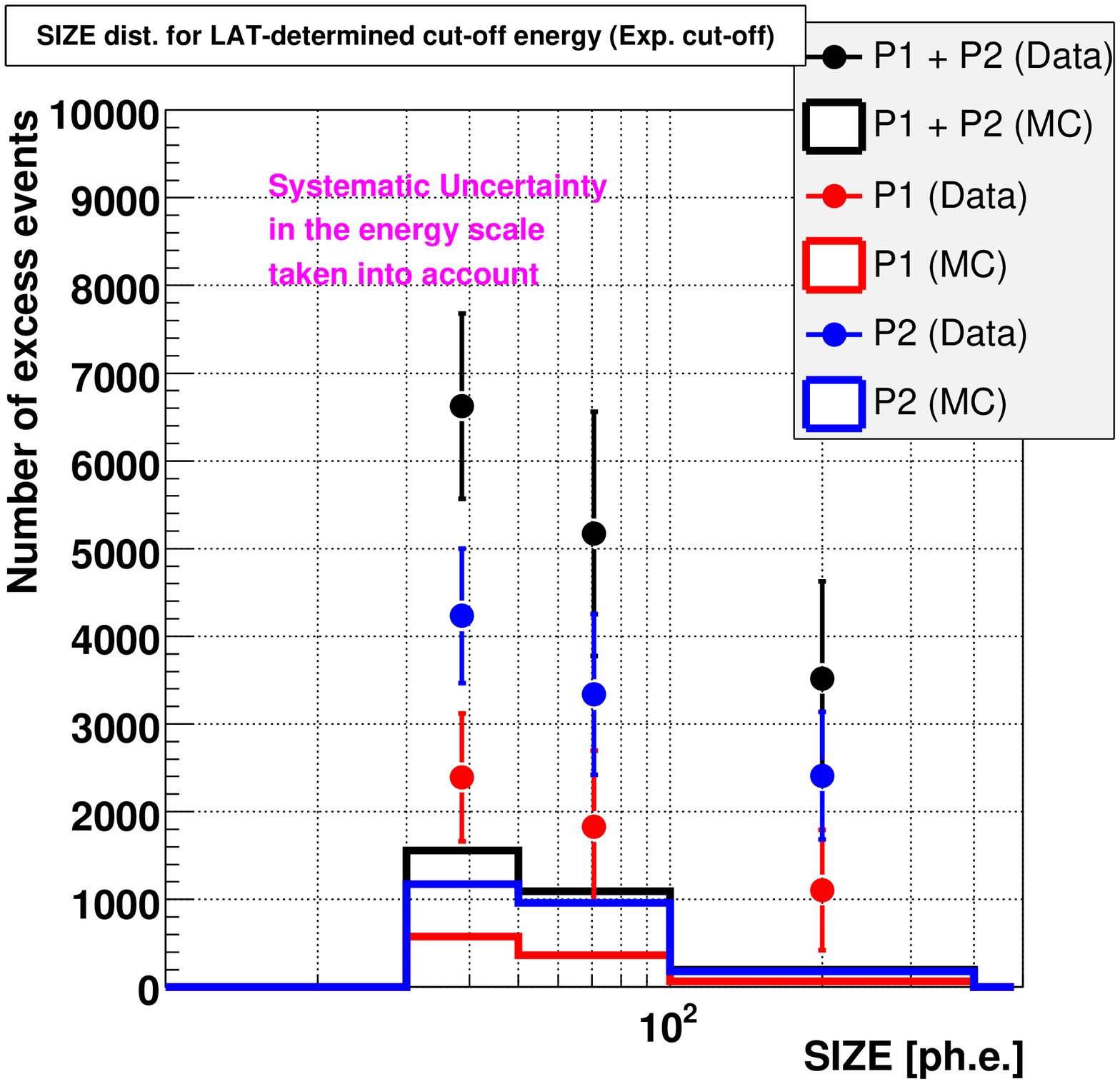}
\caption{Left: 
$\chi^2$ value based on the $SIZE$ distribution of the MC prediction and 
that of the observed data, as a function of the cut-off energy. 
The number of degree of freedom is three.
The corresponding upper probability is expressed with the corresponding Gaussian deviation
and is indicated by black dotted lines.
The black, the red and the blue points indicate P1 + P2, P1 and P2, respectively.
Green bars on the plot indicate the cut-off energy and its statistical uncertainty
obtained with the \fermi-LAT data. The energy scale of the MAGIC data are artificially
lowered by 30\% in order to examine the possibility that systematic uncertainties of 
both experiments is the reason for the discrepancy. 
Right: The comparison of the $SIZE$ distributions between the observed data and
the MC predictions for the cut-off energies obtained by \fermi-LAT data.
}
\label{FigSizeDistCompExpS0.7}
\end{figure}

 The $\chi^2$ value as a function of the cut-off energy is shown in the left panel
of Fig. \ref{FigSizeDistCompExpS0.7}.
The comparison of the $SIZE$ distributions between 
the observed data and the MC predictions for the \fermi-LAT-determined cut-off energies
 is shown in the right panel of the figure.
Even after 30\% of energy scaling, 
the discrepancies in the $SIZE$ distributions are $(5.63 \pm 0.33) \sigma$, $(2.32 \pm 0.25) \sigma$ and $(4.86 \pm 0.60)\sigma$ level for P1 + P2, P1 and P2,
 respectively.
The cut-off energies that minimize the $\chi^2$ values are 
$(8.2 \pm 0.5)$ GeV, $(6.1 \pm 0.6)$ GeV and $(10.9 \pm 0.9)$ GeV.
These values are also significantly inconsistent with the \fermi-LAT-determined values. 
These results are summarized in Table \ref{TableConsEval}

From these discrepancies, it is evident that the exponential cut-off assumption
is not valid at MAGIC energies ($>25$ GeV). The extension of the
pulsed gamma-ray emission observed by MAGIC requires a modification of the current 
pulsar models, which will be further discussed in Sect. \ref{SectElSpModi}. 



\clearpage

\subsection{How Much Do the MAGIC Measurements Deviate from an Sub-Exponential Cut-off Spectrum?}
\label{SectHowDifficult2}


As discussed in Sect.\ref{SectCutoff} and Sect. \ref{SectCutoffSteepness}, 
 nearly monoenergetic electrons (as many of theoretical models assume,
see e.g. \cite{Harding2008}, \cite{Tang2008}, \cite{Hirotani2008} and \cite{Takata2008}
) produce an exponential cut-off, i.e., 
$\Gamma_2$ in Eq. \ref{EqCutoffSpectra} is 1.
On the other hand, if $\Gamma_2$ is smaller than 1, 
the inconsistency between the \fermi-LAT measurements and the MAGIC measurements 
would become smaller
than for the exponential cut-off assumption.

Since \fermi-LAT results can also be explained by the sub-exponential cut-off assumption
($\Gamma_2 = 0.66$, see Sect. \ref{SectFermiPowerLawWithCutoff}), I examine the sub-exponential
cut-off assumption. It should be noted that $\Gamma_2 < 1$ implies that 
the responsible electrons are not monoenergetic. In this case, no theoretical
predictions for $\Gamma_2$ exist. I take the sub-exponential cut-off assumption ($\Gamma_2 = 0.66$)
as one of the most extreme assumptions 
\footnote{In the case of the Vela pulsar, 
the energy spectrum of the total pulse (phase 0 to 1) measured by \fermi-LAT
shows $\Gamma_2$ in Eq. \ref{EqCutoffSpectra} to be $0.69 \pm 0.02$ (see \cite{FermiVela}).
On the other hand, the energy spectrum from each narrow phase interval (width $\simeq 0.01$)
is still consistent with an exponential cut-off spectrum. In this analysis of the Crab
pulsar, 
the widths of phase intervals are 0.1, 0.11 and 0.21 for P1, P2 and P1~+~P2. 
Therefore, $\Gamma_2 = 0.66$ would be rather an extreme assumption. 
}


\subsubsection{Method: $\chi^2$ Test on $SIZE$ Distributions}
The method is the same as the one for the exponential cut-off assumption
(see Sect. \ref{SectHowDifficult}) except that $\Gamma_2 = 0.66$ (see Eq. \ref{EqCutoffSpectra})
is used for the steepness of the cut-off.

\subsubsection{Results of the $\chi^2$ Tests on $SIZE$ Distributions}
\begin{figure}[h]
\centering
\includegraphics[width=0.49\textwidth]{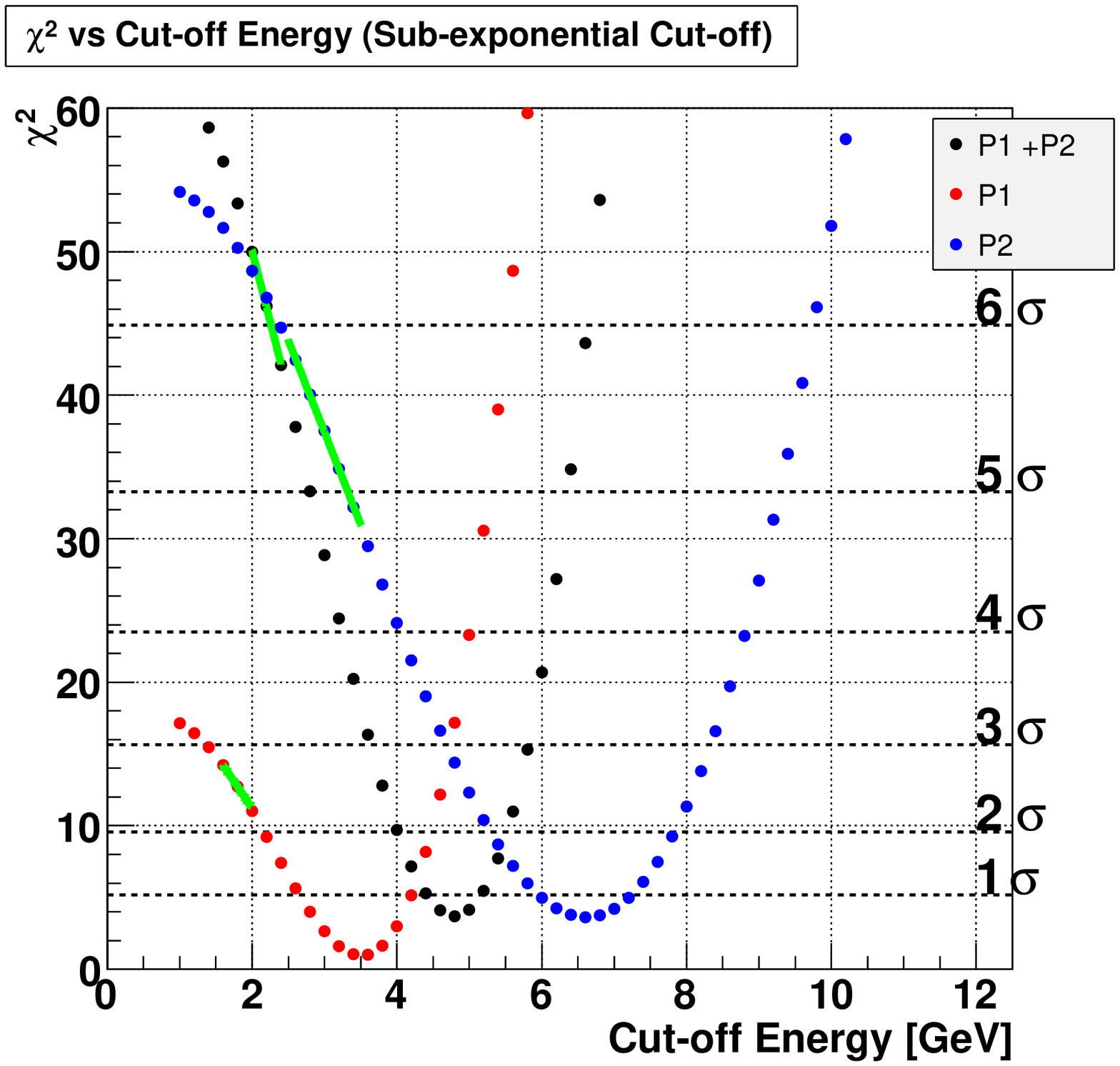}
\includegraphics[width=0.49\textwidth]{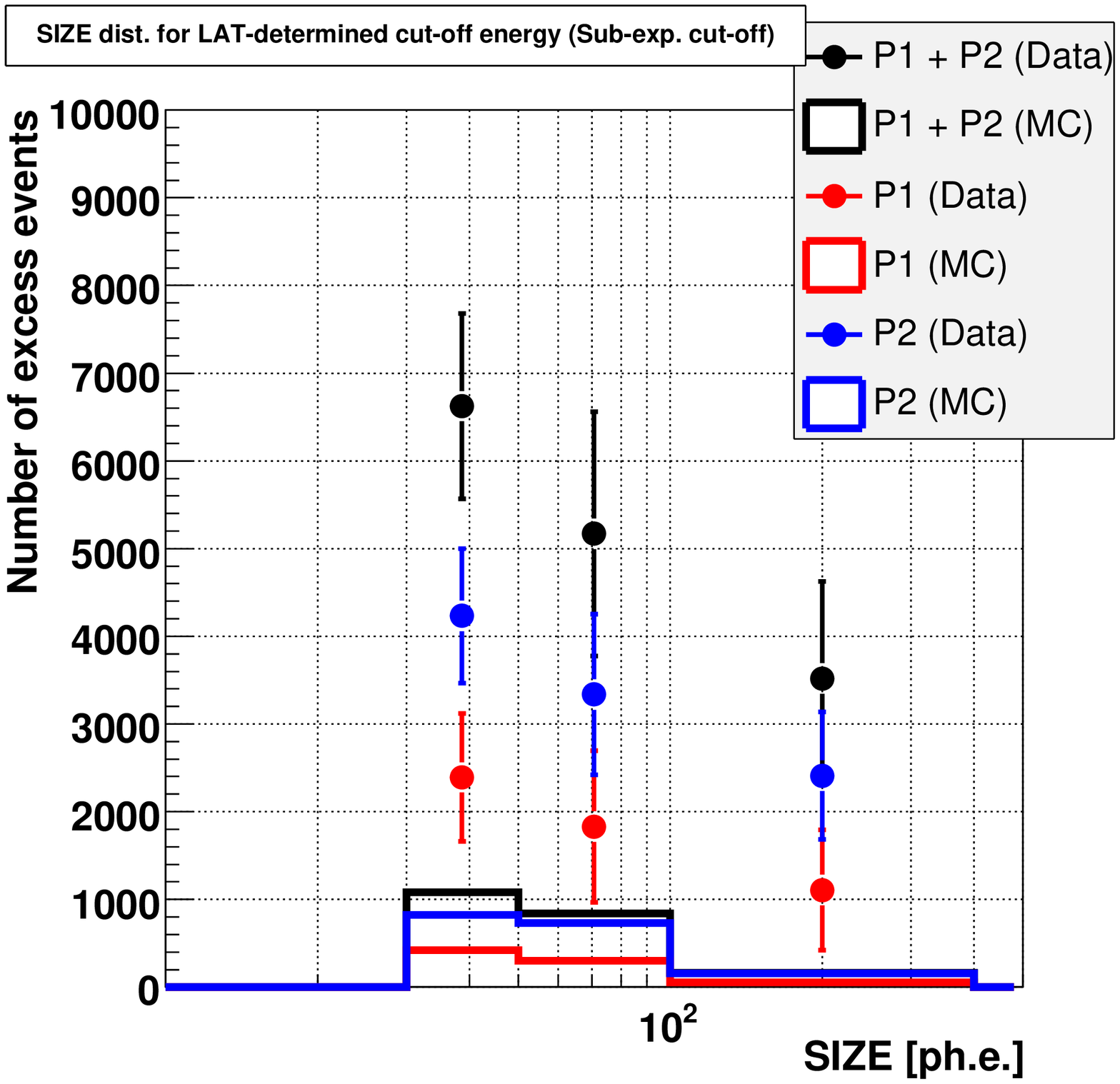}
\includegraphics[width=0.35\textwidth]{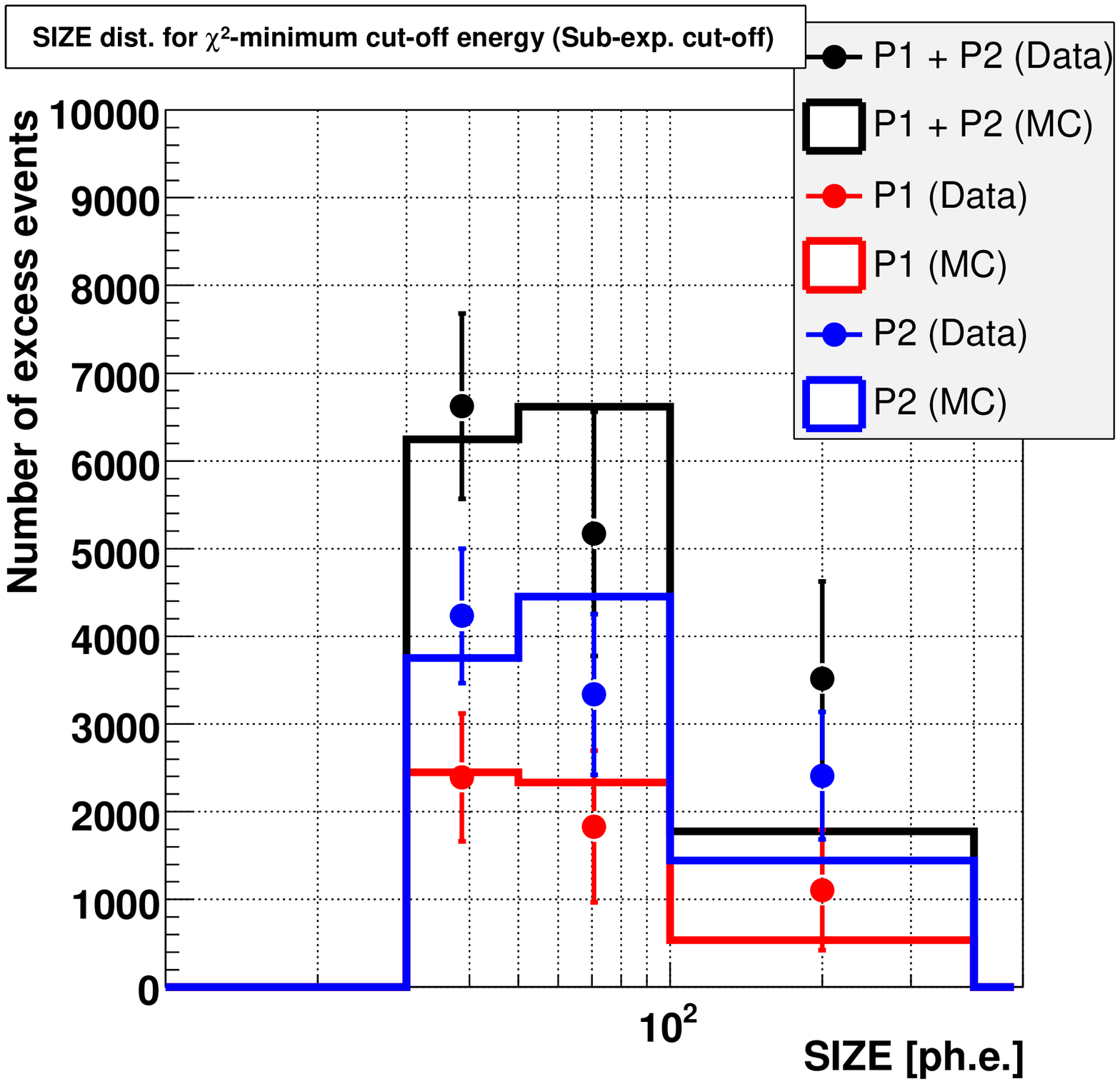}
\caption{The same as Fig. \ref{FigSizeDistComp} but for the sub-exponential cut-off assumption.}
\label{FigSizeDistComp0.66}
\end{figure}

The top left panel of Fig. \ref{FigSizeDistComp} shows the $\chi^2$ value as a function of
the cut-off energy. The number of degree of freedom is three.
The corresponding upper probability is expressed with the corresponding Gaussian deviation
and is indicated by black dotted lines.
The green lines on the plot indicate
the cut-off energies for the sub-exponential cut-off 
with statistical errors obtained from the \fermi-LAT data,
i.e. (2.20 $\pm$ 0.20) GeV , (1.79 $\pm$ 0.20) GeV and  (3.00 $\pm$ 0.51) GeV
 for P1~+~P2, P1 and P2, respectively (see Table \ref{TableFermiPulsarSpectrum}).
 The top right panel of Fig. \ref{FigSizeDistComp0.66} shows the comparison of 
$SIZE$ distributions between the observed data and 
the MC predictions for the \fermi-LAT-determined cut-off energies.
Although  
discrepancies are smaller than those for the exponential cut-off case because of the more gradual
 cut-off, 
the inconsistencies between the sub-exponential cut-off spectra determined by \fermi-LAT
and the MAGIC measurements are 
at 
$(6.04 \pm 0.26) \sigma$, $(2.55 \pm 0.23) \sigma$ and $(5.35 \pm 0.58) \sigma$ levels 
for P1~+~P2, P1 and P2, respectively. 
The cut-off energies that minimize the $\chi^2$ values are 
$(4.8 \pm 0.3)$ GeV, $(3.4 \pm 0.4)$ GeV and $(6.6 \pm 0.5)$ GeV for P1~+~P2, P1 and P2,
respectively, which are also in clear contradiction with 
the \fermi-LAT-determined cut-off energies. 
The bottom panel shows the comparison of $SIZE$ distributions between the observed data and the 
MC predictions for the cut-off energies that minimize the $\chi^2$ values.

\subsubsection{Effect of the systematic uncertainties}
The effect of the systematic uncertainties of the two experiments 
can be taken into account in the same way as for the exponential cut-off case, i.e.
the spectrum $F_{MAGIC}(E)$ used in the MC is scaled down by 30\%
(see Eq. \ref{EqEScaleDown}). 

\begin{figure}[h]
\centering
\includegraphics[width=0.48\textwidth]{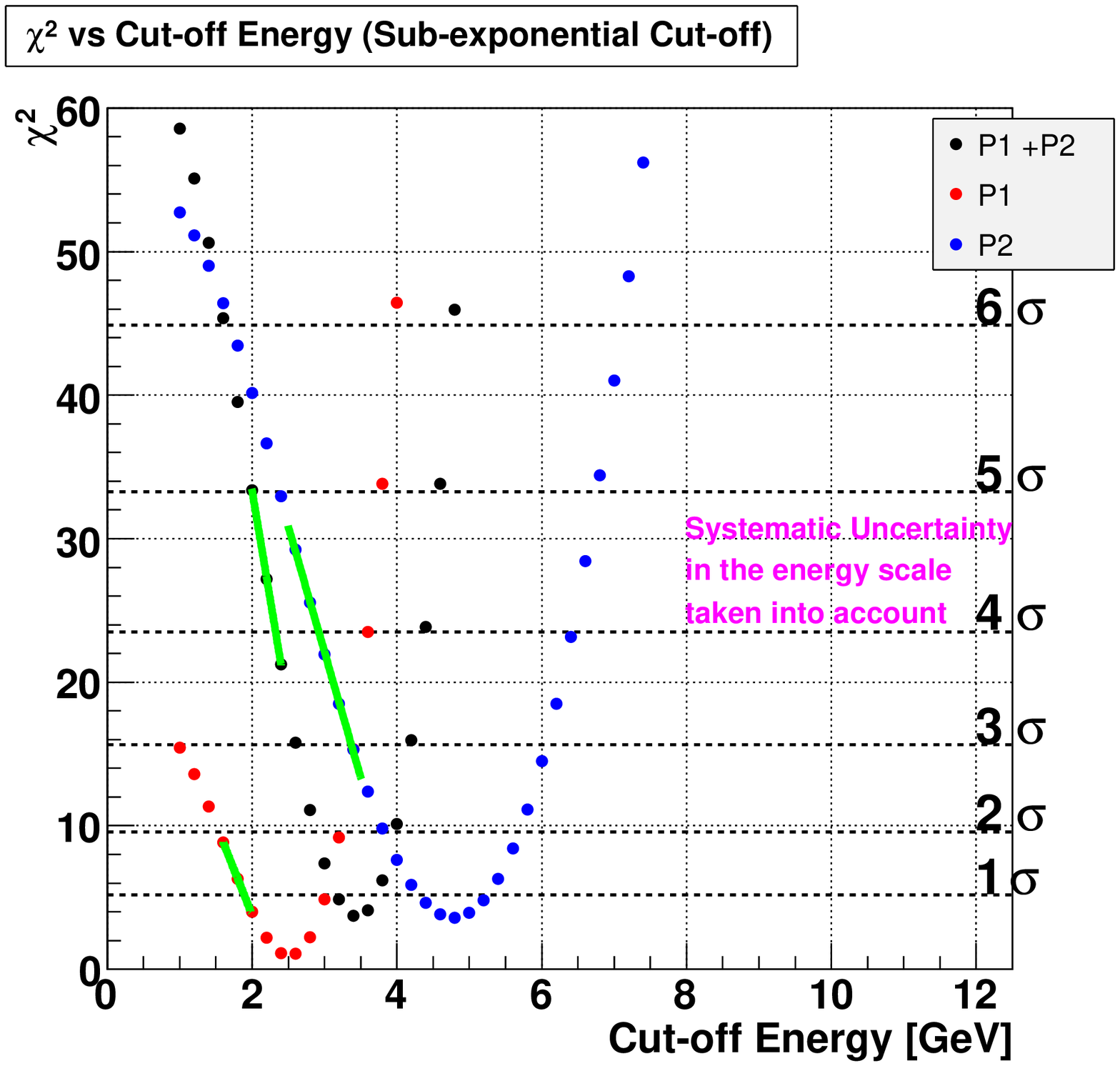}
\includegraphics[width=0.48\textwidth]{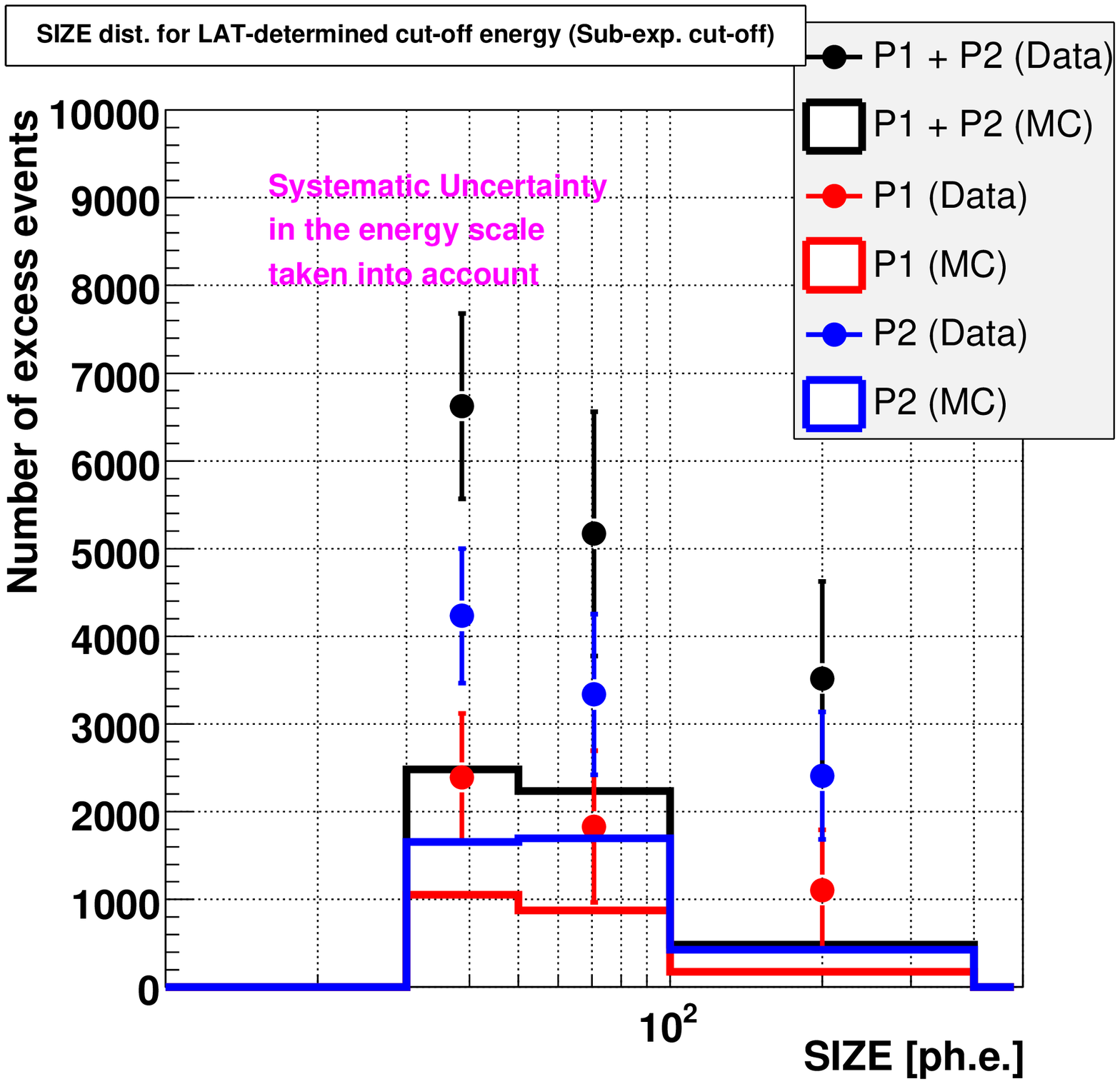}
\caption{
The same as Fig. \ref{FigSizeDistCompExpS0.7} but for the sub-exponential cut-off
assumption
}
\label{FigSizeDistComp0.66S0.7}
\end{figure}

 The $\chi^2$ value as a function of the cut-off energy
is shown in the left panel of Fig. \ref{FigSizeDistComp0.66S0.7}.
The comparison of the $SIZE$ distributions between 
the observed data and the MC predictions for the \fermi-LAT-determined cut-off energies
 are shown in the right panel of Fig. \ref{FigSizeDistComp0.66S0.7}.
After 30\% of energy scaling, 
the inconsistencies in the $SIZE$ distribution are at
$(4.37 \pm 0.65) \sigma$, $(1.25 \pm 0.62) \sigma$ and $(3.71 \pm 1.08)\sigma$ levels for P1 + P2, P1 and P2,
 respectively.
The sub-exponential cut-off spectrum might explain the energy spectrum of P1 if 
the systematic uncertainties of the two experiments are conservatively taken into account.
However, P1~+~P2 and P2 are still largely inconsistent.
The cut-off energies that minimize the $\chi^2$ values are 
$(3.39 \pm 0.23)$ GeV, $(2.40 \pm 0.26)$ GeV and $(4.77 \pm 0.38)$ GeV.
These values are still inconsistent with the \fermi-LAT-determined values.
These results are summarized in Table \ref{TableConsEval}.

The power law with a sub-exponential cut-off is not valid 
 for P1~+~P2 and P2 at the MAGIC energies ($>25$ GeV).

\clearpage

\begin{table}
\begin{threeparttable}
\begin{tabular}{|c|c|r|r|r|r|r|}
\hline
Model & Phase & $E_c^{Fermi}$ [GeV] \tnote{a}
& $E_c^{M}$ [GeV] \tnote{b}& Scale \tnote{c} & $\chi^2$ from $SIZE$ \tnote{d} 
& Rejection Power \tnote{e} \\
\hline
\hline
            & P1 + P2& 4.45 $\pm$ 0.31 & 11.68$\pm$ 0.74  & 1  & 53.4 -- 57.0  &6.6 -- 6.9 $\sigma$ \\ \cline{2-7}
Exponential & P1     & 3.68 $\pm$ 0.29 & 8.86$\pm$ 0.91  & 1  & 15.3 -- 16.4  &3.0 -- 3.1 $\sigma$ \\ \cline{2-7}
Cut-off      & P2    & 5.86 $\pm$ 0.74 & 15.39$\pm$ 1.19  & 1  & 42.1 -- 48.8  &5.8 -- 6.3  $\sigma$ \\ \cline{2-7}
\hline
\hline
Sub-        & P1 +P2& 2.20 $\pm$ 0.20 & 4.77$\pm$ 0.31  & 1  & 42.1 -- 50.2  &5.8 -- 6.4 $\sigma$ \\ \cline{2-7}
exponential & P1 & 1.79 $\pm$ 0.20 & 3.43$\pm$ 0.36  & 1  & 11.3 -- 14.2  &2.3 -- 2.8 $\sigma$ \\ \cline{2-7}
Cut-off      & P2 & 3.00 $\pm$ 0.51 & 6.59$\pm$ 0.52  & 1  & 30.8 -- 44.0  &4.8 -- 5.9 $\sigma$ \\ \cline{2-7}
\hline
\multicolumn{7}{c}{~} \\
\hline
            & P1 + P2& 4.45 $\pm$ 0.31 & 8.17$\pm$ 0.53  & 0.7  & 36.5 -- 44.4  & 5.3 -- 6.0 $\sigma$ \\ \cline{2-7}
Exponential & P1 & 3.68 $\pm$ 0.29 & 6.12$\pm$ 0.63  & 0.7  & 9.9 -- 12.8  & 2.1 -- 2.6 $\sigma$ \\ \cline{2-7}
Cut-off      & P2 & 5.86 $\pm$ 0.74 & 10.86$\pm$ 0.85  & 0.7  & 25.8 -- 38.3  & 4.3 -- 5.5 $\sigma$ \\ \cline{2-7}
\hline
\hline
Sub-        & P1 + P2& 2.20 $\pm$ 0.20 & 3.39$\pm$ 0.23  & 0.7  & 21.1 -- 33.5  & 3.7 -- 5.0 $\sigma$ \\ \cline{2-7}
exponential & P1 & 1.79 $\pm$ 0.20 & 2.40$\pm$ 0.26  & 0.7  & 4.0 -- 8.9  &  0.6 -- 1.9 $\sigma$ \\ \cline{2-7}
Cut-off      & P2 & 3.00 $\pm$ 0.51 & 4.77$\pm$ 0.38  & 0.7  & 13.2 -- 31.0  & 2.6 -- 4.8 $\sigma$ \\ \cline{2-7}
\hline
 \end{tabular}
\begin{tablenotes}
{\footnotesize
\item[a]  The cut-off energy determined by the \fermi-LAT data
\item[b] The cut-off energy estimated by the $SIZE$ distribution in MAGIC data, 
assuming that the power law part determined by the \fermi-LAT data is valid.
\item[c] The energy scaling factor. This factor is applied to the MAGIC energy. 
\item[d] $\chi^2$ value calculated from the $SIZE$ distributions of the observed
data and the MC prediction. The number of degree of freedom is three.
\item[e] The probability that the two results are consistent,
expressed in the corresponding Gaussian deviation. 
 This is calculated from the $\chi^2$ value.
}
\end{tablenotes}
\end{threeparttable}
 \caption{ Evaluation of the inconsistency between the MAGIC measurements
and the exponential/sub-exponential cut-off spectra determined by \fermi-LAT.
}
\label{TableConsEval}
\end{table}

\clearpage

\subsection{Power Law Fit above 4 GeV}
\label{SectPL4GeV}
The super-exponential cut-off assumption is ruled out by the \fermi-LAT results alone
(see Sect. \ref{SectFermiPowerLawWithCutoff}).
Moreover, 
it is now evident that neither the exponential cut-off nor the sub-exponential cut-off can explain 
the observational results of \fermi-LAT and MAGIC consistently. 

On the other hand, a power law can well describe both the \fermi-LAT results above 4 GeV
(see Sect. \ref{SectFermiPL}) and MAGIC results between 25 GeV and 100 GeV (see Sect. \ref{SectMAGICspectrum}),
as can be seen in Fig. \ref{FigSpectraComp}.
In this section, an assumption that 
the spectra above 4 GeV follow a power law is examined.

\subsubsection{Method: Combined Fit to \fermi-LAT and MAGIC Data}
A power law function is fitted to the data points of \fermi-LAT and MAGIC together.
In order to take into account a possible energy scale difference between the
two experiments, the correction factor $\kappa$ is introduced to the power law function as:
\begin{eqnarray} 
\frac{{\rm d}^3F(E)}{{\rm d}E{\rm d}A {\rm d}t} &=& f_{10} (a E/10{\rm GeV})^{\Gamma} \nonumber \\
a &=& \left\{
\begin{array}{cc}
1 &(E \leq 25 {\rm GeV}) \\
\kappa & (E > {\rm 25 GeV}) \\
\end{array}
\right.
\label{EqScaling}
\end{eqnarray} 

\fermi-LAT points are statistically independent of each other but MAGIC data points are correlated,
due to the unfolding procedure. This correlation will be taken into account when
fitting is applied. Firstly, $\kappa$ is fixed to 1 and then, secondly, $\kappa$ is treated as a 
free parameter.

\subsubsection{Results of Combined Fit to \fermi-LAT and MAGIC Data}

\begin{figure}[h]
\centering
\includegraphics[width=0.45\textwidth]{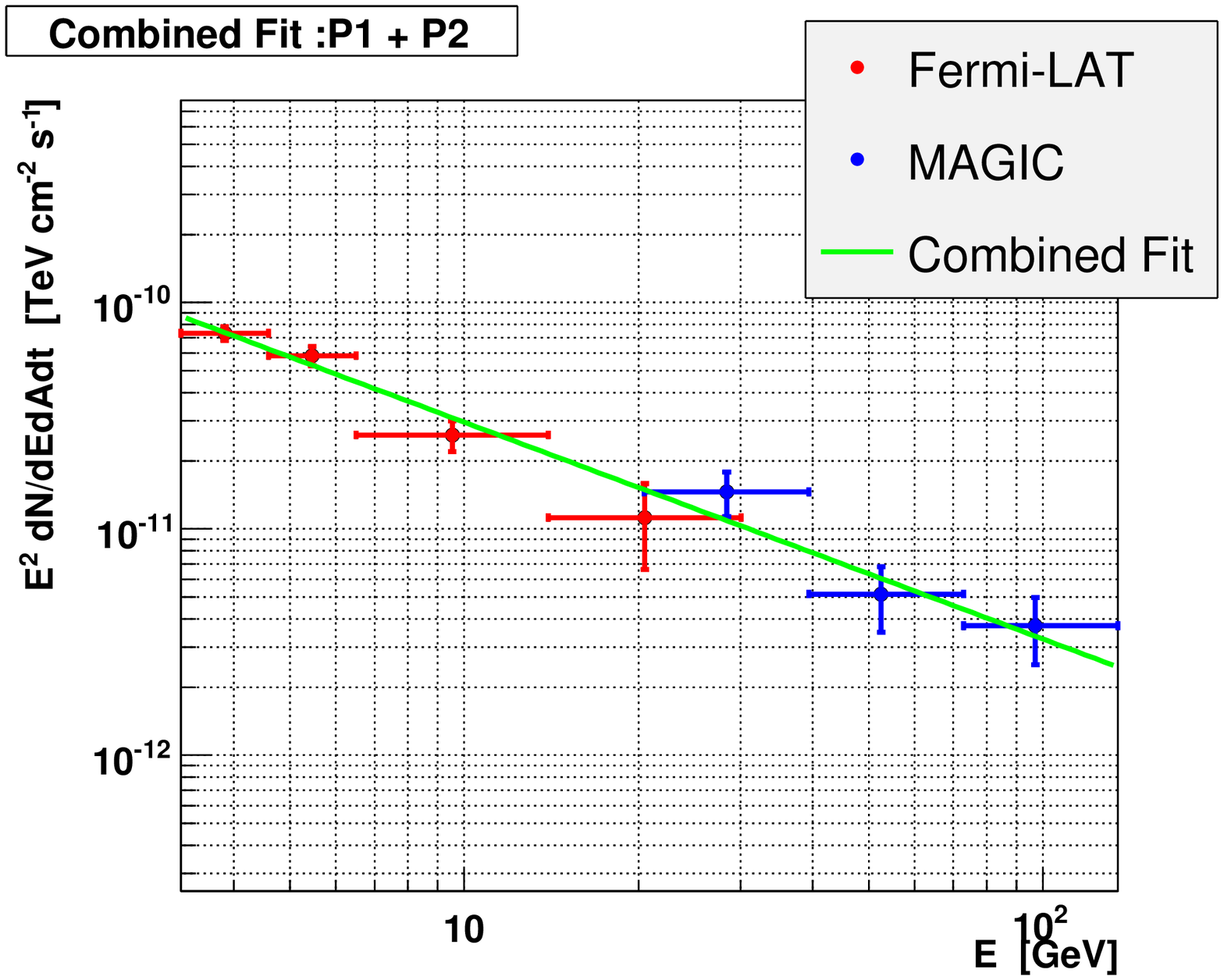}
\includegraphics[width=0.45\textwidth]{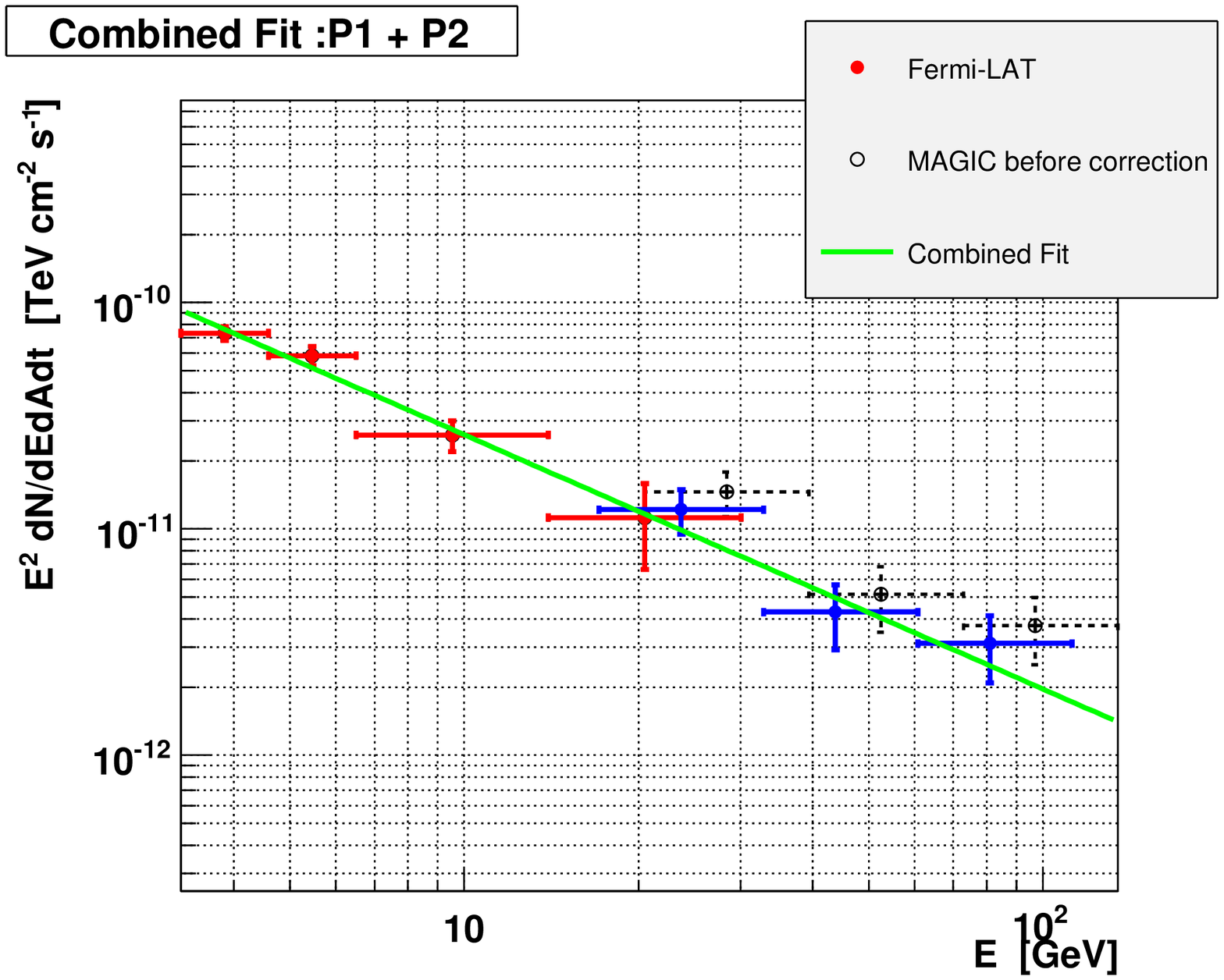}
\includegraphics[width=0.45\textwidth]{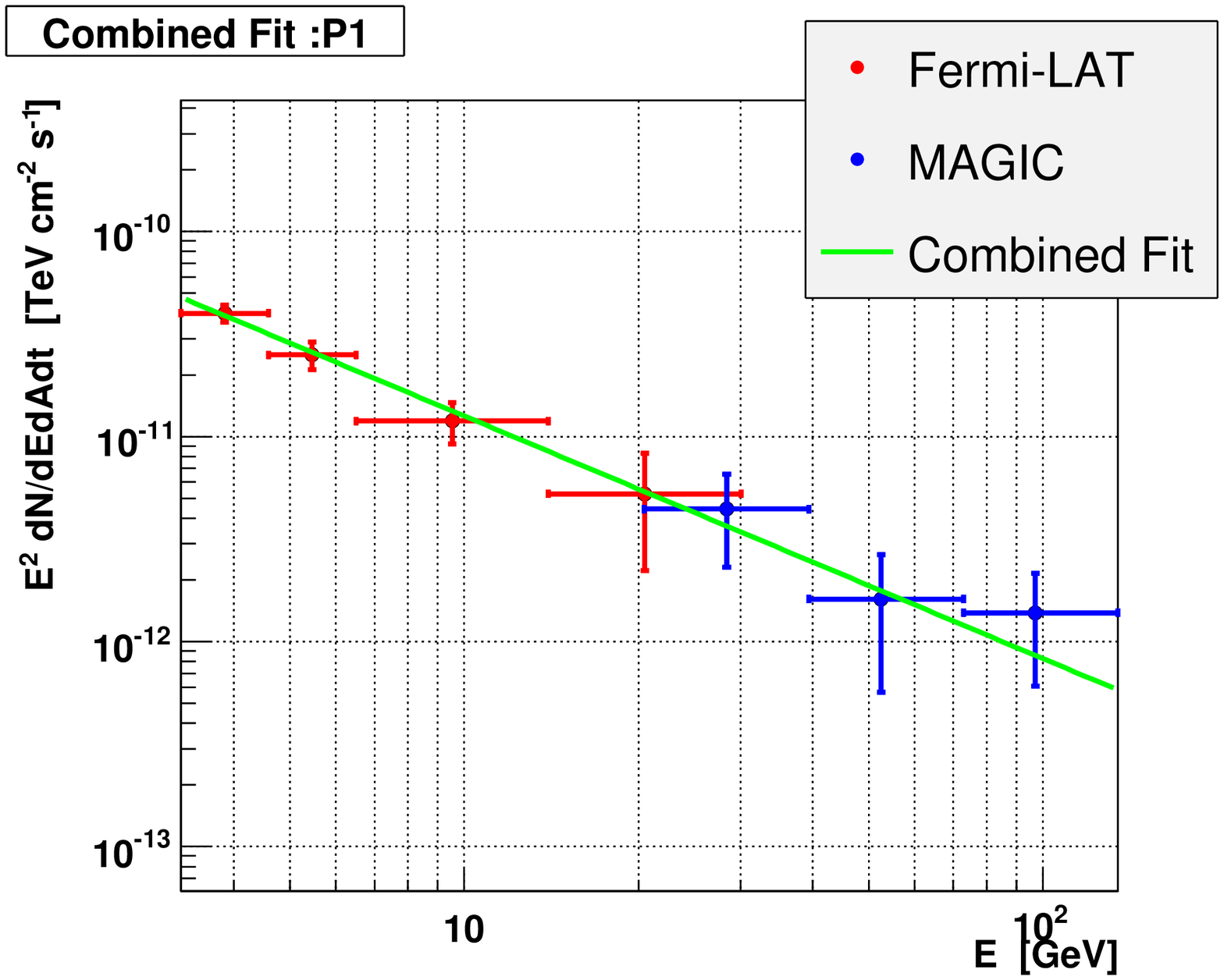}
\includegraphics[width=0.45\textwidth]{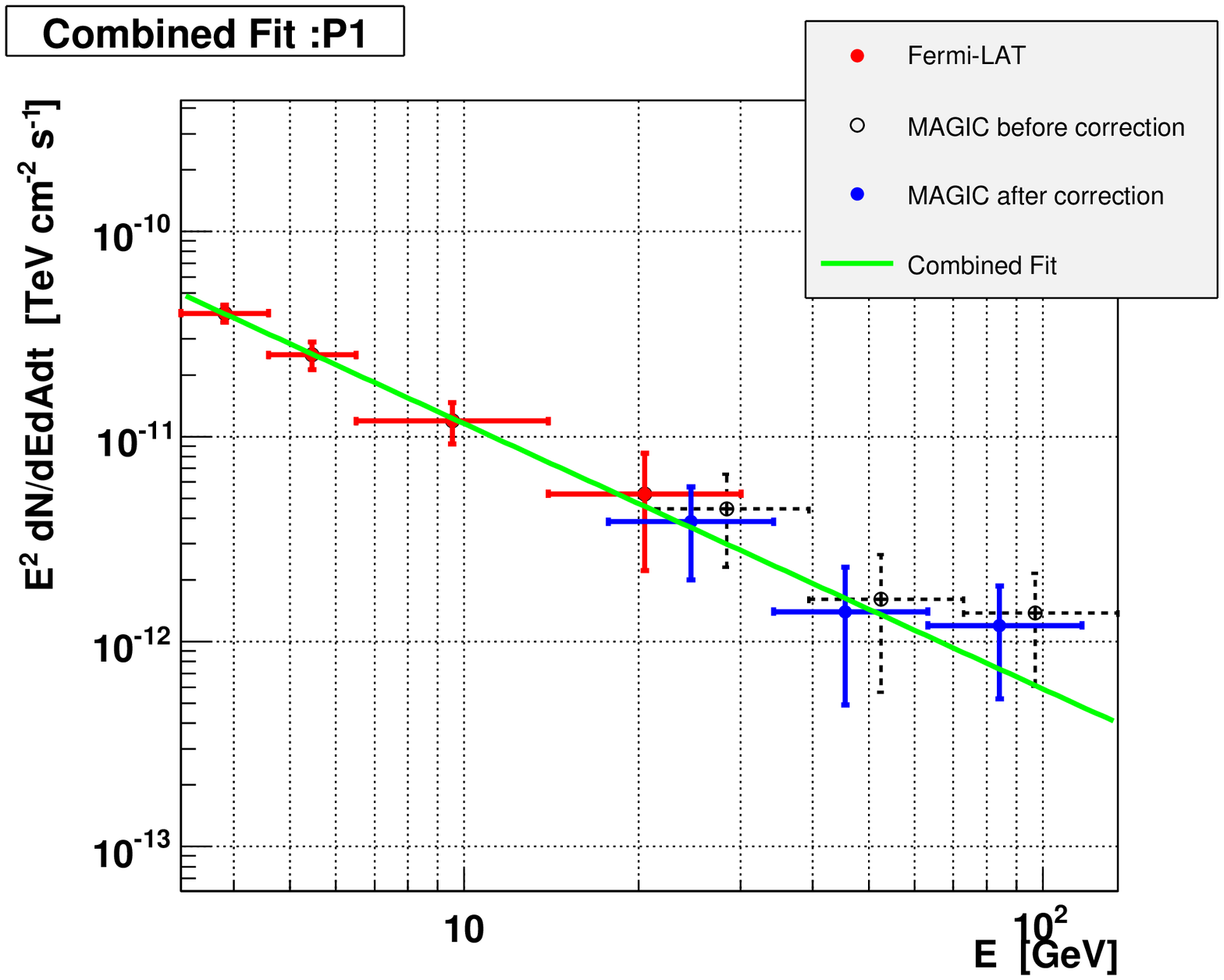}
\includegraphics[width=0.45\textwidth]{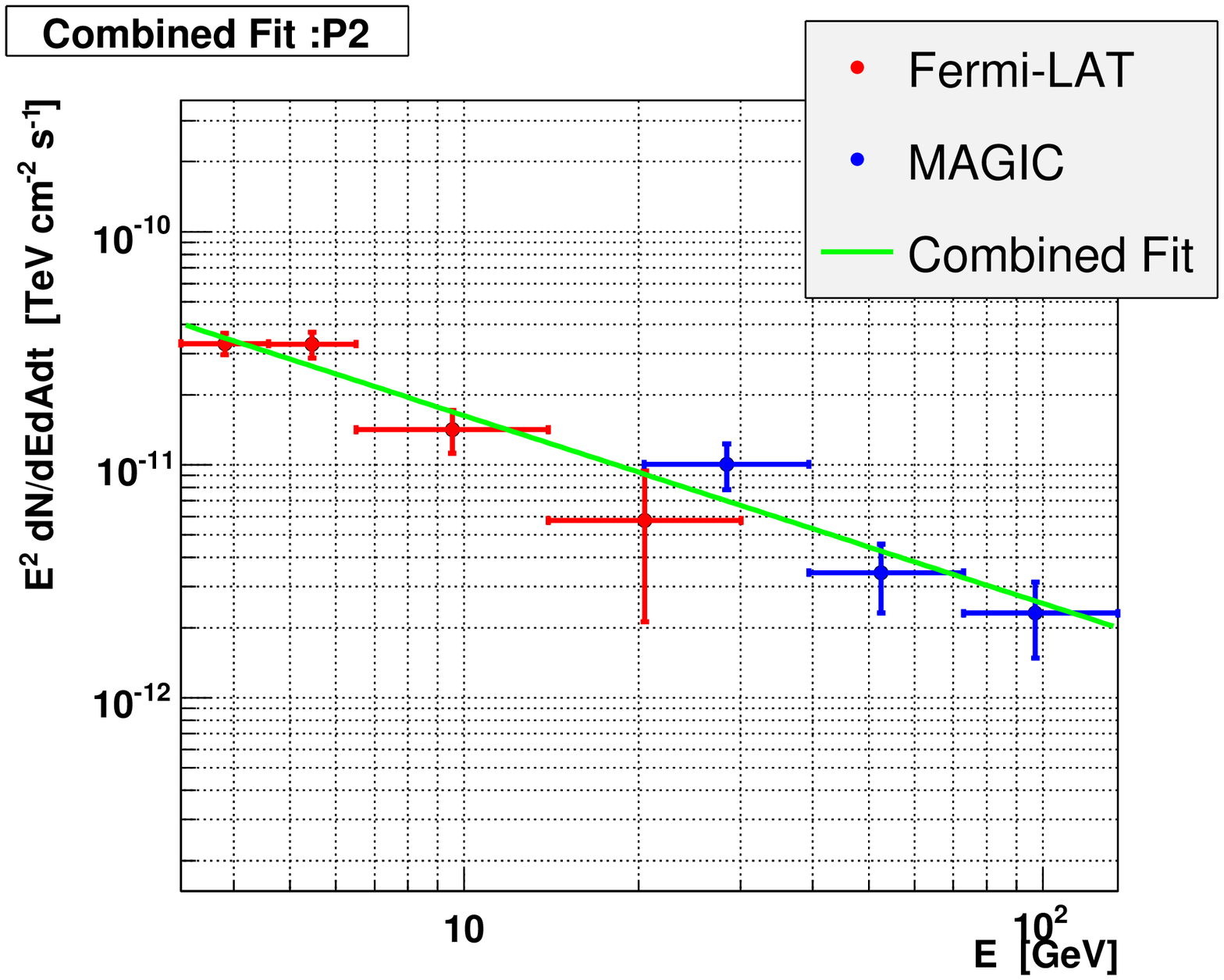}
\includegraphics[width=0.45\textwidth]{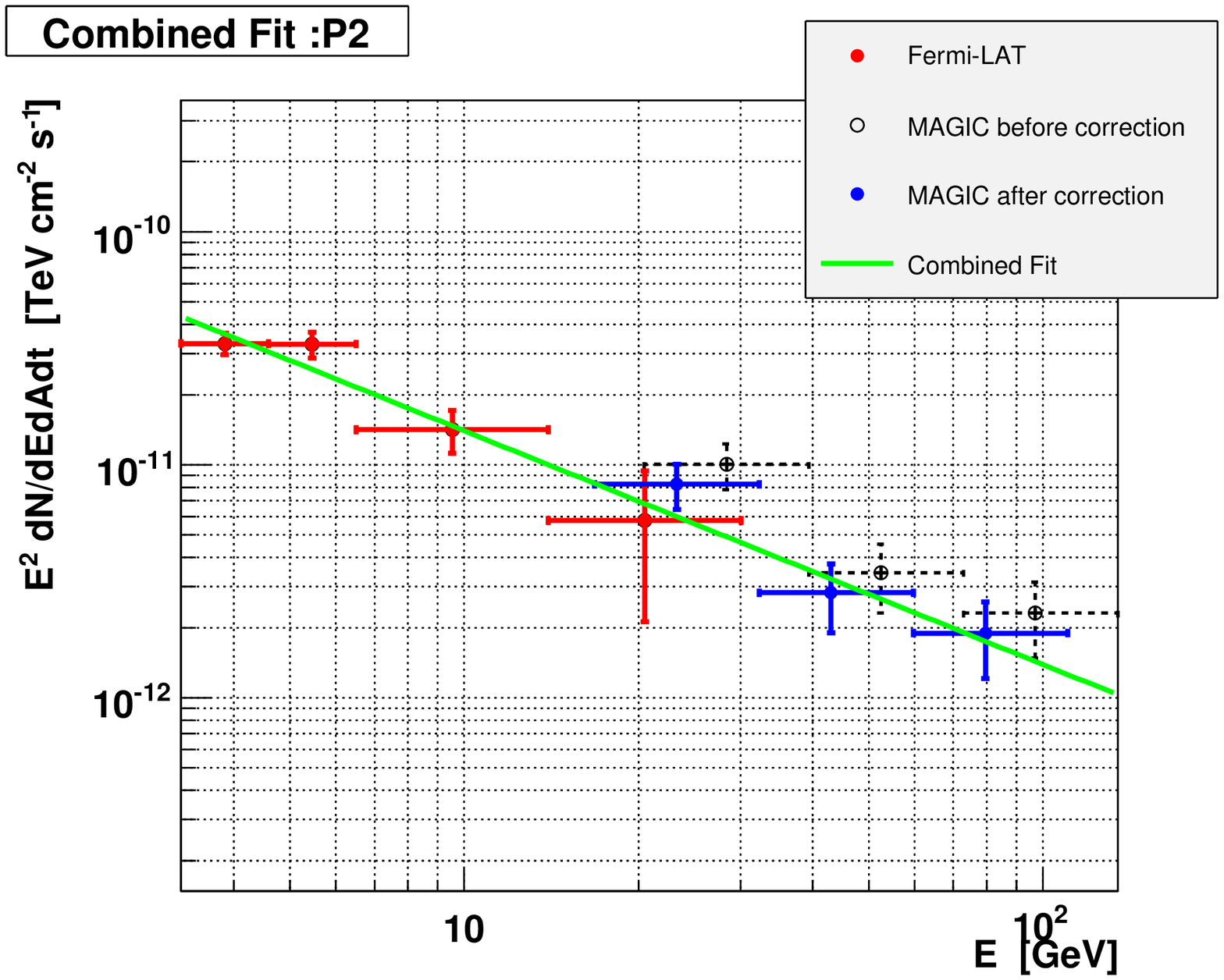}
\caption{
The power law fitting, combining the \fermi-LAT data above 4 GeV and the MAGIC data.
The correlation among the MAGIC points is taken into account.
The left panels show the case when the energy scales of both experiments are used
while the right panels show the case when a shift in 
the energy scale of MAGIC is allowed by letting 
$\kappa$ in Eq. \ref{EqScaling} be a free parameter. Top, middle and bottom panels show
P1 + P2, P1 and P2, respectively.
The best fit parameters and fitting probabilities are shown in Table \ref{TabCombFitPL}.
The fitted lines without the energy scaling (left panels) are also shown 
in Fig. \ref{FigSpectraComp} as black dotted lines.
}
\label{FigCombFitPL}
\end{figure}

The results are shown in Fig. \ref{FigCombFitPL} and Table \ref{TabCombFitPL}.
Even without relative energy correction, i.e. with $\kappa$ fixed to 1, 
the fitting probability is high enough (see the fifth column of the Table). 
By making $\kappa$ a free parameter, $\chi2$ values are only slightly reduced and 
the obtained $\kappa$s do not significantly deviate from 1. As a result, 
$f_{10}$ and $\Gamma$ do not change significantly. $\Gamma$ is about 3 for all the phase
intervals
and a significant difference is not seen. 

 The physics interpretation of this result will be discussed in Sect.\ref{SectElSpModi}.

\begin{table*}[h]
\centering
 \begin{tabular}{|c|c|c|c|c|}
\hline
\hline
Phase & $f_{10}$ [10$^{-7}$ cm$^{-2}$s$^{-1}$TeV$^{-1}]$ & $\Gamma$ & Scaling Factor $\kappa$& $\chi^2 / dof (Prob.)$ \\
\hline
P1 + P2 & 2.96 $\pm$ 0.20 & -2.96 $\pm$ 0.08 & 1 & 8.09 / 5 (0.15)\\
P1 & 1.26 $\pm$ 0.16 & -3.18 $\pm$ 0.16 &  1 & 1.98 / 5 (0.85) \\
P2 & 1.63 $\pm$ 0.13 & -2.81 $\pm$ 0.08 &  1 & 9.29 / 5 (0.10) \\
\hline
P1 + P2 & 2.61 $\pm$ 0.30 & -3.12 $\pm$ 0.14 & 0.83 $\pm$ 0.09 & 5.54 / 4 (0.24) \\
P1 & 1.16 $\pm$ 0.21 & -3.29 $\pm$ 0.22 & 0.87 $\pm$ 0.15 & 1.40 / 4 (0.84) \\
P2 & 1.40 $\pm$ 0.21 & -3.00 $\pm$ 0.18 & 0.82 $\pm$ 0.12 & 7.56 / 4 (0.11) \\
\hline
 \end{tabular}
 \caption{The power law fit combining the \fermi-LAT data above 4 GeV
and the MAGIC data}
\label{TabCombFitPL}
 \end{table*}

\clearpage

\section{P2/P1 Ratio and Bridge/P1 Ratio}
\label{SectP1P2Bridge}

\begin{figure}[h]
\centering
\includegraphics[width=0.7\textwidth]{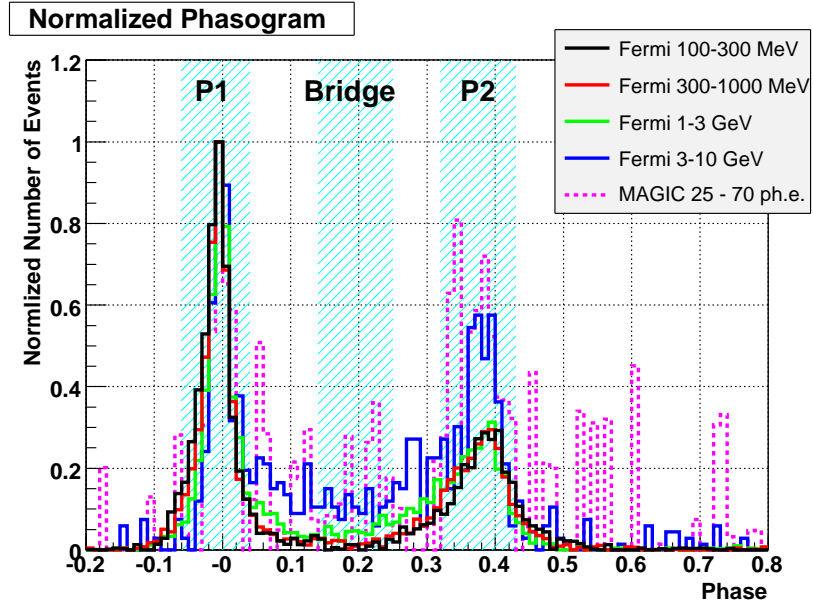}
\caption{Overlaid light curves from different energies.
The vertical values are normalized such that the heights of the P1 peak 
are equal. The energy dependence of the P2/P1 ratio and Bridge/P1 ratio 
are clearly seen.}
\label{FigEneDepRatio}
\end{figure}

\begin{table*}[h]
\centering
 \begin{tabular}{|c|c|c|c|c|}
\hline
\hline
Name & P1 & Bridge & P2 & OP \\
\hline
Phase Interval  & -0.96 to 0.04 & 0.14 to 0.25 & 0.32 to 0.43  & 0.52 to 0.88 \\
\hline
\hline
 \end{tabular}
 \caption{The Definition of Names of Pulse Phases}
\label{TabPhaseName2}
 \end{table*}

The light curves for different energies such as Fig. \ref{FigSizeDepPhasogram} 
and Fig. \ref{FigFermiPhaseEne} 
suggest that the flux ratio between P1 and P2 changes with energy.
 The fraction of 
the Bridge emission seems to be energy-dependent, too. 
Some of the light curves from different energies are overlaid in Fig. \ref{FigEneDepRatio}
in order to show the energy dependence of the P2/P1 ratio and the Bridge/P1 ratio.
The P2/P1 ratio and/or the Bridge/P1 ratio have been studied
by many authors such as \cite{Kuiper2001}, \cite{Toor1977}, \cite{Mineo1997}, \cite{Massaro1998}
and \cite{Massaro2000} for a wide energy range from optical to gamma-rays. 
On the one hand, the energy dependence of these ratios can be thought of as a consequence
 of the
 different energy spectra for different phase intervals. On the other hand, taking 
the flux ratio between two phase intervals cancels out
the systematic uncertainty of the absolute flux scale for different detectors.
Therefore, these ratios enable to study precisely
the relative spectral behavior of different phase intervals 
for a very wide energy range using the measurements
from many different detectors.
 
 Here, I calculate the P2/P1 and Bridge/P1 ratios for \fermi-LAT and MAGIC data and compare 
them with the lower energy results. 

\subsubsection{Method: Calculation of Energy-dependent P2/P1 and Bridge/P1 Ratios from Light Curves}
 \fermi-LAT data sets are divided into 5 bins in energy, 100 - 300 MeV, 
300 - 1000 MeV, 1 - 3 GeV, 3 - 10 GeV and above 10 GeV.
 MAGIC data are divided into two bins
in $SIZE$, 25 - 70 ph.e. and 70 - 500 ph.e.. 
The number of excess events for P1, P2 and Bridge
are calculated from light curves of each energy bin, by estimating background by OP phases 
(For the definition of P1, P2, Bridge and OP, see Sect. \ref{SectPhaseNaming}
and Table \ref{TabPhaseName2}.).
Then, the ratio of P2 to P1 and that of Bridge to P1 are
computed. 

\subsubsection{Results of Energy-dependent P2/P1 and Bridge/P1 Ratios }
\label{SectRatioResult}
\begin{figure}[h]
\centering
\includegraphics[width=0.45\textwidth]{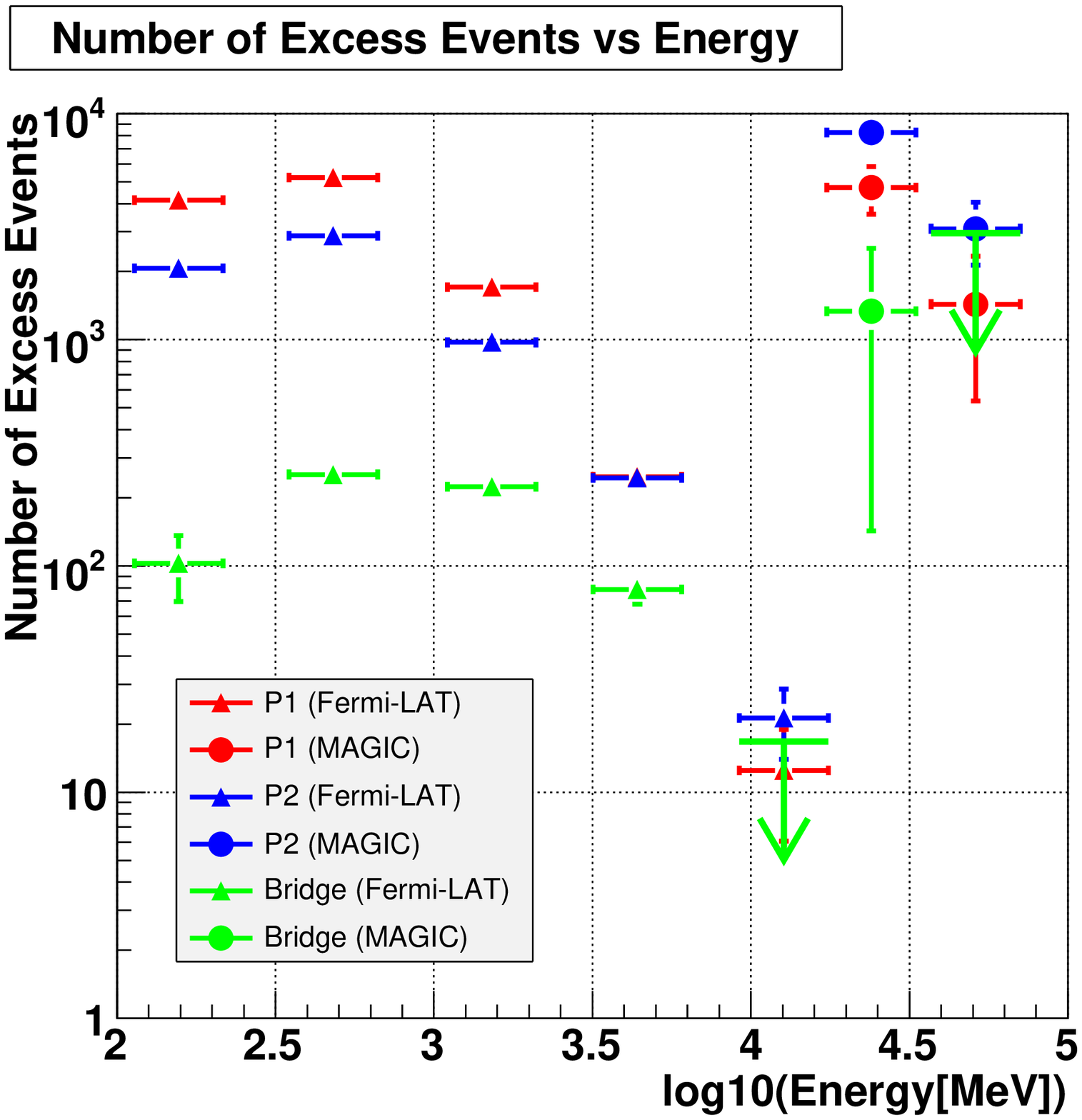}
\includegraphics[width=0.45\textwidth]{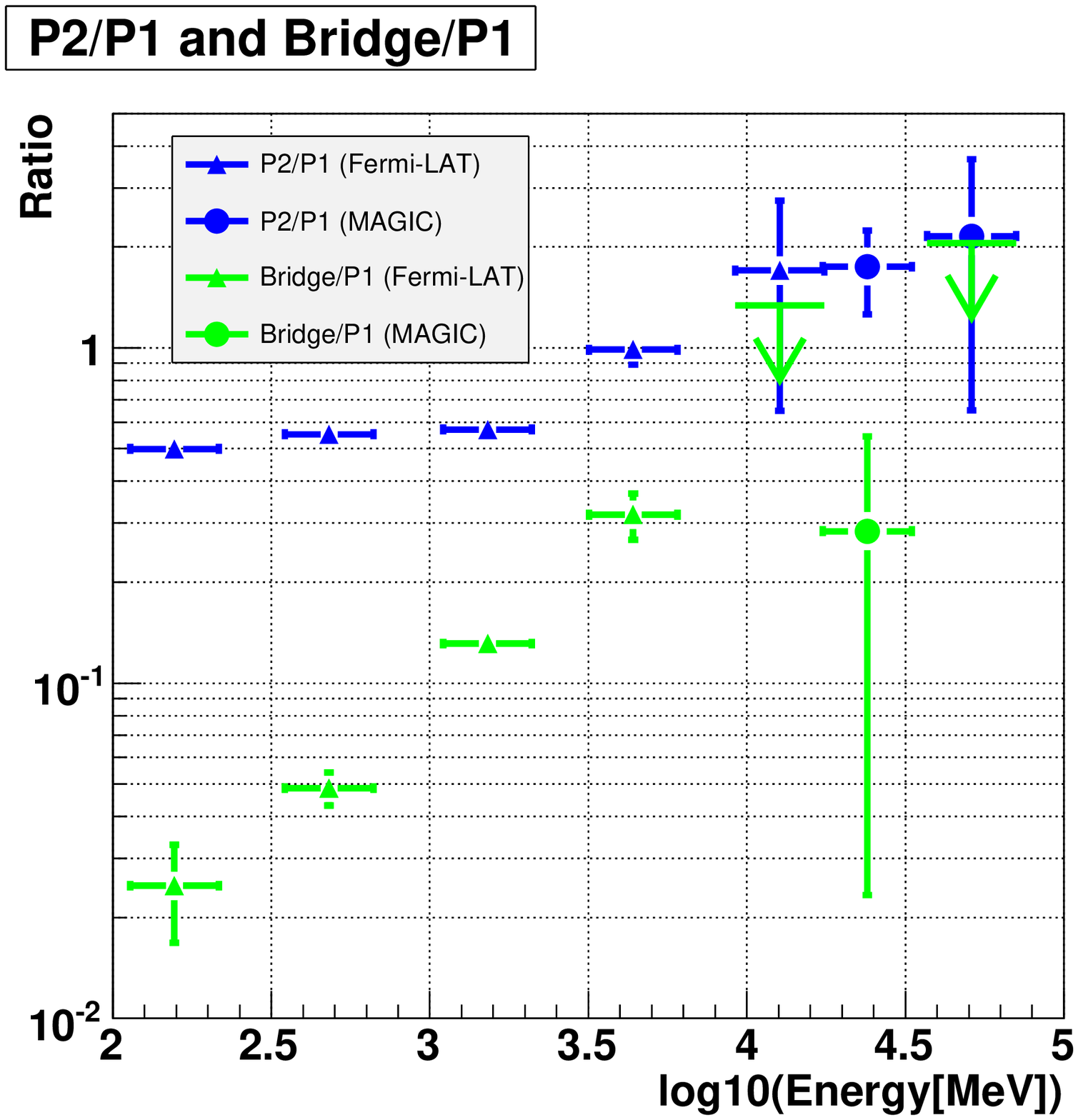}
\caption{Left: The number of excess events as a function of energy.
The red, blue, and green points denote P1, P2 and Bridge, respectively.
The last two points above 20 GeV are based on MAGIC data, while the rest are based on
\fermi-LAT data.
Right: The P2/P1 ratio (blue) and the Bridge/P1 ratio (green), as a function of energy.
The last two points above 20 GeV are based on MAGIC data, while the rest are based on
\fermi-LAT data
}
\label{FigP1P2Ratio}
\end{figure}

The numbers of excess events of P1, P2 and Bridge as a function of energy
are shown in the left panel of Fig. \ref{FigP1P2Ratio} and Table \ref{TabP1P2Ratio}. 
In the energy ranges of above 10 GeV for \fermi-LAT and 70 - 500 ph.e. for MAGIC,
no significant excess was found in Bridge phases. Therefore, 95\% confidence level
upper limits are shown for these bins.  
The horizontal central value of each point corresponds to the
log-mean energy of each bin, taking into account the detector's effective area and the 
energy spectrum
 of the pulsar, while the horizontal error bar corresponds to 30\% of the log-mean energy, 
which is a conservative estimation for the absolute energy uncertainty. 
The P2/P1 ratio stays almost constant from 100 MeV to 3GeV and rises 
at energies above 3 GeV.
On the other hand, the Bridge/P1 ratio rises from 100 MeV to 3 GeV.
Above 3 GeV, due to the lack of statistics, it is not possible
to draw a conclusion in the behavior. 

\begin{figure}[h]
\centering
\includegraphics[width=0.45\textwidth]{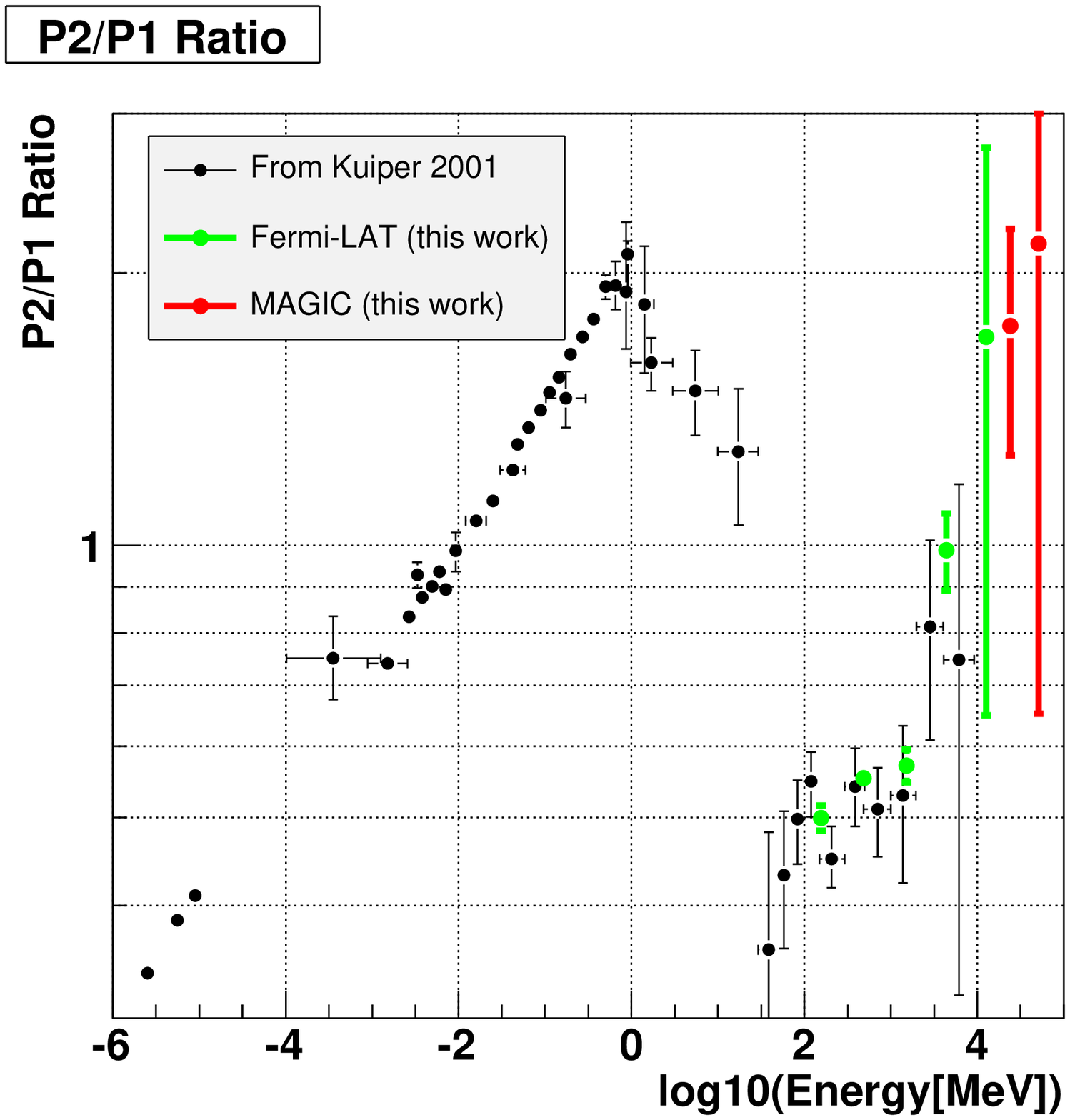}
\includegraphics[width=0.45\textwidth]{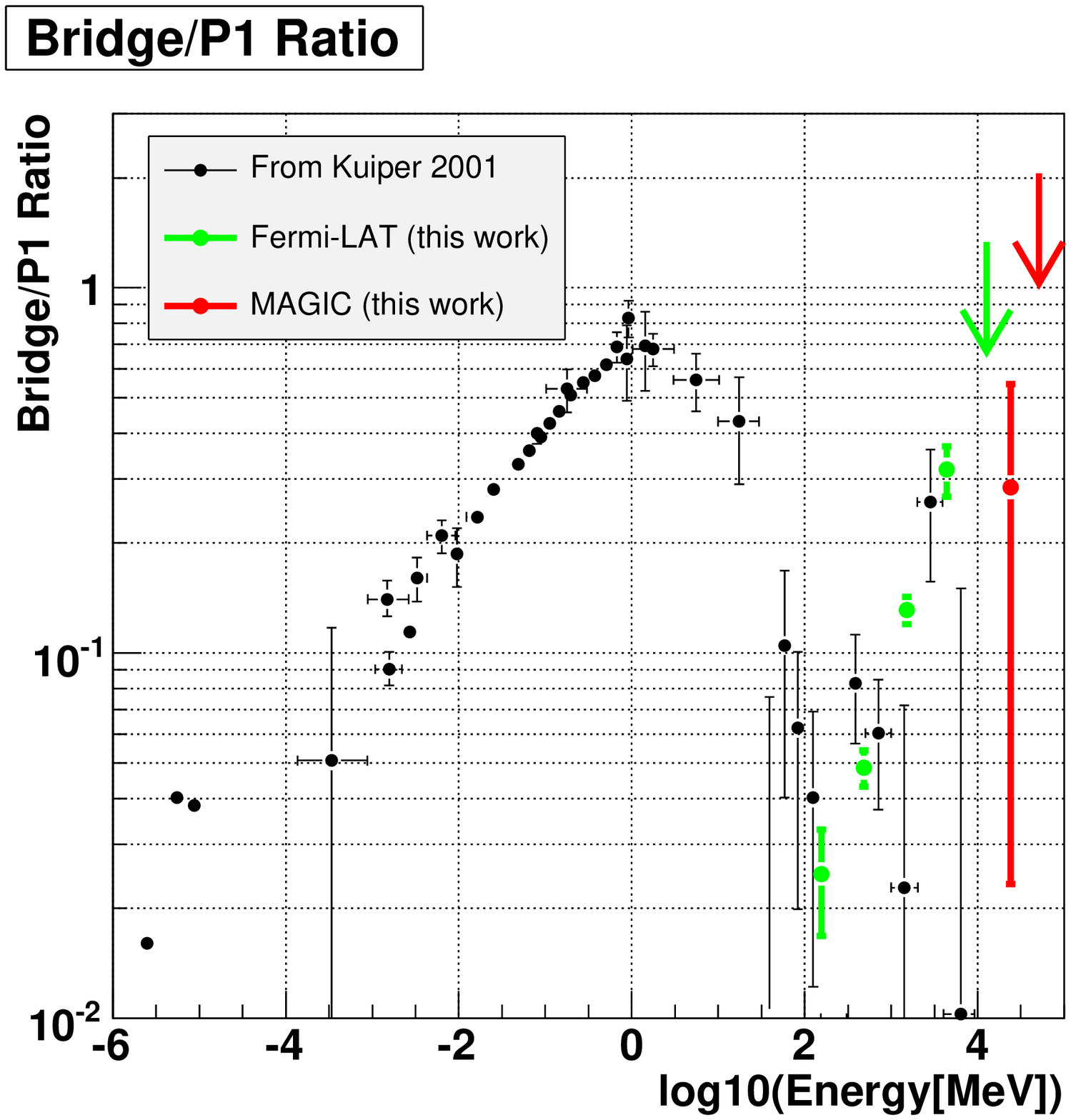}
\caption{The P2/P1 ratio (left) and the Bridge/P1 ratio (right) for a wide energy range 
from 1 eV to 100 GeV. Black points are adopted from \cite{Kuiper2001}.
Green points and a green arrow are based on \fermi-LAT data, while red points and a red arrow
are based on MAGIC data
}
\label{FigP1P2RatioWide}
\end{figure}

In Fig. \ref{FigP1P2RatioWide}, the results are compared with 
those at lower energies.
The data points of lower wavelengths are adopted from  \cite{Kuiper2001}. 
A few common features between the behavior of the P2/P1 ratio and that of the Bridge/P1 ratio 
can be seen:
From 1 eV to 1 MeV, they increase with energy with a power law.
From 1 MeV to 100 MeV, they drop rapidly and, then, rise again  above 100 MeV $-$ 1 GeV.
The ratios at $\sim 30$ GeV may be as high as those at 1 MeV, 
although the uncertainty is large due to the lack of statistics.

\begin{table*}[h]
\begin{threeparttable}
 \begin{tabular}{|r|r|r|r|r|r|r|}
\hline
 Energy & Energy \tnote{a}&  P1      & P2       & Bridge   & P2/P1 & Bright/P1 \\
  Range &    [GeV] & [counts] & [counts] & [counts] &       &  \\
\hline
\hline
Fermi-LAT & & & & & & \\
\hline
 0.1 - 0.3 GeV & 0.156& 4135 $\pm$ 71 & 2067 $\pm$ 55 & 103 $\pm$ 33 & 0.50 $\pm$ 0.02 & 0.025 $\pm$ 0.008 \\
 0.3-1.0 GeV & 0.481& 5219 $\pm$ 76 & 2885 $\pm$ 59 & 254 $\pm$ 28 & 0.55 $\pm$ 0.01 & 0.049 $\pm$ 0.005 \\
1.0-3.0 GeV & 1.52 & 1706 $\pm$ 42 & 974 $\pm$ 33 & 224 $\pm$ 18 & 0.57 $\pm$ 0.02 & 0.131 $\pm$ 0.011 \\
3.0-10.0 GeV & 4.38 & 248 $\pm$ 17 & 245 $\pm$ 17 & 79 $\pm$ 11 & 0.99 $\pm$ 0.10 & 0.318 $\pm$ 0.050 \\
$>$10 GeV & 12.7 & 12 $\pm$ 6 & 21 $\pm$ 7 & $< 17$\tnote{b}  & 1.70 $\pm$ 1.05 & $< 1.34$ \\\hline
\hline
MAGIC & & & & & & \\
\hline
 25 - 70 ph.e & 24.0 & 4711 $\pm$ 1129 & 8233 $\pm$ 1198 & 1338 $\pm$ 1195 & 1.75 $\pm$ 0.49 & 0.284 $\pm$ 0.261 \\
 70 - 500 ph.e. & 51.2 & 1437 $\pm$ 899 & 3096 $\pm$ 954 & $< 2953$\tnote{b} & 2.16 $\pm$ 1.50 & $ < 2.06$ \\

\hline
\hline
 \end{tabular}
\begin{tablenotes}
{\footnotesize
\item [a] The representative energy, taking into account the pulsar spectrum and the detector response.
\item [b] 95\% confidence level upper limit.
}
\end{tablenotes}
\end{threeparttable}
 \caption{The numbers of excess events of P1, P2 and Bridge and their ratios}
 \label{TabP1P2Ratio}
 \end{table*}


\clearpage

\section{Rising and Falling Edges}
\label{SectEdgeStudy}
The energy spectra have been calculated for the specific phase intervals. 
They are especially important to deduce the emission regions of the pulsar on a large scale.
On the other hand, as can be seen
from Fig. \ref{FigSizeDepPhasogram} and Fig. \ref{FigFermiPhaseEne}, the pulse shape
 is not the same for all energies. 
A study of the change of pulse shape with energy would also be very helpful
to understand the emission mechanism in more detail, i.e. on a small scale, 
because the pulse phase should be connected to
the geometry of the emission region, as discussed in Sect. \ref{SectLightCurve}.

Due to the lack of statistics (with respect to the large background),
 it is difficult to analyze the pulse shape with MAGIC data alone. 
 However, the fact that the pulse peak phases are very similar for all energies implies that 
the pulsations for all energies may originate from
a common physical process. Therefore, one can also expect common features in pulse shapes. 
 Once a common feature is found, one can statistically examine it in the MAGIC data.

\begin{figure}[h]
\includegraphics[width=0.495\textwidth]{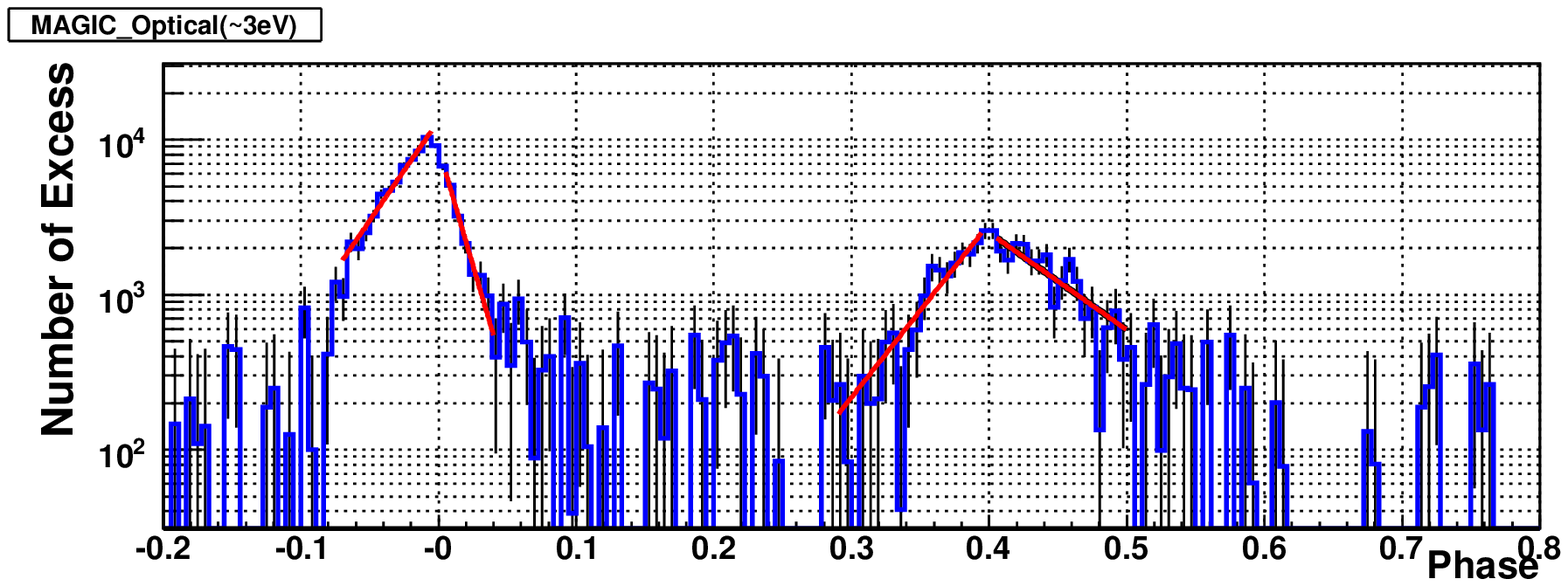}
\includegraphics[width=0.495\textwidth]{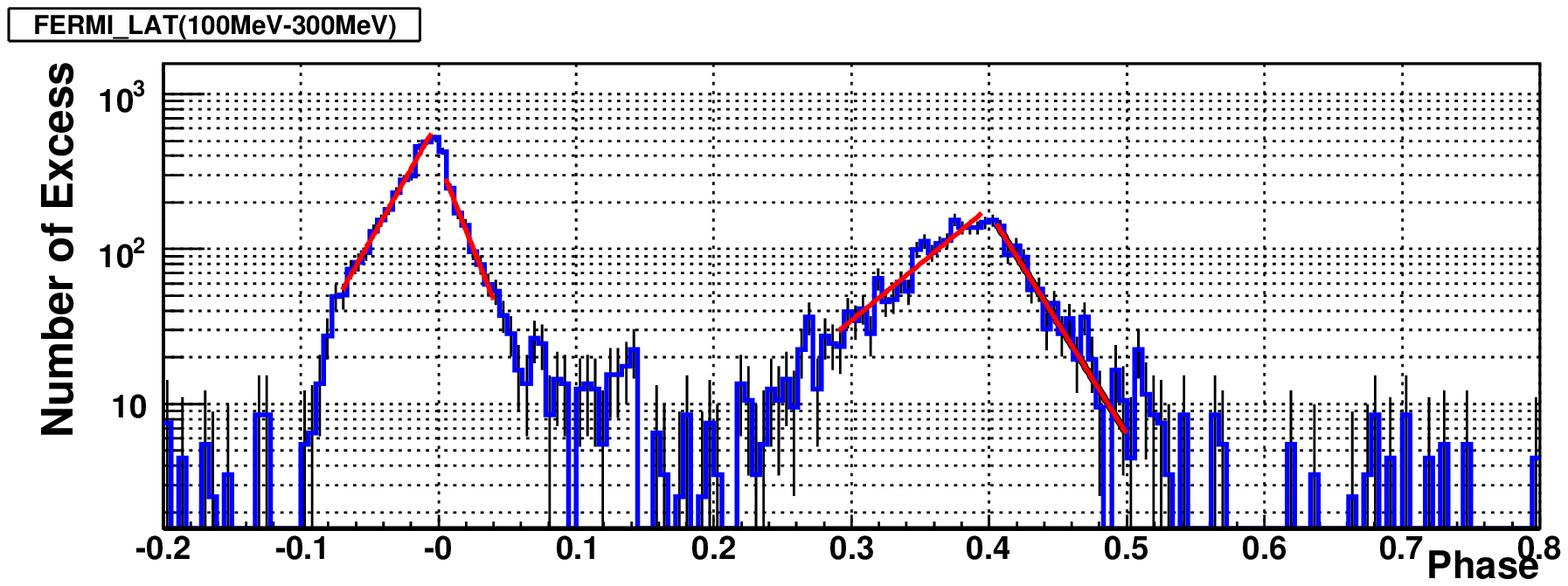}
\includegraphics[width=0.495\textwidth]{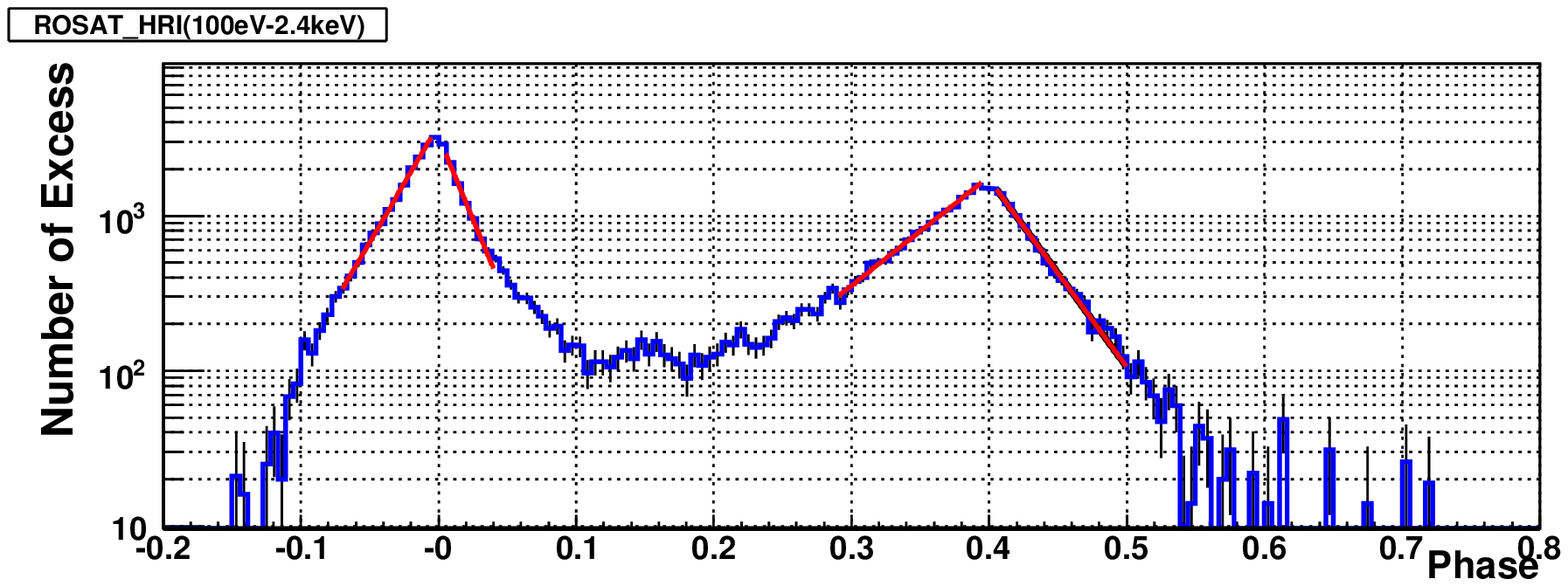}
\includegraphics[width=0.495\textwidth]{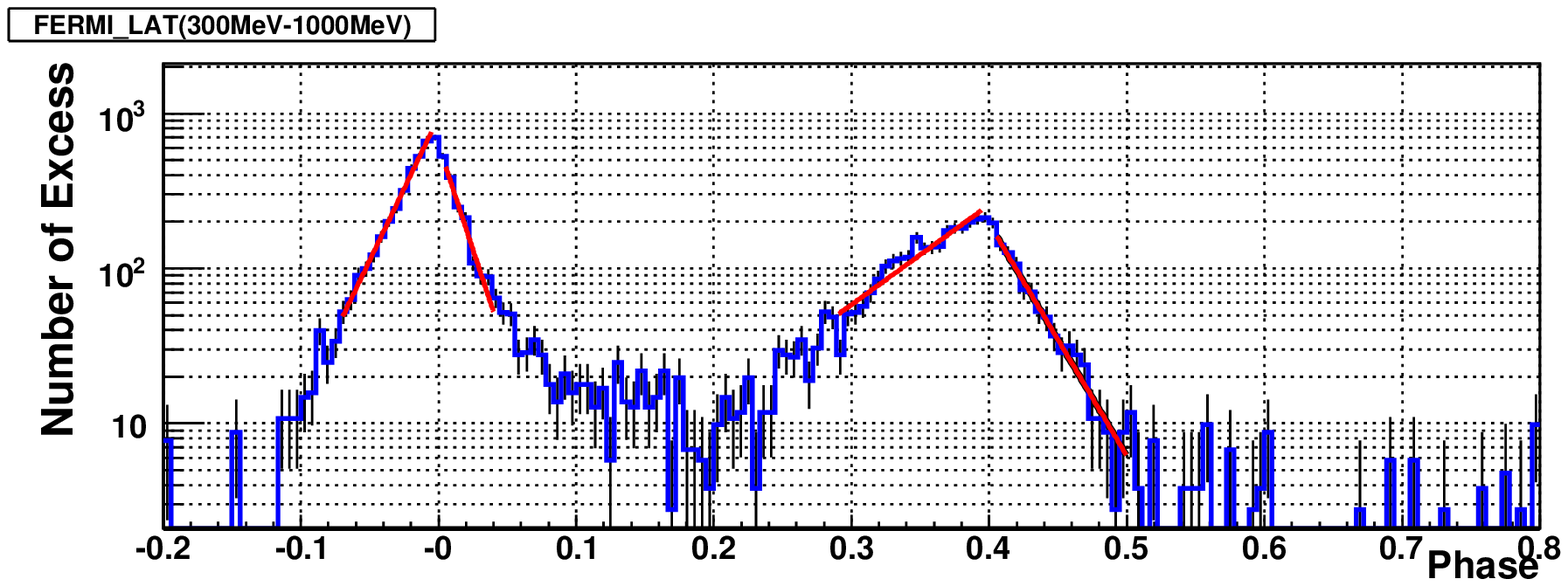}
\includegraphics[width=0.495\textwidth]{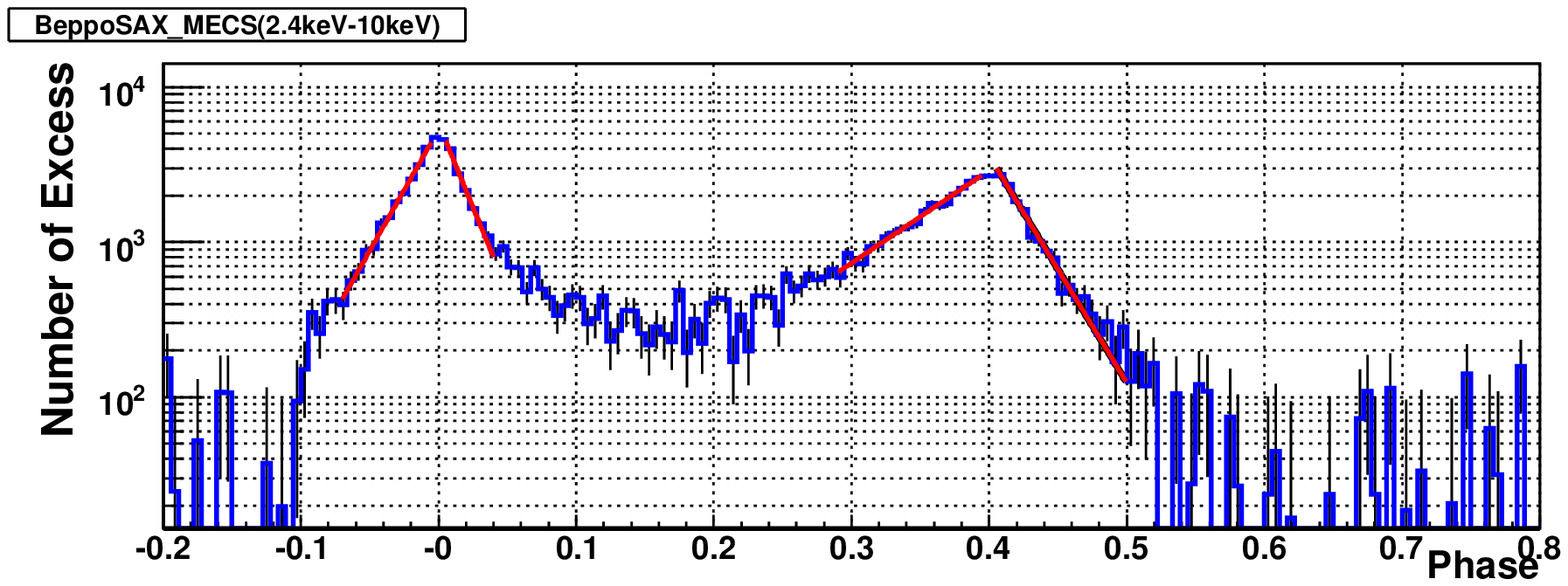}
\includegraphics[width=0.495\textwidth]{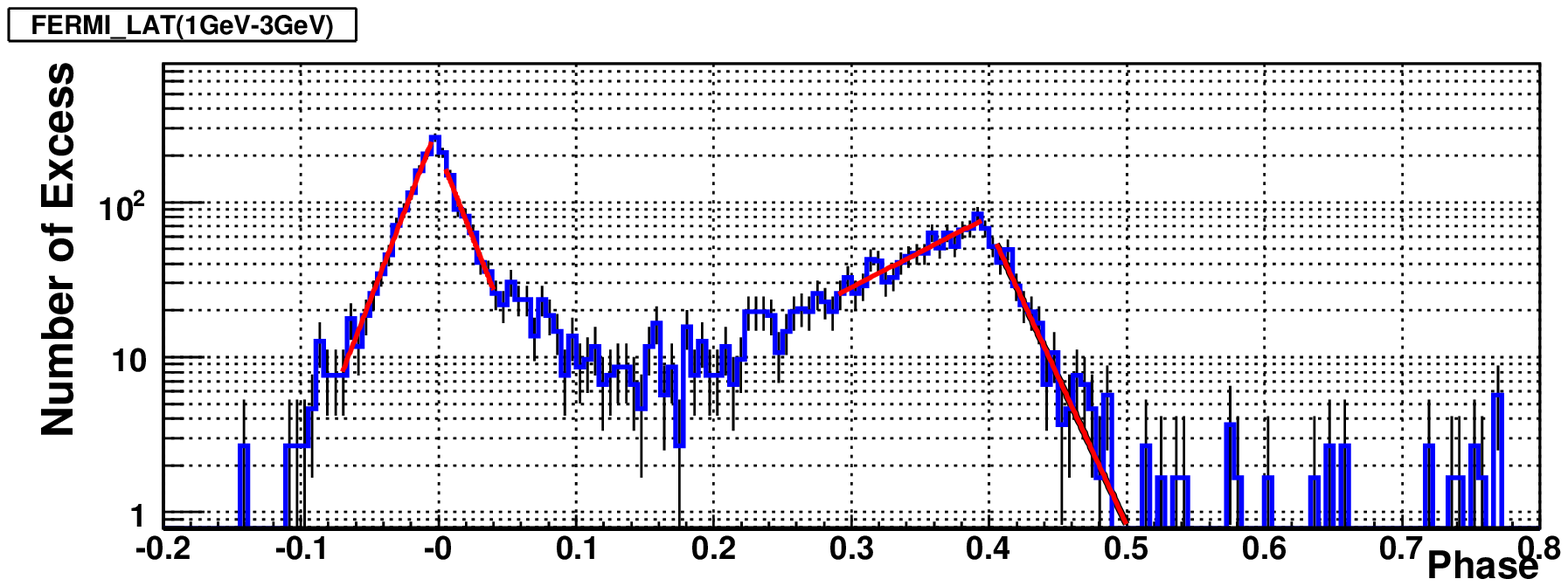}
\includegraphics[width=0.495\textwidth]{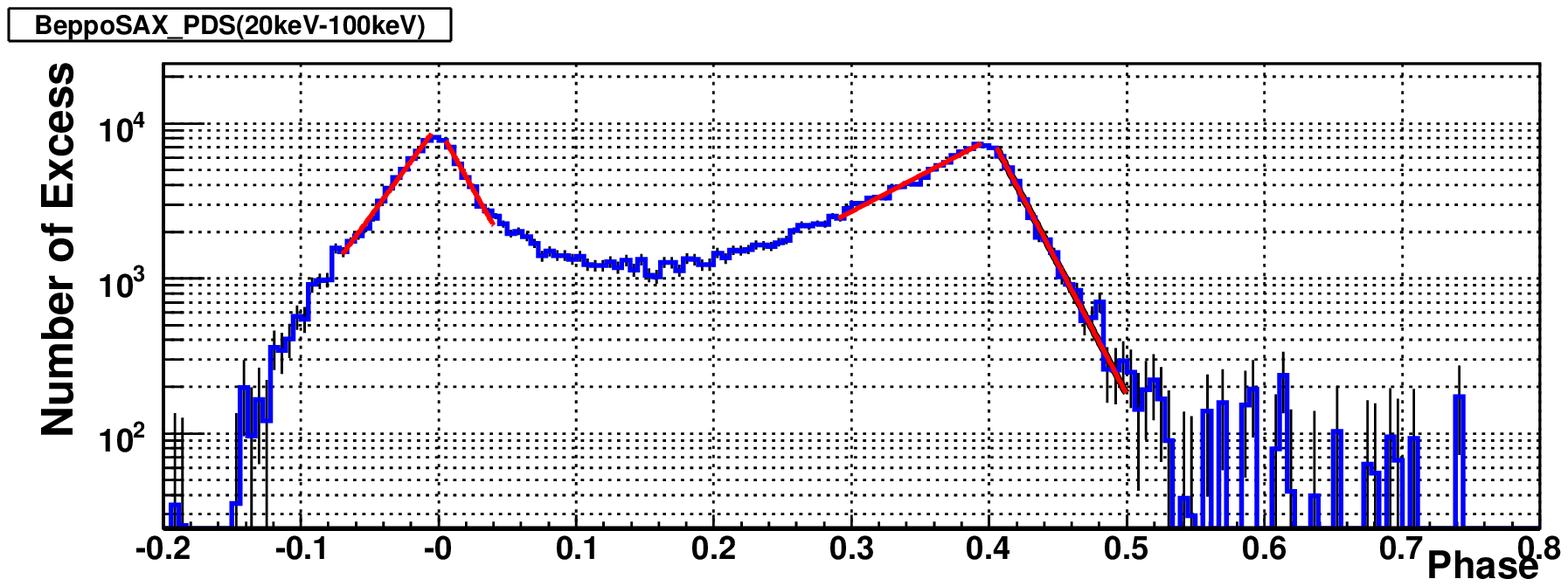}
\includegraphics[width=0.495\textwidth]{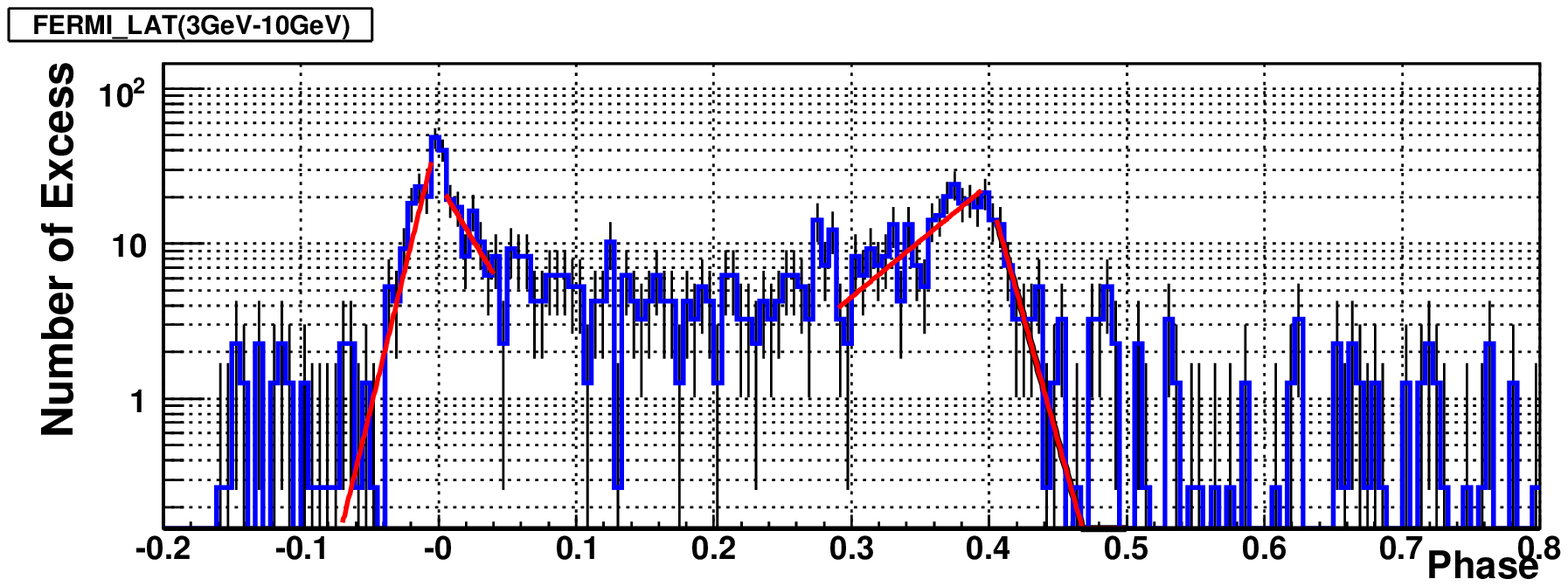}
\includegraphics[width=0.495\textwidth]{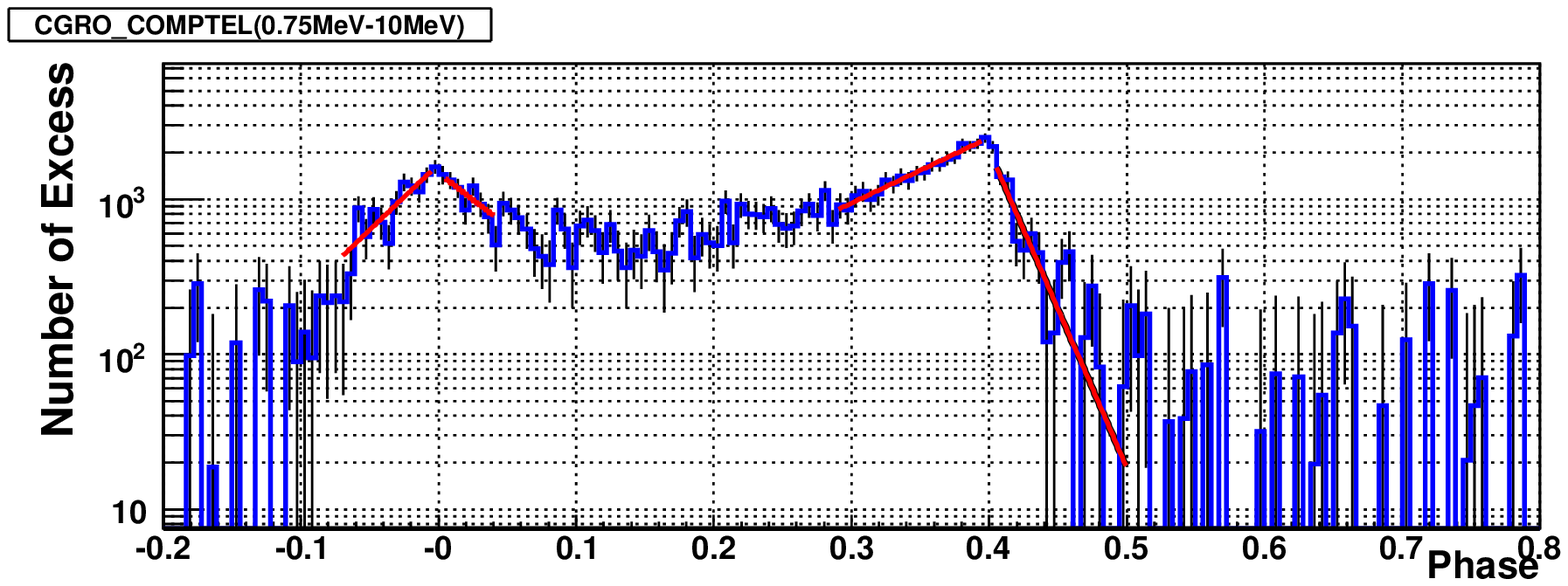}
\includegraphics[width=0.495\textwidth]{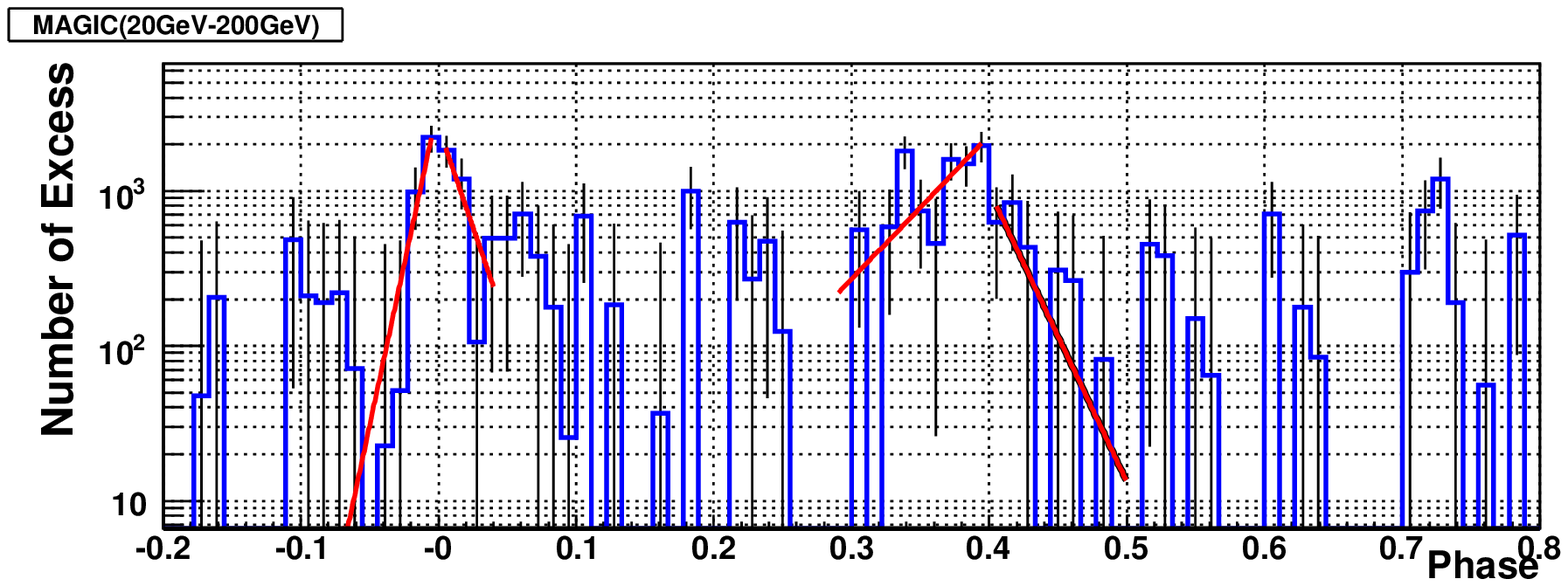}
\includegraphics[width=0.495\textwidth]{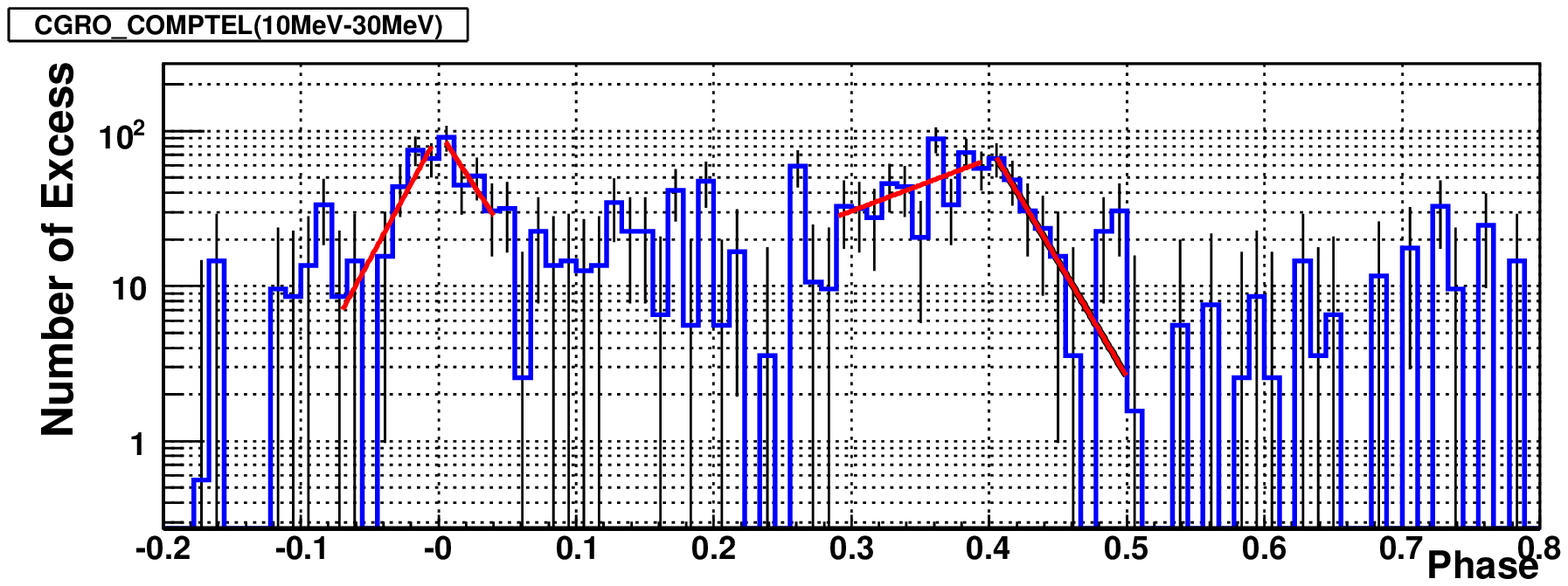}
\caption{Light curves in a logarithmic scale for different energies.
From the top left downward to the bottom left, optical measurements by the MAGIC central pixel
(see Sect. \ref{SectOptPulsation}), 100 eV to 2.4 keV by ROSAT-HRI, 
2.4 keV to 10 keV by BeppoSAX MECS, 20 keV to 100 keV by BeppoSAX PDS,
0.75 MeV to 10 MeV by COMPTEL and 10 MeV to 30 MeV by COMPTEL.
From the top right downward to the bottom right, 
100 MeV to 300 MeV, 300 MeV to 1 GeV, 1 GeV to 3 GeV, 3 GeV to 10 GeV measured by 
\fermi-LAT and 20 GeV to 200 GeV ( 25 $<SIZE <$ 500) measured by MAGIC.
}
\label{FigLogPhasogram}
\end{figure}

In fact, I found that by plotting the light curves in log scale, some interesting features become 
visible. 
In Fig. \ref{FigLogPhasogram}, light curves for different energies from optical to 
very high energy gamma-rays are shown. 
The optical light curve is obtained from the MAGIC central pixel data.
L. Kuiper provided
the X-ray light curves, which have also been  used in \cite{Kuiper2001}. 
They are based on ROSAT HRI (100 eV to 2.4 keV, see \cite{ROSAT-HRI}), 
BeppoSAX MECS (2.4 keV to 10 keV, see \cite{BeppoSAX}), BeppoSAX PDS (20 keV to 100 keV, see \cite{BeppoSAX})
and CGRO COMPTEL (750 keV to 30 MeV, see \cite{Comptel}). 
The gamma-ray light curves
from 100 MeV to 10 GeV are produced by myself based on the \fermi-LAT data.
The gamma-ray light curve at 20 to 200 GeV 
\footnote{This energy range is a rough estimation 
based on the $SIZE$ range from 25 to 500 ph.e.. No significant excess
is seen at 200 GeV.
}
 are taken from my analysis of the MAGIC data with $SIZE$ between 25 to 500 ph.e..
One can see the following features:
\begin{itemize}
\item Both rising and falling edges show an exponential behavior.  
\item Slopes are not symmetric between rising and falling edges.
\item Slopes change with energy.
\end{itemize}

 Here, I discuss the pulse edges of P1 and P2. In the previous sections and chapters,
P1 and P2 referred to
specific phase intervals, 
P1 being from -0.06 to 0.04 and P2 being from 0.32 to 0.43, while in this section 
they simply denote the first peak and the second peak.

\subsubsection{Method: Fitting Exponential Functions to Pulse Edges}

An exponential function
\begin{equation}
F(p) = A\times {\rm exp}(\pm (p-p_0)/\tau)
\end{equation}
is fitted to rising and falling edges of P1 and P2 for different energies, $A$, $p_0$ and $\tau$ 
being free parameters.
Fitting ranges were chosen such that the bridge emission and the pulse peak do not worsen 
the goodness of the fit. Fitting ranges are summarized in Table \ref{TabFittingRange}.
Then, the energy dependence of the rise time $\tau_{rise}$ and the fall time $\tau_{fall}$
for P1 and P2 are examined.

\clearpage

\begin{table*}[h]
\centering
 \begin{tabular}{|c|c|c|c|c|}
\hline
\hline
Edge & Rise of P1 & Fall of P1 &Rise of P2 & Fall of P2 \\
\hline
Phase & -0.07 to -0.005 & 0.005 to 0.04& 0.29 to 0.395 & 0.405 to 0.50 \\
\hline
\hline
 \end{tabular}
 \caption{Fitting range for the rise/fall time estimation}
 \label{TabFittingRange}
 \end{table*}




\subsubsection{Results of the Exponential Function Fitting to Pulse Edges }
\begin{figure}[h]
\centering
\includegraphics[width=0.45\textwidth]{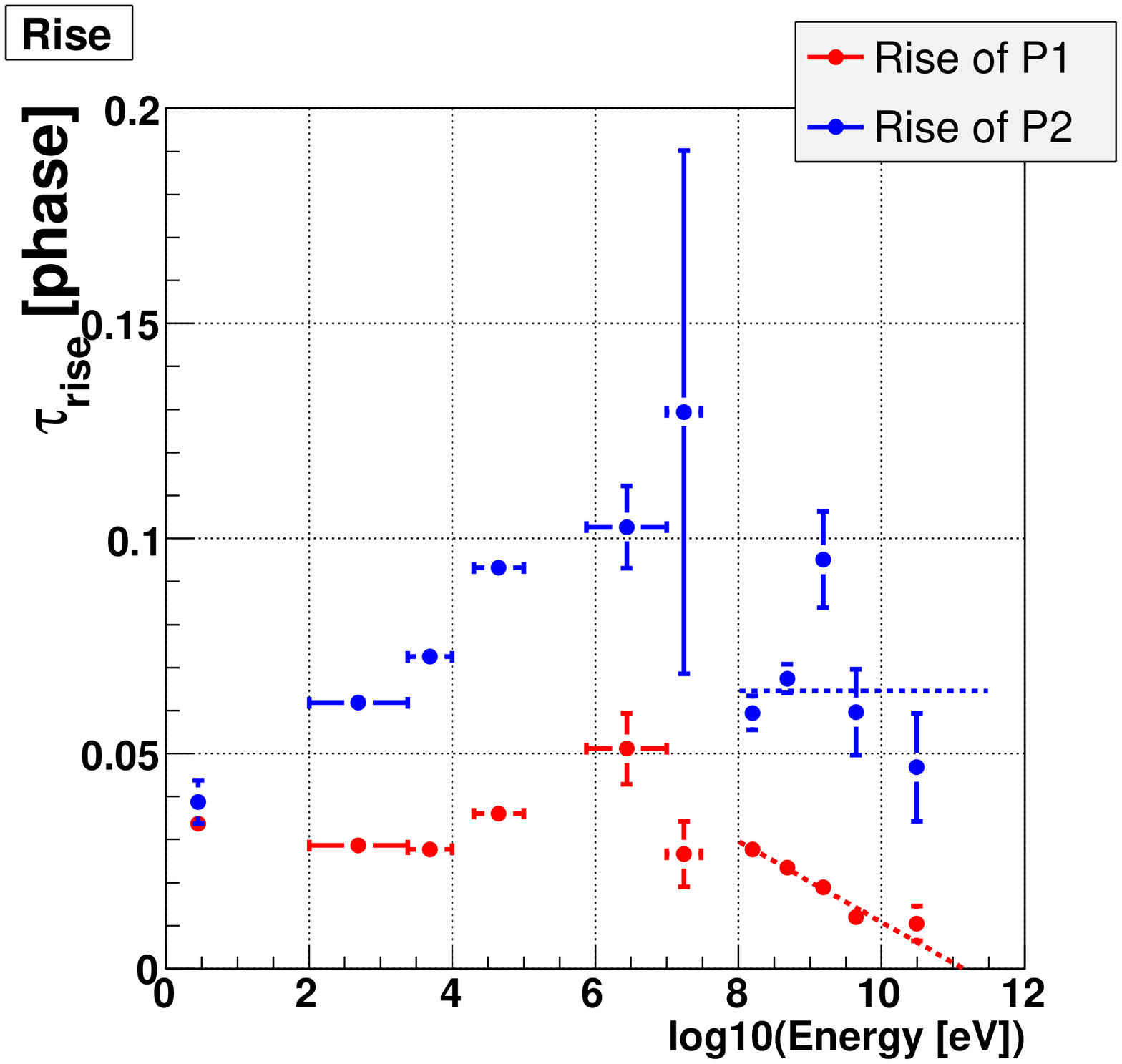}
\includegraphics[width=0.45\textwidth]{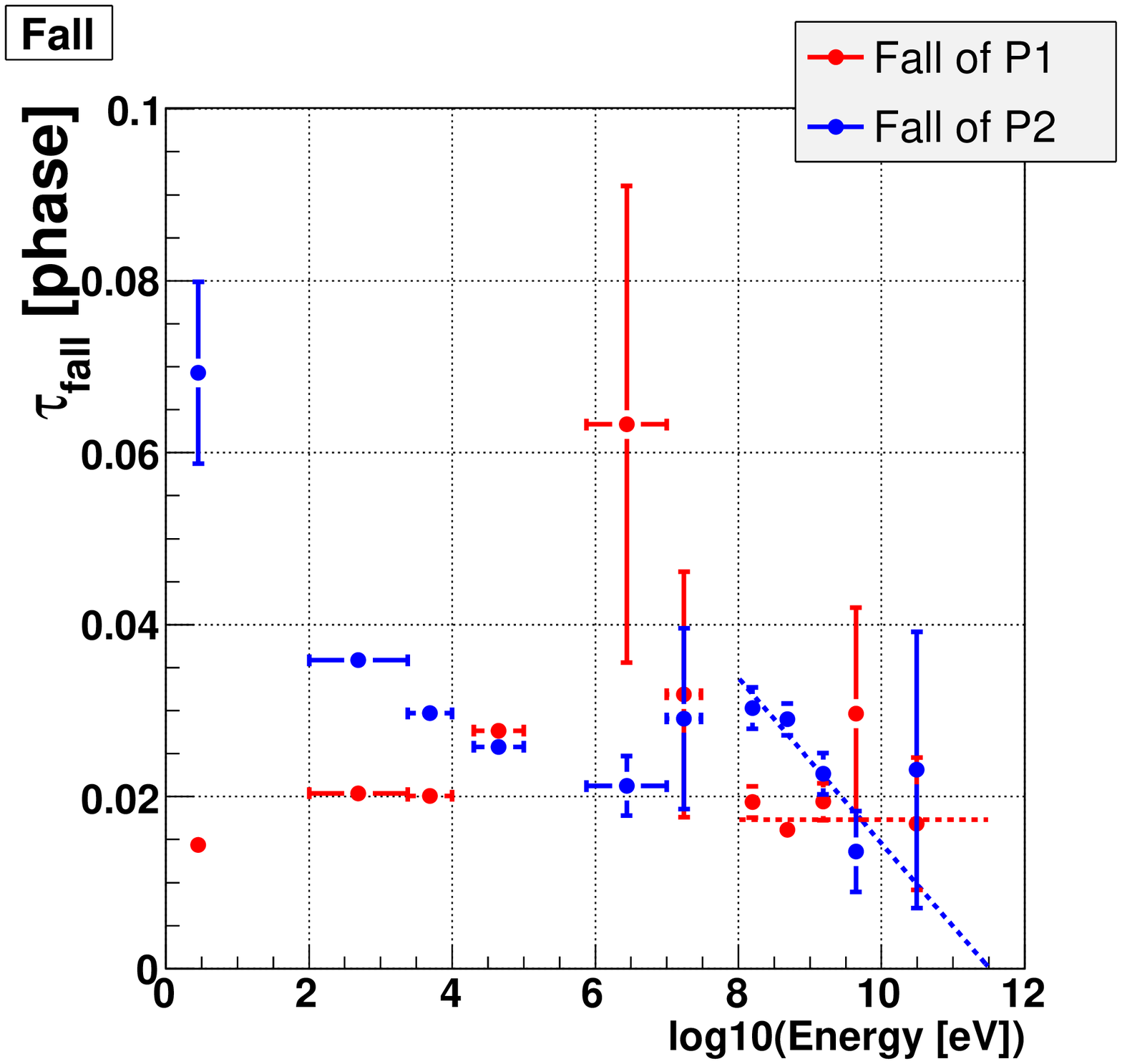}
\includegraphics[width=0.45\textwidth]{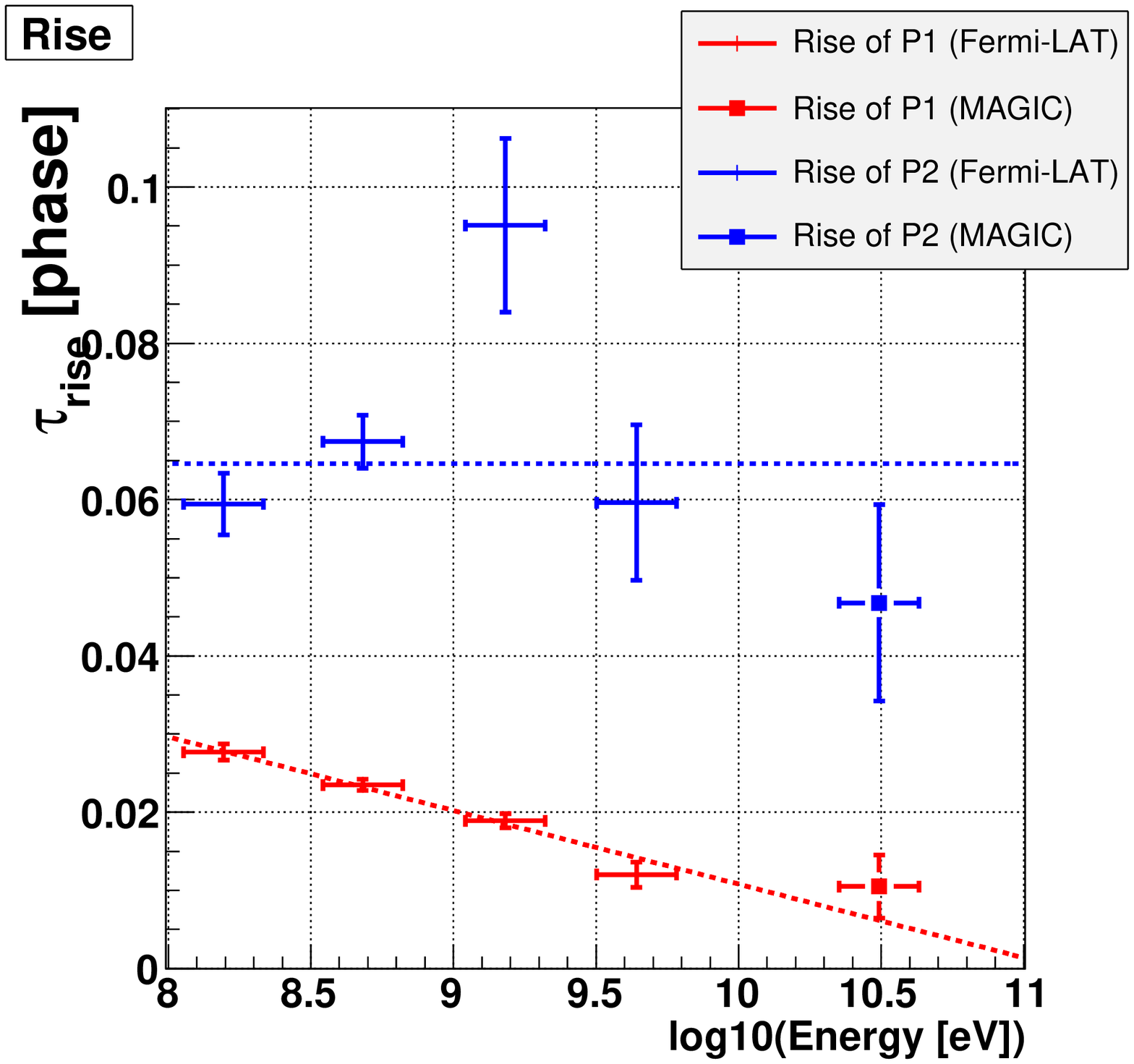}
\includegraphics[width=0.45\textwidth]{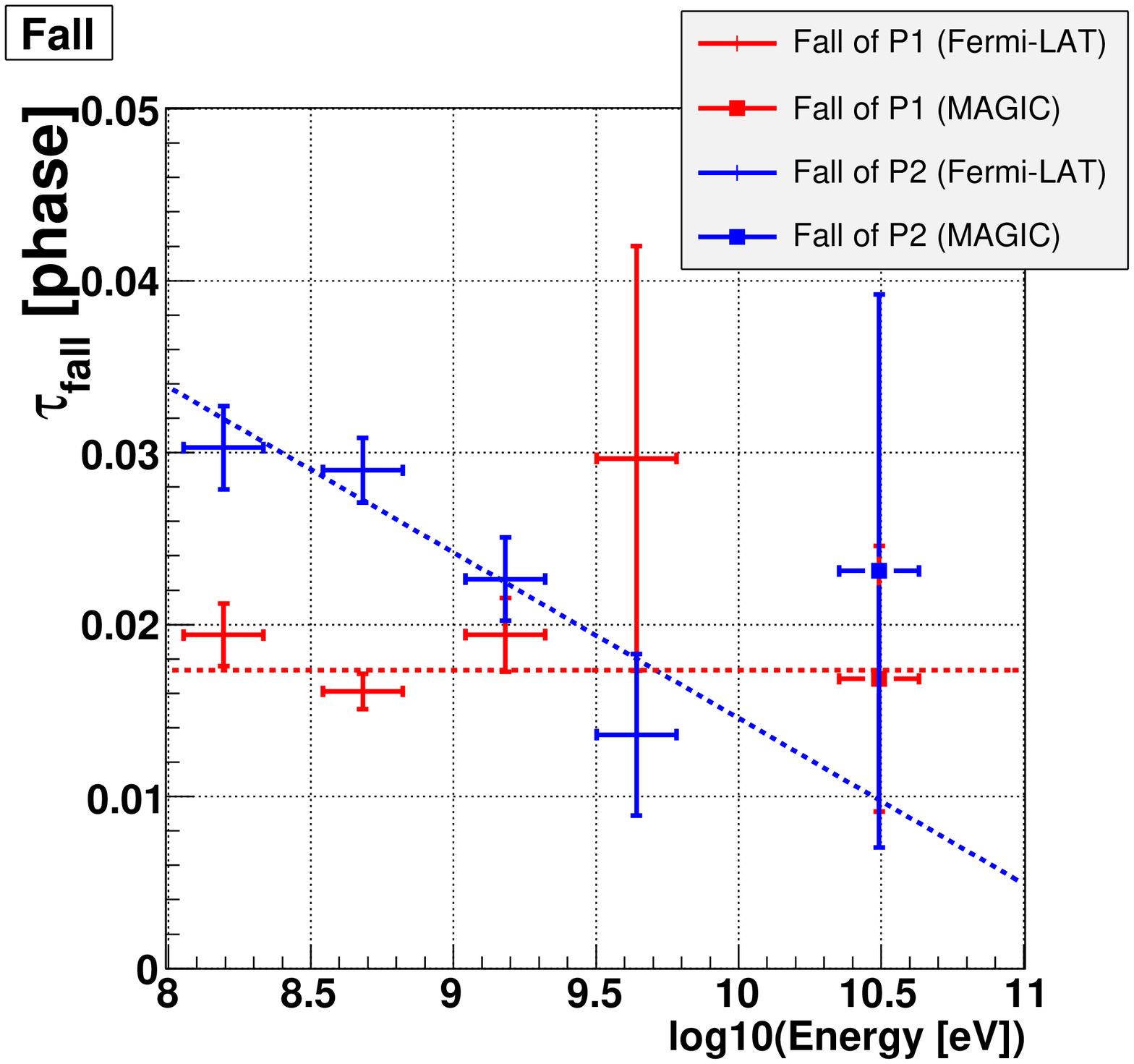}
\caption{Energy dependence of the rise and the fall times. 
Top left: The rise time of P1 (red) and P2 (blue), as a function of energy.
Top right: The fall time of P1 (red) and P2 (blue), as a function of energy.
Bottom left: The same as the top left panel but the energy range from 
100 MeV to 100 GeV is zoomed.
Bottom right: The same as the top right panel but the energy range from 
100 MeV to 100 GeV is zoomed.
}
\label{FigRiseAndFall}
\end{figure}

Fitted lines are overlaid on the light curves shown in Fig. \ref{FigLogPhasogram}.
The obtained parameters and the fitting probabilities are shown in 
Table \ref{TabRiseAndFall}. 
Fitting probabilities are reasonably good for all energies.
The energy dependence of the rise and the fall time of both peaks are graphically shown 
in the upper panels of Fig. \ref{FigRiseAndFall}.
The horizontal values in Fig. \ref{FigRiseAndFall} are determined in the following way:
Below 100 MeV, horizontal error bars show the full energy range of the data sets and 
the central values are the logarithmic center of the range. Above 100 MeV,
the horizontal central values and error bars are determined 
in the same way as the P2/P1 ratio study
(see Sect. \ref{SectRatioResult}).

General behavior of the rise time and the fall time 
from optical to gamma-rays 
can be summarized as follows:

\begin{itemize}
\item Below 10 MeV, the rise time of both pulses and the fall time of P1 are increasing,
while the fall time of P2 is decreasing.
\item Above 100 MeV, the rise time of P1 and the fall time of P2 are decreasing, while 
the rise time of P2 and the fall time of P1 do not show clear energy dependence.  
\end{itemize}

\begin{figure}[h]
\centering
\includegraphics[width=0.45\textwidth]{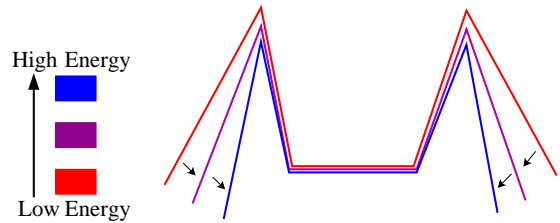}
\caption{An illustration of the energy dependence of the pulse edges above 100 MeV.
As energy increases, the outer edges get sharper, while the inner edges do not change. 
}
\label{FigEdge}
\end{figure}

Hereafter, I would like to focus on the behavior above 100 MeV, 
where I have personally analyzed all the data.
In the lower panels of Fig. \ref{FigRiseAndFall}, 
the energy dependence of the rise and the fall times above 100 MeV are shown. 
They are well described by a linear function of log$_{10}(E)$ or a constant value, 
where $E$ is the energy. 
 By fitting a linear function to the rise time of P1 and the fall time of P2 and by
fitting a constant value to the fall time of P1 and the rise time of P2, 
the following is obtained:

\begin{eqnarray}
\tau_{rise}^{P1}(E) &=& (2.02 \pm 0.08)\times 10^{-2} - (9.4 \pm 1.3)\times 10^{-3}{\rm log}_{10}(E[\rm{GeV}]) 
\label{EqTau1} \\
\tau_{rise}^{P2} &=&(6.46 \pm 0.24)\times 10^{-2}  \\
\label{EqTau2}
\tau_{fall}^{P1} &=&(1.73 \pm 0.08)\times 10^{-2}  \\
\label{EqTau3}
\tau_{fall}^{P2}(E) &=& (2.42 \pm 0.16)\times 10^{-2} - (9.6 \pm 3.1)\times 10^{-3}{\rm log}_{10}(E[\rm{GeV}]) 
\label{EqTau4}
\end{eqnarray}
$\chi2 / dof$ are 2.2/3, 12.2/4, 4.6/4 and 2.4/3 for $\tau_{rise}^{P1}$, $\tau_{rise}^{P2}$,
$\tau_{fall}^{P1}$,and $\tau_{fall}^{P2}$, respectively. \\
MAGIC points and \fermi-LAT points can be fitted consistently.
It is also interesting that
the rise time of P1 and the fall time of P2 show very similar dependence on energy.
The energy dependence of the pulse shape above 100 MeV is schematically illustrated 
in Fig.~\ref{FigEdge}.
As energy increases, the outer edges, i.e. the rising edge of P1 and the falling edge of P2 
get sharper while the inner edges, i.e. the falling edge of P1 and the rising edge of P2
do not change.

The physical interpretation of these results will be discussed in Sect. \ref{SectRiseFallDiscuss}.


\begin{table*}
\begin{threeparttable}
 \begin{tabular}{|r|r|r|r|r|}
\hline
 & \multicolumn{2}{|c|}{Rise Time} & \multicolumn{2}{|c|}{Fall Time} \\ 
\hline
Energy & $\tau_{Rise}$ [10$^{-3}$ phase]& ($\chi^2/dof, prob.$) & $\tau_{Fall}$ [10$^{-3}$ phase] & ($\chi^2/dof, prob.$) \\ 
\hline
\hline
P1 & & & &\\
\hline
2.0 - 4.0 eV  & 33.6 $\pm $ 1.2 & (16.7/10, 8.0\%) & 14.4 $\pm $ 1.6   & (2.7/4, 61.3\%)  \\
0.1 - 2.4 keV & 28.6 $\pm $ 0.5 & (7.7/10, 66.3\%) & 20.4 $\pm $ 0.7   & (5.1/4, 27.6\%)  \\
2.4 - 10 keV  & 27.7 $\pm $ 0.8 & (10.2/10, 42.5\%) & 20.1 $\pm $ 0.9  & (8.3/4, 8.0\%)  \\
20 - 100 keV  & 36.0 $\pm $ 0.7 & (17.7/10, 6.0\%) & 27.7 $\pm $ 1.0   & (10.7/4, 3.0\%)  \\
0.75 - 10 MeV & 51.1 $\pm $ 8.3 & (15.5/10, 11.3\%) & 63.3 $\pm $ 27.7 & (4.0/4, 40.5\%)  \\
10 - 30 MeV   & 26.6 $\pm $ 7.7 & (5.2/4, 26.3\%) & 31.9 $\pm $ 14.3   & (1.4/2, 48.9\%)  \\
100 - 300 MeV & 27.7 $\pm $ 1.0 & (11.0/10, 36.1\%) & 19.4 $\pm $ 1.8  & (1.3/4, 86.3\%)  \\
0.3 - 1.0 GeV & 23.5 $\pm $ 0.7 & (4.5/10, 92.0\%) & 16.1 $\pm $ 1.0   & (13.9/4, 0.8\%)  \\
1.0 - 3.0 GeV & 18.9 $\pm $ 0.9 & (3.7/10, 96.0\%) & 19.4 $\pm $ 2.1   & (3.9/4, 41.5\%)  \\
3.0 - 10 GeV  & 12.0 $\pm $ 1.6 & (16.2/10, 9.4\%) & 29.7 $\pm $ 12.4  & (4.3/4, 36.3\%)  \\
20 - 200 GeV\tnote{a} & 10.5 $\pm $ 4.0 & (0.8/4, 93.9\%) & 16.8 $\pm $ 7.7    & (1.4/2, 48.8\%)  \\
\hline
\hline

P2 & & & & \\
\hline
2.0 - 4.0 eV & 38.7 $\pm $ 5.1  & (10.7/17, 87.1\%) & 69.3 $\pm $ 10.6 & (20.1/15, 16.8\%)  \\
0.1 - 2.4 keV& 61.8 $\pm $ 1.4  & (10.5/17, 88.0\%) & 35.9 $\pm $ 0.9  & (16.9/15, 32.2\%)  \\
2.4 - 10 keV& 72.5 $\pm $ 2.7  & (14.8/17, 61.1\%) & 29.7 $\pm $ 1.1  & (20.4/15, 15.8\%)  \\
20 - 100 keV& 93.2 $\pm $ 1.9  & (39.8/17, 0.1\%) & 25.8 $\pm $ 0.6   & (29.1/15, 1.5\%)  \\
0.75 - 10 MeV& 102.6 $\pm $ 9.6 & (4.9/17, 99.8\%) & 21.2 $\pm $ 3.5   & (20.9/15, 13.9\%)  \\
10 - 30 MeV& 129.4 $\pm $ 60.8&  (11.3/8, 18.3\%) & 29.1 $\pm $ 10.5 & (8.0/7, 33.3\%)  \\
100 - 300 MeV& 59.4 $\pm $ 3.9  & (30.9/17, 2.0\%) & 30.3 $\pm $ 2.4   & (27.8/15, 2.3\%)  \\
0.3 - 1.0 GeV& 67.4 $\pm $ 3.4  & (37.2/17, 0.3\%) & 29.0 $\pm $ 1.9   & (6.4/15, 97.3\%)  \\
1.0 - 3.0 GeV& 95.1 $\pm $ 11.1 & (12.5/17, 76.8\%) & 22.6 $\pm $ 2.4  & (14.4/15, 49.4\%)  \\
3.0 - 10 GeV& 59.6 $\pm $ 10.0 & (20.9/17, 22.9\%) & 13.6 $\pm $ 4.7  & (16.6/15, 34.4\%)  \\
20 - 200 GeV \tnote{a}& 46.8 $\pm $ 12.6 & (18.0/8, 2.2\%) & 23.1 $\pm $ 16.1   & (4.2/7, 76.0\%)  \\
\hline
 \end{tabular}
\begin{tablenotes}
{\footnotesize
\item [a] This energy range is a rough estimation 
based on the $SIZE$ range in MAGIC data from 25 to 500. No significant excess
is seen at 200 GeV.
}
\end{tablenotes}
\end{threeparttable}
 \caption{Results of the rise and the fall time estimation for P1 and P2}
 \label{TabRiseAndFall}
 \end{table*}


\clearpage

\section{Peak Phases}
\label{SectPeakPhase}

\begin{figure}[h]
\centering
\includegraphics[width=0.45\textwidth]{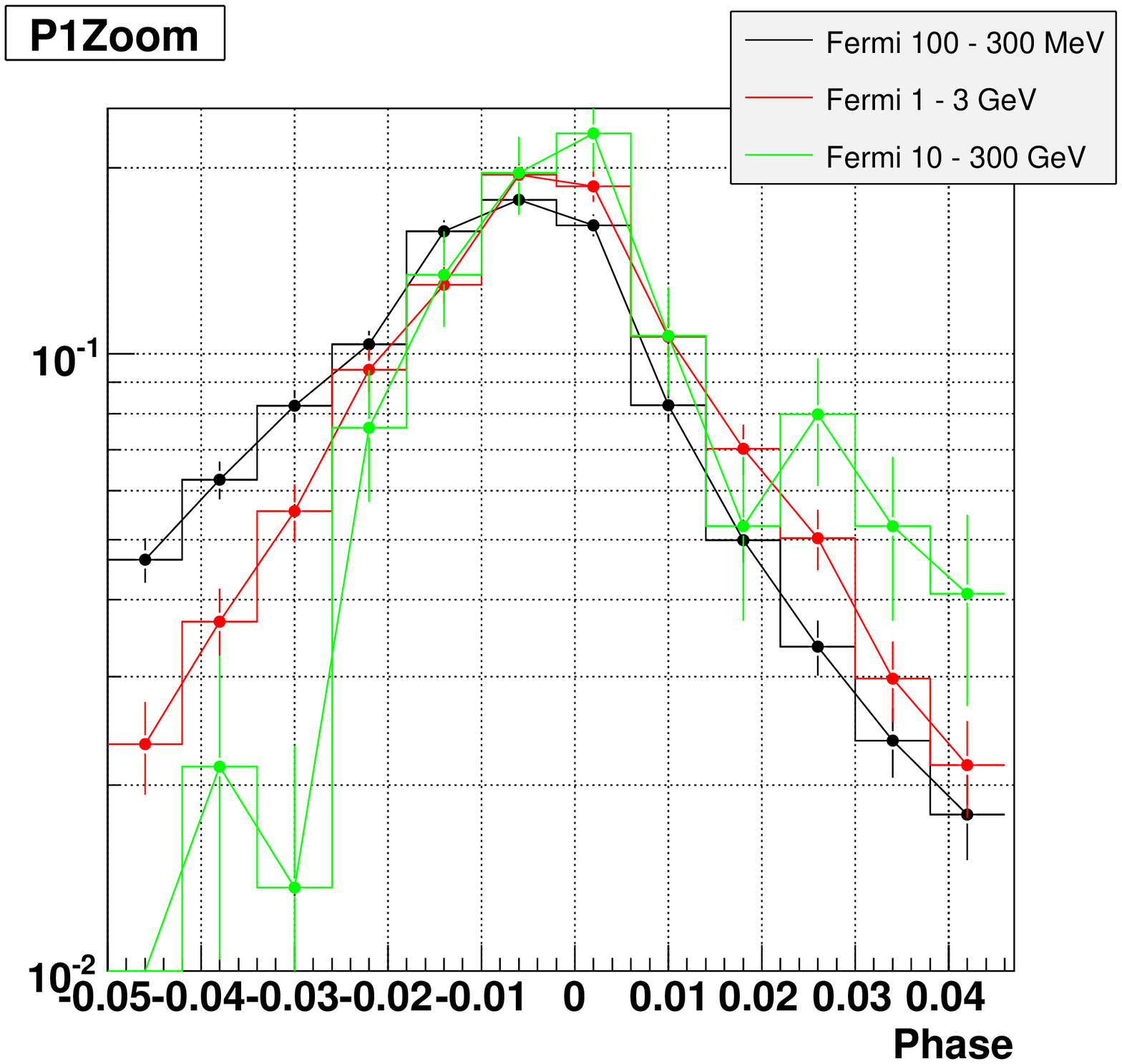}
\includegraphics[width=0.45\textwidth]{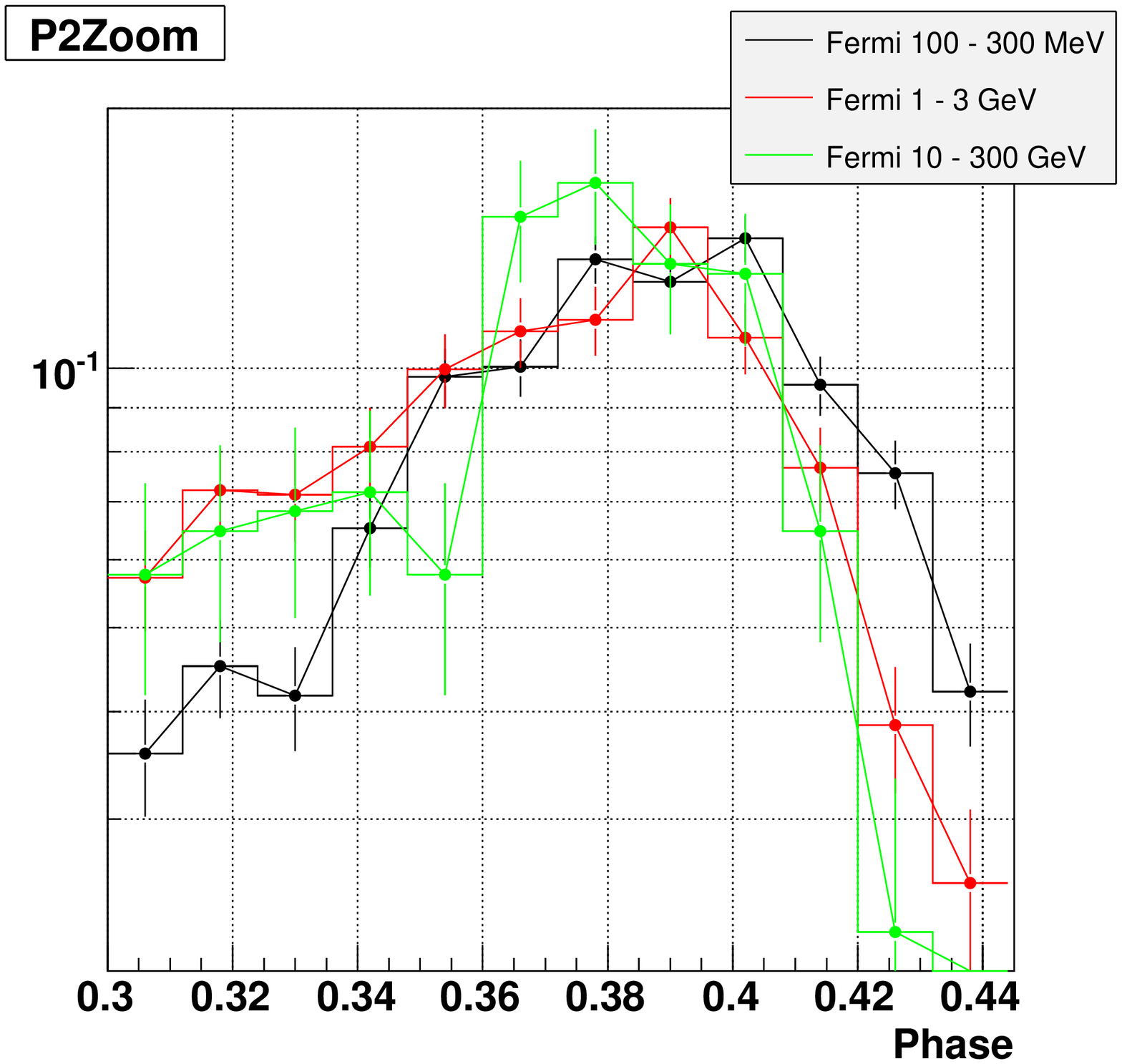}
\caption{A closer look at the peak phase of P1 (left) and P2 (right). 
The black, red and green lines indicate the light curves for 100 MeV to 300 MeV,
1 to 3 GeV and above 10 GeV, based on \fermi-LAT data.
}
\label{FigCloserLook}
\end{figure}

As described in Sect. \ref{SectCrabLC}, the pulse profiles of the Crab pulsar are very similar 
at all energies, while 
 a closer look at the light curves reveals a slight energy dependence of the peak phase.
For example, the first peak in the X-ray data (see Fig. \ref{FigCrabPulses}) and that
in optical data (see Fig. \ref{FigMAGICOptical}) precede the radio peak
by $\sim 0.01$ in phase. 
Also, above 100 MeV, there seems to be a slight shift of the peak phase,
which can be seen if the light curves based on the \fermi-LAT data are zoomed, as shown in
Fig. \ref{FigCloserLook}. 

Here, I examine the energy dependence of the peak phase above 100 MeV by using
\fermi-LAT and MAGIC data. 
 Apparently, the energy dependence of the peak phases is not very strong.
The precision of $\sim 0.003$ in phase would be required to discuss the energy dependence.
In order to fulfill this requirement with the obtained data, a sophisticated method called
the  ``Kernel density method'' is used.
\subsection{Kernel Density Method}
\label{SectKDM}
\subsubsection{The Basic Concept}
If the statistical significance of the signal is large enough,
the peak phase can be precisely determined without assuming a specific pulse shape.
However, the significance of the obtained signal is not high enough
to determine the peak phase with a precision of $\sim 0.003$, 
especially for energies above 10 GeV.
By assuming the pulse shape a priori, the fitting of the assumed shape to 
the obtained data
might improve the precision.
In such a case, in order to achieve the best possible precision, the data should not be 
converted into a binned phase diagram but should be analyzed on an event-by-event basis.
An event-by-event maximum likelihood method could have been used for that
but, in the case of MAGIC data, which
is dominated by $\sim 15$ million background events, it requires too much computational power.

The Kernel density method also uses event-by-event information but it does not require
too much computational power. Moreover,
 the assumption of the pulse shape is not needed either,
although a so-called ``kernel estimator'' must be chosen beforehand.
The kernel density method is a well established statistical method for estimating the probability 
density  function of a measured parameter, based on the observed data sample (see e.g. 
\cite{KernelDensity1}, \cite{KernelDensity2} and \cite{KernelDensity3} ).
The true pulse profile can be interpreted as a probability 
density function for the pulse phase of the signals and, hence, the kernel density method 
can be applied.

\begin{figure}[h]
\centering
\includegraphics[width=0.45\textwidth]{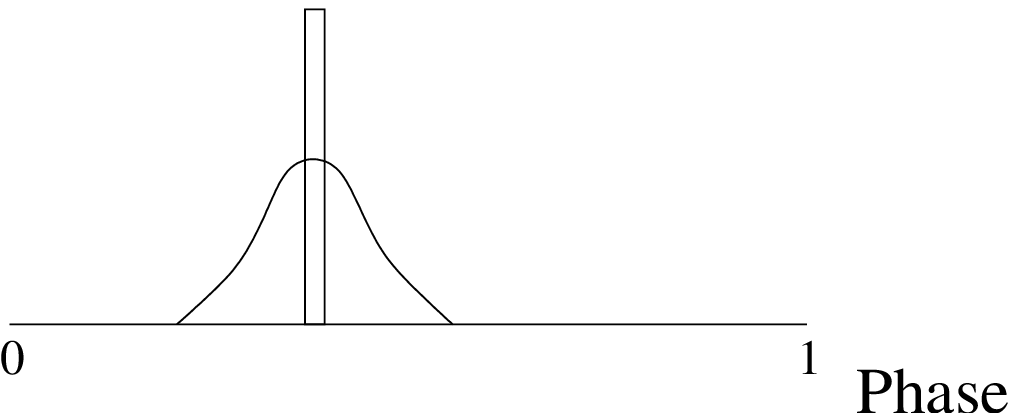}
\includegraphics[width=0.45\textwidth]{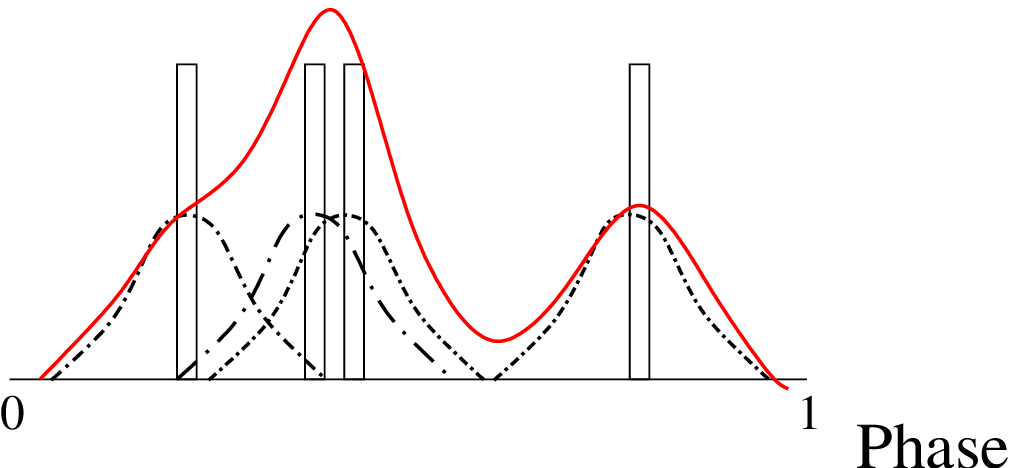}
\caption{Schematical explanation of the Kernel density method.
In a phase histogram, each event is smeared by a kernel density estimator, as shown in the left
panel. $f(p)$ is the sum of the smeared curves of all the events, as indicated by a red line
in the right panel.
}
\label{FigKernelDensityMethod}
\end{figure}

\begin{figure}[h]
\centering
\includegraphics[width=0.45\textwidth]{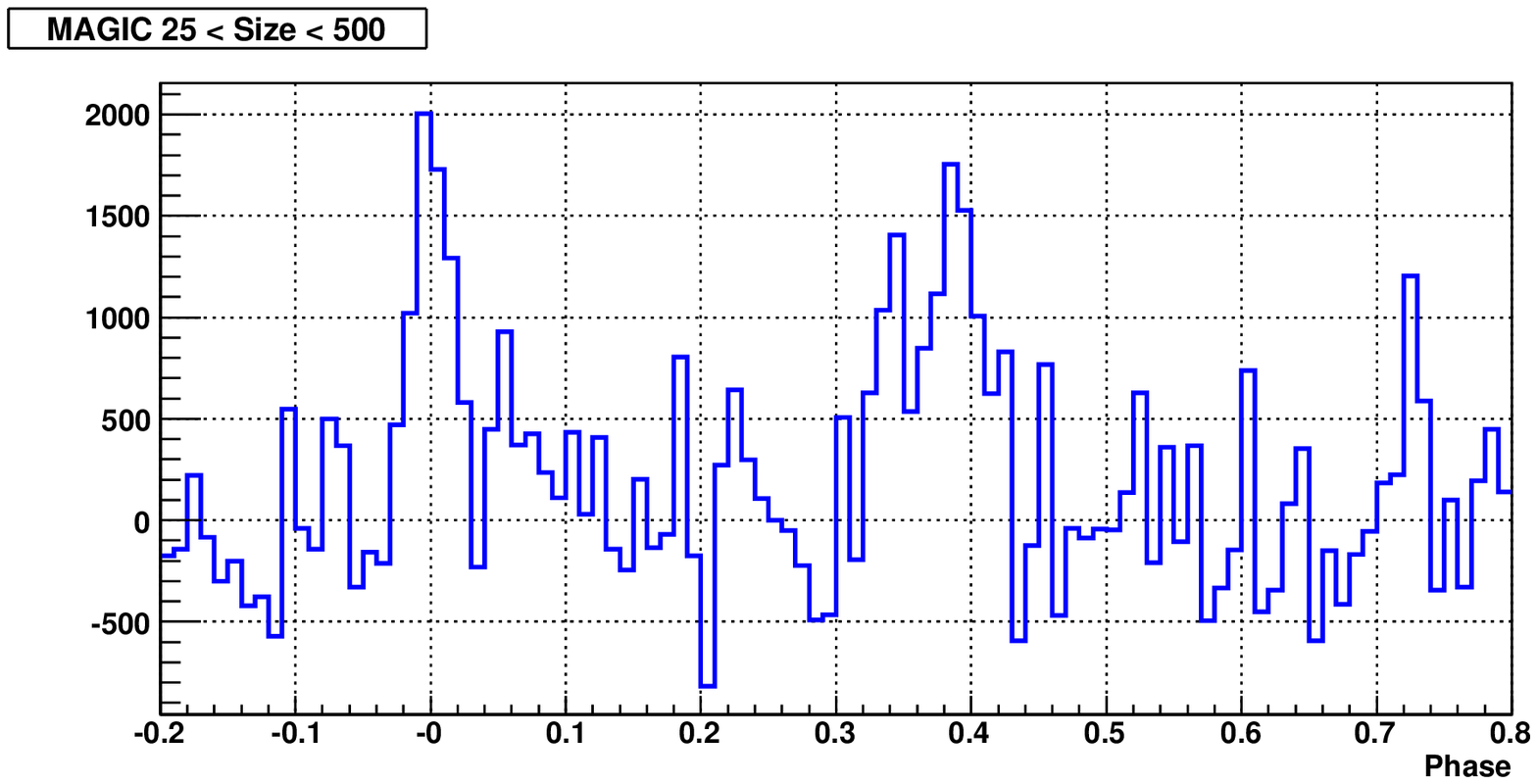}
\includegraphics[width=0.45\textwidth]{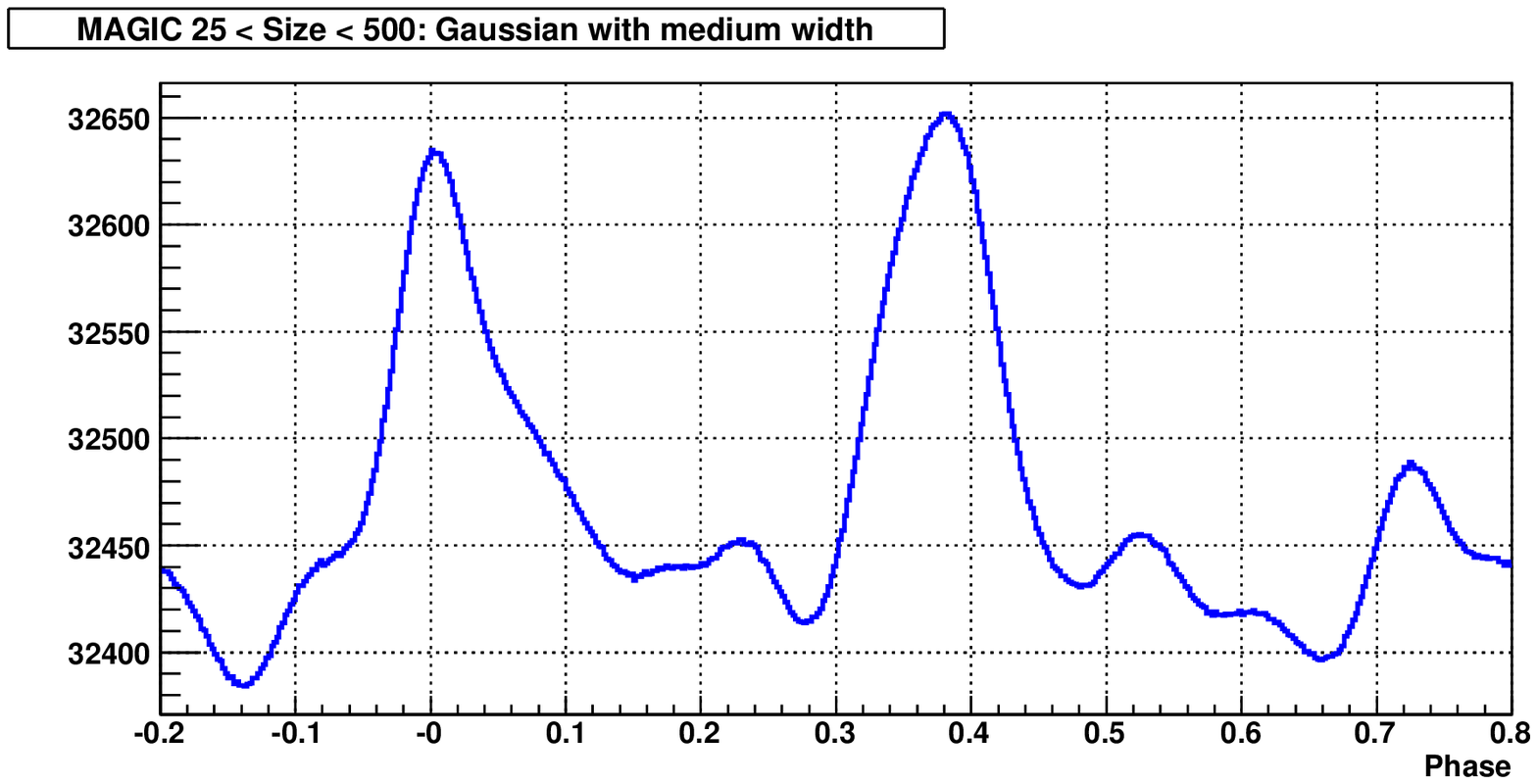}
\caption{An example of the kernel density method.
Left: The original light curve from MAGIC data with $SIZE$ between 25 and 500. 
Right: The obtained probability density function. The Gaussian kernel with $h = 0.024$ is used. }
\label{FigKernelDensityIllust}
\end{figure}

The probability density $f(p)$ as a function of phase $p$ can be estimated as follows:
\begin{eqnarray}
f(p) &=& \frac{1}{N}\sum_{i=1}^{N} K_{h}\left(\frac{p-p_i}{h}\right) 
\end{eqnarray}
 where $N$, $p_i$, $K_h(x)$ and $h$ are the total number of events, the phase of $i$th 
event, a kernel estimator and the band width of the kernel estimator $K_h$. 
The method is schematically illustrated in Fig. \ref{FigKernelDensityMethod}. 
In a light curve (a phase histogram), the phase of a given event is smeared by a kernel density 
estimator. $f(p)$ is the sum of the smeared curves of all the events.
An example of the application of the method using the MAGIC data 
is shown in Fig. \ref{FigKernelDensityIllust}.

\subsubsection{The Choice of the Kernel Estimator $K_h$ and its Width $h$}
\begin{figure}[h]
\centering
\includegraphics[width=0.6\textwidth]{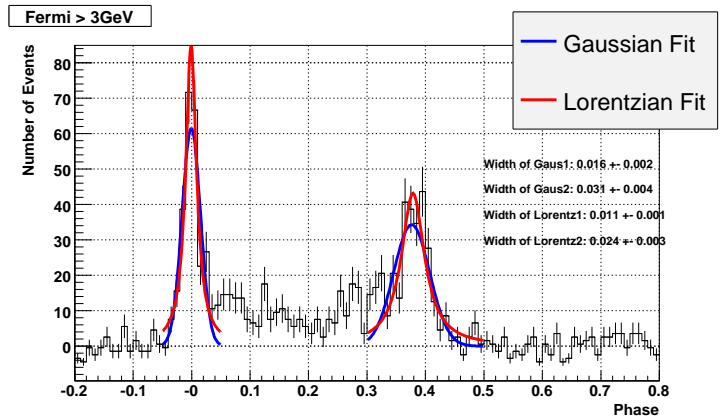}
\caption{Fitting a Lorentzian and a Gaussian for a light curve of \fermi-LAT data above 3 GeV}
\label{FigWidthDecision}
\end{figure}

 As a kernel estimator $K_h$, a normal Gaussian is commonly used in 
many different applications,
\begin{equation}
K_h\left(\frac{p-p_i}{h}\right) = \frac{1}{\sqrt{2\pi}h} e^{-\frac{ (p-p_i)^2 }{2h^2}}
\end{equation}
On the other hand, as shown in Fig. \ref{FigWidthDecision}, a Lorentzian 
represents the pulse shape better. The Lorentzian kernel estimator 
is written as
\begin{eqnarray}
K_h\left(\frac{p-p_i}{h}\right) &=& \frac{1}{\sqrt{\pi}h} \frac{1 }{1+\frac{ (p-p_i)^2 }{h^2}} \\
\end{eqnarray}
One of these estimators should be used.

 Not only the shape of the estimator but also the width $h$ must be properly chosen
(see \cite{BandWidthChoice}).
Too narrow a width will make the density function $f(p)$ too wiggly and produce many spurious 
features. 
On the other hand, too big a width will lead to too smooth a function that smears out all the 
structure. Since the pulse shape is not symmetric, as described 
in the previous section, the smearing may cause a shift in the peak phase.
  I chose $h$ based on the light curve of \fermi-LAT data above 3 GeV, which is more or less
the (logarithmic) center of the concerned energy range. First,
a Gaussian and a Lorentzian are fitted to P1 and P2 independently. Results are 
shown in Fig. \ref{FigWidthDecision}.
The best fit $h$s are $0.016\pm0.002$ (Gaussian fit to P1), 
$0.031\pm0.004$ (Gaussian fit to P2), 
$0.011 \pm 0.001$ (Lorentzian fit to P1) and $0.024\pm0.003$ (Lorentzian fit to P2).
The optimal $h$ should be close to these values. 


\begin{figure}[h]
\centering
\includegraphics[width=0.30\textwidth]{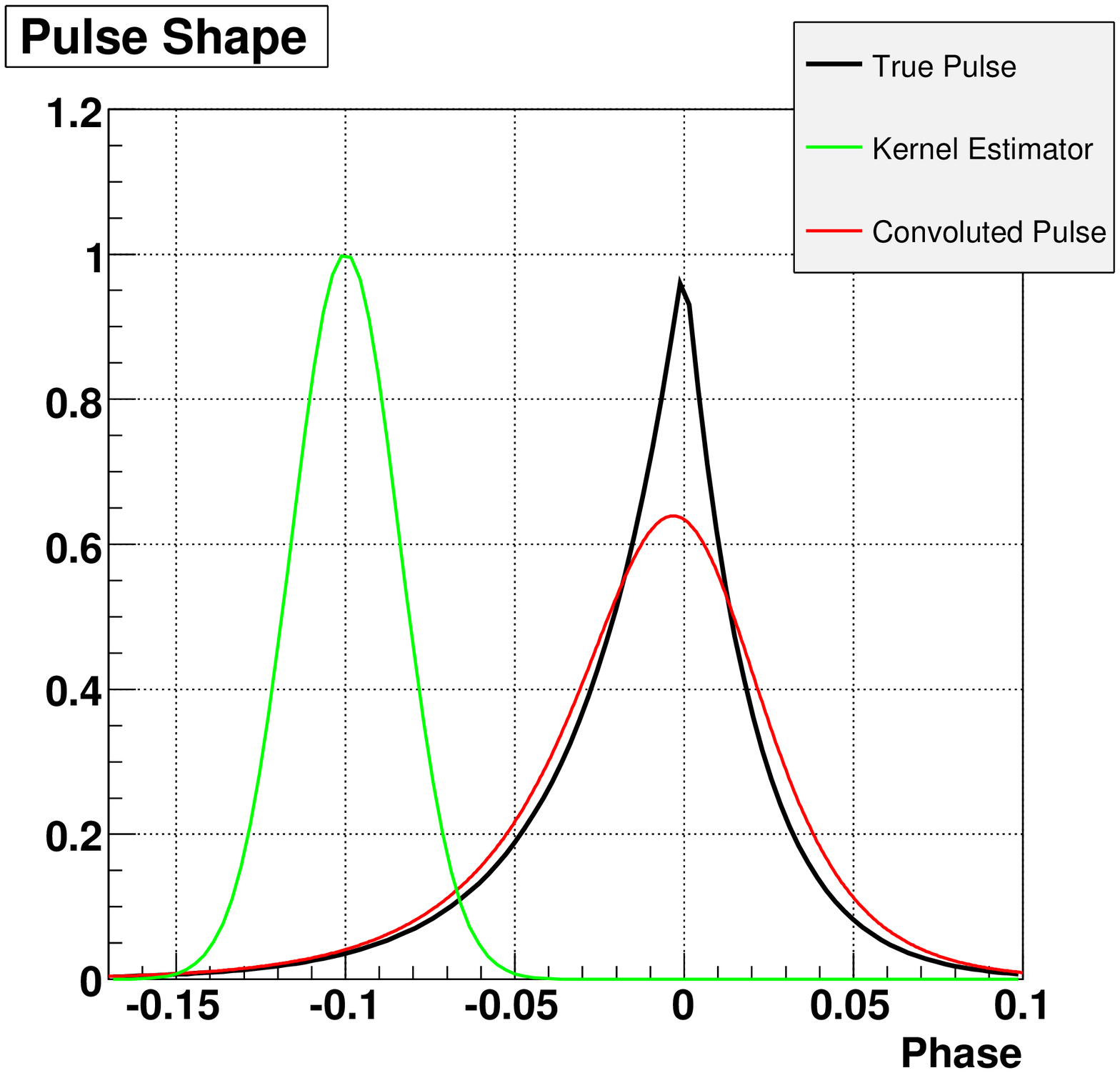}
\includegraphics[width=0.30\textwidth]{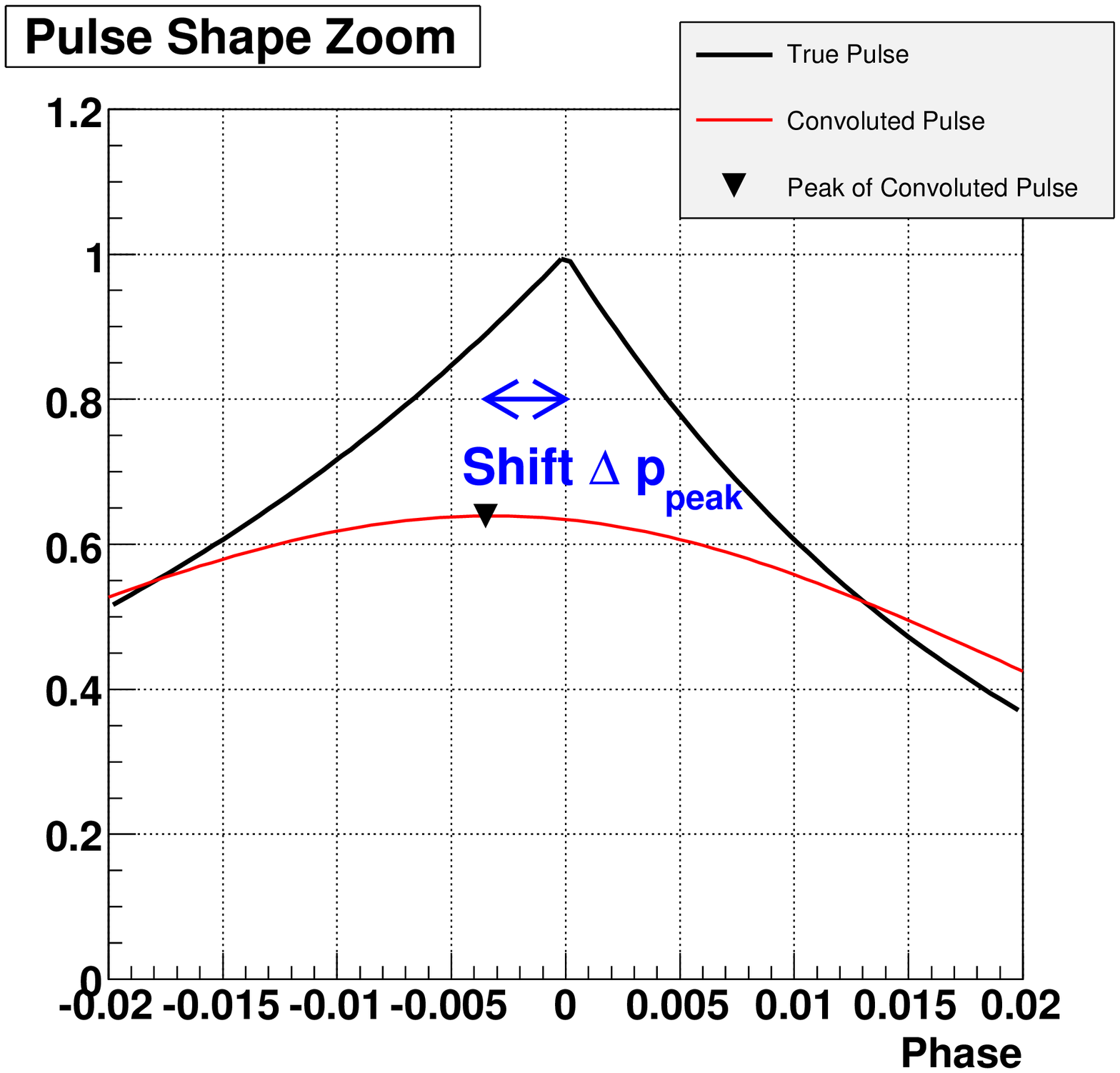}
\caption{Left: An example of the convolution of a true pulse with a kernel estimator.
$\tau_{rise} = 0.03$ and $\tau_{fall} = 0.02$ are used for the true pulse shape $f_{true}(p)$,
while a Gaussian kernel $K_h(p)$ with $h = 0.016$ is used for the kernel estimator.
A black and a red line indicate the true pulse shape $f_{true}(p)$ and the convoluted 
pulse shape $f(p)$, respectively, 
while a green line indicates the kernel estimator $K_h(p)$ 
which is scaled such that the peak height = 1.
Right: The same as the left panel but phases from -0.02 to 0.02 are zoomed. 
The analytical phase shift $\Delta p_{peak}$ is indicated by a blue arrow.
}
\label{FigPhaseShiftExampl}
\end{figure}

The final choice of the shape of $K_h$ (Gaussian or Lorentzian) and the width $h$ was made
 such that
the peak phase shift caused by the smearing effect is minimal.
The phase shift is estimated as follows:
Assuming a pulse shape to be
\begin{equation}
f_{true}(p)= \left\{
\begin{array}{cr} 
\exp (p/\tau_{rise}) & {\rm if~~~~~} p \leq 0\\
\exp (-p/\tau_{fall}) & {\rm if~~~~~} p > 0\\
\end{array}
\right.
\label{EqShapeAssumption}
\end{equation}
 $f(p)$ is calculated by convoluting the pulse shape $f_{true}(p)$ with the kernel estimator
$K_h(p)$
\begin{equation}
f(p)= \int ^{0.25}_{-0.25} f_{true}(p^\prime) K_h(p-p^\prime) dp^\prime
\end{equation}
Then, the difference in the peak phase $\Delta p_{peak}$ between $f_{true}(p)$ and $f(p)$ is estimated.
Hereafter, $\Delta p_{peak}$ is referred to as the ``analytical phase shift''.
An example of the analytical phase shift 
is shown in Fig. \ref{FigPhaseShiftExampl}.

\begin{figure}[]
\centering
\includegraphics[width=0.45\textwidth]{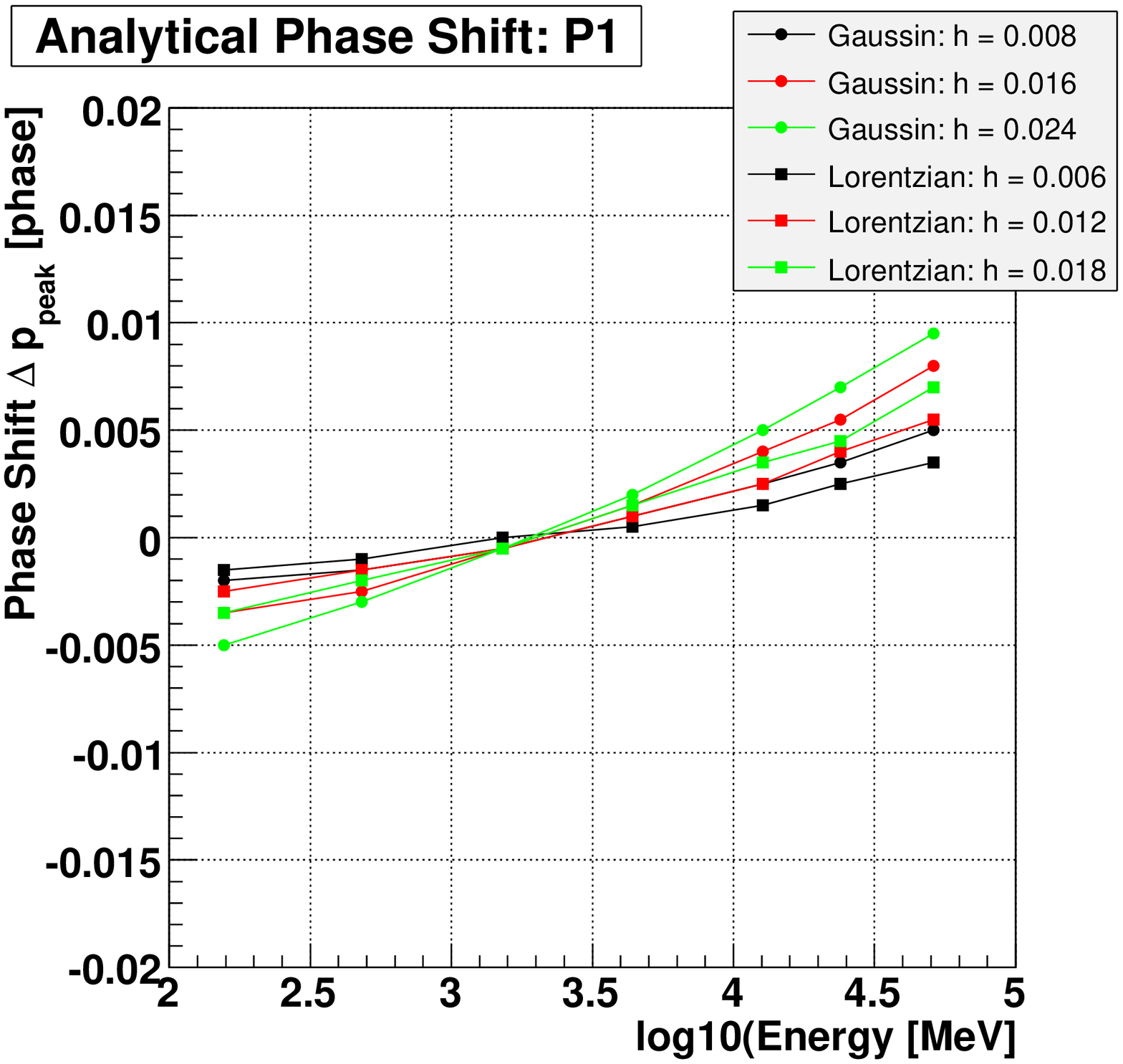}
\includegraphics[width=0.45\textwidth]{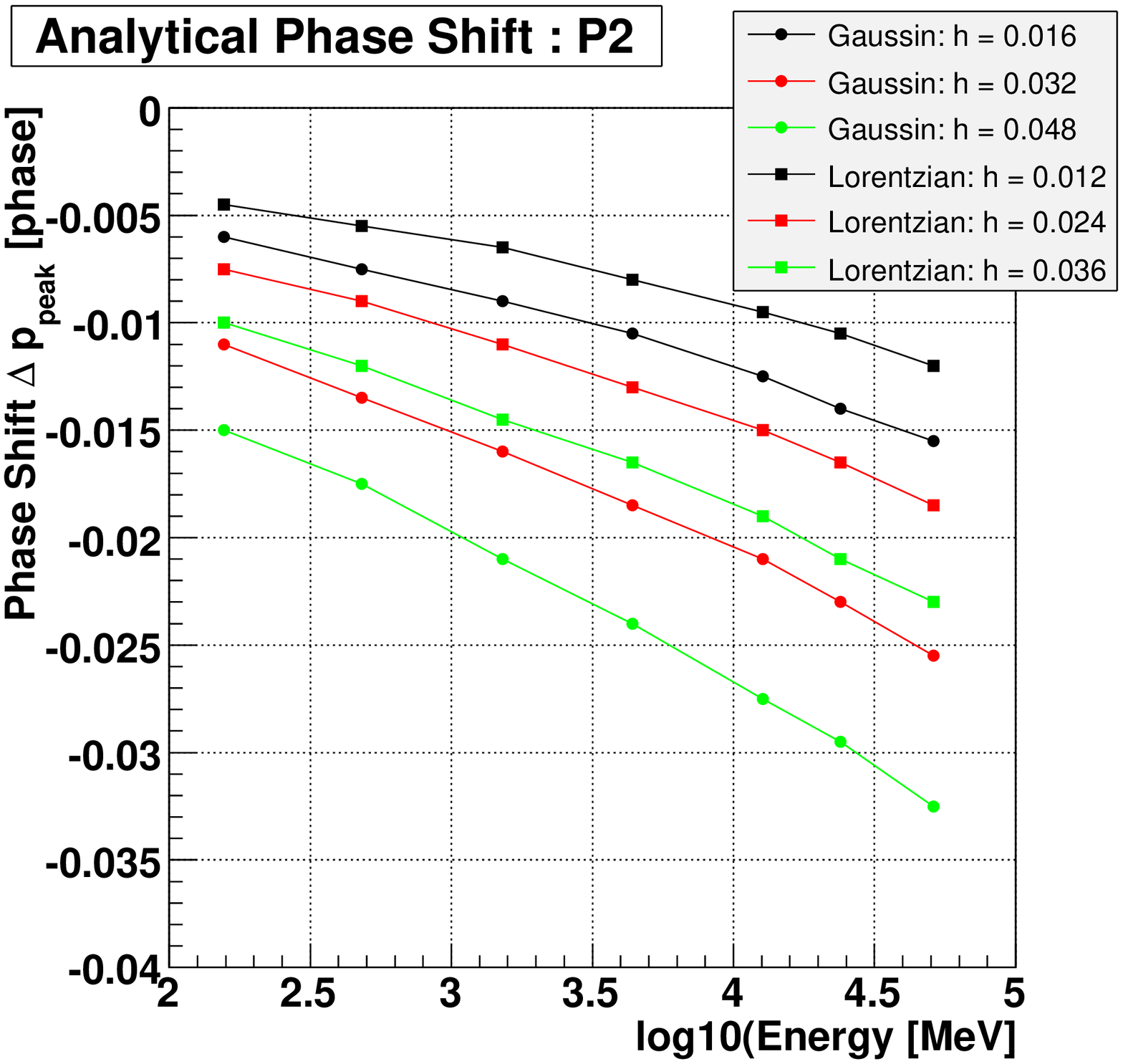}
\caption{The energy and the estimator dependence of the analytical phase shift $\Delta p_{peak}$ 
for P1 (left) and P2 (right)}
\label{FigAnaPhaseShift}
\end{figure}

By substituting Eqs. \ref{EqTau1} $-$ \ref{EqTau4} for $\tau_{rise}$ and $\tau_{fall}$ in 
Eq. \ref{EqShapeAssumption}, I estimated the energy dependence of the analytical phase shift
$\Delta p_{peak}$
for different $K_h$s,
as shown in Fig. \ref{FigAnaPhaseShift}.
The Lorentzian kernel estimators with $h = 0.006$ for P1 and $h = 0.012$ for P2,
which are the half of the fitted values (see Fig. \ref{FigWidthDecision})
 have the smallest effect. Even smaller $h$ might reduce the effect
further. However, if $h$ is too small compared to the fitted values, 
$f(p)$ would, in turn, produce spurious structures, as mentioned before. 
Therefore, in the following analysis,  I use Lorentzian for $K_h$ 
with $h = 0.006$ for P1 and $h = 0.012$ for P2. The residual $\Delta p_{peak}$ will be
subtracted from the obtained results.



\subsubsection{Statistical Uncertainty Estimation by the Bootstrapping Method}
One can estimate a probability density function $f(p)$
from observed data by the kernel density method and, then, determine the peak position from it.
 However, it does not give the statistical uncertainty.

The statistical uncertainty of the result can be estimated by the bootstrapping method 
which is well established and is used in many statistical treatments
(see e.g. \cite{Bootstrap1} and \cite{Bootstrap2}).
The procedure is explained as follows:
Let $N$ be the total number of observed events.
One randomly chooses $N$ events out of the observed $N$ events. The 
same events can be chosen multiple times. 
Then, the same kernel density method is applied to the
chosen data sample. By repeating this procedure $M$ times, one obtains 
$M$ different $f(p)$s. The RMS of the peak phase distribution from them is used
as the statistical uncertainty of the peak phase.
In this analysis, $M = 900$ is used.

\clearpage
\subsection{Example of the Kernel Density for Different Energies}
\begin{figure}[h]
\centering
\includegraphics[width=0.45\textwidth]{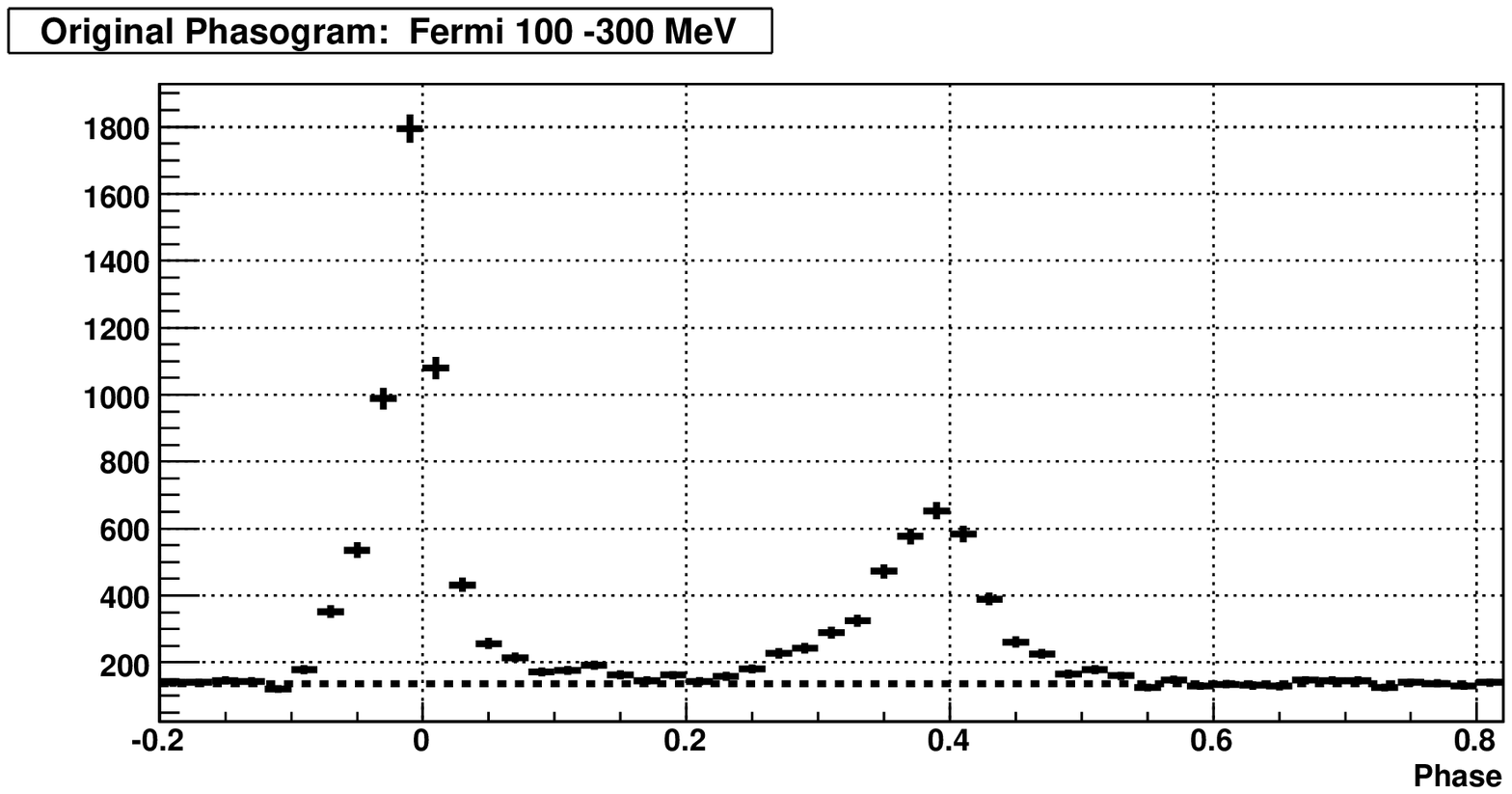}
\includegraphics[width=0.45\textwidth]{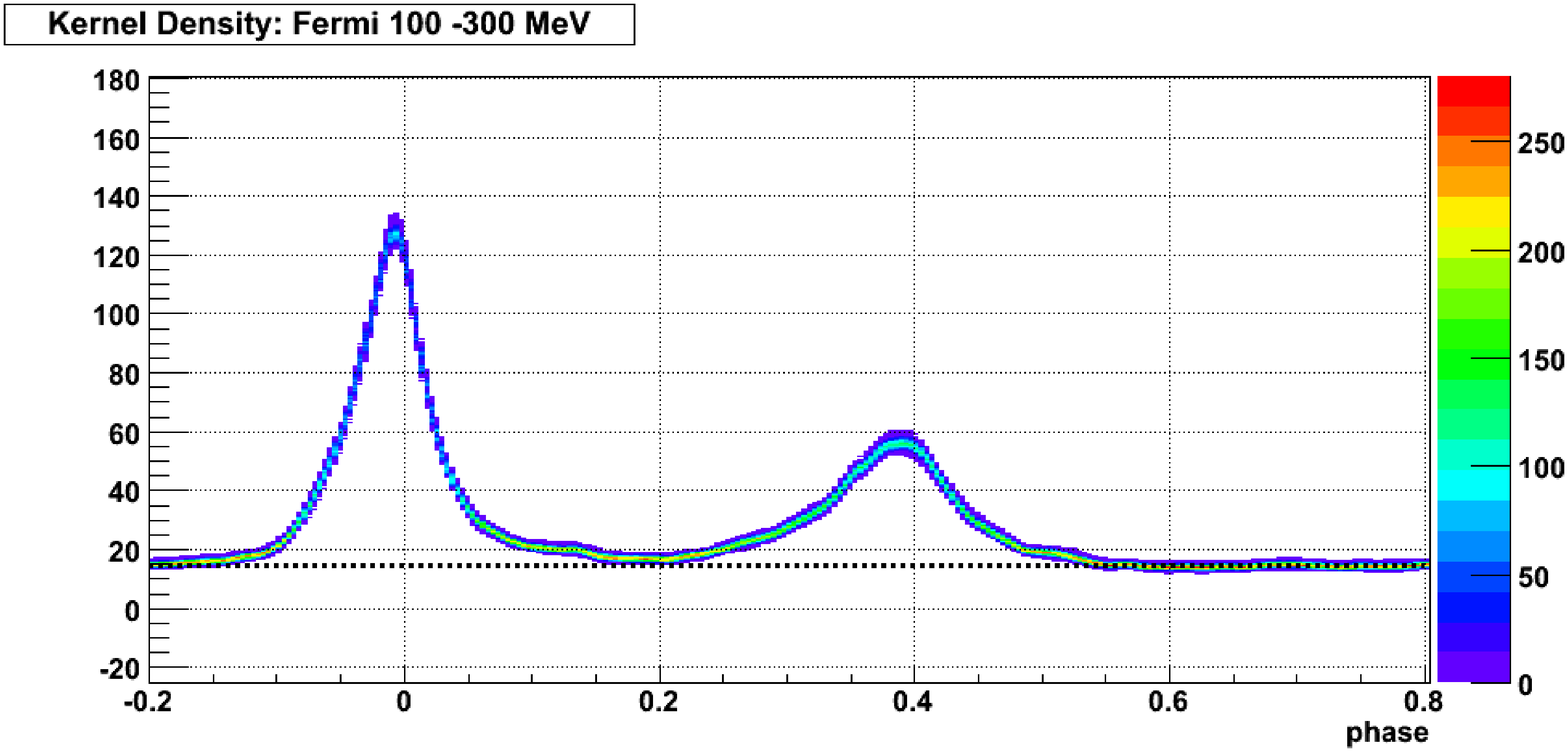}
\includegraphics[width=0.45\textwidth]{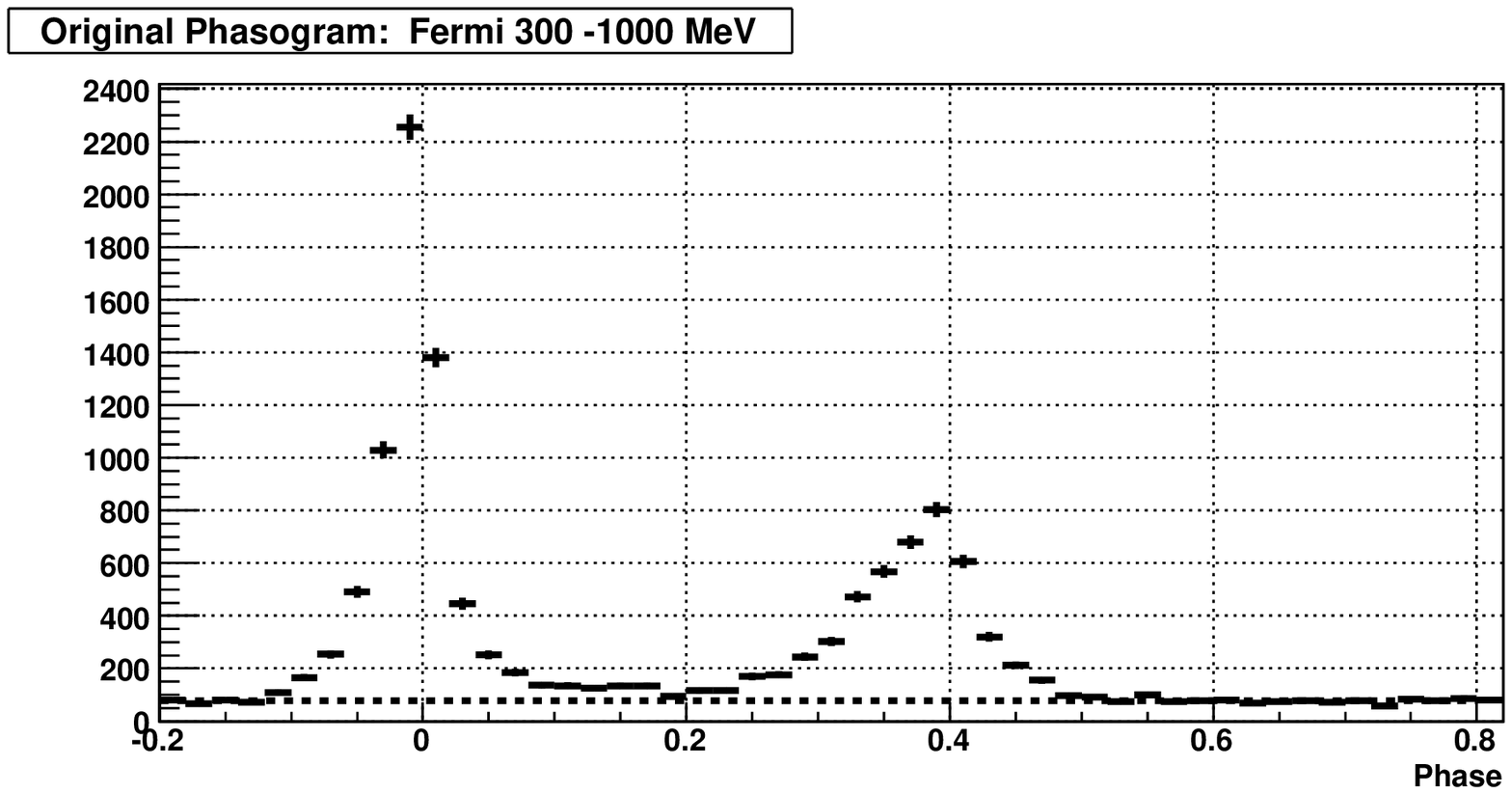}
\includegraphics[width=0.45\textwidth]{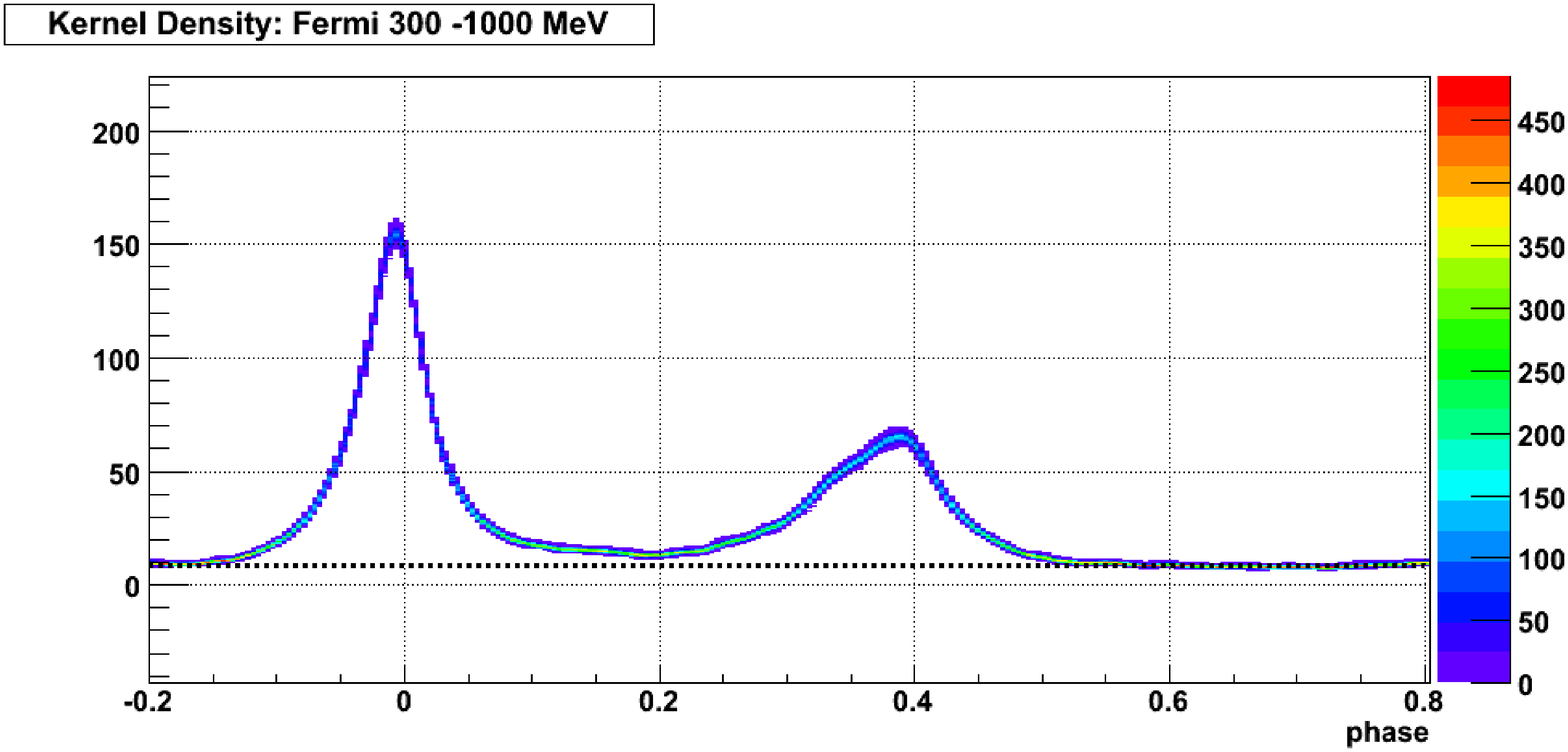}
\includegraphics[width=0.45\textwidth]{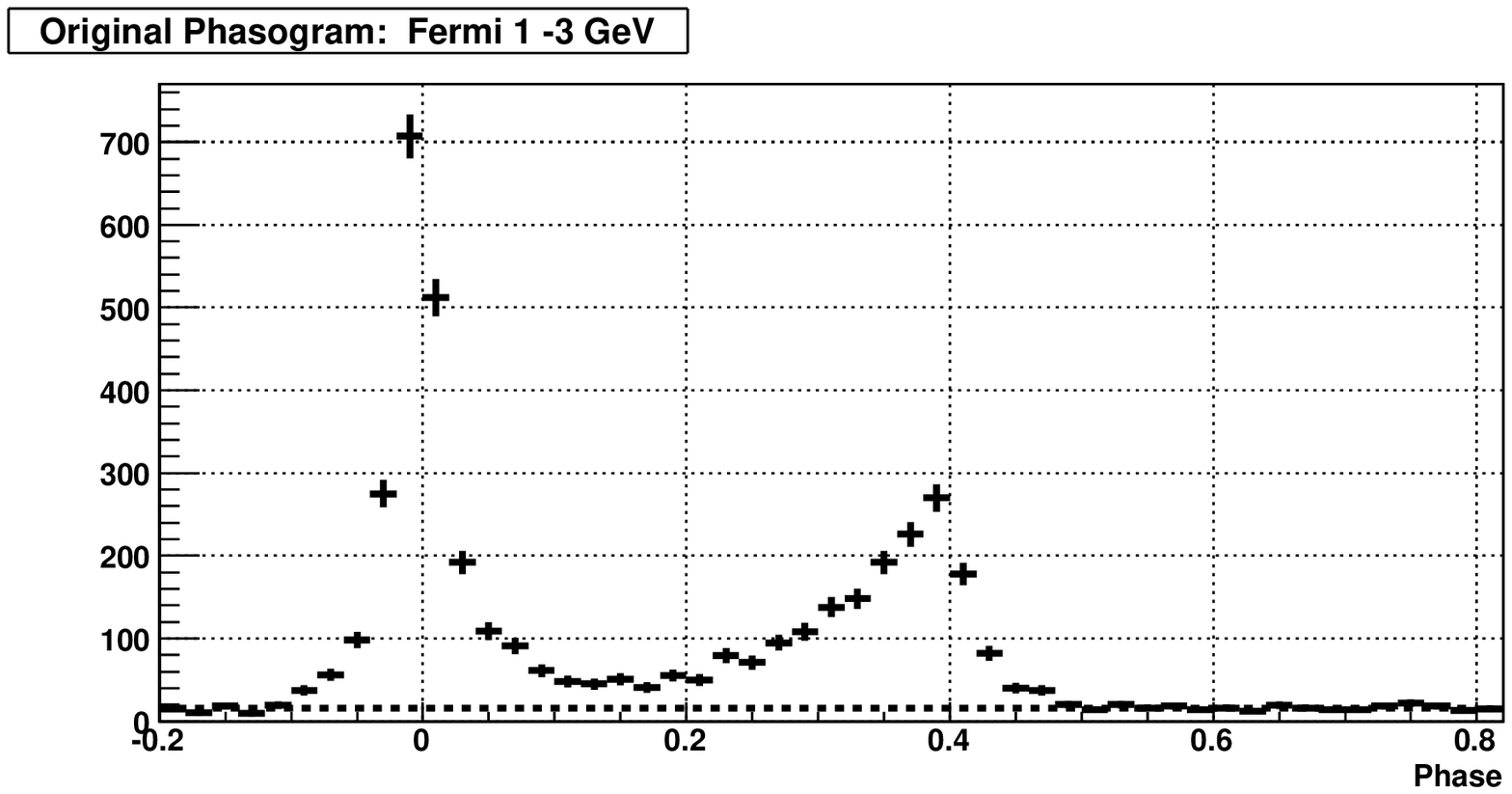}
\includegraphics[width=0.45\textwidth]{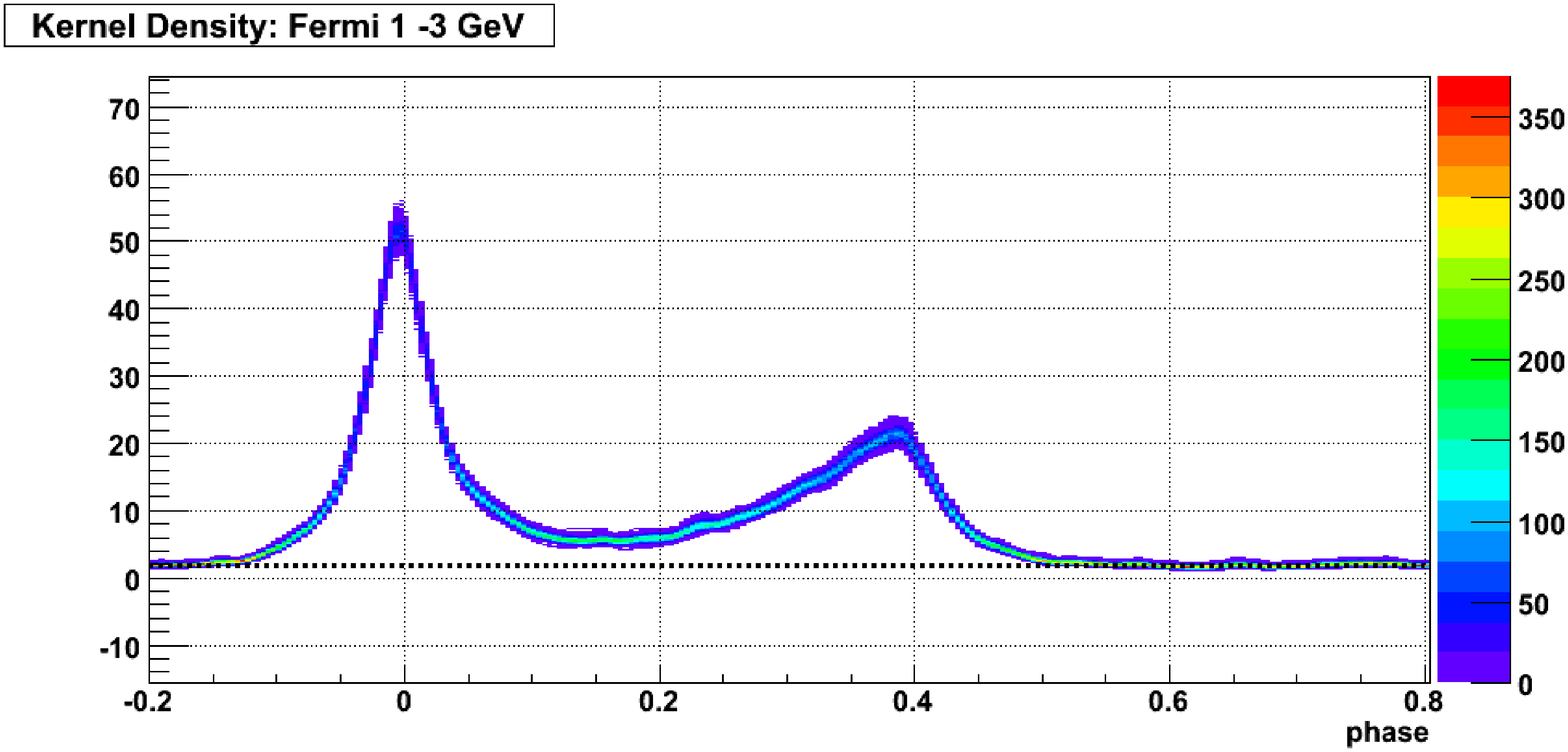}
\caption{The original light curves (left)
and the resulting probability density function (kernel density) by using Lorentzian kernel with $h = 0.012$ (right).
This kernel density is used for the P2 peak determination,
 while $ h = 0.006$ is used for the P1 peak. 
Black dotted lines indicate the background level.
900 curves obtained by the bootstrapping method are overlaid in the left panels
and colors indicate their density.
From the top: 100~-~300 MeV, 0.3~-~1 GeV and 1~-~3 GeV are shown, 
all of which are based on \fermi-LAT data.
}
\label{FigKernelDensityExample}
\end{figure}

\begin{figure}[h]
\centering
\includegraphics[width=0.45\textwidth]{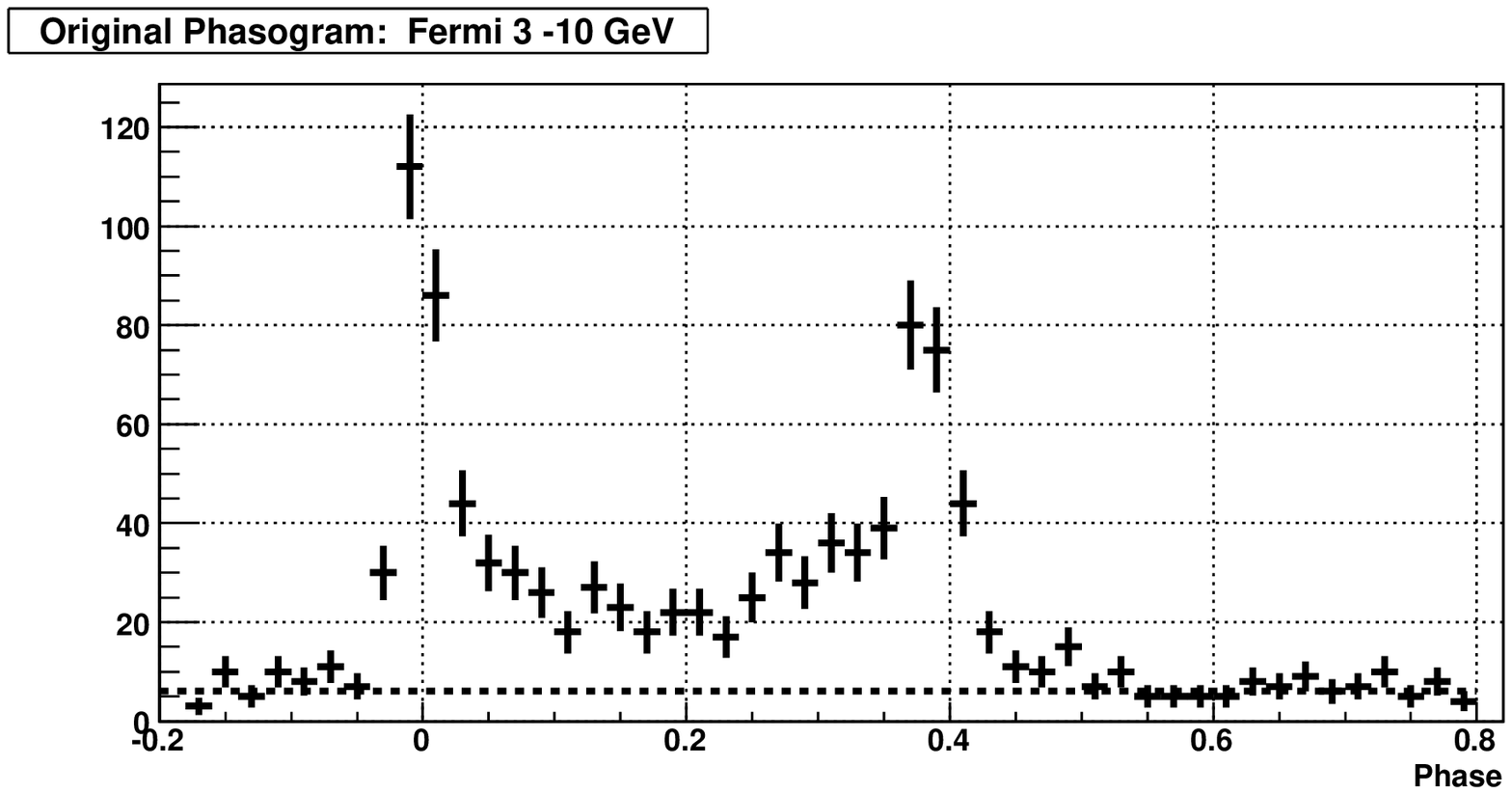}
\includegraphics[width=0.45\textwidth]{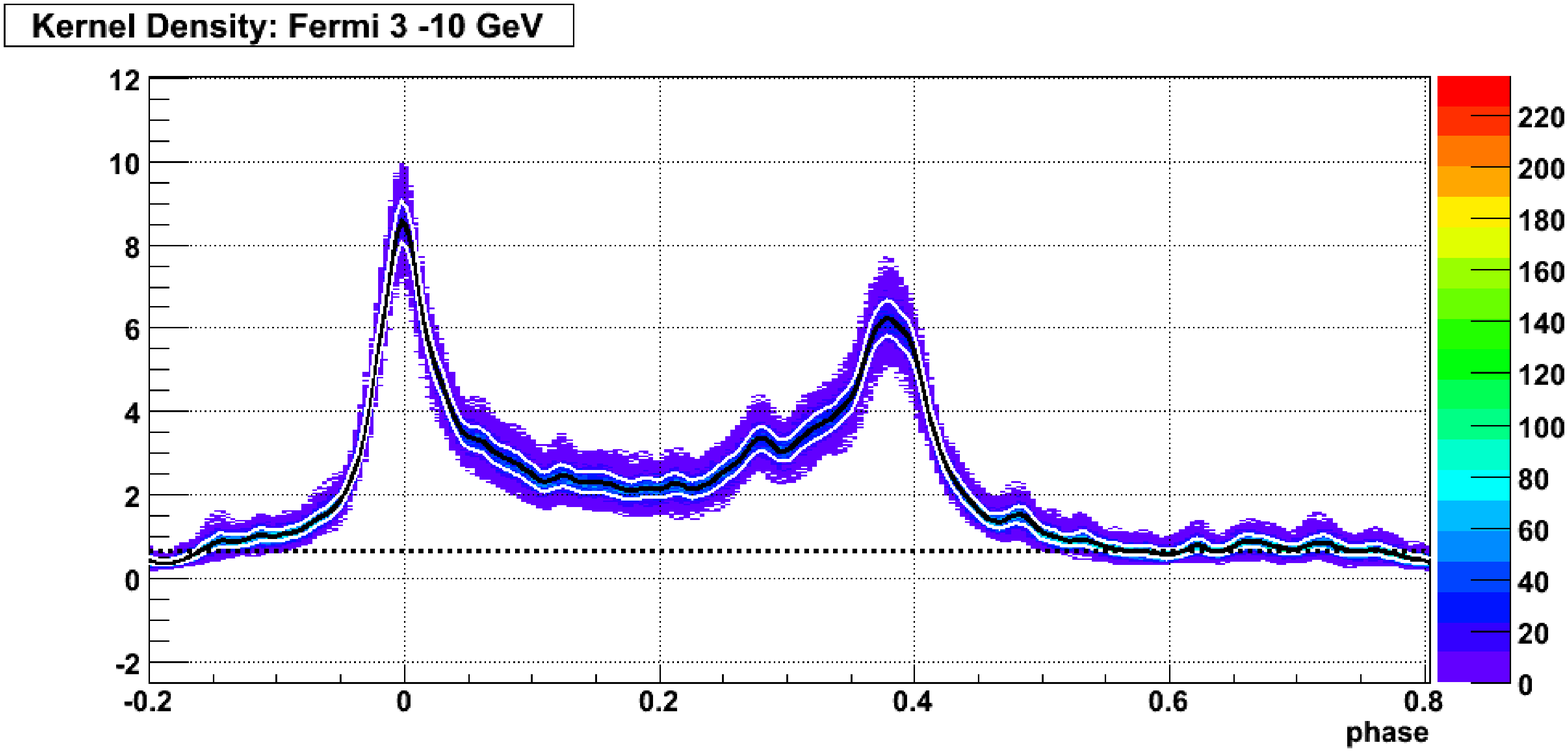}
 \includegraphics[width=0.45\textwidth]{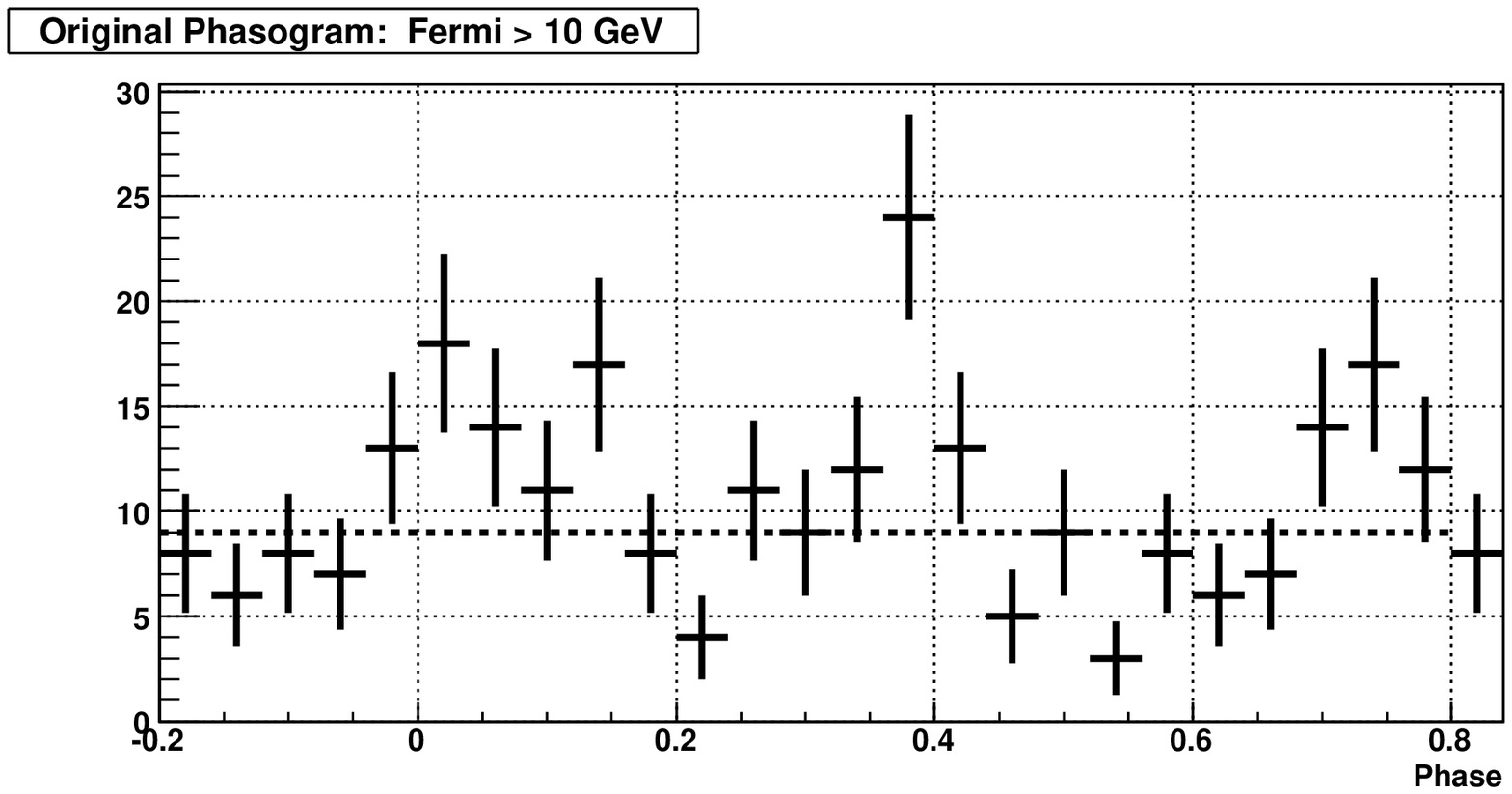}
 \includegraphics[width=0.45\textwidth]{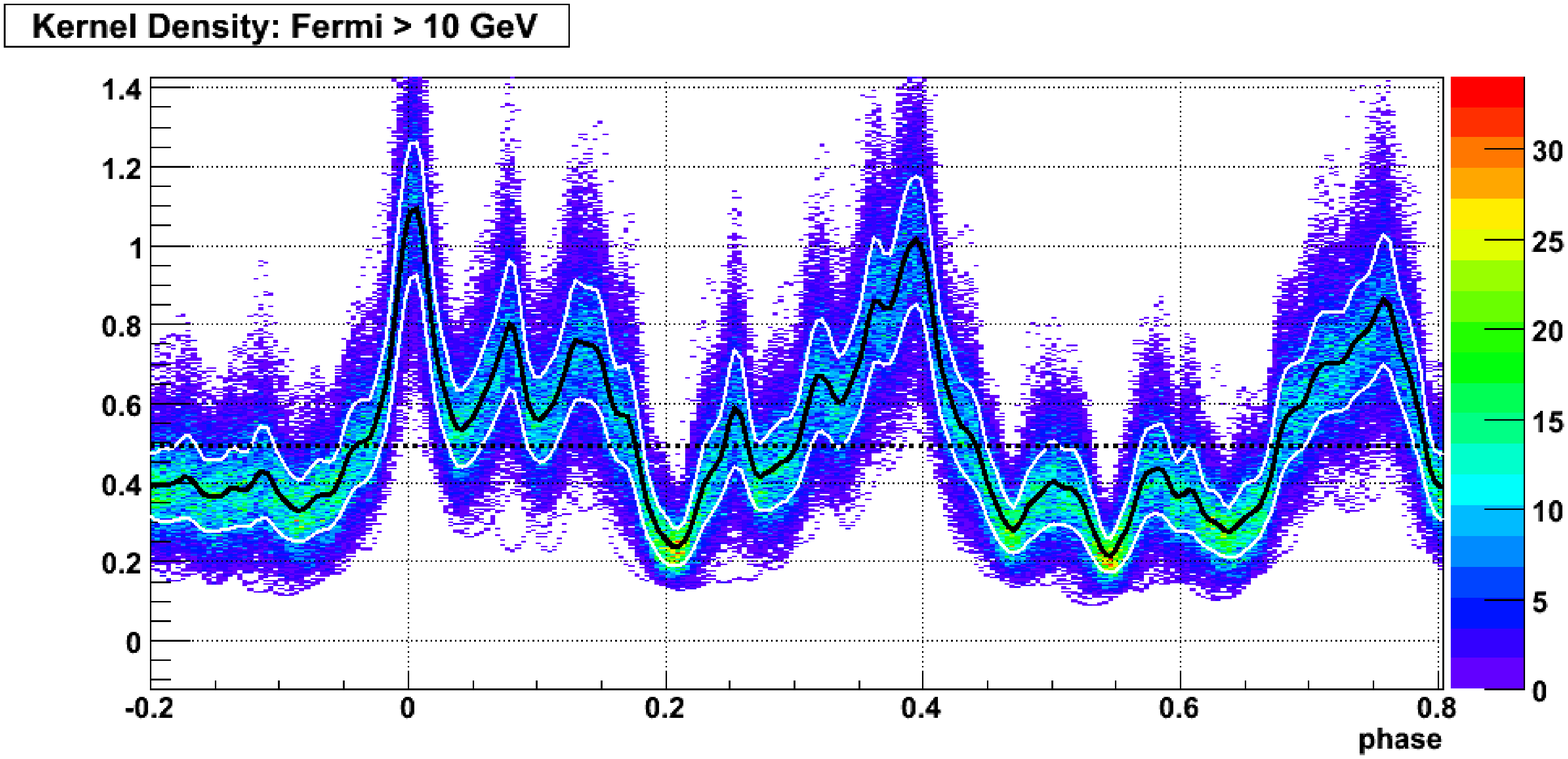}
 \includegraphics[width=0.45\textwidth]{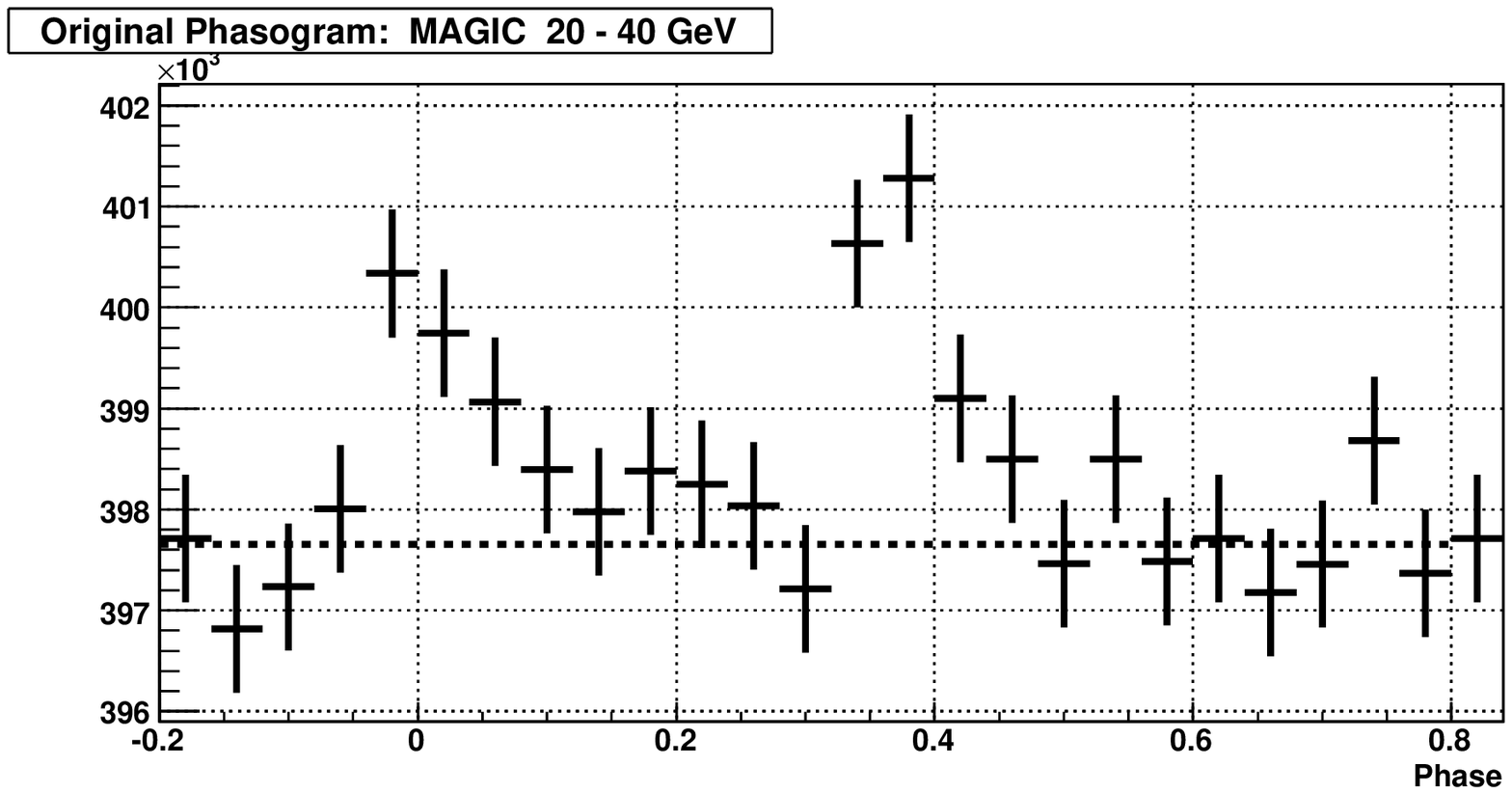}
 \includegraphics[width=0.45\textwidth]{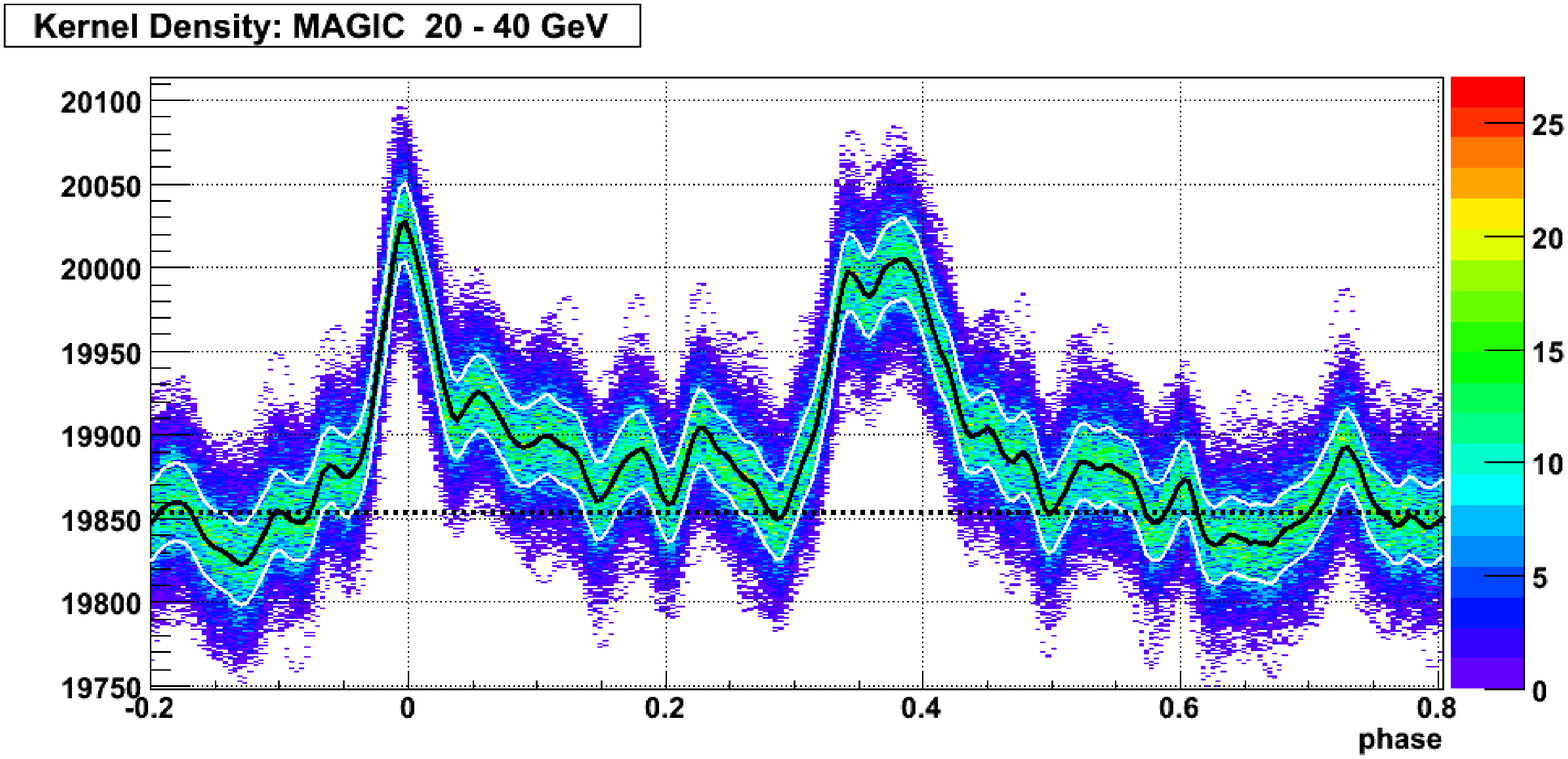}
 \includegraphics[width=0.45\textwidth]{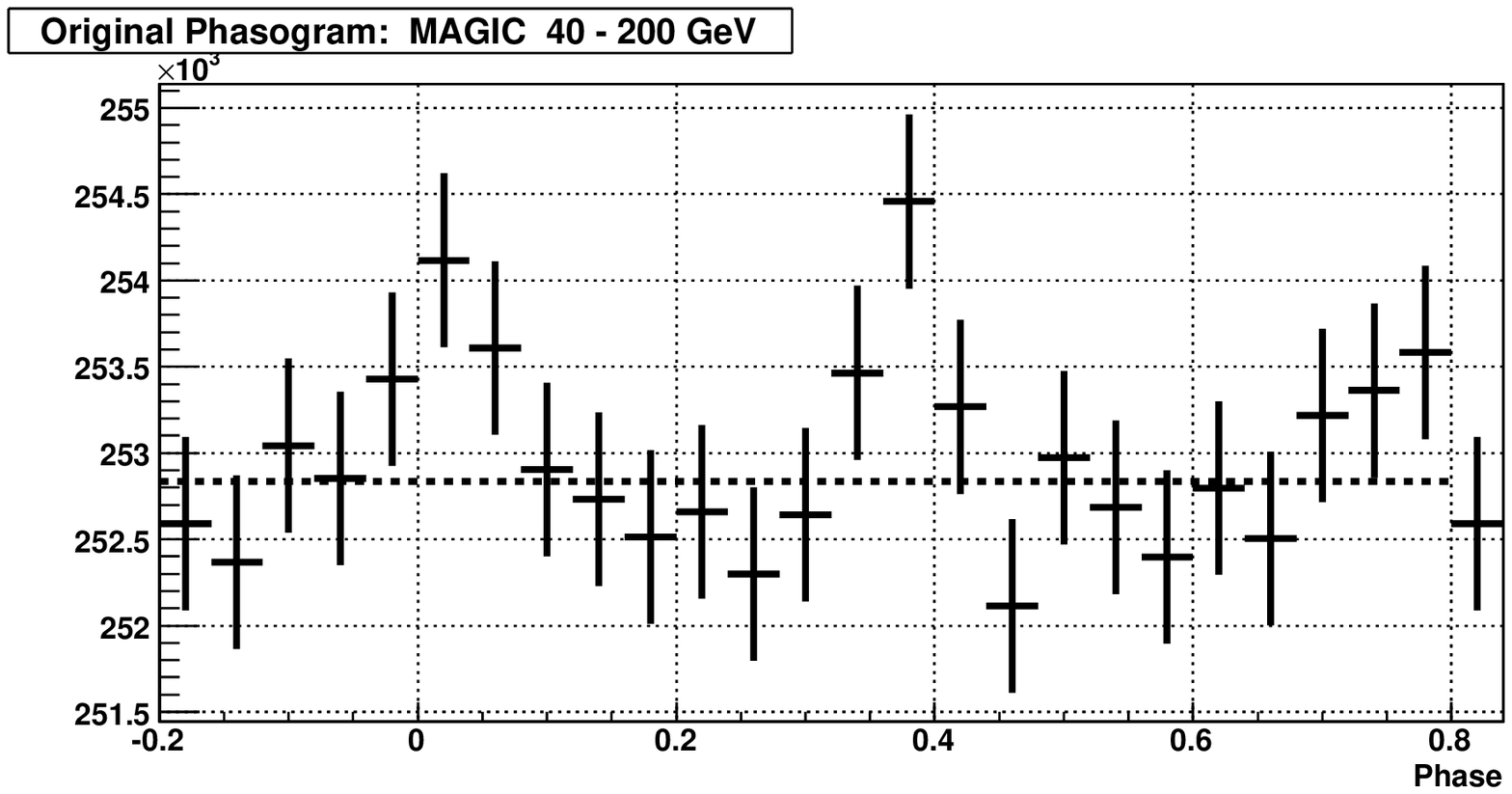}
 \includegraphics[width=0.45\textwidth]{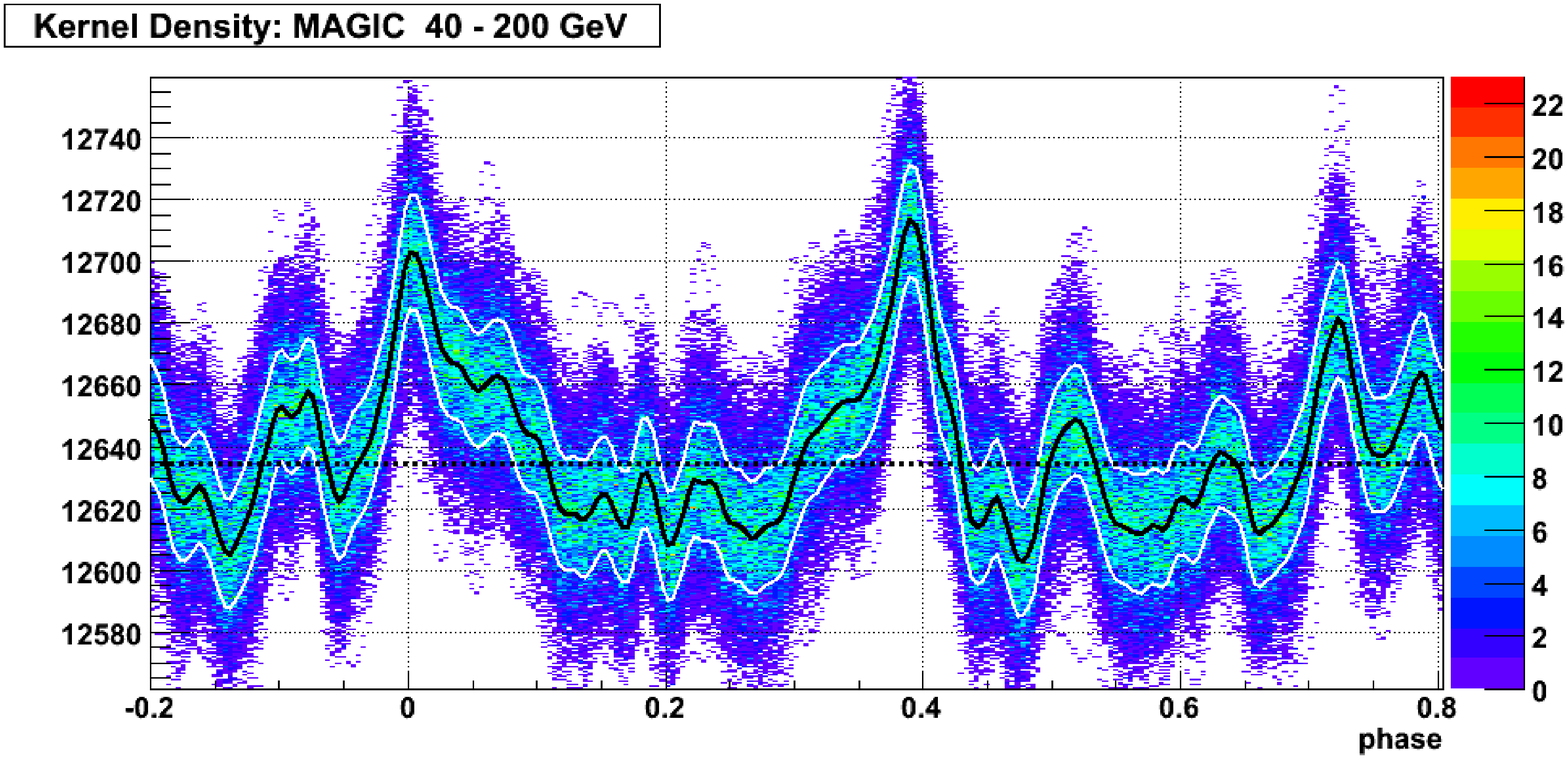}
\caption{The same as Fig \ref{FigKernelDensityExample} but for higher energies.
From the top: 3~-~10 GeV from \fermi-LAT data, above 10 GeV from \fermi-LAT data, 
20~-~40 GeV ( 25~-~70 in $SIZE$) from MAGIC data 
and 40~-~200 GeV ( 70~-~500 in $SIZE$) from MAGIC data. 
These energy ranges for MAGIC data are rough estimation based on the $SIZE$ range
and no significant excess is seen at 200 GeV.
Because of the relatively low significance of the signals,
an (acceptably) wiggly structure is visible. By using larger $h$ in the kernel estimator,
the structure will become less pronounced, while it leads to a larger analytical bias.
Black lines in the left panels indicate the resulting probability density functions,
while white lines are their (1 $\sigma$) uncertainty.
}
\label{FigKernelDensityExampleHigh}
\end{figure}

 In Fig. \ref{FigKernelDensityExample} and Fig. \ref{FigKernelDensityExampleHigh},
 the original light curves and the resulting probability functions (kernel densities) are shown. 
The statistical uncertainty of the function estimated by the bootstrapping method
are indicated by colors.




\clearpage

\subsection{Results}
\begin{figure}[h]
\centering
\includegraphics[width=0.45\textwidth]{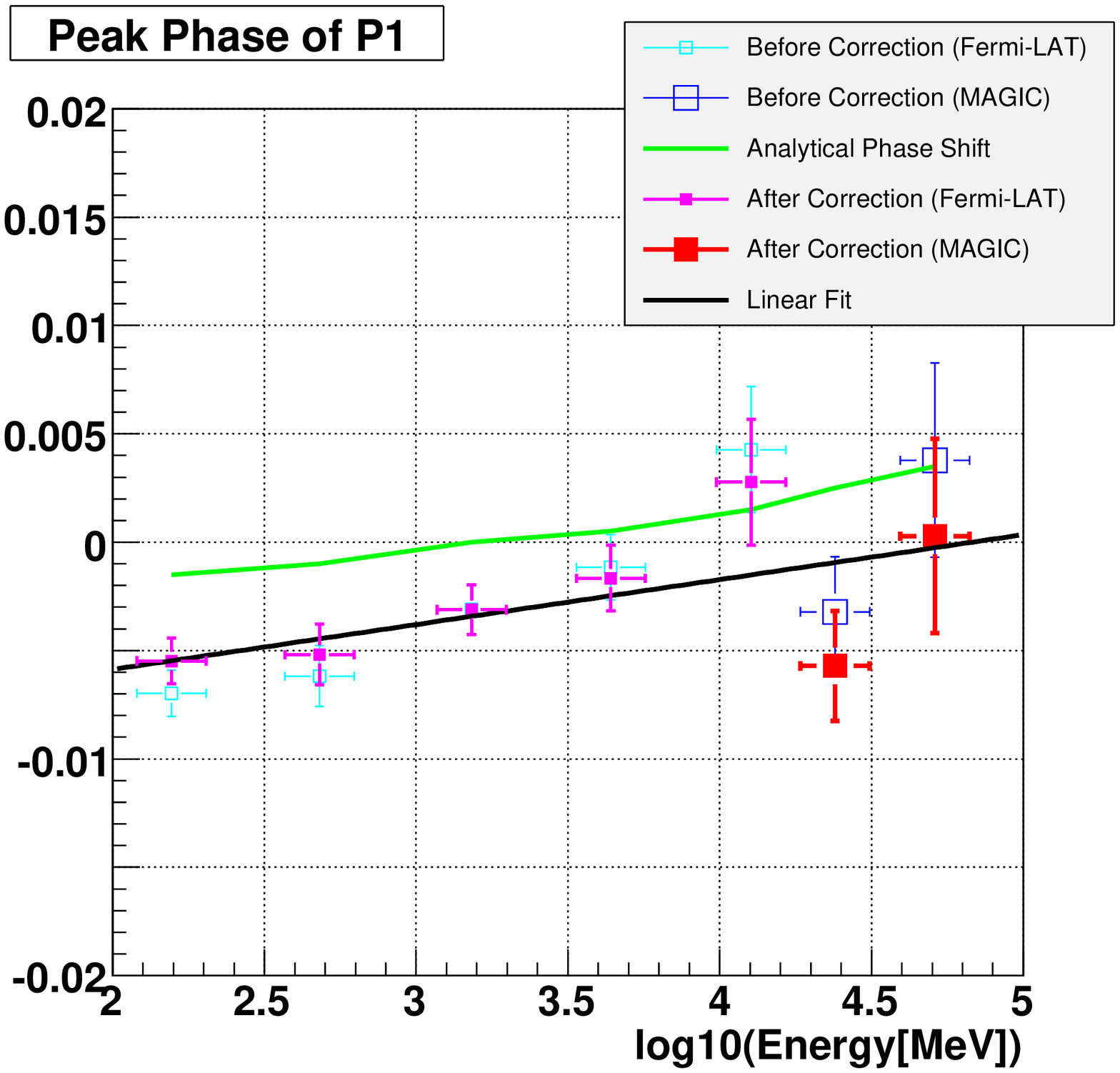}
\includegraphics[width=0.45\textwidth]{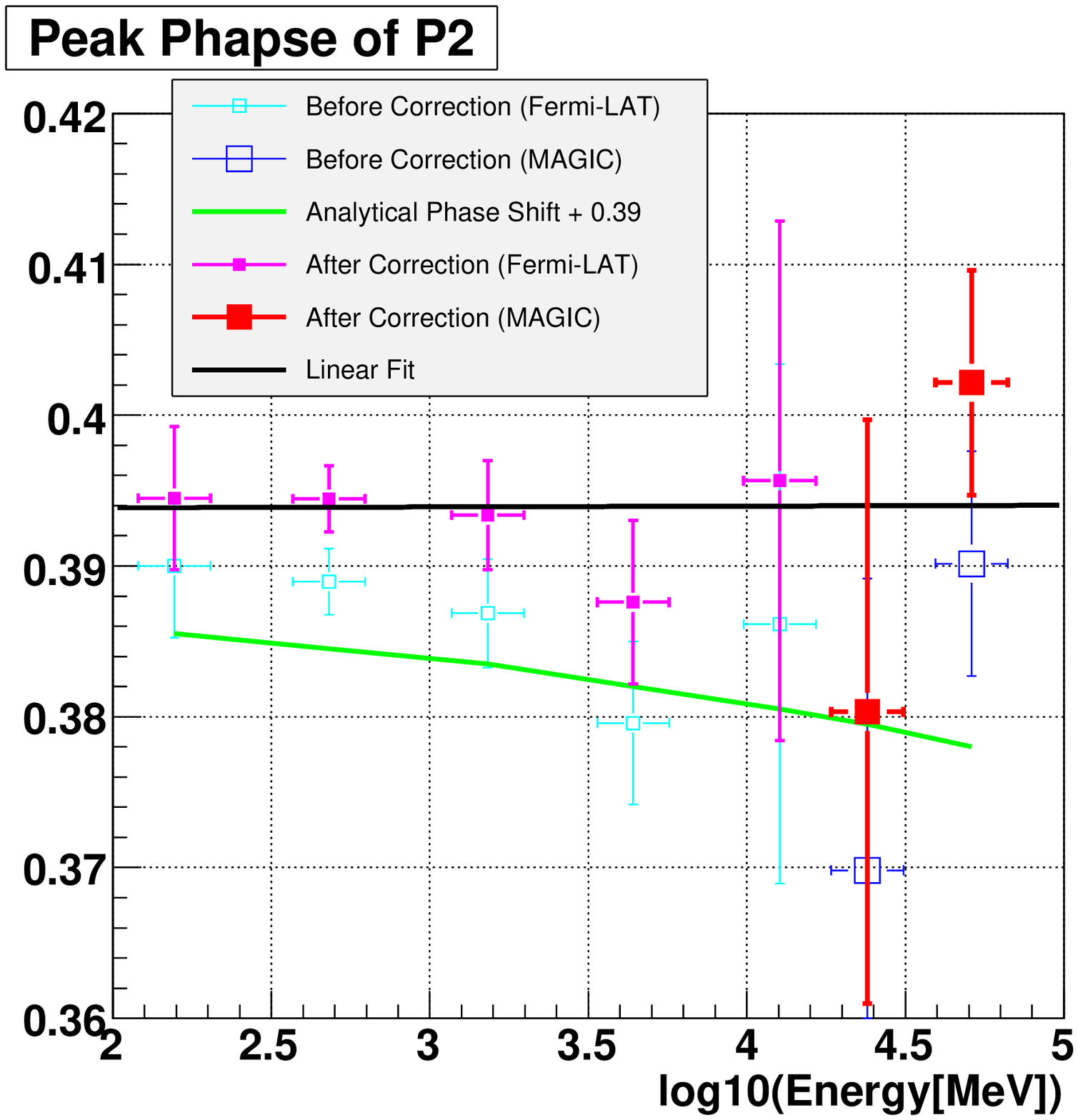}
\caption{The energy dependence of the peak phase of P1 (left) and P2 (right).
Open and filled squares denote the peak phase before and after correcting the analytical
phase shift $\Delta p_{peak}$. Green lines indicate the $\Delta p_{peak}$ 
(see Fig. \ref{FigAnaPhaseShift}). The peak phases after the correction as a function of
energy are fitted by a linear function, as shown by the black lines.
}
\label{FigEneDepPeak}
\end{figure}

 Results are shown in Fig. \ref{FigEneDepPeak}. 
Open and filled squares indicate the results before and after correcting the analytical phase shift
$\Delta p_p$, respectively.
Horizontal values and errors are determined in the same way as the P2/P1 ratio study
(see Sect. \ref{SectRatioResult}).
Blue lines indicate the energy dependence of $\Delta p_p$ estimated in Sect. \ref{SectKDM} 
(see Fig. \ref{FigAnaPhaseShift}). 


After the correction (filled squares), the energy dependence is clearly seen for P1
but it is not the case for P2, due to the large uncertainties. 
This difference comes mainly from the pulse width. 
Since P2 has twice as large a width as P1, the precision of the peak phase becomes worse.
The results after the correction (filled squares) are fitted by a linear function and  
\begin{eqnarray}
Peak1(E) &=& (-3.8 \pm 0.6)\times 10^{-3} + (2.1 \pm 0.9)\times 10^{-3}{\rm log}_{10}(E[\rm{GeV}]) \\
\label{EqPeakPhase1}
Peak2(E) &=& 0.39+(3.9 \pm 1.6)\times 10^{-3} - (0.05 \pm 2.8)\times 10^{-3}{\rm log}_{10}(E[\rm{GeV}]) 
\label{EqPeakPhase2}
\end{eqnarray}
are obtained.  $\chi2 / dof$ are 6.26/5 and 3.18/5,  for $Peak1$ and $Peak2$, respectively.  

The physical interpretation of the results will be discussed in Sect. \ref{SectEDepPeak}.

\clearpage
\section{ Possible Existence of the Third Peak}

As mentioned in Sect. \ref{SectFermiLightCurve}, a possible third peak is seen 
above 10 GeV in \fermi-LAT data. Judging from the light curve (see Fig. \ref{FigFermiPhaseEne}),
the flux might be as high as P1 and P2, although the large statistical uncertainty 
does not permit a solid argument. Here, I examine the existence of the 
possible third peak by using both \fermi-LAT data and MAGIC data. 

\subsection{Definition of the ON and OFF Phases for the Third Peak P3}

The third peak has not been detected in other energies, 
except for specific frequencies in radio, where fourth and fifth peaks are also seen 
(see Sect. \ref{FigCrabPulses}). Therefore, it is not possible to define the third peak phase
interval (P3) a priori. 
I define P3 to be phases from 0.7 to 0.8. 
It is based on the observed result itself, which leads to the overestimation of the 
flux and the statistical significance, while
 no fine tuning of the bin-edges is carried out in order to reduce the effect.
In all the previous analyses, background level had been estimated by OP phases
0.52 to 0.88 (0.52 to 0.87 for the \fermi-LAT nebula analysis). For P3,
phases from 0.5 to 0.65 and from 0.85 to 0.9 are used for the background estimation. 

\subsection{MAGIC and \fermi-LAT above 10 GeV}

\begin{figure}[h]
\centering
\includegraphics[width=0.75\textwidth]{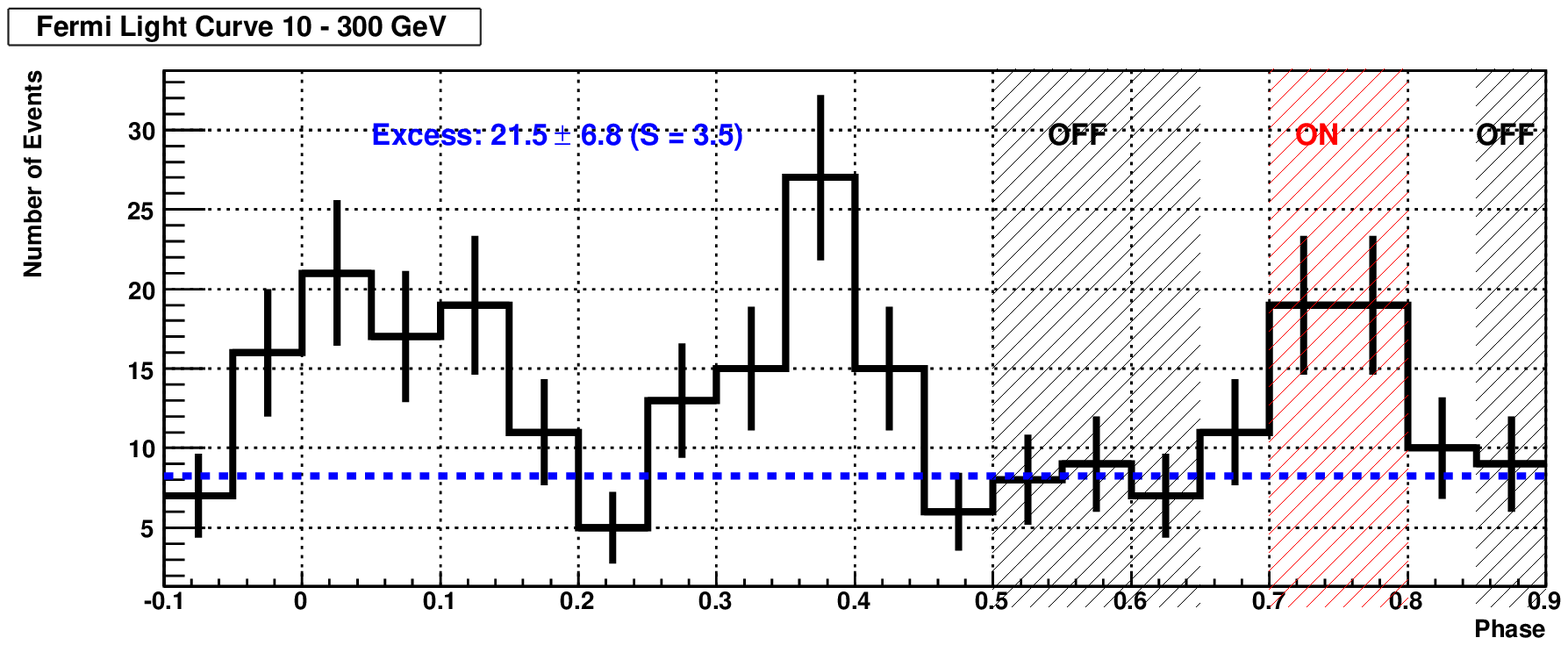}
\includegraphics[width=0.75\textwidth]{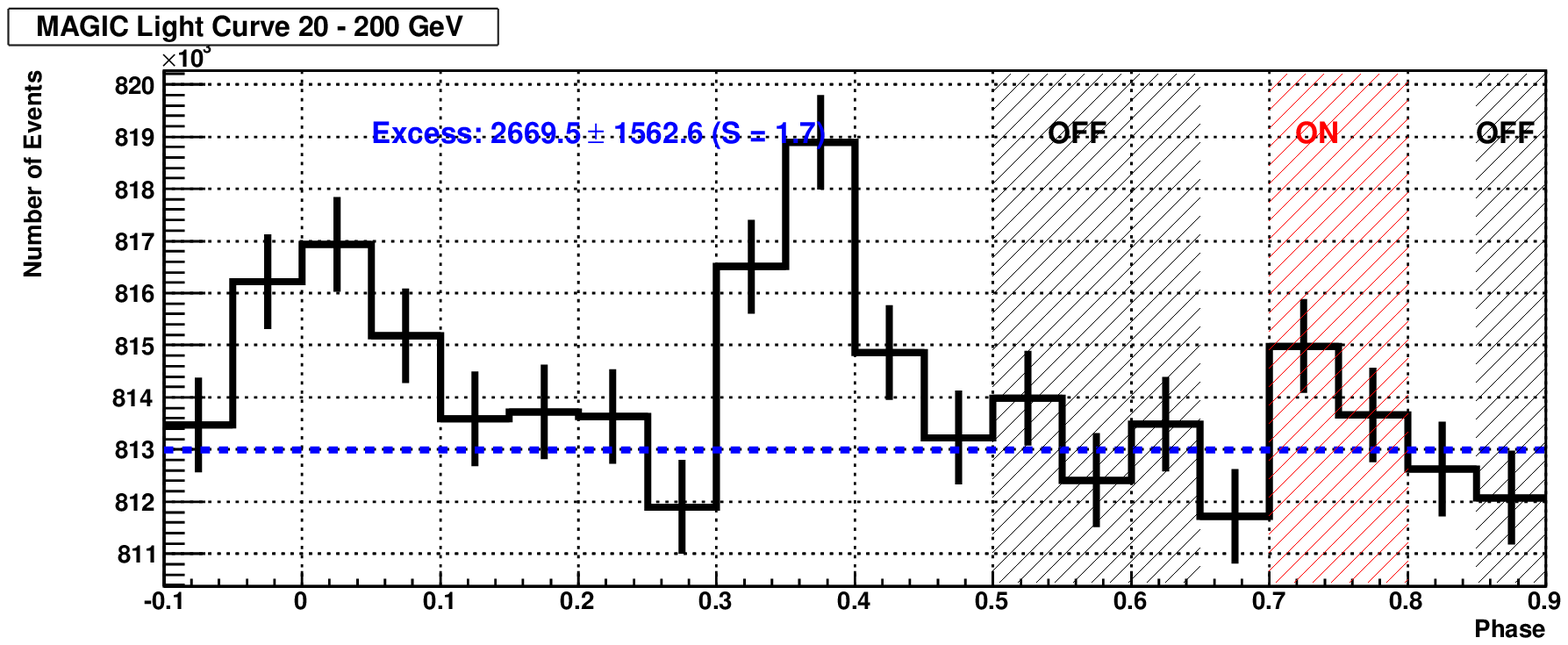}
\caption{The light curves of the \fermi-LAT data above 10 GeV (top) and that 
of the MAGIC data with $SIZE$ from 25 to 500. P3 phases and background estimation phases
are indicated by red and black shadows, respectively. 
}
\label{FigThirdPeakAll}
\end{figure}

The light curves of \fermi-LAT data above 10 GeV and MAGIC data with $SIZE$ from 25 to 500 are
shown in Fig. \ref{FigThirdPeakAll}. \fermi-LAT data show  21.5 $\pm 6.8$ excess events
corresponding to $3.5 \sigma$, while MAGIC data show $2670 \pm 1563$ excess events corresponding
to $1.7 \sigma$. More statistics are required in order to verify 
or refute the presence of the signal. 

\begin{figure}[h]
  \centering
  \includegraphics[width=0.5\textwidth]{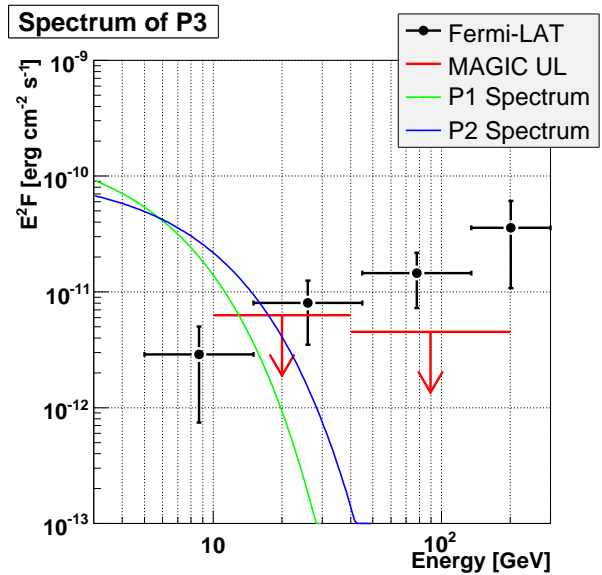}
  \caption{The energy spectrum of P3. Black points and red arrows indicate the measurements
by \fermi-LAT and the upper limits based on MAGIC data, respectively. The flux 
is estimated by simply dividing the number of excess events by the observation time and the effective area. A green and a blue line indicate the exponential 
cut-off energy spectra of P1 and P2, respectively.}
 \label{FigP3Spect}
 \end{figure}

 Although the existence of the signal is not clear, I estimated the energy spectrum
of P3 based on \fermi-LAT data. Based on MAGIC data, the differential flux upper limit
was also
calculated  with a 95\% confidence level . They are shown in Fig. \ref{FigP3Spect}. 
The estimation is done in the following way:
 Instead of using the likelihood method for \fermi-LAT data and the unfolding method
for MAGIC data, the number of excess events (or the upper limit on the number of excess events) 
in each energy bin was simply divided by 
the effective area
and the observation time. Therefore, the spill-over effect from the adjacent bins are not taken
into account. 
The discrepancy between
MAGIC upper limits and \fermi-LAT measurements may imply an upward bias in the 
\fermi-LAT data
analysis, probably because the P3 phases are defined based on the observed data themselves.
The possibility that there is some signal, whose excess was enhanced by the background fluctuation,
cannot be excluded. The time variability of the P3 excess is also another explanation. 

\section{Concluding Remarks}
Neither an exponential cut-off nor a sub-exponential cut-off can explain the energy spectra
measured by MAGIC and \fermi-LAT consistently, even if the possible systematic uncertainties
of both experiments are taken into account.
 Assuming that the energy spectrum 
does not roll off exponentially but extends with a power law after the cut-off energy, 
they can be consistently explained with an index of $\sim -3.0 \pm 0.1$ (after the cut-off).
The physics interpretation of these spectral features will be discussed in Sect
\ref{SectElSpModi}. 

The P2/P1 ratio stays constant between 100 MeV and 3 GeV and
rises above 3 GeV. On the other hand, the Bridge/P1 ratio increases by a power law
between 100 MeV and 3 GeV, while  the behavior above 3 GeV
cannot be analyzed, due to the lack of statistics. The edges of the two pulses 
have an exponential shape. The rise time of P1 and the fall time of P2 have a clear energy
dependence while the fall time of P2 and the rise time of P2 do not. The physics
interpretation of this will be discussed in Sect. \ref{SectRiseFallDiscuss}.
The peak phase of P1 has a slight energy dependence while this is not clear for P2, due to the
difficulty in determining the P2 peak phase. The physics discussion on this will also be presented
 in 
Sect. \ref{SectEDepPeak}. The third peak seen in \fermi-LAT data above 10 GeV with
3.5 $\sigma$ is not clearly visible in MAGIC data. 
In order to verify or refute the existence of the P3 signal, more statistics
 is required.

 \chapter{Discussion}
\label{ChapDiscussion}

Several interesting features 
of the pulsed gamma-ray radiation from the Crab pulsar above 100 MeV have been newly found
from the MAGIC and \fermi-LAT data analyses.
Quite a few results are not in agreement with model predictions and extrapolations.
 Especially the following findings are remarkable:

\begin{itemize}
\item None of the super-exponential cut-off, the exponential cut-off and 
the sub-exponential cut-off 
can describe the measured energy spectrum.
\item The measured energy spectrum extends at least up to 100 GeV.
\item A power law with an index of $-3.0 \pm 0.1$ can well explain
the measured energy spectrum between 4 GeV and 100 GeV.
\item The edges of the two pulses have a clear exponential behavior. The rise time of P1 and 
the fall time of P2 have a clear energy dependence while the fall time of P1 and the rise time
of P2 do not. 
\item The peak phase of P1 has a small but significant energy dependence. 
\end{itemize}
In this chapter, I discuss the new constraints on the pulsar emission models
and possible modifications of 
the standard model, based on these findings. 
In addition, the radiation efficiency in gamma-rays above 100 MeV
is discussed.

\section{Constraints on the Emission Region}

As discussed in \ref{SectRegionDet}, there are mainly two approaches to infer the emission region,
namely, by the steepness of the cut-off and by the highest energy of the observed photons. 


\subsection{By the Steepness of the Cut-off}
As described in \ref{SectCutoffSteepness},
 if the emission region is close to the star surface, a superexponential cut-off is expected.
$\Gamma_2$ in Eq. \ref{EqCutoffSpectra} is typically 2 (see e.g. \cite{Nel1995} and \cite{Jagar2002}). 
However, the analysis of Fermi-LAT data revealed that superexponential assumption ($\Gamma_2 =2$)
is less likely than exponential cut-off ($\Gamma_2 =1$) by 4.8$\sigma$, 7.7$\sigma$,
5.0$\sigma$ and 4.3$\sigma$, for TP, P1 + P2, P1, and P2 respectively.
 This indicates that emission region is far from the star surface and that
gamma-rays do not cause magnetic pair creation. In essense, the modest steepness
of the spectral cutoff allow one to reject the PC model.

\subsection{By the Highest Energy of the Observed Photons}
\label{Sect7.8R0}
More quantitative estimation of the emission region can be made by the highest energy of
the observed photons. As discussed in Sect. \ref{SectCutoff},
the highest energy of photons $E_{max}$ 
which can escape from a given height $r$ can be estimated
as 
\begin{eqnarray}
E_{max}(r) &\simeq&  40 \sqrt{P} \left(\frac{r}{R_0}\right)^{7/2} \frac{B_{cr}}{B_0} {\rm ~~~~ MeV} 
\label{EqMaxHeight}
\end{eqnarray}
where $P$, $R_0$, $B_0$, and $B_{cr}$ are the period of the pulsar in second, the radius of the neutron star, the magnetic field strength at the stellar surface and the critical magnetic field
($B_{cr} = 4.4 \times 10^{13}$ G). 

MAGIC detected gamma-rays up to 100 GeV. For $P = 33.6$~ms, $B_0 = 3.8\times 10^{12}$~G 
$=0.086B_{cr}$ and $E_{max} = 100$ GeV, one obtains $r/R_0$ = 7.8. The height of the emission 
region must be more than 7.8 times the pulsar radius, which is too large for the PC model. 

\section{Estimates of the Electron Spectrum and \\ Constraints on the Acceleration Electric Field}
\label{SectElSpModi}

From the argument in the previous section, it is clear that the emission region
is free from magnetic pair creation process, i.e. the emission region
should be in the outer magnetosphere.  
Since magnetic pair creation is a basic physics process, this conclusion is 
rather robust.

Even if the emission region is assumed to be in the outer magnetosphere,
the energy spectrum observed by MAGIC contradicts the most favored theoretical models,
which predict an exponential cut-off.  
Below, I will discuss possible modifications of the standard outer magnetosphere
model. After briefly reviewing the basic equations for the electron energy spectrum 
within the pulsar magnetosphere and the curvature radiation spectrum from
these electrons in Sect. \ref{SectBasicEq}, I will deduce the electron spectrum 
based on the observed gamma-ray spectrum of P1~+~P2 in Sect. \ref{SectSpecSpec}. 
Then, constraints on the strength of the acceleration electric field
will be discussed in Sect. \ref{SectECons}. 
Possibilities to explain the deviation 
of the observational results from the standard model 
by a imperfect dipole magnetic field will be briefly
described in Sect. \ref{SectMagMod}. 


\subsection{Basic Equations for the Electron Energy and the Curvature Radiation Photon Energy}
\label{SectBasicEq}
As discussed in Sect. \ref{SectRadCooling},
the energy of the electron is set by the equilibrium between 
the gain in energy due to the acceleration electric field and the energy loss via the curvature radiation.
The Lorentz factor of the electron $\Gamma$ can be written as a function 
of the acceleration electric field strength $E_\parallel$ 
and the magnetic field curvature $R_{curv}$:
\begin{eqnarray}
\Gamma = 2.8\times 10^7 \left(\frac{eE_\parallel}{\rm (10^7~eV/cm)}\right)^{1/4} \sqrt{R_{curv}/{\rm 1000~km}}
\label{EqGammaMax2}
\end{eqnarray}
See Sect. \ref{SectCutoff}, for the derivation of this equation, 

As described in Sect. \ref{SectCurv},
the curvature radiation spectrum from a single electron with an energy $\Gamma m_ec^2$ is 
written as

\begin{eqnarray}
\frac{dN_\gamma}{dE_\gamma dt}_{mono} &\simeq& 
KE_\gamma^{-0.7}{\rm exp}(-E_\gamma/E_c) \\
E_c &=& \frac{3}{2} \Gamma^3 h\nu_{curv} = \frac{3}{2} \Gamma^3 \frac{hc}{2\pi R_{curv}}
\label{EqGEne} 
\end{eqnarray}



From Eq. \ref{EqGammaMax2} and Eq. \ref{EqGEne}, 
the cut-off energy $E_c$ 
of the curvature radiation is written as a function of $E_\parallel$ and $R_{curv}$:
\begin{eqnarray} 
E_c = 6.5 \left(\frac{eE_\parallel}{\rm (10^7~eV/cm)}\right)^{3/4} \sqrt{R_{curv}/{\rm 1000~km}} {~~~~~} {\rm [GeV]}
\label{EqPhotonEne}
\end{eqnarray} 



\clearpage

\subsection{Estimates of the Electron Spectrum\\ Based on the Measured Gamma-ray Spectrum}
\label{SectSpecSpec}
Here, I deduce the 
electron spectrum (the $\Gamma$ spectrum) based on the measured gamma-ray spectrum of P1~+~P2,
assuming that gamma-rays above 4 GeV are generated by the curvature radiation. 
For simplicity, the curvature of the magnetic field is assumed to be fixed at $R_{curv} = 1000$ km
\footnote{The different $R_{curv}$ for different emission region 
can be taken into account
by replacing $\Gamma$ with $\Gamma (R_{curv}/1000 {\rm [km]})^{-1/3}$ (see Eq. \ref{EqGEne})}. 

\subsubsection{0) Standard Model: Nearly Monoenergetic Electron Spectrum}
In general, it is considered that $E_\parallel$ and $R_{curv}$ of the last closed
field line do not change extremely 
largely over 
the emission region
(see e.g. \cite{Takata2007}). In addition, 
the dependency of $\Gamma$ on $E_\parallel$ and $R_{curv}$
 is rather weak (see Eq. \ref{EqGammaMax2}). These are the reasons why 
a nearly monoenergetic spectrum is derived for accelerarted electrons 
in most of pulsar models based on the outer magnetosphere emission hypothesis.

\begin{figure}[h]
\centering
\includegraphics[width=0.33\textwidth]{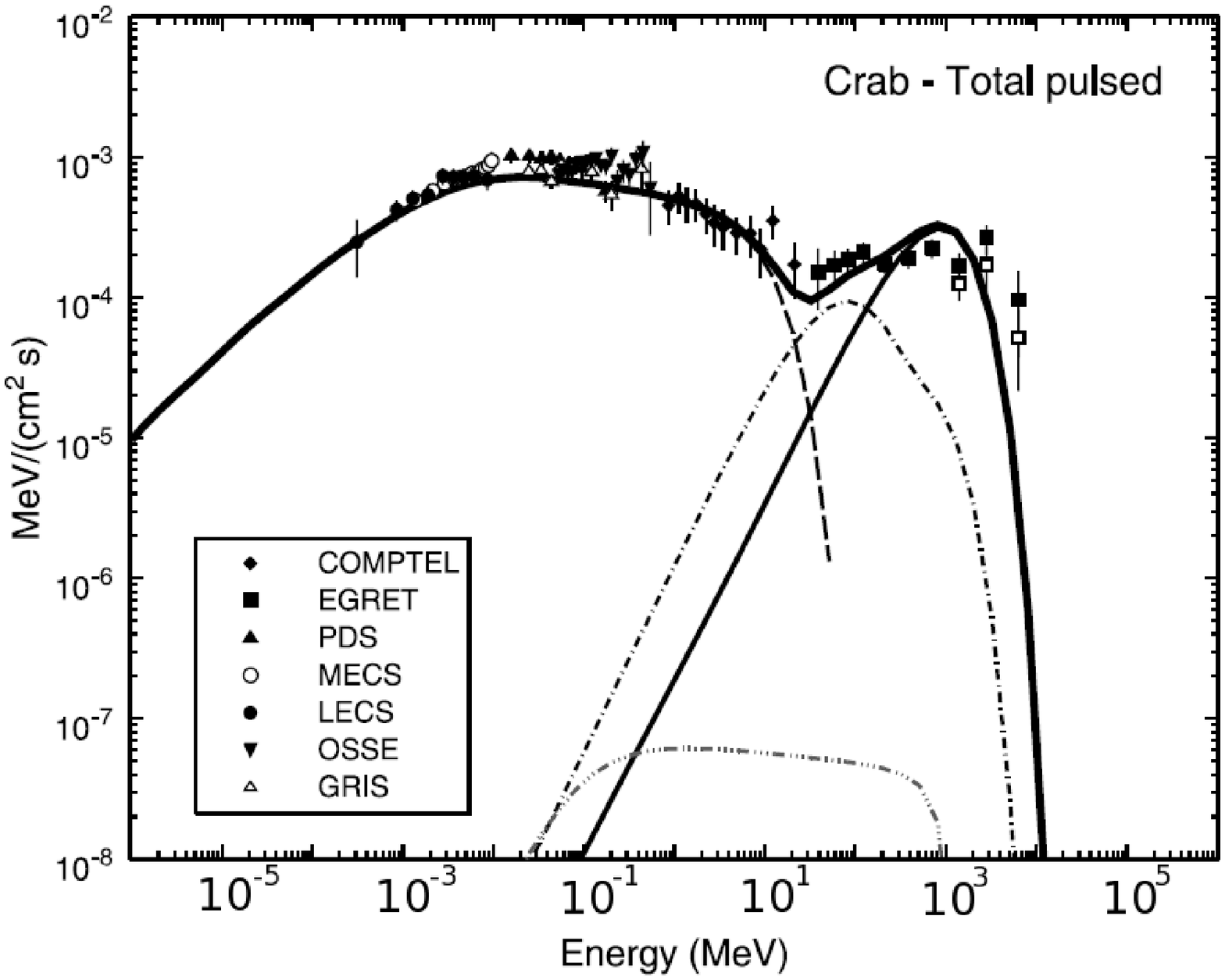}
\includegraphics[width=0.33\textwidth]{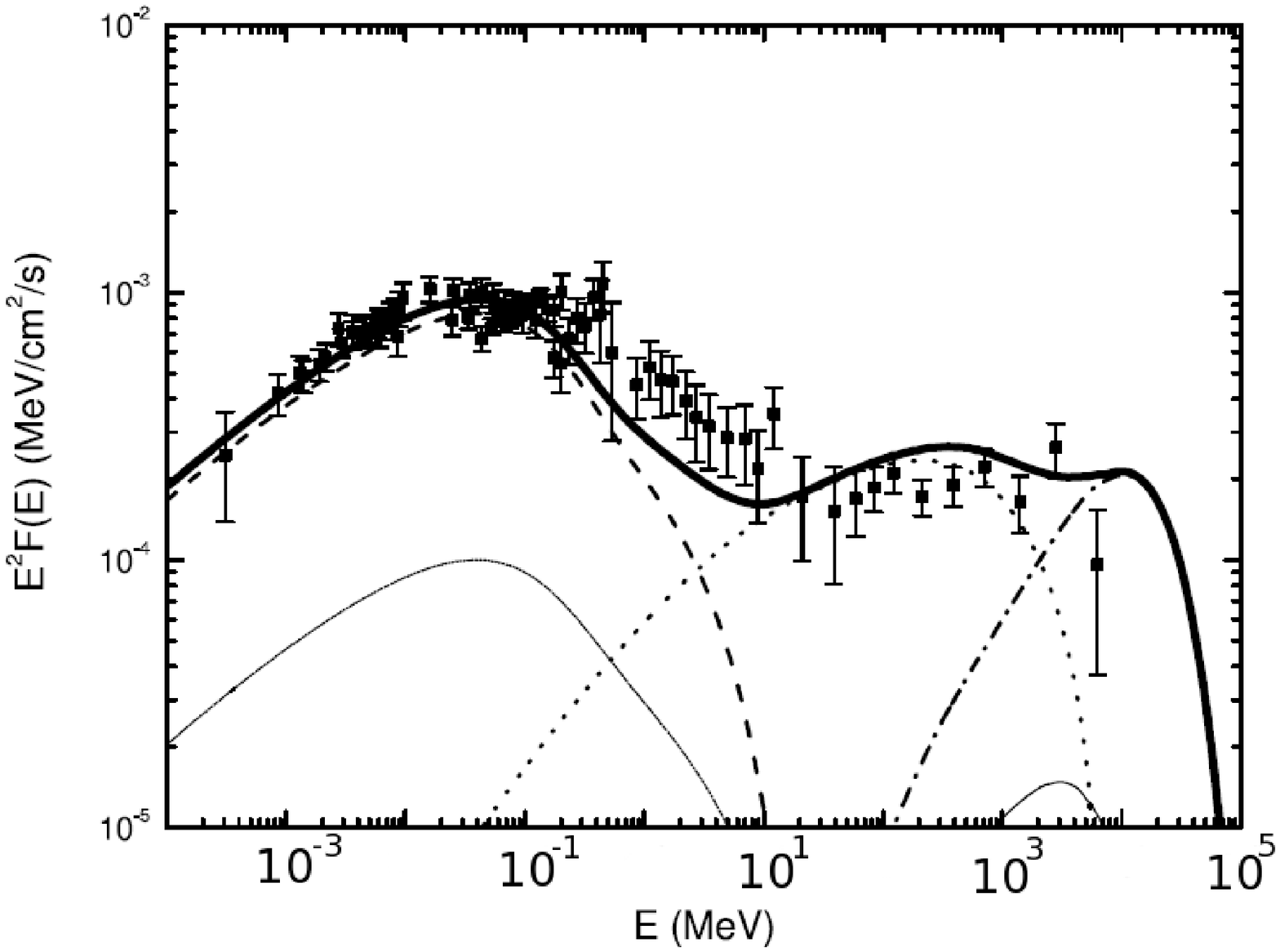}
\caption{
Theoretical explanations of the energy spectrum of the Crab pulsar
assuming that the emission region is in the outer magnetosphere. 
Two models, namely Harding et al. (left, see \cite{Harding2008}) and Tang. et al (right, see \cite{Tang2008}) are shown. The highest end of the spectrum are explained by the curvature radiation from 
nearly monoenergetic electrons as indicated by a solid line (left) and a dashed-dotted line
(right). The emission between 100 MeV and 1 GeV, where a power law spectrum with an index 
of $\sim -2.0$ is observed, is explained by the synchrotron radiation (a dash-dotted line
in the left panel) or inverse Compton emission (a dotted line in the right panel).
Below 10 MeV, the emissions from secondary electron-positron pairs
 created by high energy gamma-rays are considered to be responsible for the observed spectrum.
}
\label{FigOuterSpectrum2}
\end{figure}

The cut-off in the high energy gamma-ray spectrum 
is determined by the curvature radiation spectrum from these nearly monoenergetic electrons,
in turn leading to the exponential cut-off (see Sect. \ref{SectBasicEq}). 
Examples of the theoretical explanations of the Crab pulsar spectrum observed before 2007 are shown
in Fig. \ref{FigOuterSpectrum2}, which are the same as Fig. \ref{FigOuterSpectrum}.
The highest end of the spectrum is explaned by the curvature radiation from
nearly monoenergetic electrons.
In order to explain the power law spectrum with an index of $\sim 2$ between $\sim 10$ MeV and $\sim 1$ GeV, the synchrotron radiation (left panel) or the inverse Compton scattering (right panel) 
is considered. Below 10 MeV, the emissions from secondary electron-positron pairs
 created by high energy gamma-rays are considered to be responsible for the observed spectrum. 

\begin{figure}[h]
\centering
\includegraphics[width=0.4\textwidth]{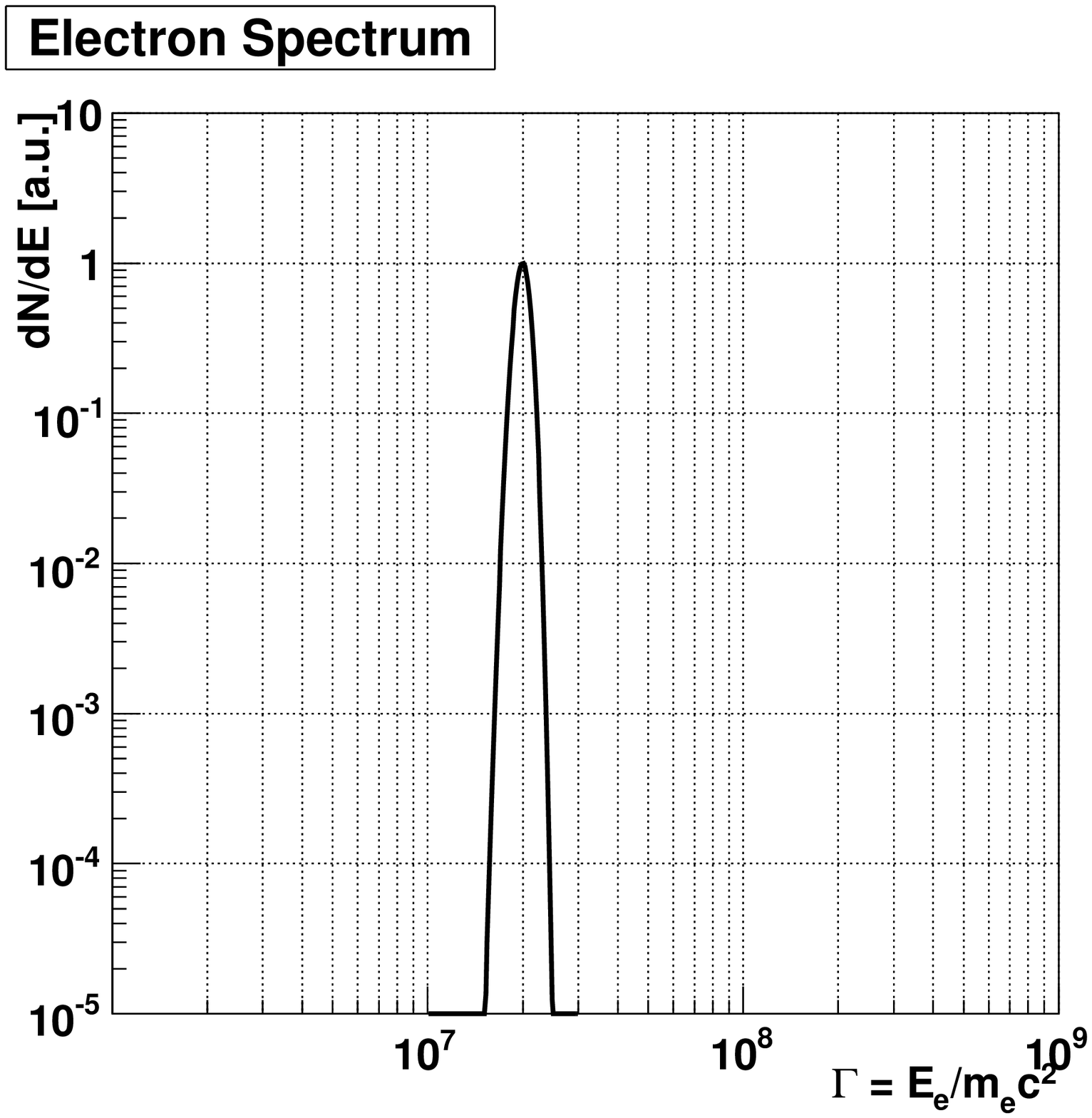}
\includegraphics[width=0.4\textwidth]{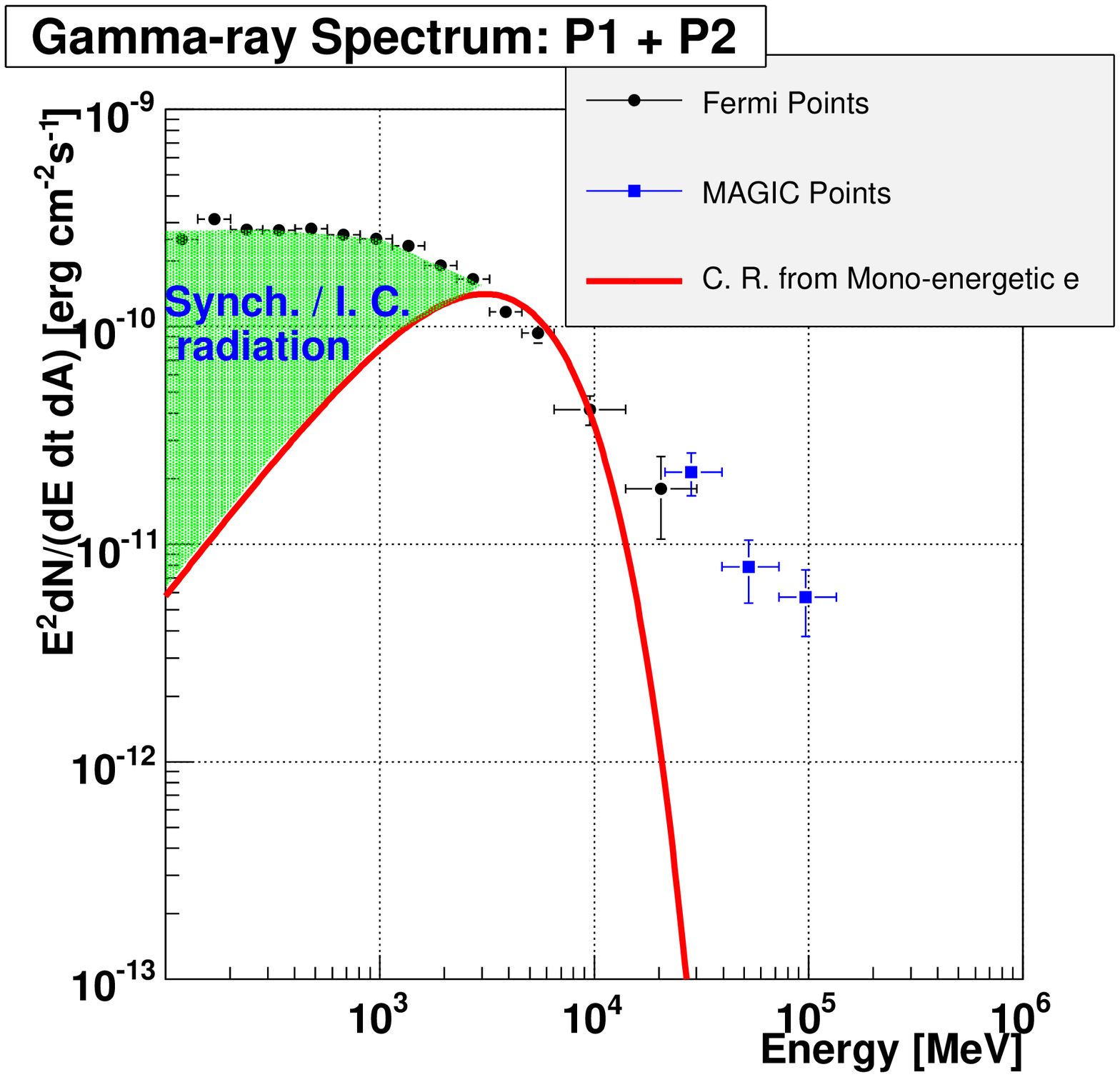}
\caption{
Applying the standard scenario 
to the highest end of the observed spectrum for P1~+~P2. 
Left: Assumed $\Gamma$ factor spectrum for electron. 
It has a Gaussian shape with 
the mean of $2\times10^7$ and the RMS of $10^6$. 
Right: The energy spectrum of the curvature radiation from the assumed electron spectrum,
overlaid with the observed results.  
MAGIC measurements are largely deviated from the expected spectrum.  
The discrepancy between the curvature radiation spectrum and observed data below 3 GeV
(a green shaded region) could be explained by either the synchrotron radiation or
the inverse Compton scattering (see. Fig. \ref{FigOuterSpectrum2}).
}
\label{FigElGStd}
\end{figure}

Applying the standard scenario to the observed spectrum of P1 + P2 
around the cut-off energy is shown in Fig. \ref{FigElGStd}.
I made a simple calculation assuming that the $\Gamma$ spectrum of 
the electrons has a Gaussian shape with 
the mean of $2\times10^7$ and the RMS of $10^6$.
This corresponds to $E_\parallel = 3 \times 10^6$ [V/cm], which is consistent
with, e.g. \cite{Takata2007}.
The absolute flux scale was chosen such that the predicted gamma-ray spectrum 
matches with the measurements. 

As can be clearly seen in the figure, the measured spectrum requires modification of the
standard models. 



\clearpage



\subsubsection{1) Steep Power Law Tail Assumption}
Given the fact that the power law with an index of $\sim -3$ 
describes well the observed gamma-ray spectrum above 4 GeV, the simplest assumption 
for the electron spectrum would be a power law with an index of $\sim -8$,
because if the electron has a power law spectrum with an index of $-p$, 
the resulting curvature radiation spectrum should show the power law with an index
of $-q = -(p+1)/3$, as discussed in Sect \ref{SectCurv}.
I assumed a power law with an index of $-8$ between $1.7\times 10^7$ and $10^8$ 
in the $\Gamma$ spectrum. 
For $\Gamma <10^7$, a Gaussian with the mean of $1.7\times 10^7$ and the RMS of
 $10^6$ is assumed so that the modification with respect to the standard model is only on higher
 energy side.
 The measured spectrum of P12 can well be explained by the curvature radiation from 
these electrons as expected (the top panels of Fig. \ref{FigElGSpec}).
Obviously, the main difficulty is to find a convincing argument for the shape of the
electron spectrum.
 
\subsubsection{2) Log-Gaussian Assumption}
It might also be possible that the observed power-law-like behavior above the cut-off energy is part of a curved
spectrum, whose curvature is not visible due to the statistical uncertainty, the 
limited energy resolution 
and the limited energy coverage of the measurements.

\begin{figure}[h]
\centering
\includegraphics[width=0.4\textwidth]{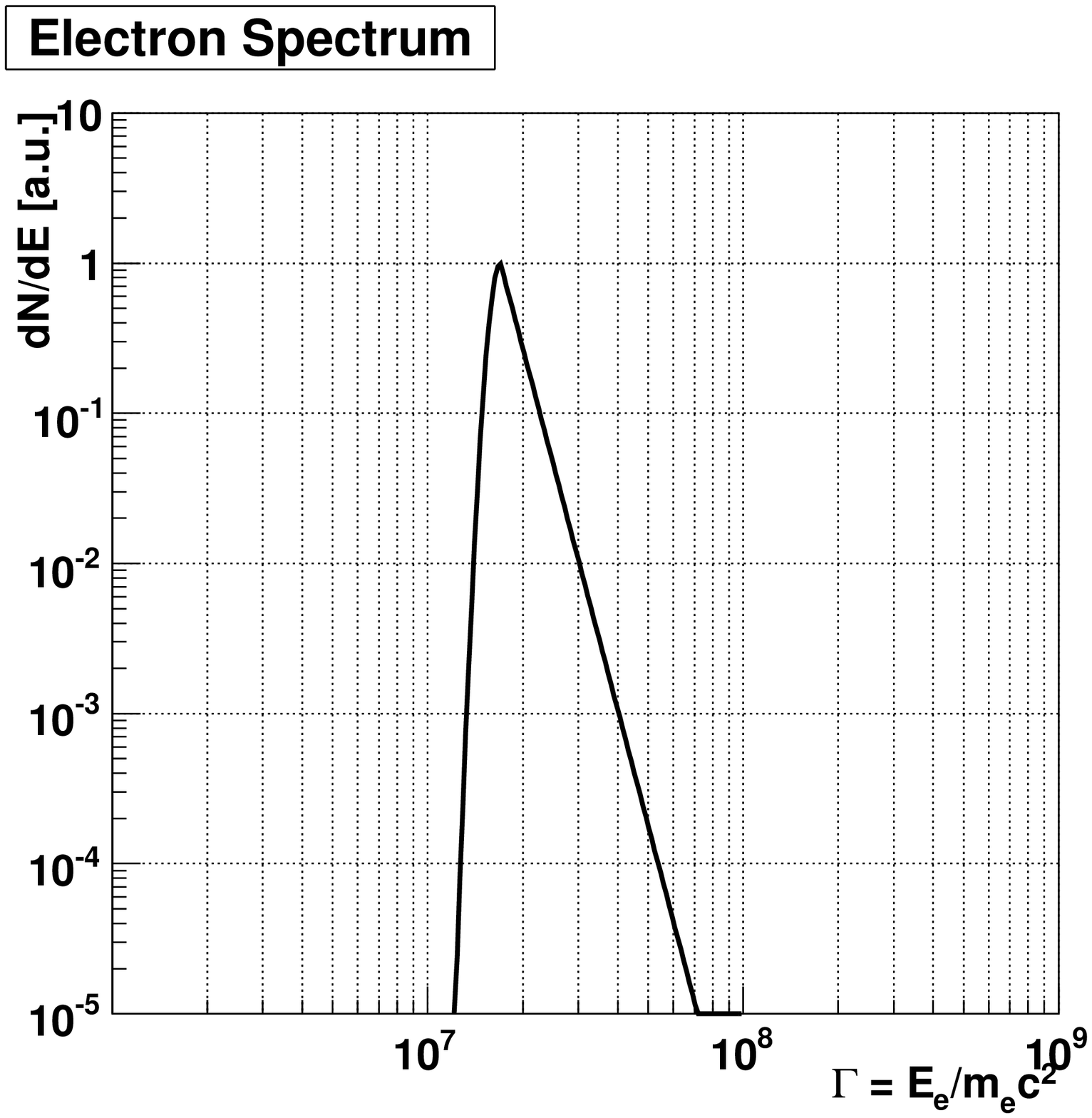}
\includegraphics[width=0.4\textwidth]{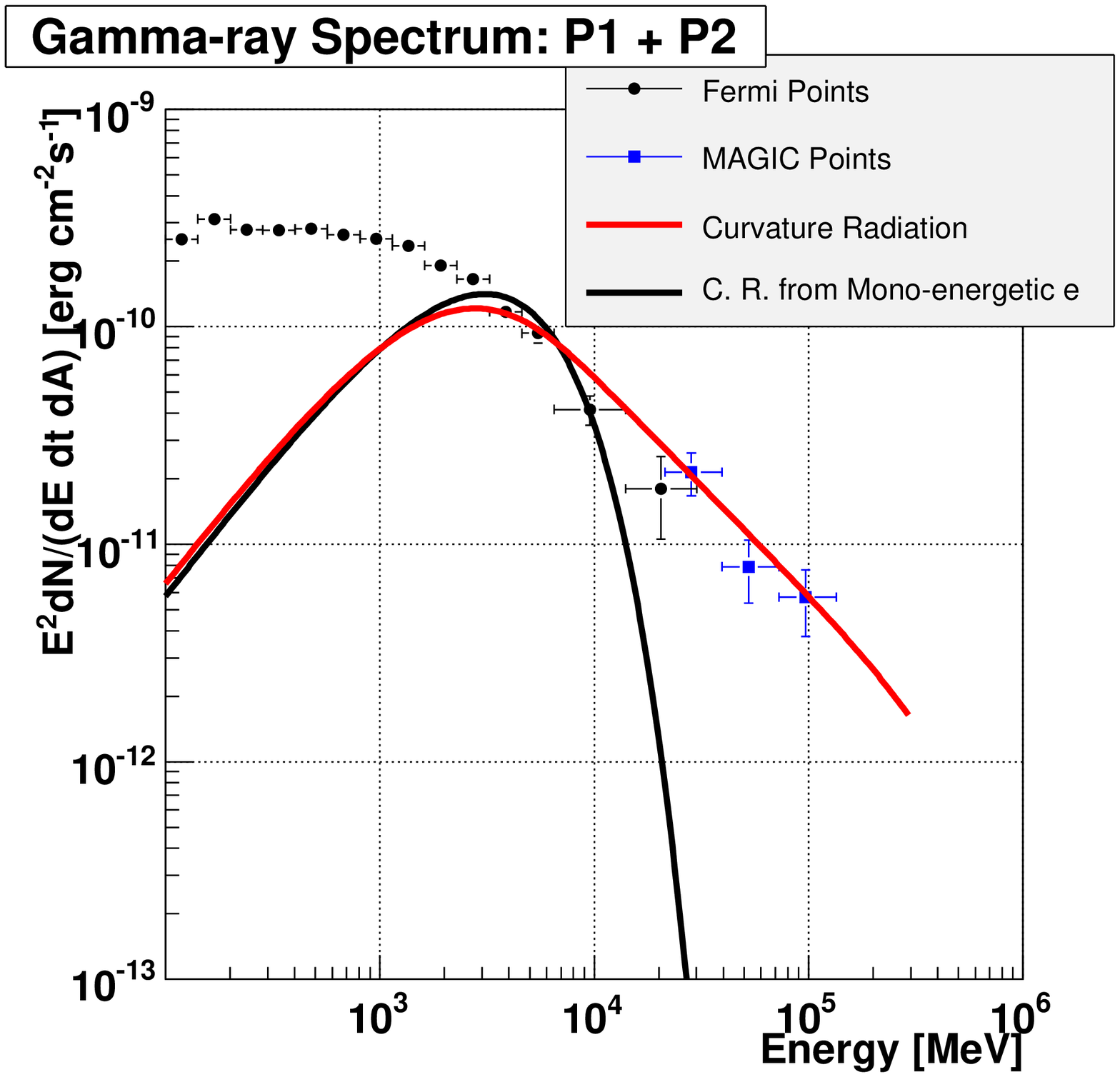}
\includegraphics[width=0.4\textwidth]{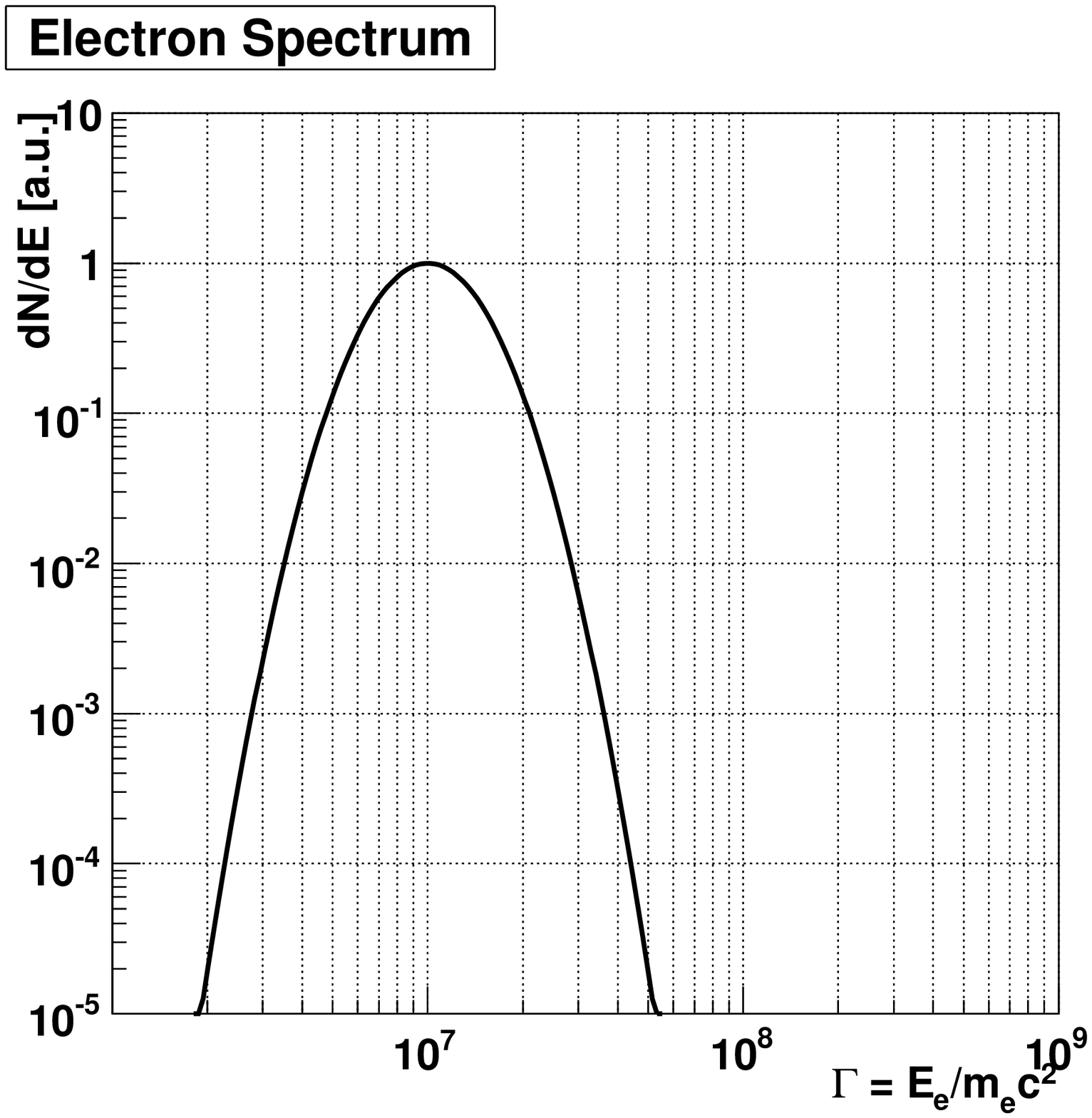}
\includegraphics[width=0.4\textwidth]{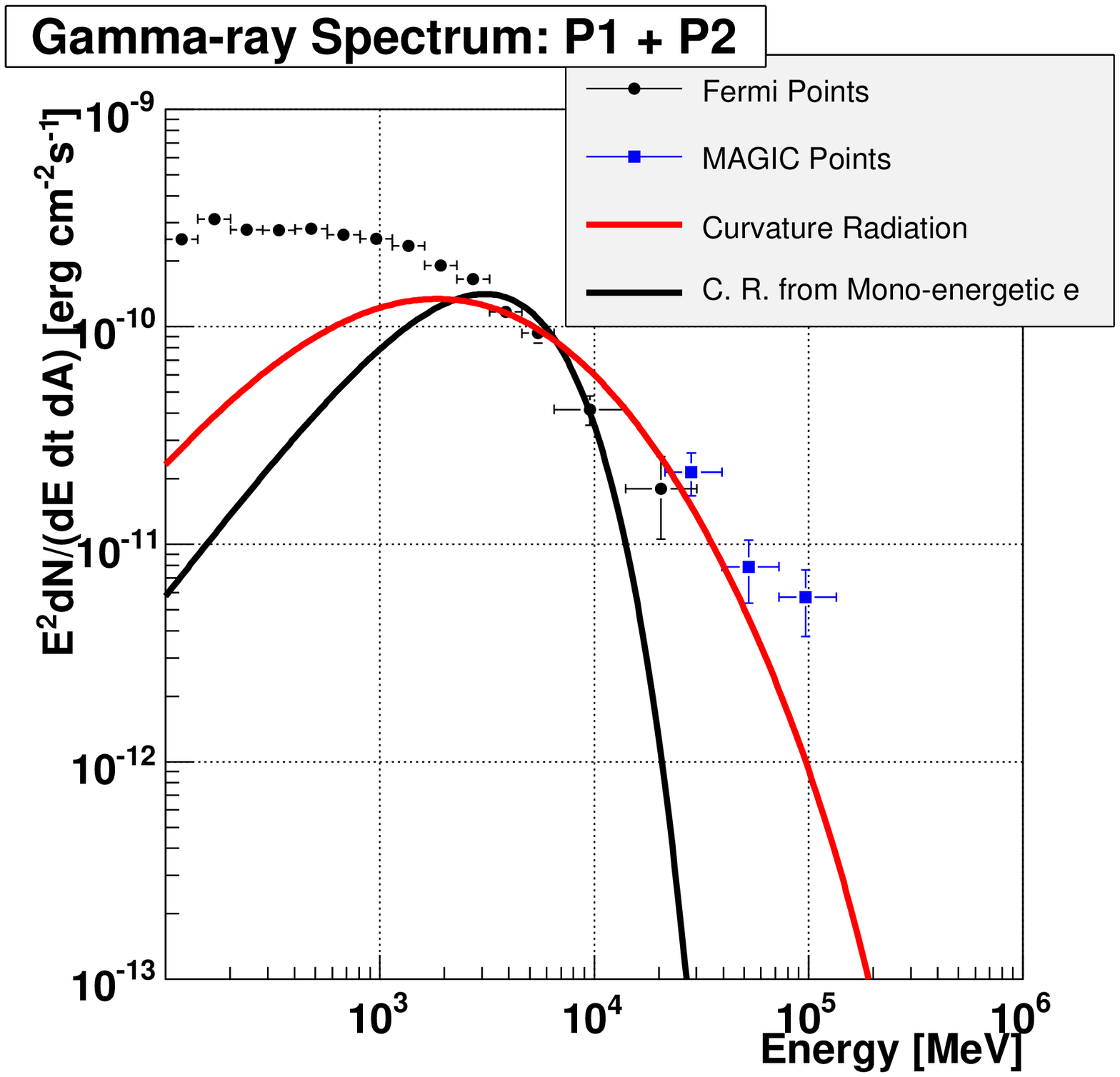}
\caption{The same as Fig. \ref{FigElGStd} but for different $\Gamma$ spectrum.
Top: A power law tail after the peak is assumed for the $\Gamma$ spectrum.  
At $\Gamma <10^7$, a Gaussian with the mean of $1.7\times 10^7$ and the RMS of
 $10^6$ is assumed, while a power law spectrum with an index of $-8$ is assumed  
between $1.7\times 10^7$ to $10^8$. The resulting curvature radiation 
reproduces the observed results very well.  
Bottom: A Log-Gaussian spectrum is assumed for the $\Gamma$ spectrum.
The mean of log$_{10}(\Gamma)_{mean} = 7 $ and the RMS of log$_{10}(\Gamma)_{RMS} = 0.15$ is used.
The resulting curvature radiation 
reproduces the observed results reasonably well
taking into account the possible energy scale difference 
between MAGIC and \fermi-LAT up to $\sim 30$\%.
}
\label{FigElGSpec}
\end{figure}

For example, a log-Gaussian spectrum of the electrons produce a gamma-ray spectrum extending to
the MAGIC energies. In the bottom panels of Fig. \ref{FigElGSpec}, the log-Gaussian spectrum
of the electron with the mean of log$_{10}(\Gamma)_{mean} = 7 $ and the RMS of log$_{10}(\Gamma)_{RMS} = 0.15$ is assumed. 
\begin{eqnarray}
\frac{dN_e}{d\Gamma} \propto \exp\left(- \frac{(\log_{10}\Gamma - 7)^2}{2\cdot 0.15^2})\right)
\end{eqnarray}

The spectrum above the cut-off can be reasonably explained, taking into
account that MAGIC and \fermi-LAT may have a relative energy scale difference of up to $\sim 30$\%.
The idea behind the log-Gaussian spectrum of $\Gamma$ is that 
the electrons are not as monoenergetic as the standard model predicts.
This might originate from a small distortion of the pulsar magnetosphere structure
from the standard model.


\clearpage

 \subsubsection{3) Two Population Assumption}							      
\begin{figure}[h]
\centering
\includegraphics[width=0.4\textwidth]{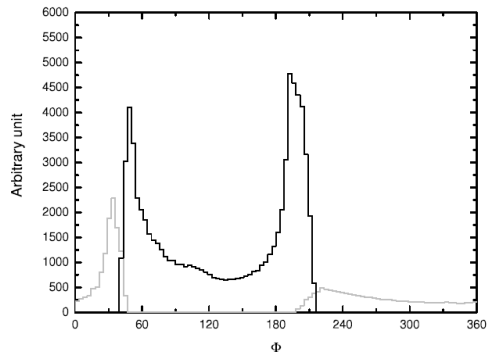}
\caption{A theoretical explanation of the light curve assuming that
emissions from the two poles contribute to the observed pulsation.
The inclination angle of the dipole axis with respect to the rotation axis is
assumed to be 50 degrees and the viewing angle is assumed to be 76 degrees.  
Contributions from each pole are overlaid with different intensity of the lines.
Figure adopted from \cite{Tang2008}}
\label{FigTwoPolLC}
\end{figure}

Several authors consider the possibility that the observed pulsation is 
the sum of the emissions from	
the two poles (see e.g. \cite{Takata2007} and \cite{Tang2008}). The two contributions
to the light curve are shown in Fig. \ref{FigTwoPolLC}.
 Even though their calculations do not predict an energy spectrum extending to 100 GeV, 
the possibility of the contributions from the two 
 isolated places make the two population assumption intriguing. 

In fact, the model of two population of electrons can explain the measurements as well, as shown   
in the top panels of Fig. \ref{FigP12TwoComp}.
Here I assumed two Gaussian spectra
with the mean $\Gamma$ being $2\times 10^{7}$ and $4\times 10^7$. 
$R_{curv}=1000$ km is used. 
The RMSs of the two Gaussians
are 20 times smaller than its mean.
The peak flux of the second
population is 200 times lower than that of the first.

\subsubsection{4) Power Law + Exponential Cut-off Assumption}

\begin{figure}[h]
\centering
\includegraphics[width=0.4\textwidth]{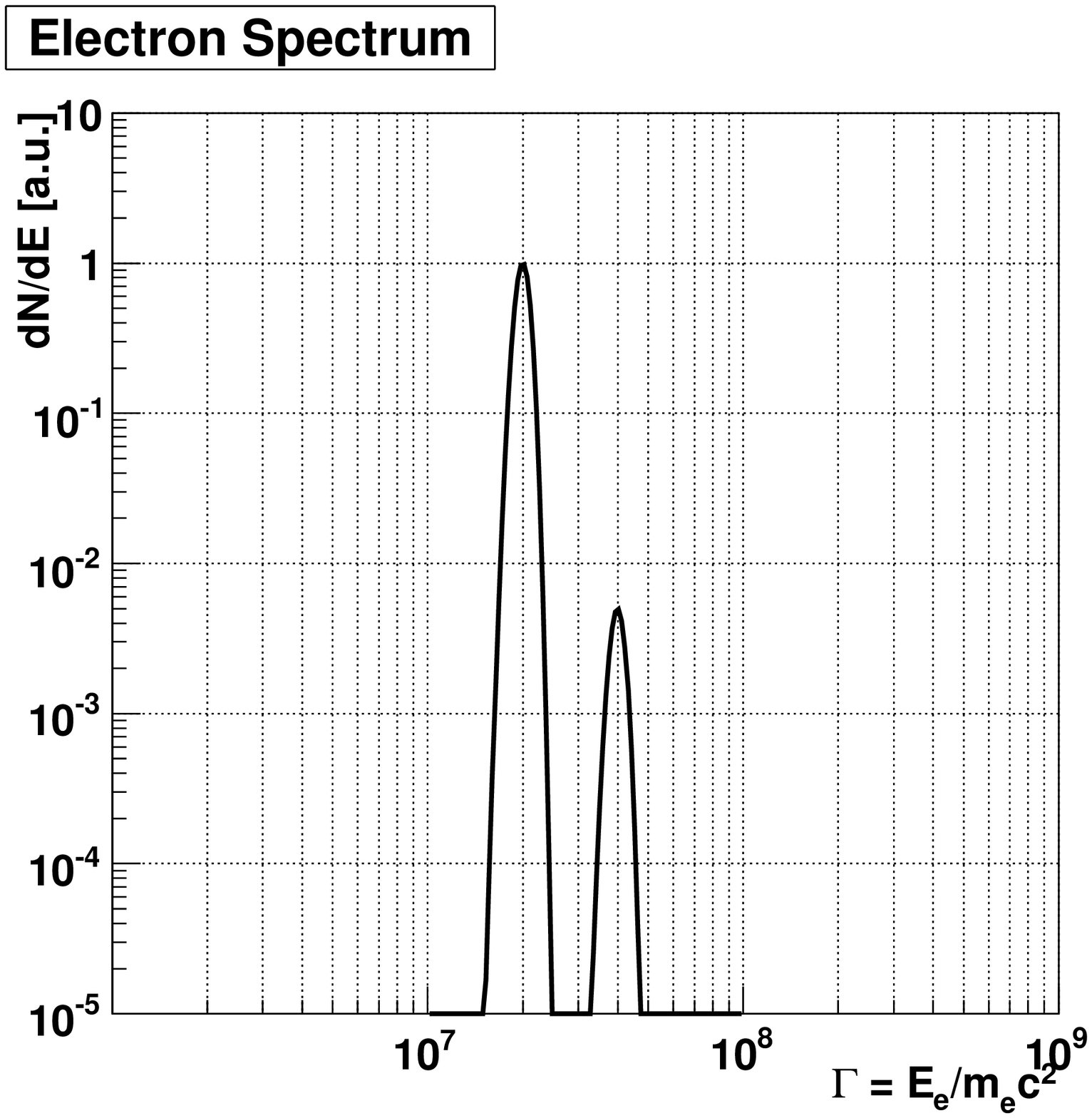}
\includegraphics[width=0.4\textwidth]{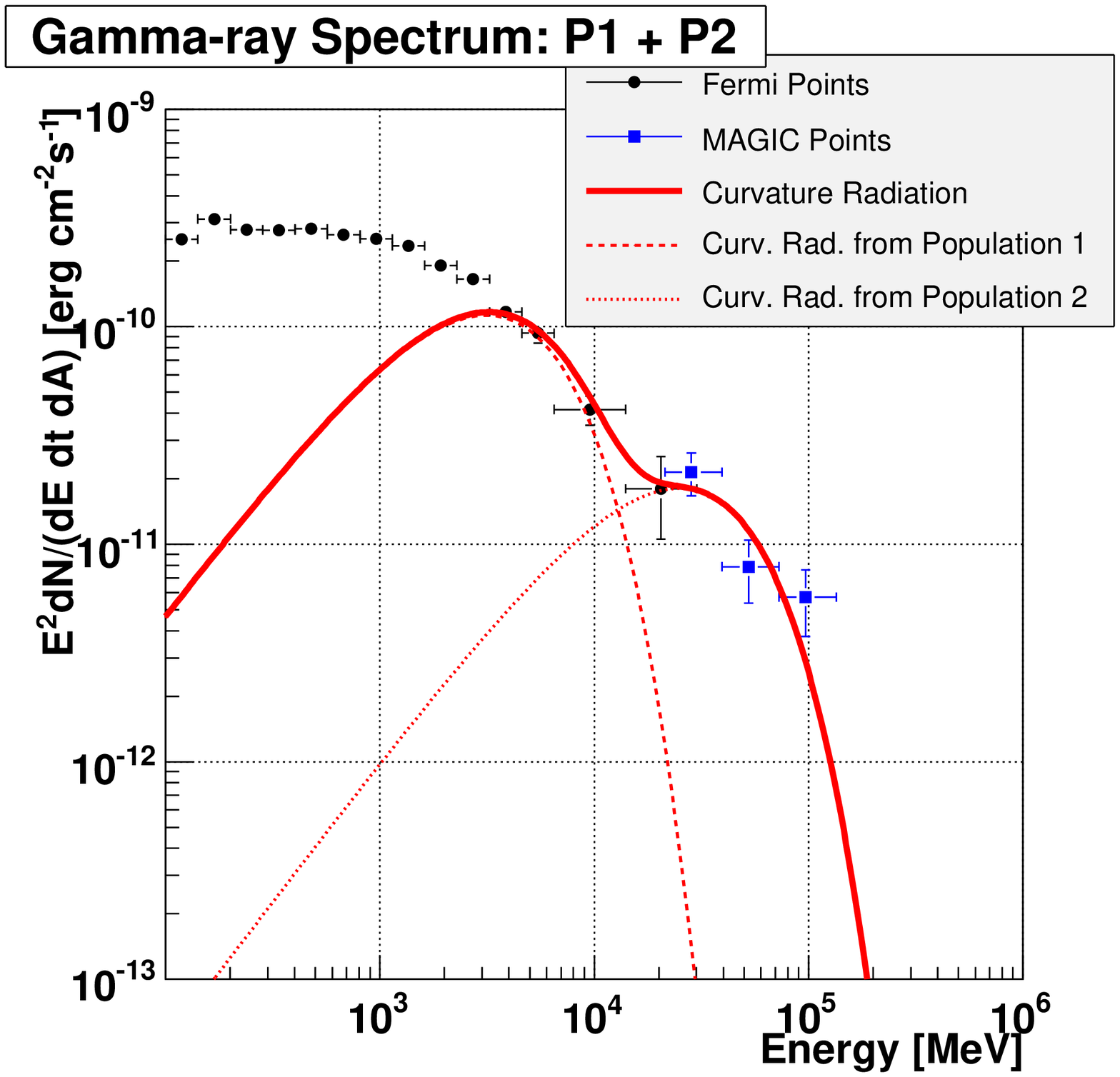}
\includegraphics[width=0.4\textwidth]{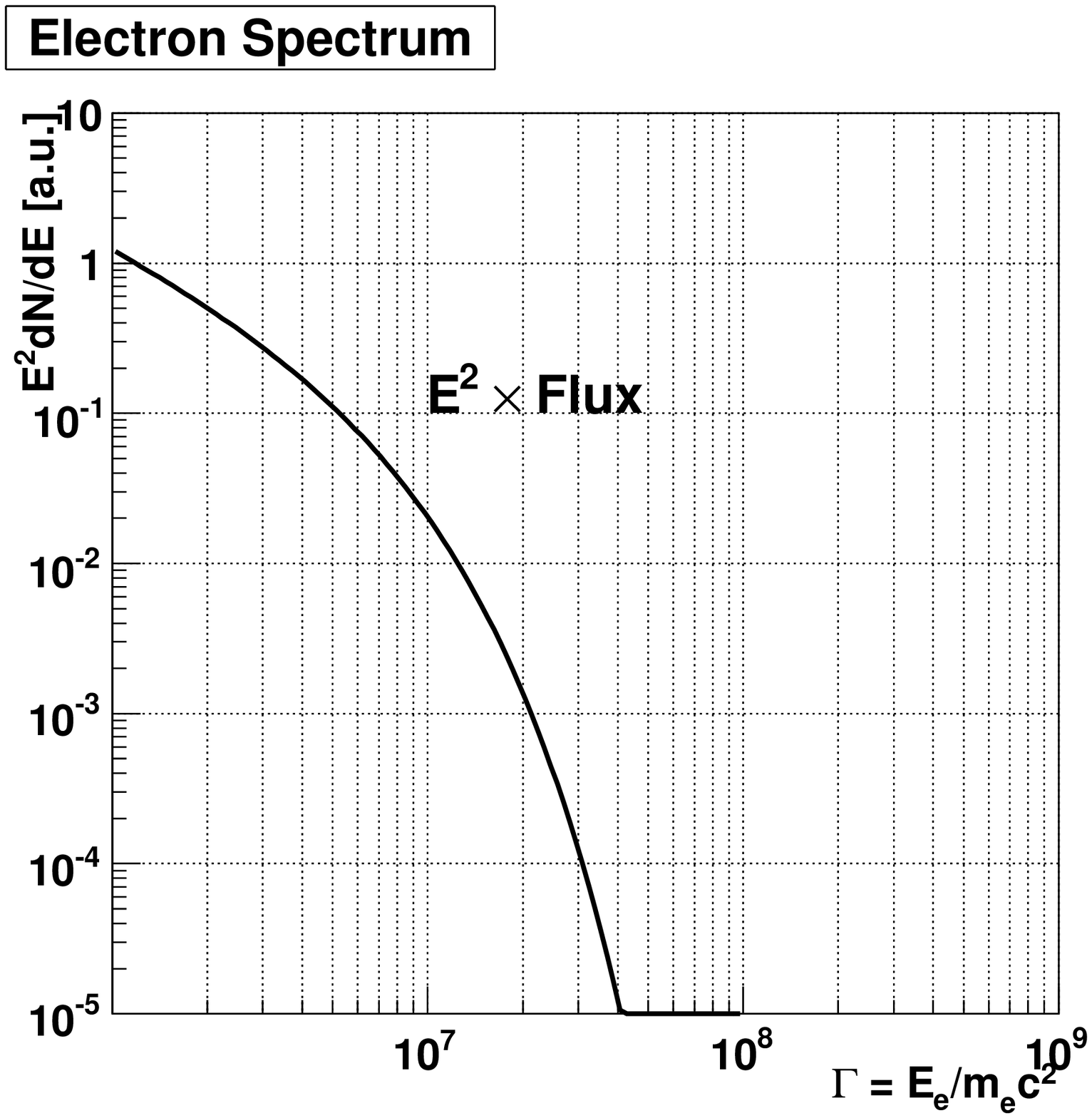}
\includegraphics[width=0.4\textwidth]{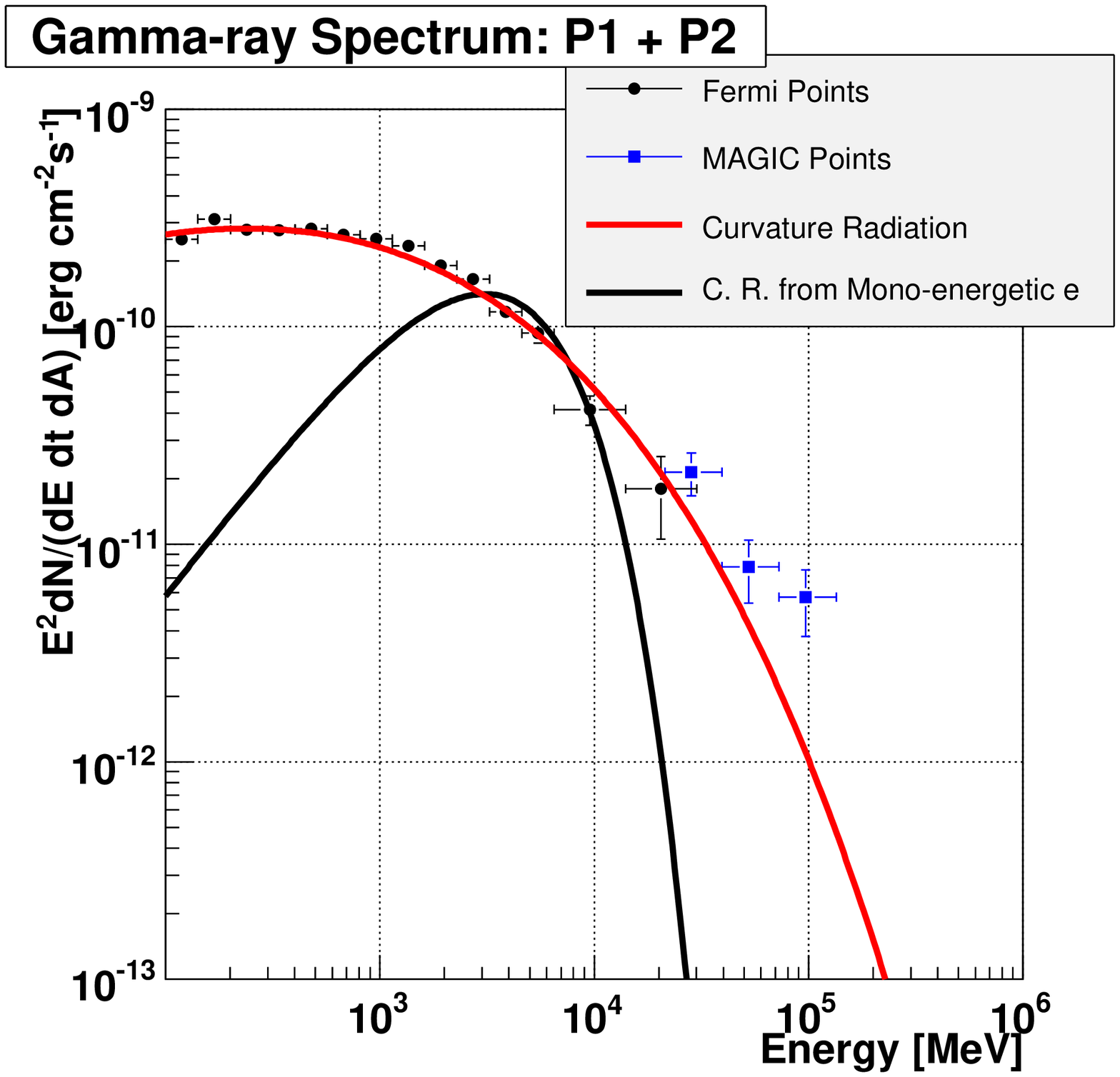}
\caption{The same as Fig. \ref{FigElGStd} but for different $\Gamma$ spectrum.
Top: Two population of the $\Gamma$ spectrum is assumed. 
Two Gaussian spectra
with the mean $\Gamma$ being $2\times 10^{7}$ and $4\times 10^7$ are used. 
The RMSs of the two Gaussians
are 20 times smaller than its mean. The peak flux of the second
population is 200 times lower than that of the first.
The resulting curvature radiation 
reproduces the observed results very well.  
Bottom: 
The power law with an exponential cut-off is assumed for $\Gamma$ spectrum.
The power law index of $-3$ and cut-off at $\Gamma = 0.5 \times 10^7$ are used.
The resulting curvature radiation can explain the observed spectrum
very well from 100 MeV to 100 GeV, if 
the relative energy scale difference of up to $\sim 30$\% is taken into account. 
This assumption does not require additional emission such as the inverse Compton scattering or 
the synchrotron radiation between 100 MeV to 3 GeV.
}
\label{FigP12TwoComp}
\end{figure}

The power law with an exponential cut-off for the electron spectrum 
produces an interesting results. 
By assuming a power law with an index of $-3$ and cut-off at $\Gamma = 0.5 \times 10^7$
for electrons, the resulting curvature radiation can explain the observed gamma-ray spectrum
very well from 100 MeV to 100 GeV, if 
the relative energy scale difference of up to $\sim 30$\% is taken into account
(see the bottom panels of Fig. \ref{FigP12TwoComp}). 
This assumption does not require additional emission mechanisms 
such as the inverse Compton scattering and the
synchrotron radiation in order to explain the measured spectrum between 100 MeV to 3 GeV.

It is known that the electron energy spectrum may exhibit a power law with an exponential cut-off
in the case of shock acceleration in a supernova remnant taking into account 
either the acceleration-time limits (see \cite{Drury1991}) or radiative-loss limits
(see \cite{Webb1984}). 
In order to apply a similar scheme to a pulsar, the acceleration mechanism in the pulsar
magnetosphere needs to be reconsidered from scratch.


\clearpage

\subsection{Constraints on the Acceleration Electric Field in an Ideal Dipole Magnetic Field}
\label{SectECons}

\begin{figure}[h]
\centering
\includegraphics[width=0.6\textwidth]{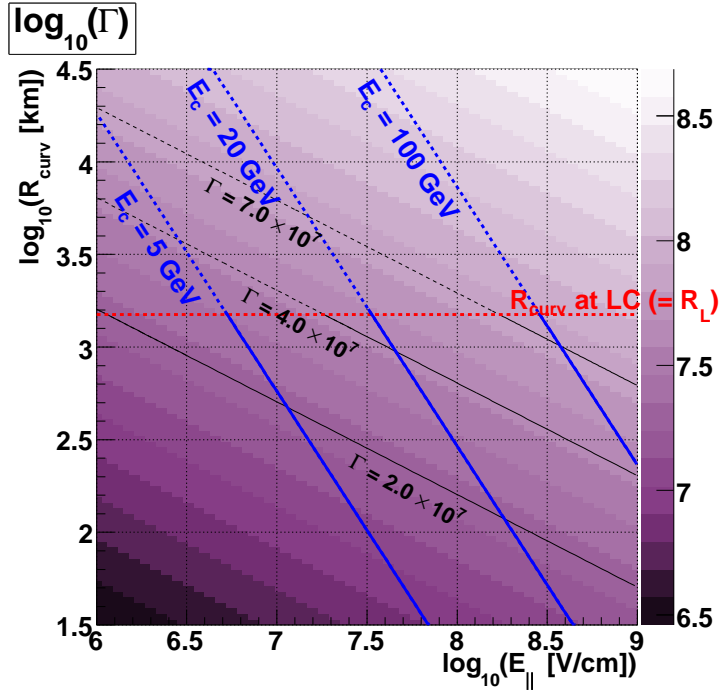}
\caption{The electron Lorentz factor $\Gamma$ as a function of
the field curvature $R_{curv}$ and the acceleration field strength $E_\parallel$,
expressed with a color scale.
Black lines indicate the contor lines for $\Gamma = 2.0 \time 10^7$, $4.0\times 10^7$
and $7.0\times 10^7$.
Blue lines indicate the corresponding cut-off energy of curvature radiation photons
for 5 GeV, 20 GeV and 100 GeV (see Eq. \ref{EqGEne}.).
A red dotted line indicate the co-rotation radius, which should be the
upper limit in $R_{curv}$.
}
\label{FigGammaVsRandE}
\end{figure}

Based on Eq. \ref{EqGammaMax2}, $\Gamma$ as a function of $R_{curv}$ and $E_\parallel$
is graphically shown in Fig.~\ref{FigGammaVsRandE}.
$R_{curv}$ as a function of $E_\parallel$ for $E_c = 5$ GeV,
20 GeV and 100 GeV are shown by blue lines in the same figure.  

It is known that $R_{curv}$ of the dipole magnetic field is well approximated 
as $R_{curv}(r) = \sqrt{R_Lr}$ (see e.g. \cite{Baring2004}),
 where $R_L$ is the co-rotation radius (see Sect. \ref{SectLC})
and $r$ is the distance from the center of the neutron star.  
Since the emission region must be within the light cylinder, $R_{curv} < R_L$ would be the
reasonable upper limit on $R_{curv}$. 
Assuming that the (true) gamma-ray spectrum is extending as a power law up to 100 GeV
without a cutoff,
as is the case for {\bf 1) Steep Power Law Tail Assumption} in the previous subsection, 
there must be a place
where $E_\parallel > 2.9 \times 10^8$~[V/cm]. This is $\sim 100$ times larger than 
the value used in the standard model (see e.g. \cite{Takata2007}).
Assuming that the gamma-ray spectrum observed at around $\sim 100$ GeV 
is basically the tail of the curvature radiation spectrum with a 20 GeV cut-off,
as is the case for {\bf 3) Two Population Assumption} in the previous subsection, the lower limit 
in $E_\parallel$ will be $E_\parallel > 3.3 \times 10^7$~[V/cm].
This is $\sim 10$ times larger than the standard value. 

\subsection{Possible Explanations for the High Energy Tail
of the Gamma-ray Spectrum in an Imperfect Dipole Magnetic Field}
\label{SectMagMod}

\begin{figure}[h]
\centering
\includegraphics[width=0.35\textwidth]{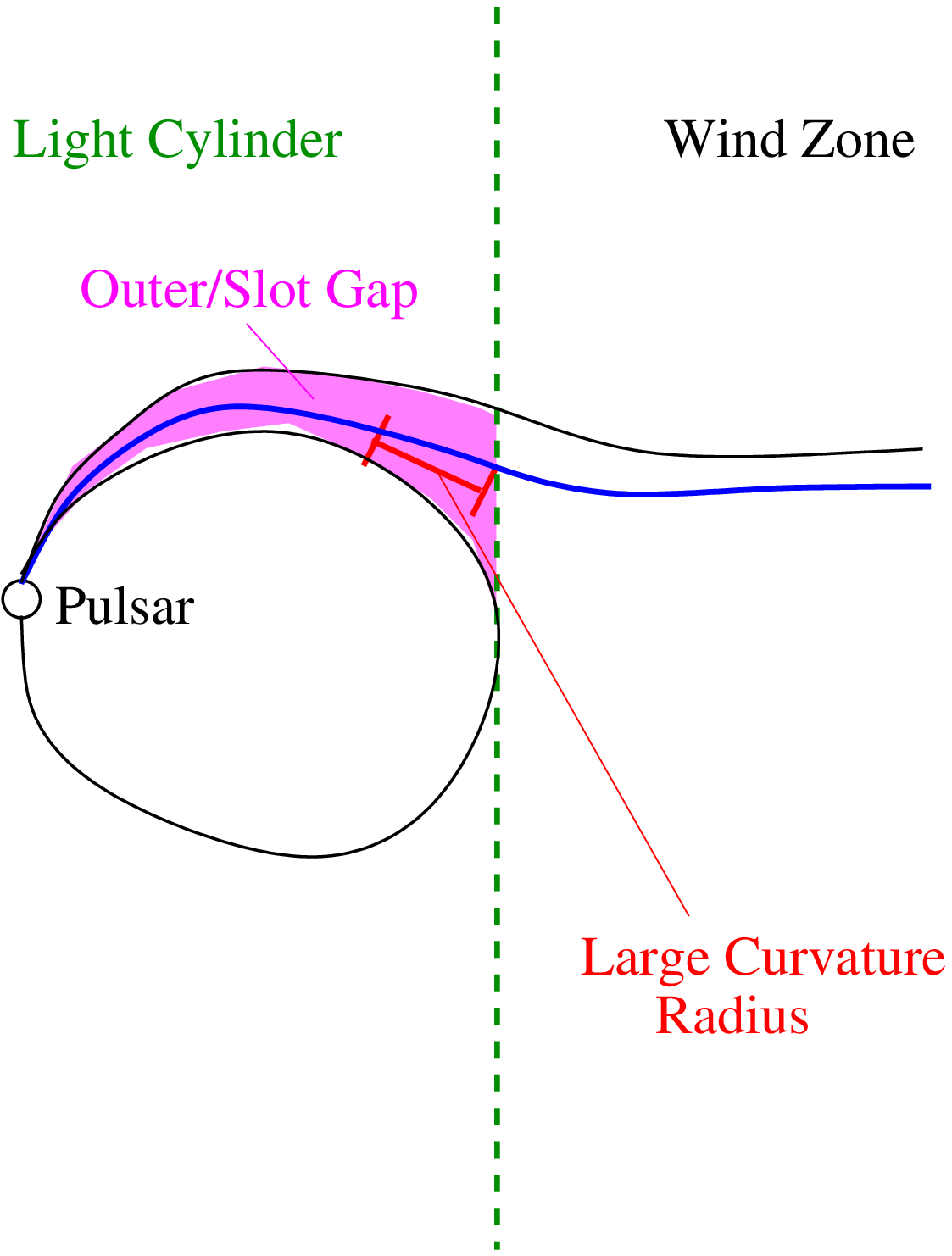}
\includegraphics[width=0.6\textwidth]{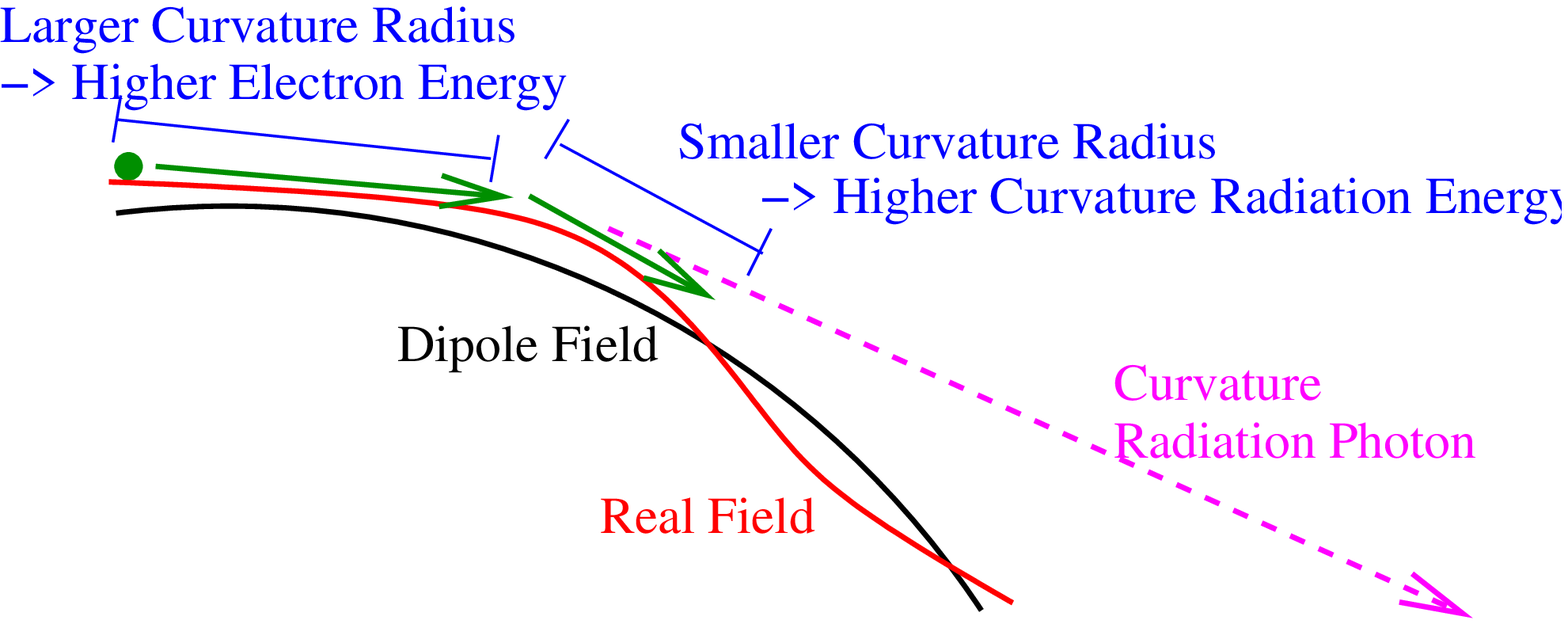}
\caption{Schematical explanations of two ideas for the high energy tail
of the observed gamma-ray spectrum.
Left: The magnetic field curvature near the light cylinder can be much larger than that of 
the dipole fields, because some field lines
in the gap should connect to the wind zone. 
Such a large field curvature
may produce a high energy tail in the gamma-ray spectrum as can be understood
from Eq. \ref{EqPhotonEne}.
Right: If the magnetic field lines are slightly wiggling along the dipole structure, 
there may be two radially connected regions the inner one of which
has a larger-than-dipole field curvature and the outer one of which
has a smaller-than-dipole field curvature.
Since the curvature radiation cooling is weaker in the larger curvature field, 
electrons can be accelerated to higher energies. 
Then, these electrons subsequently enter the smaller curvature field,
generating the unusually high energy curvature photons.
}
\label{FigWiggle}
\end{figure}

The high energy tail of the observed gamma-ray spectrum could be explained by the
imperfect dipole structure of the magnetic field.

The thickness of the Outer Gap or the Slot Gap can be as much as 0.1$R_L$
at the light cylinder.
The magnetic field lines in the middle of the gap should 
not close within the light cylinder and should be connected to
the wind zone. Therefore, there is a possibility that the magnetic field curvature
$R_{curv}$
near the light cylinder in the middle of the gap 
is significantly larger than that of the dipole structure
(see the left panel of Fig.~\ref{FigWiggle}).
As can be seen from Eq. \ref{EqGammaMax2}, the larger the curvature ($R_{curv}$),
the higher the electron energy ($\Gamma$). Alghouth the energy of the curvature radiation photon
is proportinal to $R_{curv}^{-1}$, it is also proportional to $\Gamma^3$
(see Eq. \ref{EqGEne}), resulting
in the higher photon energy from the larger curvature (see Eq.~\ref{EqPhotonEne}).

A more efficient mechanism for producing the high energy tail could be as follows:
Let us assume that the magnetic fields are slight wiggling along the dipole field 
in a small scale ($1 - 10$ km). 
There may be two radially connected regions the inner one of which
has a larger-than-dipole field curvature and the outer one of which
has a smaller-than-dipole field curvature (see the right panel of Fig.~\ref{FigWiggle}). 
Since the curvature radiation cooling is weaker,
electrons can be accelerated to higher energies in the larger curvature field. 
These electrons subsequently enter the smaller curvature field,
generating the anormally high energy curvature photons.

\clearpage

\section{Inverse Compton Scattering as a Second Radiation Component}

As shown in the top panels Fig. \ref{FigP12TwoComp}, a second emission component 
can explain the deviation of the MAGIC measurements from the standard model. 
In the previous section, a second population of electrons is assumed.  
By introducing the inverse Compton scattering, a single monoenergetic electron 
population might 
also explain the MAGIC measurements. 

 As discussed in Sect. \ref{SectIC}, the energy of the radiated photon via
the inverse Compton scattering
is
(see Eq. \ref{EqICE})
 \begin{eqnarray}
   E_\gamma  &\simeq& \epsilon \Gamma^2 
 \end{eqnarray}
where $\epsilon$ is the energy of the target photon and $\Gamma$ is the Lorentz factor
of the electron. 
In order to have a radiation peak at 30 GeV for an electron population 
with $\Gamma = 2 \times 10^7$, 
$\epsilon$ should be peaked at 
 \begin{eqnarray}
\epsilon \simeq   E_\gamma / \Gamma^2 \simeq 10^{-4} \rm ~~~ [eV]
\label{EqTargetPhE}
 \end{eqnarray}

For a thermal radiation, $10^{-4}$ eV corresponds to 1 K, which is by far lower than
the stellar surface temperature ($\sim 10^6$ K, see Sect \ref{SectSurfaceTemp}). 
On the other hand, the CMB radiation corresponds to 2.7 K in temperature, 
which is actually close to 
$\sim 10^{-4}$ eV. However, 
it is known that the number density of CMB photons (400 cm$^{-3}$) is much lower than the 
radio photons emitted within the pulsar magnetosphere ($> 10^{7}$ cm$^{-3}$) 
(see \cite{Hirotani2001}). 

Therefore, in order to reproduce the inverse Compton scattering spectrum peaked at around 30 GeV, 
there must be a particular mechanism to produce a soft photon spectrum peaked at
$\sim 10^{-4}$ eV, which is hard to imagine. 
Therefore, an effect of a simple inverse Compton scattering 
for the extended spectrum is highly unlikely.

\section{Radiation Efficiency}

The total energy loss of a pulsar, i.e. the spin down luminosity $\dot E$ 
can be estimated from the period and the time derivative of it, as described in
Sect \ref{SectSpinDown}. For the Crab pulsar, it is $\dot E = 4.6 \times 10^{38}$ erg/s. 

Here, I add an estimate of the radiation efficiency, 
which is a fraction of the spin down luminosity
deposited in a given energy range. The estimation method is adopted from \cite{FermiCrab}.
When the observed flux is $F_{obs}$, the luminosity $L$ can be calculated 
as $L = 4\pi f_\Omega F_{obs} D^2$, where $f_\Omega$ and $D$ are the beaming angle factor
and the distance from the pulsar to the Earth. 
$f_\Omega$ depends on the inclination angle $\alpha$ of the magnetic dipole axis and 
the viewing angle $\xi$ (see Sect \ref{SectLightCurve}). 
For the Crab pulsar, assuming the SG model or the OG model, $\alpha \sim 70$ degrees and  
$\xi \sim 60$ degrees
well explain the light curves (see e.g. \cite{Cheng2000}), the nebula torus structure in X-ray
(see \cite{Ng2008}),
and the polarization of the optical pulsation (see \cite{Slowikovska2009}). 
For $\alpha$ and $\xi$ near these values, $f_\Omega \simeq 1.0$ is the good approximation
according to \cite{Watters2009}. The distance is known to be $D = 2.0 \pm 0.2$ kpc 
(see e.g. \cite{Trimble1973}). From the measurements, $F_{obs}$ above 100 MeV is calculated
to be $(1.31 \pm 0.03) \times 10^{-9}$~erg/s. 

Using $\dot E = 4.6 \times 10^{38}$ erg, $D = 2.0$ kpc and $f_\Omega = 1.0$, the efficiency $\eta$ above 100 MeV is calculated as
\begin{eqnarray}
\eta =  \frac{L}{\dot E} = \frac{4\pi f_\Omega F_{obs} D^2}{\dot E} = (1.36 \pm 0.04) \times 10^{-3}
\end{eqnarray}
Only 0.13\% of the total energy loss is due to the radiation above 100 MeV. 
As discussed in Sect. \ref{SectDipoleAccel} and Sect. \ref{SectMagDipoleRadiation},
the rotation energy is carried away mostly by the pulsar wind. 

\begin{figure}[h]
\centering
\includegraphics[width=0.45\textwidth]{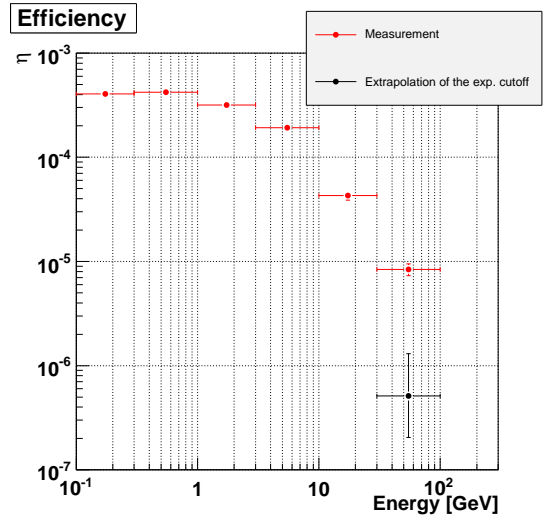}
\caption{The radiation efficiency $\eta$ for different energy intervals.
Red points indicate the observed results. 
Below 10 GeV, TP spectrum with the exponential cut-off assumption is used
while for the bin of 10 GeV to 30 GeV, a power law assumption above 4 GeV is used
(see Sect. \ref{SectFermiSpectrum}). For the bin of 30 GeV to 100 GeV, 
combined analysis with an power law assumption for P1~+~P2 is used
(see Sect. \ref{SectPL4GeV}).  
The black point indicates the extrapolation of the exponential cut-off spectrum
of TP based on \fermi-LAT measurement.
}
\label{FigRadEff}
\end{figure}

Fig. \ref{FigRadEff} shows $\eta(E)$ as a function of energy. 
Based on the combined analysis described in Sect. \ref{SectPL4GeV}, $\eta$ for energies
between 30 and 100 GeV is
$ (8.4 \pm 1.1) \times 10^{-6}$. If the spectrum had shown the pure exponential cut-off,
 it would have been
$5.1_{- 2.1}^{+ 2.8} \times 10^{-7}$, which is a factor of $\sim 10$ lower than
the measurement. 
The discrepancy between the
exponential cut-off and the MAGIC measurement corresponds to
$(8 \times 10^{-6}) / (1.4 \times 10^{-3}) = 6 \times
10^{-3}$ = 0.6\% of the radiation energy above 100 MeV.

\clearpage
\section{Remarks on the High Energy Tail of the Crab Pulsar Energy Spectrum}
\label{SectRemark}
 \fermi-LAT measured the energy spectrum of the Crab pulsar 
with good precision up to $\sim10$ GeV, which is 
only a factor of $\sim 2$ higher than the exponential cut-off energy
($E_c$~in~Eq.~\ref{EqCutoffSpectra}).
The \fermi-LAT-measured spectrum is consistent with the standard model.  
Therefore, the standard model is successful in explaining the energy spectrum
of the Crab pulsar in the energy region where a vast majority of the 
gamma-ray radiation energy is deposited.

On the other hand, MAGIC observed the energy spectrum of the Crab pulsar
above 25 GeV, which is a factor of $\sim 5$ higher than the exponential cut-off energy.
Only a small fraction of the radiation energy is deposited above 25 GeV.
The deviation from the standard model detected by MAGIC requires
only a higher order correction for the standard model.

It should also be mentioned that, 
all the dedeuced electron spectra discussed in Sect.~\ref{SectSpecSpec}
eventually need to be corrected by the ``smearing'' of the energy
due to the limited and not-perfectly-known 
energy resolution. While the simulations clearly showed that the observed
spectrum up to 100 GeV is not just an artifact of the energy resolution
(see Sect. \ref{SectHowDifficult}),
I could not determine the precise partition of genuine high energy gamma-rays 
and lower energy gamma-rays mismeasured with higher energy assignment.
It is obvious that clarification can only come from better measurements.


\section{Energy Dependence of the Rising and Falling Edges in the Light Curve}
\label{SectRiseFallDiscuss}

The basic shape of the light curve is explained by the structure of the last closed field lines,
 as discussed in Sect. \ref{SectLightCurve}. 
In that explanation,
electrons are assumed to move parallel to the field lines, i.e. the pitch angle $\phi$
of electrons is assumed to be 0. Since the beaming angle of the emitted photons 
 is $\sim 1/\Gamma \sim 10^{-7}$, all the photons are considered to be emitted tangentially to
the field lines.  However, this scenario explains neither the exponential behavior of the edges
 nor the energy dependence of the rise/fall time of the edges, 
which are clearly visible in the observed data.

\begin{figure}[h]
\centering
\includegraphics[width=0.25\textwidth]{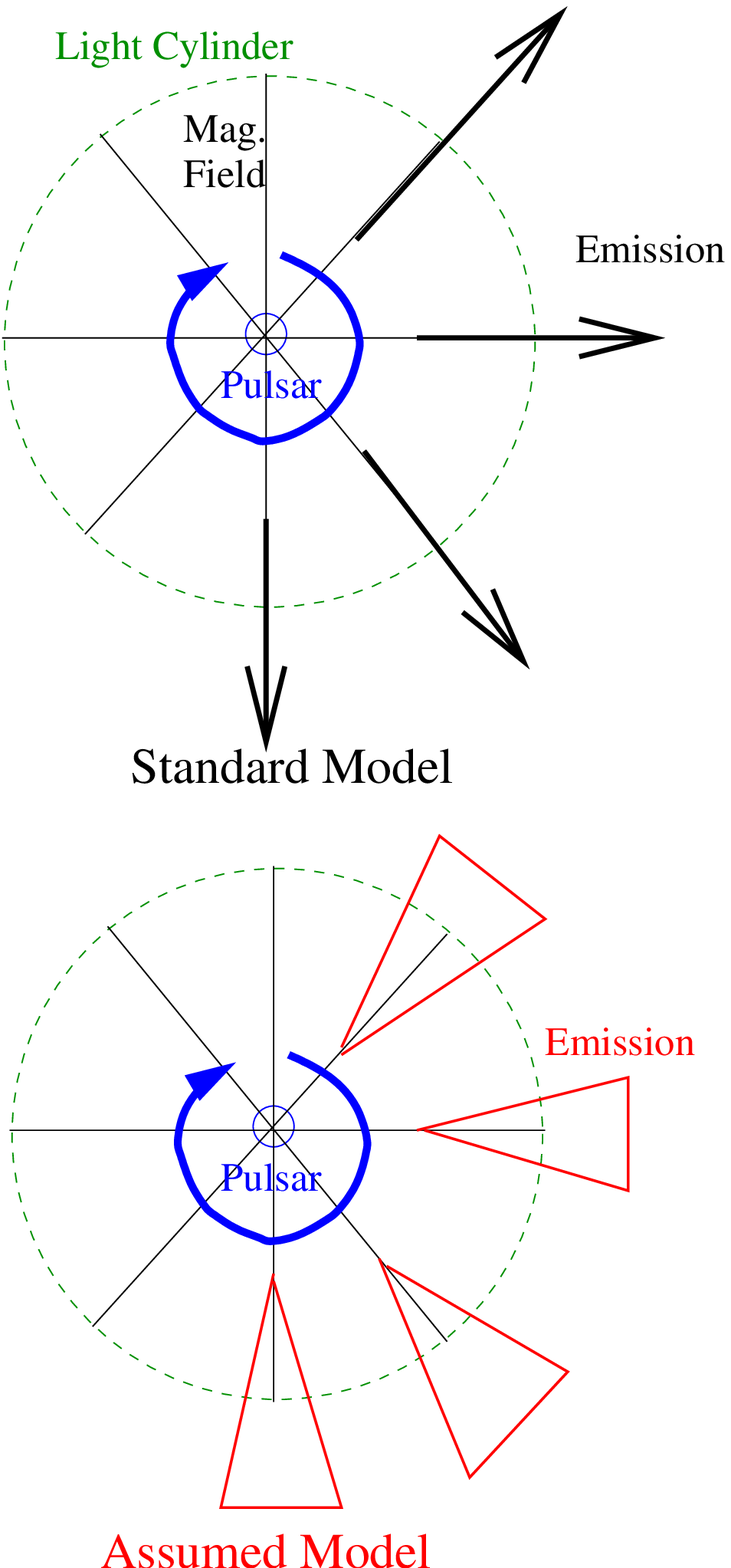}
\includegraphics[width=0.55\textwidth]{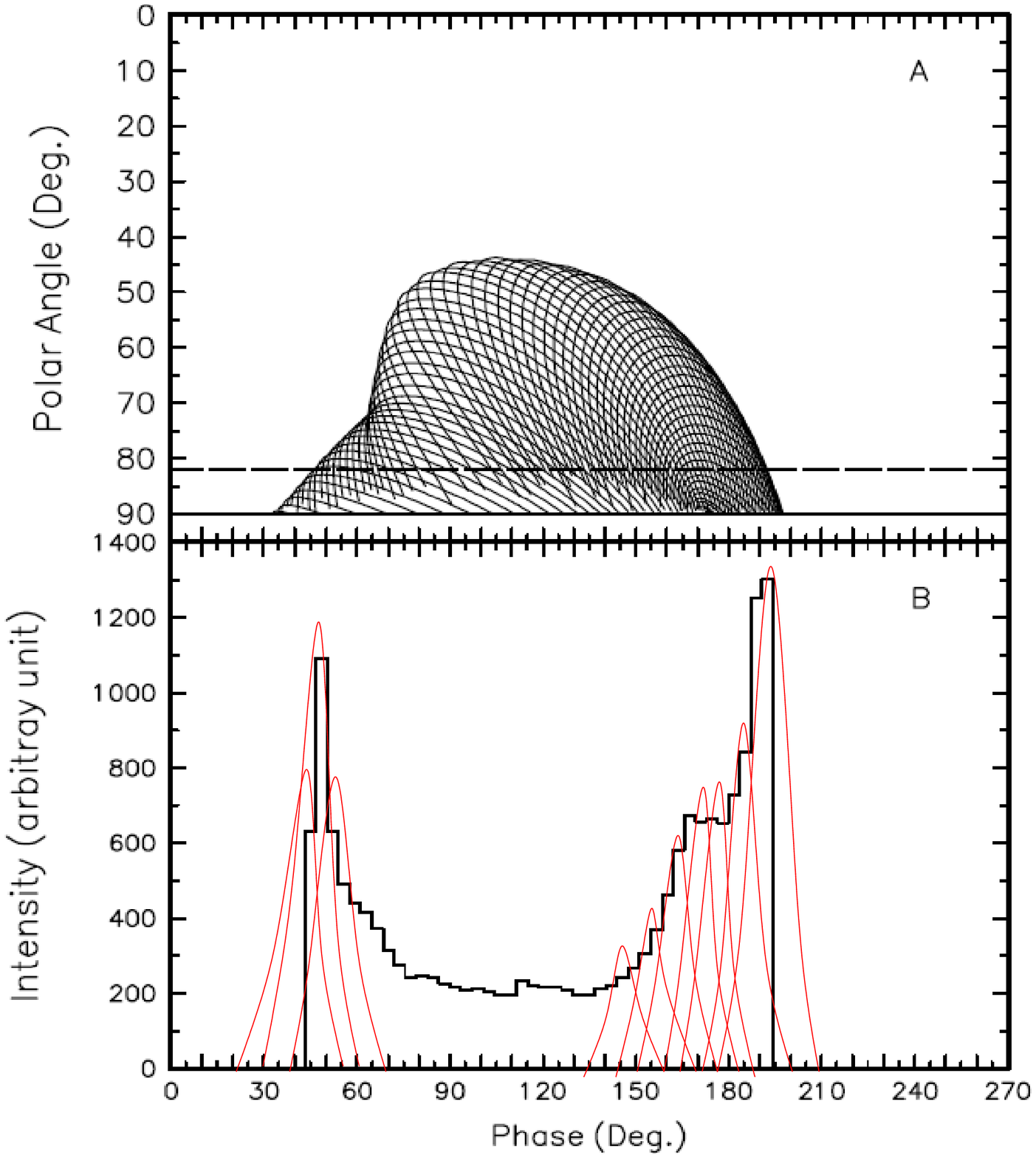}
\caption{Top left: A simplified top view of a pulsar of the standard model.
The emission is strongly beamed along the field lines.
Bottom left: A simplified top view of a pulsar of the assumed model.
The emission has an exponential angular distribution along the field lines. 
Top right:
 The emission profile map of the standard model assuming that inclination angle between
the rotation axis and the dipole axis is 65 degrees. Figure adopted from \cite{Cheng2000}.
Bottom right: The hypothetical light curve.
A black histogram shows the basic structure of the light curve explained 
by the emission profile map shown in the top panel. Red lines show the hypothetical broadening
effect taking into account the angular distribution of the emission
Original figure is adopted from \cite{Cheng2000} and red lines are added by myself.
 }
\label{FigBroadening}
\end{figure}

 Observed results can be explained by, for example,
 assuming that the emission from each field line 
is not beamed with an angle of  $\sim 1/\Gamma$, 
but has an exponential angular distribution characterized by 
its decay constant $\theta_c$:
\begin{eqnarray}
F(\theta) = F_0\exp(-\theta/\theta_c)
\end{eqnarray}
where $\theta_c$ is dependent on the photon energy. 
Then, the resulting light curve should be broadened
compared with the one without a sizable emission angle
, as shown in Fig. \ref{FigBroadening}.

Here, I additionally assume that the shape of the the outer edges (the rising edge of P1
and the falling edge of P2) reflect the angular distribution of the emission
along a single field line, while the shape of the inner edges
(the falling edge of P1 and the rising edge of P2) are governed by 
the overall field line structure (see Fig.~\ref{FigBroadening}).
From Eq. \ref{EqTau1} and Eq. \ref{EqTau4}, the rise time of P1 and the fall time of P2
as a function of energy are:
\begin{eqnarray}
\tau \simeq (0.022 \pm 0.0018) - (0.0095 \pm 0.0034) \log_{10}(E {\rm [GeV]})
\label{EqTauApp}
\end{eqnarray}
where the simple mean of $\tau_{rize}^{P1}$ and $\tau_{fall}^{P2}$
is adopted.
Since the emission angle $\theta$ can be translated into the pulse phase $p$ 
as $p = \theta$/$2\pi$, Eq. \ref{EqTauApp} leads to
\begin{eqnarray}
\theta_c &\simeq& (0.14 \pm 0.01) - (0.06 \pm 0.02)\log_{10}(E {\rm [GeV]})  {~~~~~~ \rm[rad]}\\
&=& (7.9 \pm 0.6)  - (3.4 \pm 1.2) \log_{10}(E {\rm [GeV]}){~~~~~~ \rm[deg]}
 \label{EqTauTheta}
\end{eqnarray}
This is by far larger than $1/\Gamma \sim 10^{-7}$.
The large emission angle should be the consequence of the large pitch angle of 
the high energy electrons. These electrons with a spiral orbit should emit 
synchrotron radiation, which should explain the observed gamma-rays at least below 3 GeV
(where the energy dependence of the edges is clearly seen as shown in Fig. \ref{FigRiseAndFall}.)
The strength of the magnetic field at 1000 km from  the neutron star is $> 10^6$G and 
high energy gamma-rays are therefore reasonably expected (see Sect. \ref{SectSynch}). 

 However, in order to realize such a large pitch angle orbit, there must exist a process
which nearly instantly provides a large perpendicular momentum, because a
gradual increase of perpendicular momentum cannot occur due to the strong energy loss
by the synchrotron radiation itself. (This is actually the reason why the pitch angle
of electrons is normally considered to be 0.)


A possibility that the pitch angle $\phi$ can has a large
value has been discussed by, for example,  S. A. Petrova 
(see \cite{Petrova2002} and references therein).
The basic idea is as follows:
The accelerated electrons efficiently absorb radio photons with 
a frequency corresponding to the cyclotron frequency when it is converted into the electron 
rest frame.
This cyclotron resonance absorption of radio photons is so efficient that it can cause
a large pitch angle even though electrons are continuously losing energy 
by the synchrotron radiation.
This scenario is used in \cite{Harding2008} to reproduce the Crab pulsar energy spectrum
at around 100 MeV (see Fig. \ref{FigOuterSpectrum2}). 
However, this process produces neither the exponential angular distribution  
nor the electron-energy dependent pitch angle (see \cite{Petrova2002}). 

\section{Energy Dependence of the Peak Phase}
\label{SectEDepPeak}

\begin{figure}[h]
\centering
\includegraphics[width=0.45\textwidth]{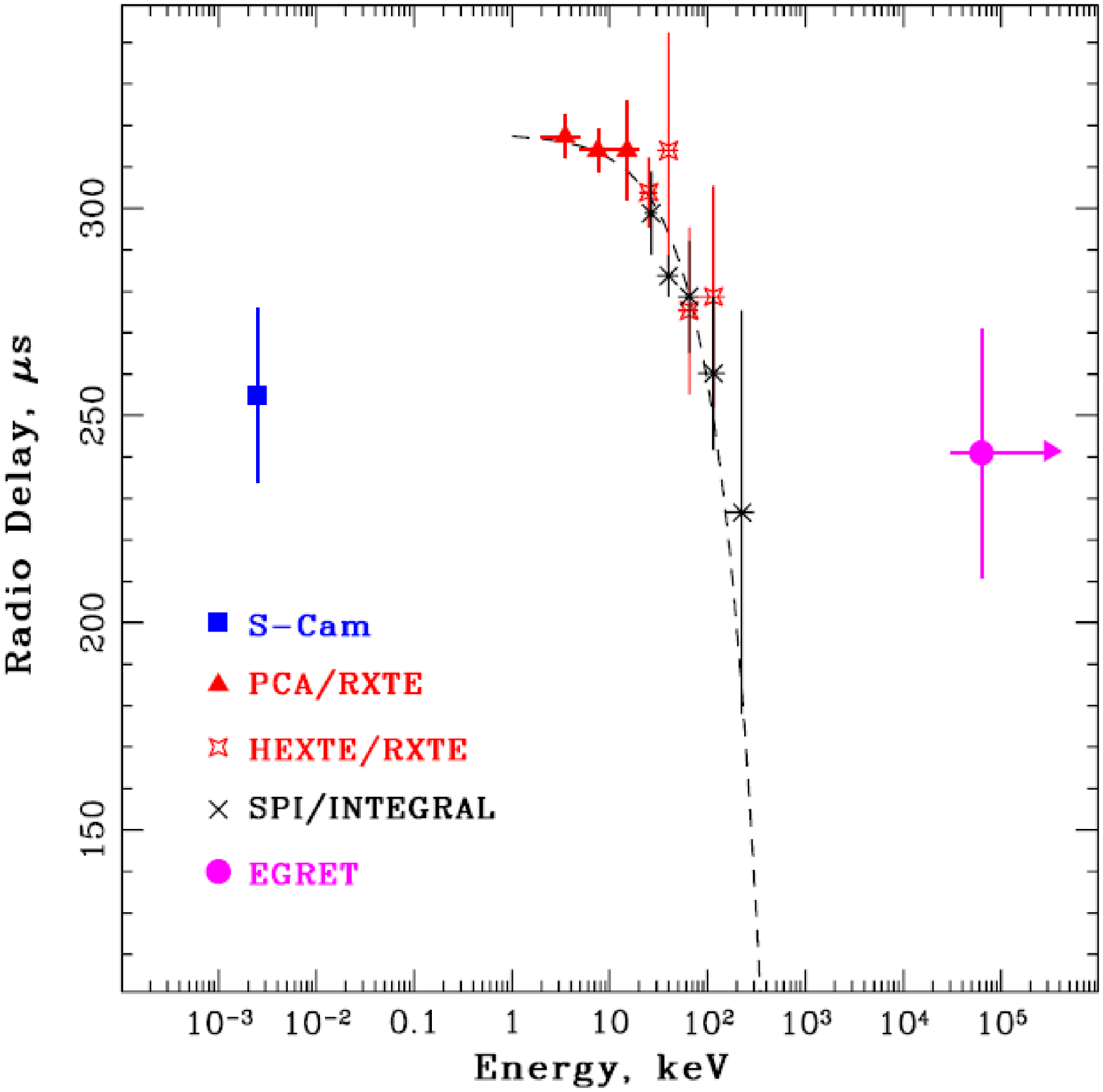}
\includegraphics[width=0.45\textwidth]{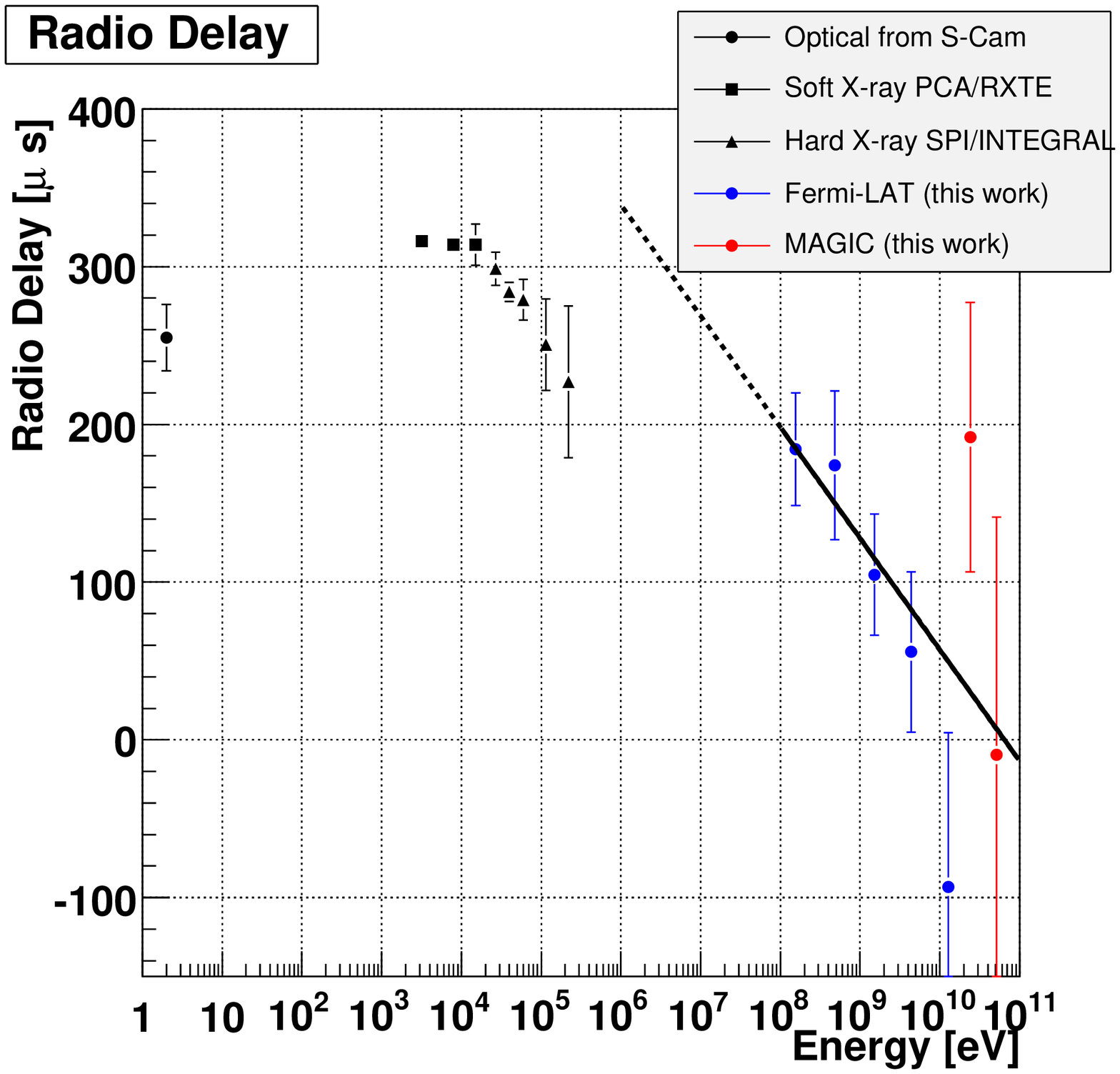}
\caption{Left: Radio delay as a function of energy. 
A simple linear function fit to X-ray range is indicated by a dashed line.
Figure adopted from \cite{Molkov2010}
Right: The same as the left panel but \fermi-LAT and MAGIC results analyzed by myself
 are shown. The energy dependence calculated based on Eq. \ref{EqPeakPhase1}
is shown as a solid line above 100 MeV. The dotted line is an extrapolation 
of it down to 1 MeV.
The points below 1 MeV are copied from the left panel by eye.
 }
\label{FigRadioDelay}
\end{figure}

In Fig. \ref{FigRadioDelay}, the first peak (P1) position as a function of energy 
is shown from optical to high energy gamma-rays. It is expressed as ``radio delay'',
which tells how much earlier the peak at a given energy occurs compared to the radio peak
(Jodrell Bank, at 610 MHz). 
The energy dependence of the radio delay has been studied in the X-ray range.
As discussed by many authors 
(see e.g. \cite{Kuiper2003}, \cite{Rots2004} and \cite{Molkov2010}),
the simplest explanation for the radio delay and its energy dependence is 
the difference in the emission region within the magnetosphere. i.e.,
the higher the energy is, the more inner of the magnetosphere the emission 
region is. 
In \cite{Molkov2010}, a simple linear function was fitted to the data,
deriving a shift of $0.6 \pm 0.2 \mu$s keV$^{-1}$ (see the left panel of the figure).  
It corresponds to 
\begin{eqnarray}
L(E) &=& (180 \pm 60)\times(E{\rm [keV]}) \rm {~~~}[m]
\label{EqPLX}
\end{eqnarray}
where $L$ indicate the path length difference with respect to the radio emission region.
 Eq. \ref{EqPeakPhase1} can be rewritten as
$(130 \pm 20) - (70 \pm 30){\rm log}_{10}(E/\rm{GeV}) {~~~} [\mu{\rm s}] 
$ in radio delay, which is shown as a black line in the left panel of the figure. 
In path length, this corresponds to
\begin{eqnarray}
L(E) &=& (40 \pm 6) - (20 \pm 9){\rm log}_{10}(E\rm{[GeV]}) {~~~} [km]
\label{EqPLG}
\end{eqnarray}
Considering the size of the light cylinder ($R_L = 1500$ km), Eq. \ref{EqPLX} and Eq. \ref{EqPLG}
are reasonable. 
On the other hand Eq. \ref{EqPLX} and Eq. \ref{EqPLG} are apparently not consistent
as can be seen from Fig. \ref{FigRadioDelay}. 
Actually, the shape of the energy spectrum of the Crab pulsar (see Fig. \ref{FigCrabOpticaltoGamma})
and the theoretical model (see Fig. \ref{FigOuterSpectrum})
suggests different mechanisms for X-ray and high energy gamma-ray
radiation.
Therefore, it may be natural that the energy dependence of the radio delay is different in
the two well-separated energy ranges. 


It should be noted that the definition of the peak phase is different for different 
types of analyses. For example, A. A. Abdo et al. (see \cite{FermiCrab}),
which is the official publication from the \fermi~collaboration, 
 determined the peak phase by fitting
an asymmetric Lorentzian function to the binned light curves.
In their analysis, the peak phase above 100 MeV is constant and  $-0.008 \pm 0.001$,
which corresponds to $280 \pm 30$ $\mu$s in radio delay, though
their highest energy bin of 3 to 10 GeV shows marginal deviation of
$-0.005 \pm 0.002$, corresponding to $170 \pm 70$ $\mu$s in radio delay.


 \chapter{Development of High Quantum Efficiency Hybrid Photodetector  HPD R9792U-40}

see http://wwwmagic.mppmu.mpg.de/publications/theses/TSaito.pdf
 \chapter{Conclusions and Outlook}
\section{Conclusions}

Before 2007, the energy spectrum of the Crab pulsar had been measured only up to $\sim 5$ GeV
by the satellite-borne detector, EGRET,
while  IACTs had set flux upper limits only above 100 GeV.
There existed no sufficient measurement at around the cut-off energy, i.e, at energies between 
a few GeV and a few tens of GeV, while the spectral shape at around the cut-off energy
is essential to constraining the emission region of the pulsation, 

The MAGIC telescope with the newly implemented SUM trigger
successfully detected emission from the Crab pulsar above 25 GeV
during the observations between October 2007 and February 2008
thanks to the collective efforts of my colleagues, T. Schweizer, M. Lopez, 
A. N. Otte, M. Rissi and M. Shayduk.
However, an in-depth analysis and detailed discussion 
in comparison with the adjacent energy range had not yet been performed.
Also, a new satellite-borne detector, \fermi-LAT, became operational in August 2008 
and the observational data were made public in August 2009. 

In this thesis, the Crab pulsar 
has been studied in detail
in the previously (almost) unstudied energy gap between 5 GeV and 100 GeV.
For the analysis, I used the data from both the upgraded MAGIC telescope and 
the public \fermi-LAT data.
The main results are summarized as follows:

\begin{itemize}
\item MAGIC observations between October 2007 and January 2009
resulted in the detection of the Crab pulsar above 25 GeV
with a statistical significance of
4.3~$\sigma$, 7.4~$\sigma$ and 7.5~$\sigma$ for the first peak (P1), the second peak (P2),
and the sum of the two peaks (P1~+~P2), respectively.
\item 
{\bf The energy spectrum of the Crab pulsar is 
consistent with a power law with an index of $\sim -3.5 \pm 0.5$ between 25 GeV and 100 GeV
for P1, P2 and P1~+~P2. At 30 GeV, the flux of P2 is twice as high as that of P1.} 

\item A variation of the flux and the light curve 
of the Crab pulsar
on a yearly time scale were not found in the MAGIC data.

\item One year of \fermi-LAT data showed a clear detection ($> 100$ $\sigma$) 
of the Crab pulsar from 100 MeV to $\sim 30$ GeV. 
Between 100 MeV and $\sim 30$ GeV, the energy spectrum 
is consistent with a power law with an exponential cut-off, 
for total pulse (TP), P1, P2 and P1~+~P2.
The cut-off energies are estimated to be $6.1 \pm 0.5$ GeV, $3.7 \pm 0.3$ GeV, $5.9 \pm 0.7$ GeV and $4.5 \pm 0.3$ GeV for TP, P1, P2 and P1~+~P2, respectively.
Due to the small detector area of $\sim 1$ m$^2$, 
the statistical uncertainty of the spectrum above $\sim 10$ GeV is rather large and 
it is not possible to detect any pulsed signal above 30 GeV.
\item {\bf From the \fermi-LAT observations,  the superexponential cut-off assumption ($\Gamma_2 = 2.0$ in Eq. \ref{EqCutoffSpectra})
is ruled out by 4.8 $\sigma$, 5.0$\sigma$, 4.3$\sigma$ and 7.7$\sigma$ for 
TP, P1, P2 and P1~+~P2, respectively.
}

\item {\bf The combination of the results from MAGIC and \fermi-LAT revealed that
 the exponential cut-off spectra determined by \fermi-LAT are inconsistent with MAGIC
results above 25 GeV by $> 2.1 \sigma$, $>4.3 \sigma$ and $> 5.3 \sigma$ for P1, P2 and P1~+~P2,
respectively, even if the possible absolute energy scale difference
between the two experiments is carefully taken into account (up to 30\%)}.

\item The flux ratio of P2 to P1 and that of  Bridge to P1
increase rapidly with energy between 100 MeV and 100 GeV. 
This behavior is similar to that in the energy range 
below 1 MeV but contrary to that in the energy range between 1 MeV to 100 MeV. 

\item {\bf Both edges of the two peaks show a clear exponential behavior. 
In addition, the outer edges, i.e.,
the rising edge of P1 and the falling edge of P2 become sharper 
as the energy 
increases, while the inner edges, i.e. the falling edge of P1 and the rising
 edge of P2 have no energy dependence. }
The rise time of P1 ($\tau^{P1}_{rise}$) and the fall time of P2 ($\tau^{P2}_{fall}$)
can be expressed as \\
$\tau_{rise}^{P1}(E) = (2.02 \pm 0.08)\times 10^{-2} - (9.4 \pm 1.3)\times 10^{-3}{\rm log}_{10}(E[\rm{GeV}])$ 

$\tau_{fall}^{P2}(E) = (2.42 \pm 0.16)\times 10^{-2} - (9.6 \pm 3.1)\times 10^{-3}{\rm log}_{10}(E[\rm{GeV}])$

\item {\bf The phase of the first peak has a slight but significant energy dependence. }
This shift can be expressed as \\
$Peak1(E) = (-3.8 \pm 0.6)\times 10^{-3} + (2.1 \pm 0.9)\times 10^{-3}{\rm log}_{10}(E[\rm{GeV}]) $.
As the energy increases, 
the peak position shifts to a later time in the light curve. 
For the second peak, because of the broader width, the peak phase is determined with 
a worse precision, and no significant energy dependence has been found.

\item In the \fermi-LAT data above 10 GeV, a hint of the third peak is seen at phase $\sim 0.75$ with a significance of 3.5 $\sigma$. However, in the MAGIC data, only a 1.7 $\sigma$ excess
 has been found
and the flux upper limit based on the MAGIC data is in marginal contradiction with the \fermi-LAT
results. 

\item Aiming for better observations of pulsars and other sources below 100 GeV with MAGIC, 
I participated in the development of a new photodetector, the Hamamatsu hybrid photodetector
HPD R9792U-40. 
Compared to the currently used PMTs, 
its photodetection efficiency is twice higher and its 
ion-feedback rate is 500 times lower. Its charge resolution is excellent, too.
The lifetime of the photocathode was proven to be long enough to allow a ten year observation
time without significant degradation. 
A compensation circuit for the correction of the temperature dependence of the gain and
 safety circuits against the strong light were also successfully developed.
\end{itemize}

Based on these results, the following physics conclusions have been drawn:

\begin{itemize}

\item The extension of the pulsed gamma-ray emission up to 100 GeV 
observed by MAGIC sets a lower limit in 
height of the emission region at 7.8 times the neutron star radius.  
This rules out the inner magnetosphere emission scenario, i.e. the Polar Cap model,
 for the pulsation mechanism.
Strong rejection of the super-exponential cut-off assumption by \fermi-LAT 
also favors the outer magnetosphere
emission scenario, i.e. the Slot Gap model or the Outer Gap model.

\item The rejection of the exponential cut-off assumption by the combined analysis 
of \fermi-LAT results and MAGIC results requires modifications of the standard
outer magnetosphere model. If the magnetic field has an ideal dipole structure,
there must be a place where the acceleration electric field is more than 10 times
larger than that of the standard model ($>3.3\times 10^7$ [V/cm]). 
A distorted dipole structure of the magnetic field
is another possible explanation.

\item It is unlikely that the contribution of the inverse Compton scattering is 
the reason for the discrepancy between the standard outer magnetosphere model and 
the observed results, considering the energy of accelerated electrons and 
that of possible target photons. 

\item The radiation efficiency above 100 MeV is estimated to be 
$(1.36 \pm 0.04) \times 10^{-3}$ from the \fermi-LAT measurement while that above 30 GeV is 
estimated to be $(8.4 \pm 1.1) \times 10^{-6}$ from the MAGIC measurement.
The discrepancy in the radiation energy above 30 GeV between the standard outer magnetosphere
model and
the MAGIC measurement amounts to 0.6\% of the radiation energy above 100 MeV.

\item 
The exponential behavior of the pulse edges can be explained by
assuming that the emission angle with respect to the magnetic field line 
has an exponential distribution. 
Under this assumption, the energy dependence of the 
exponential decay constant $\theta_c$ can be expressed as 
$\theta_c = (7.9 \pm 0.6) - (3.4 \pm 1.2) \log_{10}(E {\rm [GeV]})$ [deg], $E < 100$ GeV.

\item The simplest explanation for the energy dependence of the peak phase is that 
the emission region shifts inward toward the neutron star.
The energy-dependent difference in path length 
$L(E)$ with respect to the radio emission region can be written as \\
$L(E) = (40 \pm 6) - (20 \pm 9){\rm log}_{10}(E/\rm{GeV})$ [km], $E < 100$ GeV.

\end{itemize}

\clearpage
\section{Future Prospect: Observations of Other Gamma-ray Pulsars}
The power-law-like extension of the gamma-ray energy spectrum of the Crab pulsar
 beyond the cut-off energy 
is a new discovery and currently there is no concrete theoretical explanation for it,
except the discussions presented in this thesis. 
It would be necessary to check if this feature is unique for the Crab pulsar or
 common for all/some of the other pulsars. 
Although \fermi-LAT detected 46 gamma-ray pulsars,
it is not possible to study the spectral behavior well beyond the cut-off energy
with \fermi-LAT data due to the limited effective area of the detector. 
 MAGIC is currently the only detector that can study pulsars 
at energies well beyond the
cut-off energy. 
The next target of observation for MAGIC could be the Geminga pulsar, 
which is the second  brightest above 1 GeV (next to the Crab pulsar)
 among the pulsars in the sky region which MAGIC can observe.
Above 10 GeV, its flux is comparable to that of the Crab pulsar, according to the \fermi-LAT observations. 

\section{Future Prospect: Improvement of the Telescope Performance}

The Crab pulsar is the only pulsar that has up to now been detected by an IACT.
Other pulsars have never been detected from ground despite tremendous efforts. 
Even for the Crab pulsar, the energy spectrum 
could be determined only with the moderate statistical significance 
after 59.1 hours of observations, which is relatively long for IACT
observations. 
The measured energy spectrum of the Crab pulsar is consistent with a power law
above 25 GeV. 
However, if the statistical uncertainty of the
measurements is smaller and the energy resolution is better, 
a possible curvature of the spectrum might become visible, which surely helps to understand the reason for the
spectral extension after the cut-off. 

In order to detect more pulsars and determine the energy spectrum with higher precision,
one needs to meet the following requirements:

\begin{itemize}
\item A lower energy threshold.
\item A larger effective area below 100 GeV
\item A better  (hadron+muon)/gamma separation below 100 GeV
\item A better angular resolution below 100 GeV
\item A better  energy resolution below 100 GeV
\end{itemize}

For meeting these requirements, several improvements might be explored, as described 
in the following subsections.

\subsection{Installation of the HPD R9792U-40}
The replacement of the PMTs in the camera with the HPDs described in
Chapter \ref{ChapterHPD} would improve the telescope performance.
The HPDs will double the number of detected Cherenkov photons from air showers.
One can record shower images with higher precision, which will lead to
a better (hadron+muon)/gamma separation, a better angular resolution
 and a better energy resolution.
The energy threshold will also be lowered. 
The effective area should also increase largely, not only thanks to the higher photodetection
efficiency
but also due to the lower ion-feedback rate, allowing a more efficient trigger.
However, the effect of the fast-and-huge afterpulsing,
which may be attributed to the generation of characteristic X-rays inside the HPD, 
should be carefully studied.


\subsection{Stereoscopic Observation with the MAGIC Stereoscopic System}

\begin{figure}[h]
\centering
\includegraphics[width=0.45\textwidth]{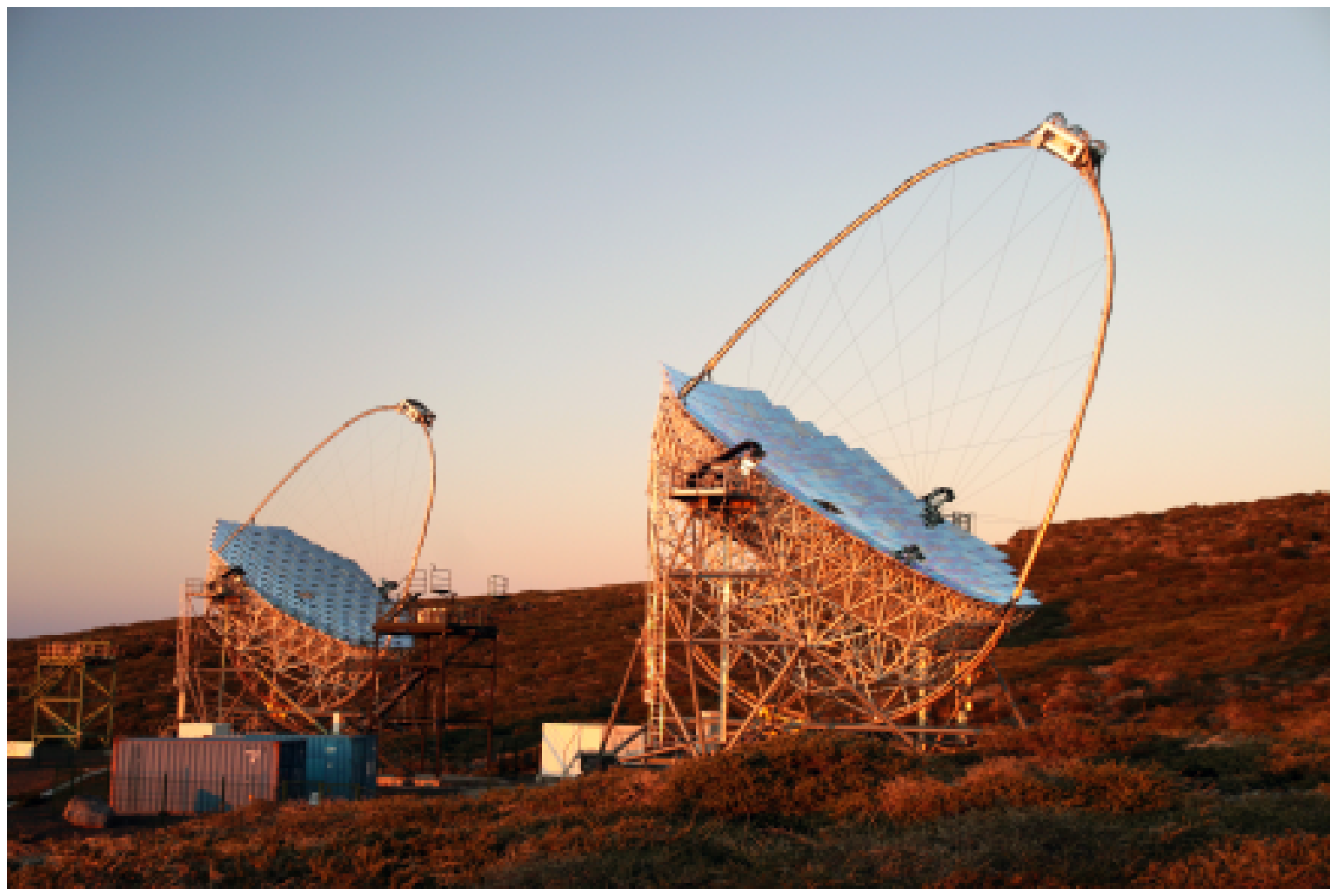}
\includegraphics[width=0.5\textwidth]{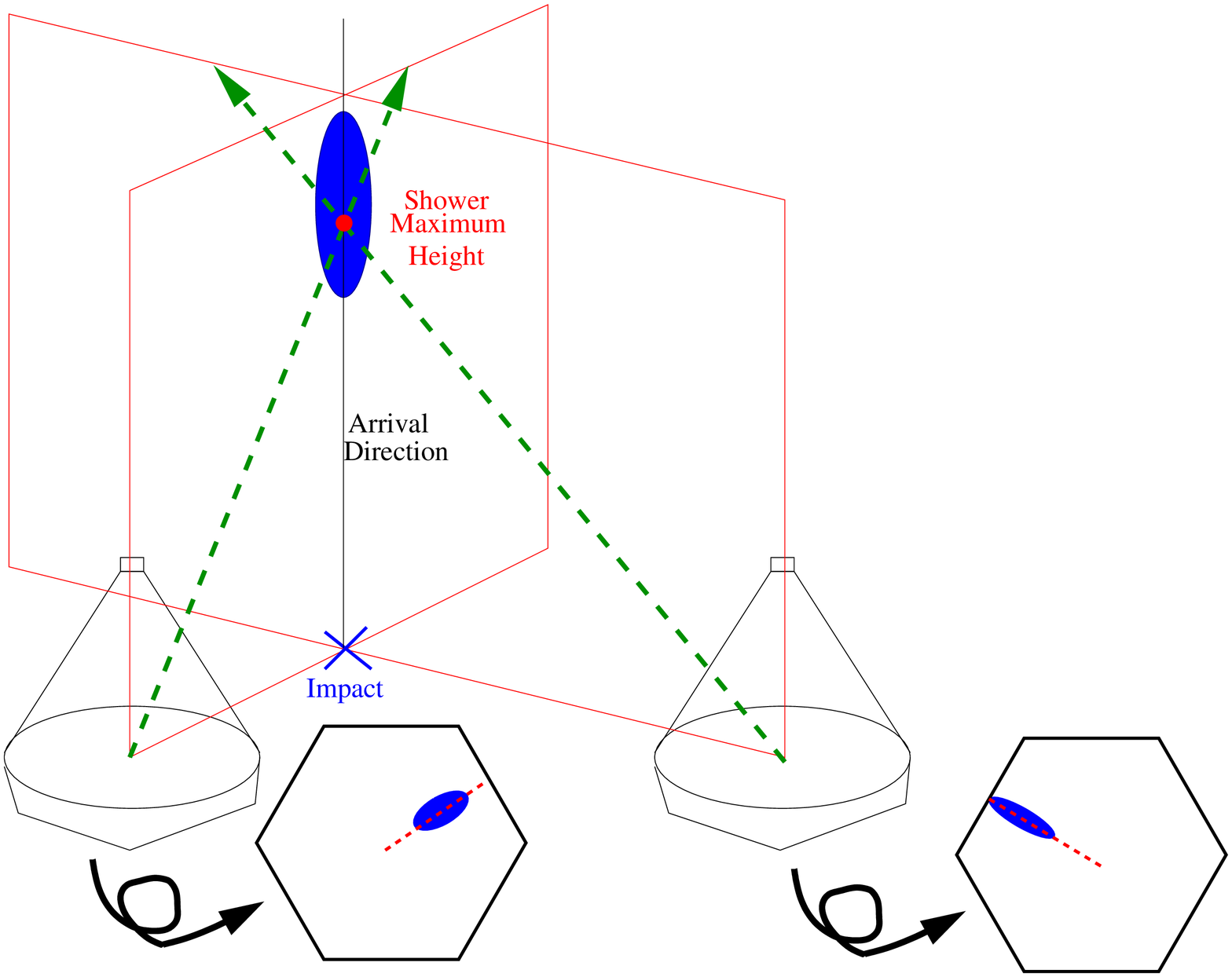}
\caption{Left: A photograph of the two MAGIC telescopes adopted from \cite{MAGICGal}.
Right: Concept for the stereo observations. Using two images recorded by 
the two telescopes, the arrival direction, the shower maximum height and the
impact point of an air shower can be reconstructed with a much higher precision
than with a single telescope.
}
\label{FigStereo}
\end{figure}

\begin{figure}[h]
\centering
\includegraphics[width=0.3\textwidth]{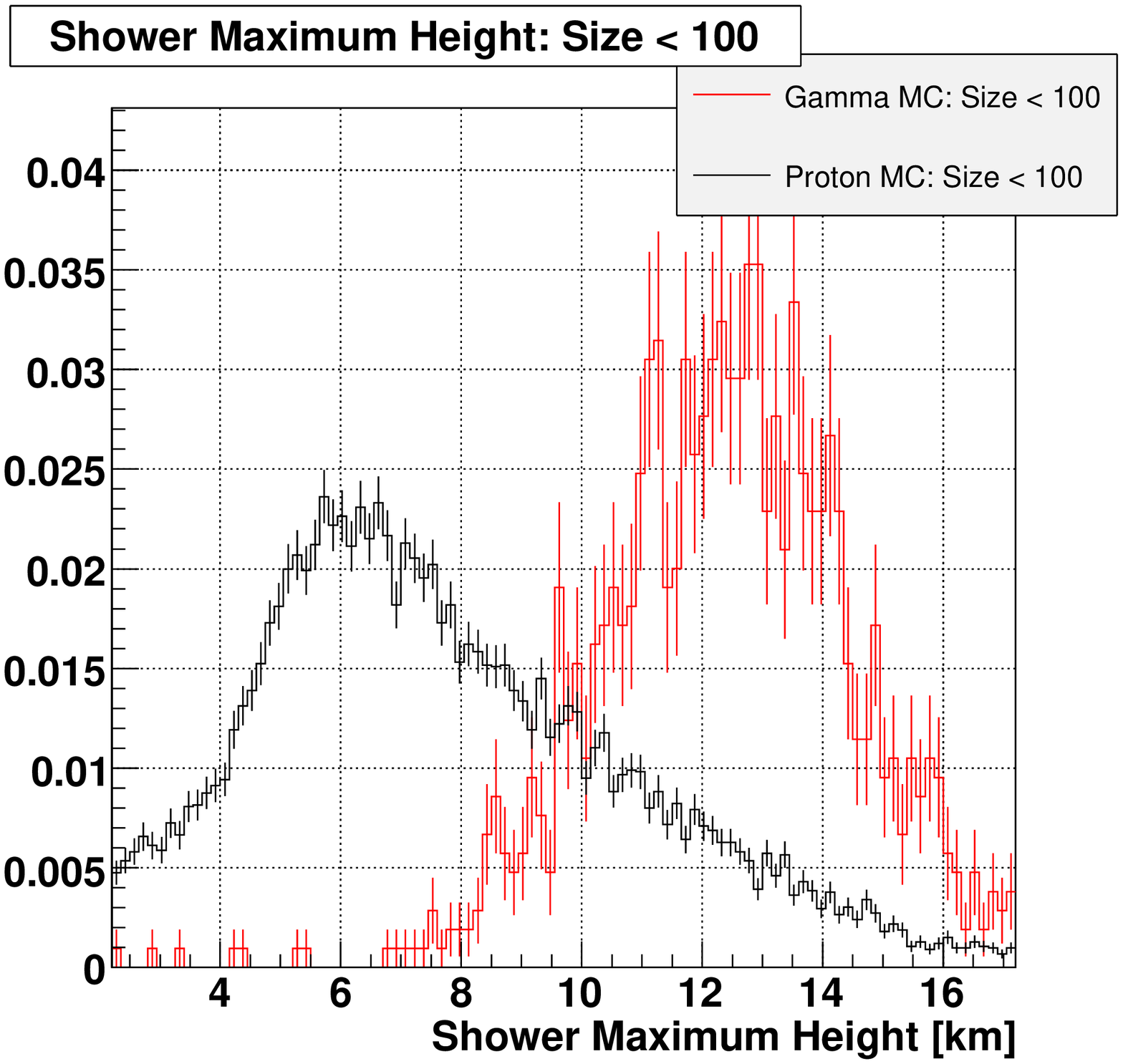}
\includegraphics[width=0.3\textwidth]{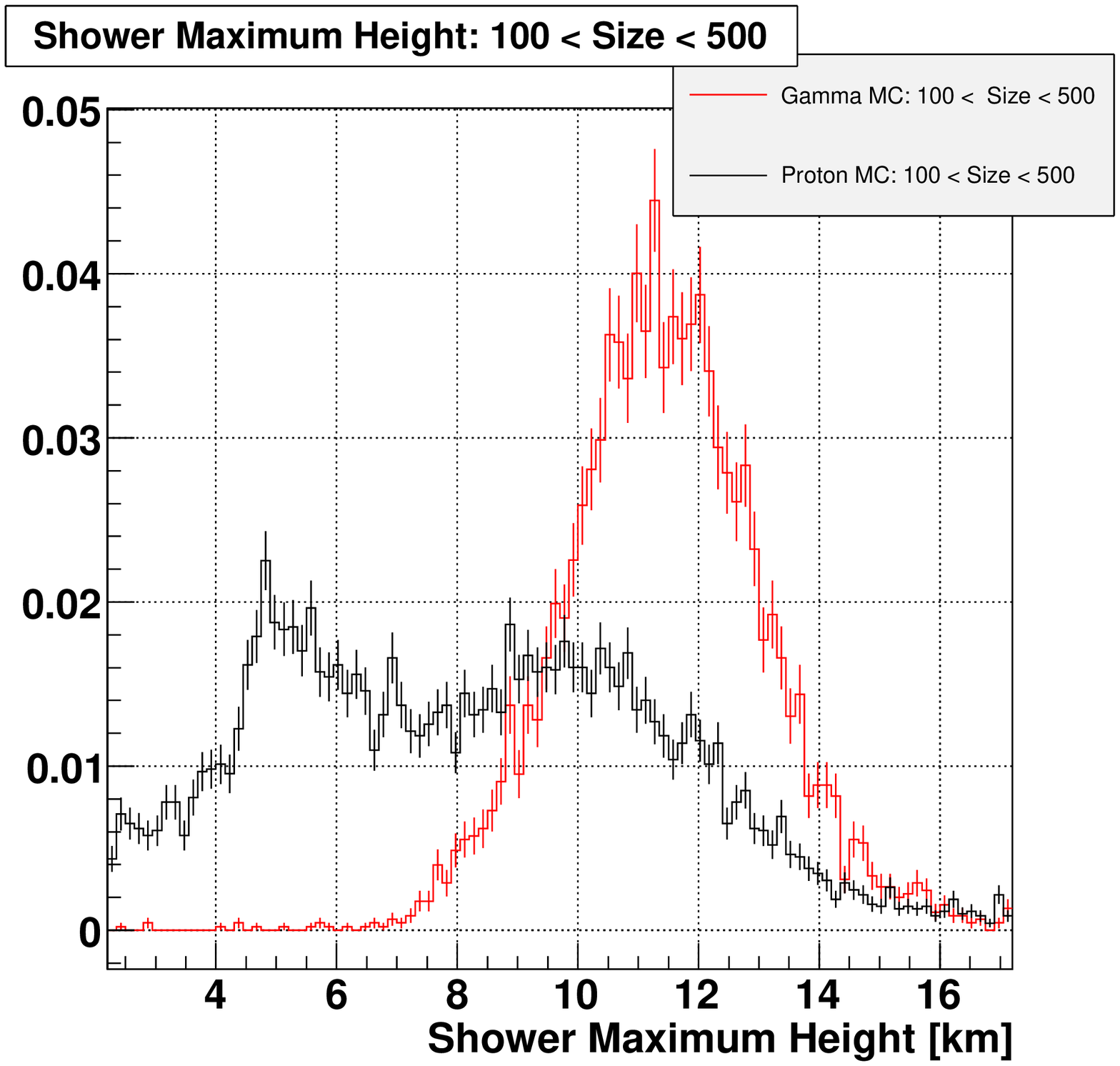}
\includegraphics[width=0.3\textwidth]{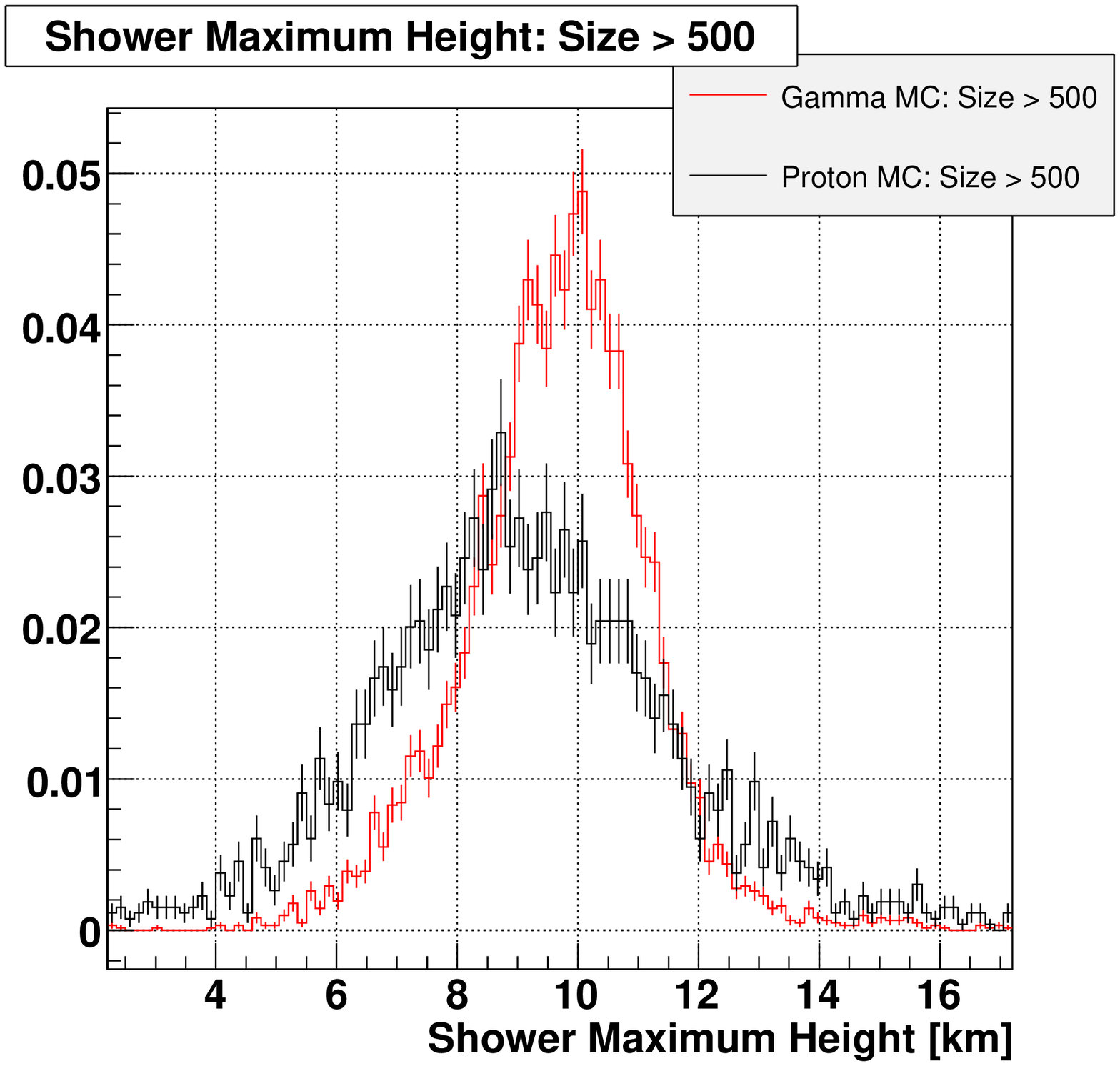}
\caption{The shower maximum height distributions for observed data (black histograms)
and gamma-ray MC (red histograms). 
$SIZE < 100$ (left), $100 < SIZE < 500$ (middle) and $SIZE > 500$ (right) are shown.
At $SIZE >500$, 
two distributions are not clearly separated.
At $100 < SIZE < 500$, two peaks can be seen for the observed data. 
The first peak at $\sim 5$ km can be explained by the 
the contribution from large-impact-distance muons 
(see text for detailed explanation).
At $SIZE < 100$, only one peak can be seen at $\sim 6$ km for the observed data, 
suggesting that most of the events are 
large-impact-distance muons.
Consequently, the distribution of the observed data
is  clearly separated from that of the gamma-ray MC.
Therefore, the shower maximum height, which can precisely be estimated only 
with the stereo observations,
is a powerful parameter for the gamma-ray/hadron separation for very low energies.
 }
\label{FigHMax}
\end{figure}

In October 2009, MAGIC started stereoscopic observations with two telescopes. 
As schematically explained in the right panel of Fig. \ref{FigStereo},
if an air shower image is recorded by the two telescopes, the arrival direction can be
reconstructed better than by a single telescope. In addition, 
the impact point of the shower and the shower maximum height can 
be determined with higher precision.
This leads to a better energy resolution. 
The information of the shower maximum height has another advantage.
A muon image may look like a low energy gamma-ray image if 
the impact distances from both telescopes are large ($> 80$ m). 
However, the majority of muons can be identified by 
the reconstructed shower maximum height (a muon does not produce a shower but can create
a shower-like image). In the case of muon images, 
the height should be reconstructed to be $\sim 5$ km, which is unusually low 
for low energy gamma-rays (see Fig. \ref{FigHMax}). 
Also, Hadron/gamma-ray separation will improve simply due to the double amount of information
from the two telescopes. 

 However, the requirement for a coinciding signal from the two telescopes reduces
the effective area especially for low energies. Below 50 GeV, the reduction
can be as large as a factor of $\sim 5$. The energy threshold also increases. 
The trigger condition, i.e. whether or not the coincidence condition is fulfilled,
  must be carefully studied 
taking into account the advantage and the disadvantage of stereoscopic observation. 
\\
\\
 
\subsection{Pulsar Observation with CTA}

\begin{figure}[h]
\centering
\includegraphics[width=0.45\textwidth]{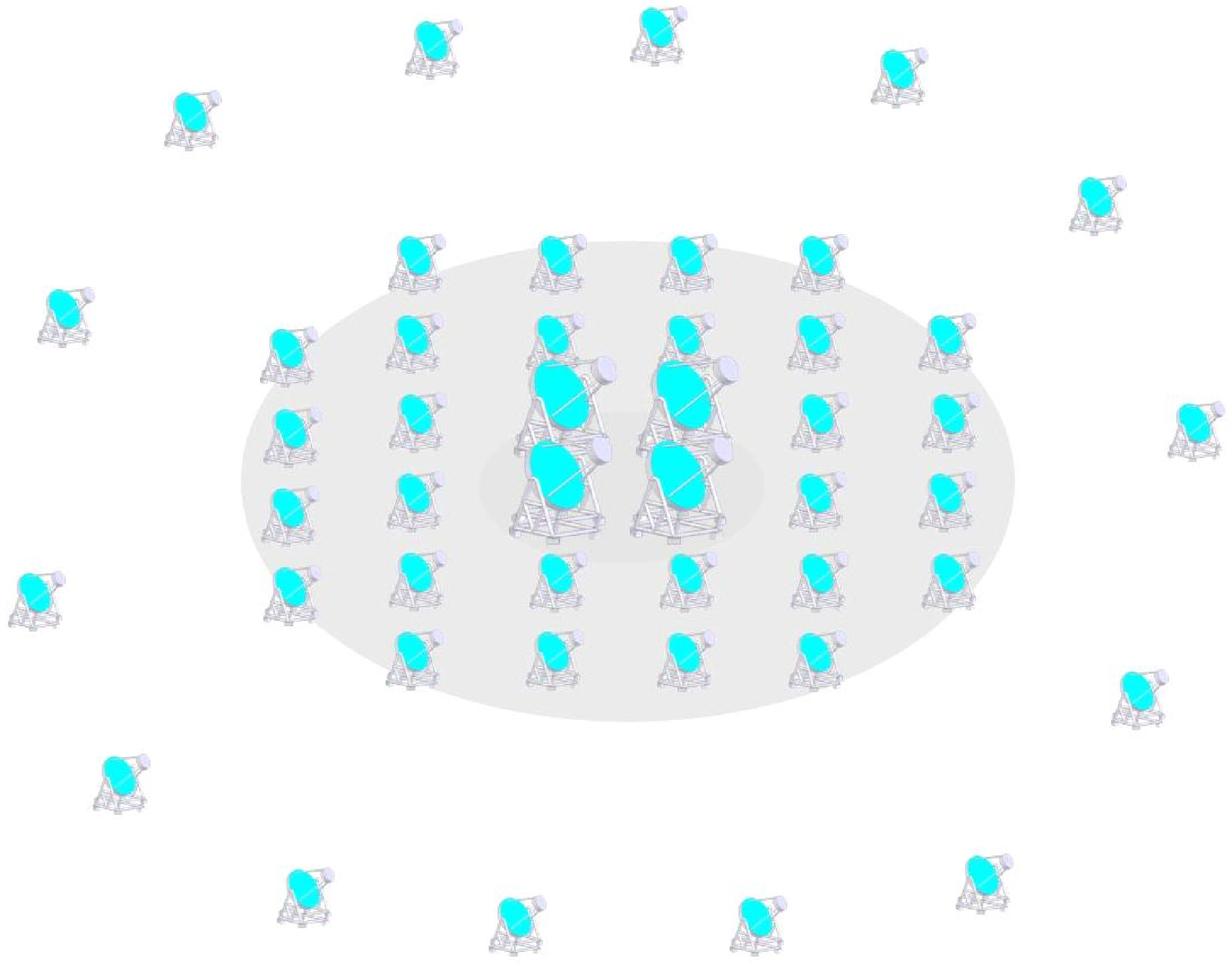}
 \includegraphics[width=0.45\textwidth]{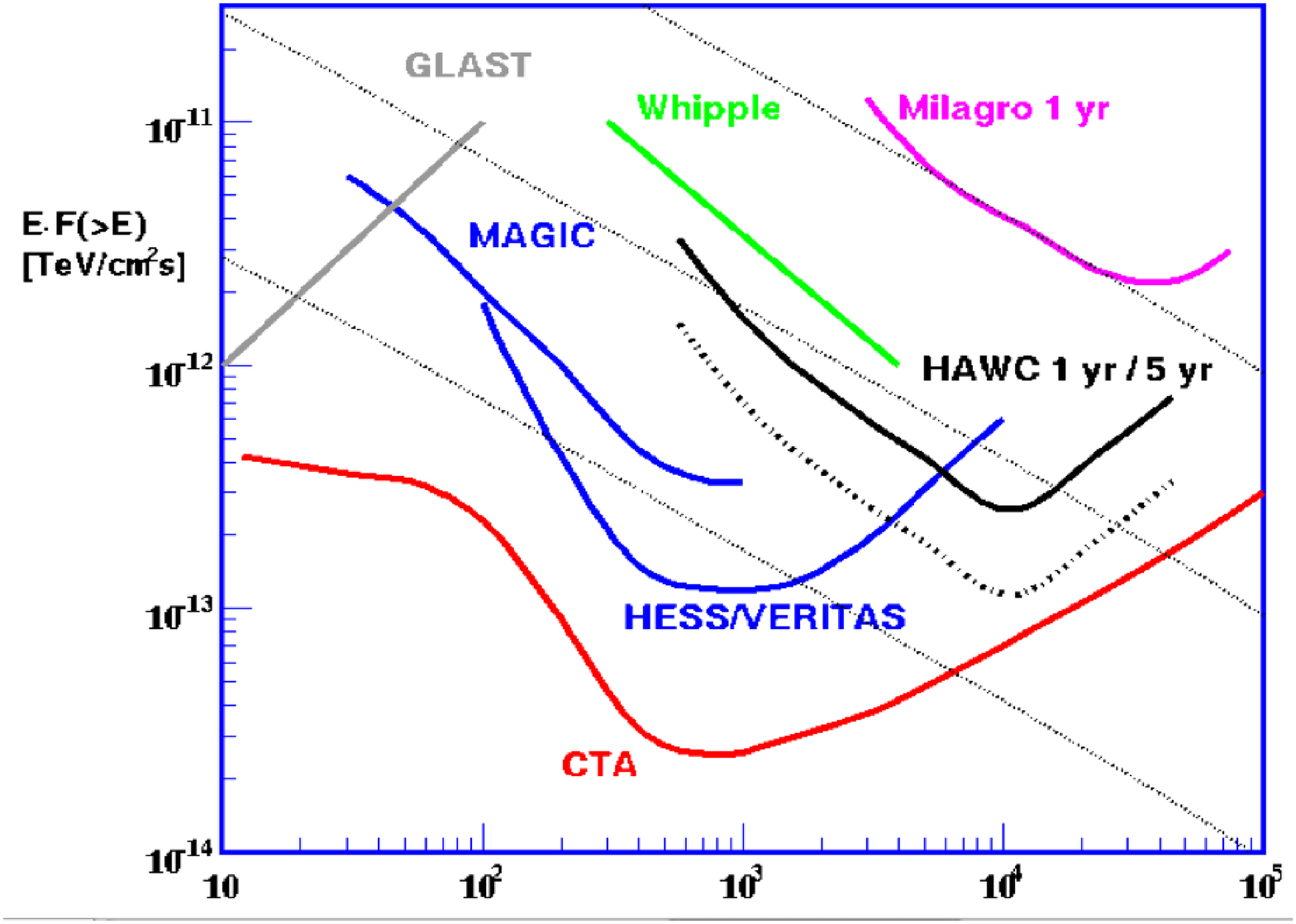}
\caption{Left:An illustration of the CTA project. Figure adopted from \cite{CTA}.
Right: Sensitivity curves for currently operational experiments (GLAST(\fermi-LAT),
MAGIC, HESS, VERITAS and Milagro) and expected sensitivity curves for future experiments
(HAWC and CTA).Figure adopted from \cite{HAWC}. 
 }
\label{FigCTA}
\end{figure}

A new project in VHE gamma-ray astronomy named Cherenkov Telescope Array (CTA) 
was recently proposed and a large international collaboration was formed (see \cite{CTA}). 
The basic concept of CTA is an array of a large number of IACTs spread over a
square kilometer area.  

 The telescope configuration would be as follows;
\begin{itemize}
\item The array consists of 3 different sizes of telescopes, $3 - 5$ large size ones,
$20 - 30$ middle size ones, and $20 - 30$ small size ones.
\item The large size telescopes have a reflector diameter of $\sim 23$ m with a FoV of 5 degree.
\item The middle size telescopes have a reflector diameter of $\sim 12$ m with a FoV of 8 degree.
\item The small size telescopes have a reflector diameter of $\sim 6.5$ m with a FoV of 10 degree.
\item The distance between telescopes is around the order of 100 m.
\item Two arrays will be constructed, a larger one in the southern hemisphere and a smaller
one comprising only 23 m and 12 m telescopes in the northern hemisphere.  
\end{itemize}
The parameters described above such as the numbers of telescopes, the diameters of reflectors
 and the angle of FoVs are not yet fixed. 
The primary purpose of the CTA project is 
the observation of VHE gamma-ray sources in the energy range between 
100 GeV and 10 TeV with a sensitivity $5 - 10$ times higher than that of 
current IACTs. The accessible energy range will also be extended
down to $10-20$ GeV and up to 100 TeV. 

For pulsar observations, a sensitivity below 100 GeV is essential.
Below 100 GeV, the sensitivity is mainly determined by the large size telescopes,
whose reflector area will be $\sim 1.7$ times larger than that of MAGIC.  
If the SUM trigger system is used in CTA, and if a coincidence between multiple telescopes
is not required for a trigger, 
then the energy threshold of the CTA may be roughly estimated to be $\sim 15$ GeV
by scaling with the reflector diameter.
Since there will be 3 - 5 telescopes, the effective area would increase correspondingly.
The coincidence requirement might increase the threshold energy, but the reduction
of the effective area may not be as much as in the case of the MAGIC-stereo system.
The gain in sensitivity below 100 GeV should be intensely studied before all the designs
are fixed and the construction of telescopes starts.

  \include{anhang}
   \backmatter
    \bibliographystyle{jkthesis}
\bibliography{literatur}

%

  \addcontentsline{toc}{chapter}{\protect Acknowledgement}

\chapter*{Acknowledgement}
\small
First of all, I would like to thank Prof. Christian Kiesling 
for being my academic advisor at the Ludwig-Maximimlian-Universit\"at, M\"unchen,
and for supporting my PhD works. I also appreciate Prof. Masahiro Teshima 
for providing me with the opportunity to join the MAGIC collaboration
and to do research in the Max-Planck-Institut f\"ur Physik.

I am also deeply grateful to Dr. Thomas Schweizer, Dr. Nepomuk Otte, Dr. Michael Rissi
and Dr. Maxim Shayduk for the greatest achievement of lowering the energy threshold 
of the MAGIC telescope by developing and installing the SUM trigger. 
Without their tremendous efforts and ability, this Ph.D. work would not have
been realized. 

 I am very happy to have been supervised by Prof. Masahiro Teshima, 
Dr. Thomas Schweizer, Dr. Maxim Shayduk, Dr. Razmick Mirzoyan, and Dr. Eckart Lorenz. 
Discussions with them were always constructive, educative, and very inspiring.
Whenever I got stuck in my research, Prof. Teshima gave me accurate advice,
which immediately brought things forward.
Thomas's deep knowledge, wide experiences, 
and agile mind helped me to accomplish this work. 
Maxim, with whom I spent all my 
Ph.D. time in the same office, was always taking care of my work and checking reasonability
of all kinds of  results based on his rigorous knowledge and 
stupendous intelligence. 
I learned many details of the IACT technique and photosensors from Dr. Mirzoyan. 
His explation was always with concrete numbers, which gave me very clear understanding
of IACT technique and photosensors. 
Dr. Lorenz guided my work to the right direction 
by giving me advice from a long-term perspective, which he can do better than anyone
because of his long experience always standing close to students. 

I would also like to thank all my friends in the Max-Planck-Institut f\"ur Physik,
Daniela Borla Tridon, 
Emiliano Carmona, 
Pierre Colin, 
Antonios Dettlaff, 
Toni Engelhardt, 
Christian Fruck, 
Florian G\"obel,
Dennis H\"afner, 
Jürgen Hose, 
Tobias Jogler, 
Hanna Kellermann, 
Julian Krause, 
Matthias Kurz, 
Hiroko Miyamoto, 
Koji Saito, 
Maxim Shayduk, 
Julian Sitarek, 
Burkhard Steinke, 
Hajime Takami, 
Robert Wagner, 
and many others. I enjoyed with them having many parties, playing many different sports, 
chatting during lunch and break time and having many physics discussion. 
Among them, I had the best time with Julian Sitarek and Maxim Shayduk. 
It was a great fun to stay with them because they are not only intelligent
but also very cheerful. 
I am also very grateful to other collaborators
 in different institutes. Drinking alcohol with them 
after the collaboration meeting in many different cities was the best pleasure  
during my Ph.D time. 

I thank my parents, my brother and my sister who 
supported me from my home country. Their letters and their e-mails encouraged me very much. 
In addition, nothing could have made me happier than Japanese foods they regularly sent me. 
Last but not least, I thank and will thank for all my life my wife, Emi, who came over to 
Munich to live together with me. 
I could overcome many difficulties doing Ph.D. work here in Munich
just because she was always with me.

\end{document}